\documentclass[twocolumn]{aastex63}

\newcommand{\HI}{H\,{\sc i} }
\newcommand{\HInospace}{H\,{\sc i}}

\newcommand{\Htwo}{H$_{2}$ }

\shorttitle{VERTICO: The Virgo Environment Traced In CO Survey}
\shortauthors{Brown et al.}

\usepackage{subfiles} 
\def\biblio{\bibliographystyle{aasjournal}\bibliography{refs}}  

\begin{document}
\def\biblio{}

\title{VERTICO: The Virgo Environment Traced In CO Survey}

\correspondingauthor{Toby Brown}
\email{tobiashenrybrown@gmail.com}

\author[0000-0003-1845-0934]{Toby Brown}\affiliation{Herzberg Astronomy and Astrophysics Research Centre, National Research Council of Canada, 5071 West Saanich Rd, Victoria, BC, V9E 2E7, Canada}\affiliation{Department of Physics \& Astronomy, McMaster University, 1280 Main Street W, Hamilton, ON, L8S 4M1, Canada}

\author{Christine D. Wilson}\affiliation{Department of Physics \& Astronomy, McMaster University, 1280 Main Street W, Hamilton, ON, L8S 4M1, Canada}

\author{Nikki Zabel}\affiliation{Kapteyn Astronomical Institute, University of Groningen, PO Box 800, NL-9700 AV Groningen, the Netherlands}

\author{Timothy A. Davis}\affiliation{School of Physics \& Astronomy, Cardiff University, Queens Buildings, The Parade, Cardiff CF24 3AA, UK}

\author{Alessandro Boselli}\affiliation{Aix-Marseille Universit\'{e}, CNRS, CNES, LAM, Marseille, France}

\author{Aeree Chung}\affiliation{Department of Astronomy, Yonsei University, 50 Yonsei-ro, Seodaemun-gu, Seoul 03722, South Korea}

\author{Sara L. Ellison}\affiliation{Department of Physics \& Astronomy, University of Victoria, Finnerty Road, Victoria, BC,   V8P 1A1, Canada}

\author{Claudia D.P. Lagos}\affiliation{International Centre for Radio Astronomy Research, The University of Western Australia, 35 Stirling Hwy, 6009 Crawley, WA, Australia }\affiliation{ARC Centre of Excellence for All Sky Astrophysics in 3 Dimensions (ASTRO 3D), Australia}

\author{Adam R.H. Stevens}\affiliation{International Centre for Radio Astronomy Research, The University of Western Australia, 35 Stirling Hwy, 6009 Crawley, WA, Australia }

\author{Luca Cortese}\affiliation{International Centre for Radio Astronomy Research, The University of Western Australia, 35 Stirling Hwy, 6009 Crawley, WA, Australia }\affiliation{ARC Centre of Excellence for All Sky Astrophysics in 3 Dimensions (ASTRO 3D), Australia}

\author{Yannick M.~Bah\'{e}}\affiliation{Leiden Observatory, Leiden University, PO Box 9513, 2300 RA Leiden, The Netherlands}

\author{Dhruv Bisaria}\affiliation{Department of Physics, Engineering Physics and Astronomy, Queen’s University, Kingston, ON K7L 3N6, Canada}

\author{Alberto D. Bolatto}\affiliation{Department of Astronomy, University of Maryland, College Park, MD, 20742, USA)}\affiliation{Visiting Scholar at Flatiron Institute, Center for Computational Astrophysics (NY 10010, USA) }

\author{Claire R. Cashmore}\affiliation{E.A. Milne Centre for Astrophysics, The University of Hull, Cottingham Road, Kingston-Upon-Hull, HU6 7RX, UK}

\author{Barbara Catinella}\affiliation{International Centre for Radio Astronomy Research, The University of Western Australia, 35 Stirling Hwy, 6009 Crawley, WA, Australia }\affiliation{ARC Centre of Excellence for All Sky Astrophysics in 3 Dimensions (ASTRO 3D), Australia}

\author{Ryan Chown}\affiliation{Department of Physics \& Astronomy, McMaster University, 1280 Main Street W, Hamilton, ON, L8S 4M1, Canada}

\author{Benedikt Diemer}\affiliation{Department of Astronomy, University of Maryland, College Park, MD 20742, USA}

\author{Pascal J. Elahi}\affiliation{Pawsey Supercomputing Centre, Commonwealth Scientific and Industrial Research Organisation}

\author{Maan H. Hani}\affiliation{Department of Physics \& Astronomy, McMaster University, 1280 Main Street W, Hamilton, ON, L8S 4M1, Canada}

\author{Mar\'{i}a J. Jim\'{e}nez-Donaire}\affiliation{Observatorio Astronómico Nacional (IGN), C/Alfonso XII, 3, E-28014 Madrid, Spain}\affiliation{Centro de Desarrollos Tecnológicos, Observatorio de Yebes (IGN), 19141 Yebes, Guadalajara, Spain}

\author{Bumhyun Lee}\affiliation{Kavli Institute for Astronomy and Astrophysics, Peking University, Beijing 100871, China}

\author{Katya Leidig}\affiliation{Department of Astronomy, University of Maryland, College Park, MD, 20742, USA)}

\author{Angus Mok}\affiliation{Department of Physics \& Astronomy, The University of Toledo, Toledo, OH 43606, USA}

\author{Karen Pardos Olsen}\affiliation{Department of Astronomy and Steward Observatory, University of Arizona, Tucson, AZ 85721, USA}

\author{Laura C. Parker}\affiliation{Department of Physics \& Astronomy, McMaster University, 1280 Main Street W, Hamilton, ON, L8S 4M1, Canada}

\author{Ian D. Roberts}\affiliation{Leiden Observatory, Leiden University, PO Box 9513, 2300 RA Leiden, The Netherlands}

\author{Rory Smith}\affiliation{Korea Astronomy and Space Science Institute (KASI), 776 Daedeokdae-ro, Yuseong-gu, Daejeon 34055, Korea}\affiliation{University of Science and Technology (UST), Daejeon 34113, Korea}

\author{Kristine Spekkens}\affiliation{Royal Military College of Canada, PO Box 17000, Station Forces, Kingston, ON, Canada K7K 7B4}

\author{Mallory Thorp}\affiliation{Department of Physics \& Astronomy, University of Victoria, Finnerty Road, Victoria, BC, V8P 1A1, Canada}

\author{Stephanie Tonnesen}\affiliation{CCA, Flatiron Institute, 162 5th Ave, New York NY 10010}

\author{Evan Vienneau}\affiliation{Department of Physics \& Astronomy, McMaster University, 1280 Main Street W, Hamilton, ON, L8S 4M1, Canada}

\author{Vicente Villanueva}\affiliation{Department of Astronomy, University of Maryland, College Park, MD 20742, USA}

\author{Stuart N. Vogel}\affiliation{Department of Astronomy, University of Maryland, College Park, MD, 20742, USA)}

\author{James Wadsley}\affiliation{Department of Physics \& Astronomy, McMaster University, 1280 Main Street W, Hamilton, ON, L8S 4M1, Canada}

\author{Charlotte Welker}\affiliation{Department of Physics and Astronomy, Krieger School of Art and Science, The Johns Hopkins University, Baltimore, MD, USA}

\author{Hyein Yoon}\affiliation{Sydney Institute for Astronomy, School of Physics, University of Sydney, NSW 2006, Australia}\affiliation{ARC Centre of Excellence for All Sky Astrophysics in 3 Dimensions (ASTRO 3D), Australia}


\begin{abstract}
	We present the Virgo Environment Traced in CO (VERTICO) survey, a new effort to map $^{12}$CO($2-1$), $^{13}$CO($2-1$), and C$^{18}$O($2-1$) in 51 Virgo Cluster galaxies with the Atacama Compact Array, part of the Atacama Large Millimeter/submillimeter Array (ALMA). The primary motivation of VERTICO is to understand the physical mechanisms that perturb molecular gas disks, and therefore star formation and galaxy evolution, in dense environments. This first paper contains an overview of VERTICO’s design and sample selection, $^{12}$CO($2-1$) observations, and data reduction procedures. We characterize global $^{12}$CO($2-1$) fluxes and molecular gas masses for the 49 detected VERTICO galaxies, provide upper limits for the two non-detections, and produce resolved $^{12}$CO($2-1$) data products (median resolution $= 8\arcsec \approx 640~{\rm pc}$). Azimuthally averaged $^{12}$CO($2-1$) radial intensity profiles are presented along with derived molecular gas radii. We demonstrate the scientific power of VERTICO by comparing the molecular gas size--mass scaling relation for our galaxies with a control sample of field galaxies, highlighting the strong effect that radius definition has on this correlation. We discuss the drivers of the form and scatter in the size--mass relation and highlight areas for future work. VERTICO is an ideal resource for studying the fate of molecular gas in cluster galaxies and the physics of environment-driven processes that perturb the star formation cycle. Upon public release, the survey will provide a homogeneous legacy dataset for studying galaxy evolution in our closest cluster.
\end{abstract}

\section{Introduction}\label{sec:introduction}

Beginning with the pioneering works of \citet{Gunn1972}, \citet{Cowie1977}, and \citet{Larson1980}, the last 50 years have seen a steady stream of work demonstrating that galaxies' cold gas and star formation properties are influenced by the environment in which they reside \citep[see reviews by][and references therein]{Haynes1984, Boselli2006, Cortese2021}. 

Much of this work has focused on local galaxy clusters as laboratories for studying the role environment plays in galaxy evolution. Containing hundreds or even thousands of galaxies, clusters are characterized by high internal velocity dispersion ($v_{\rm disp}\sim10^2 - 10^3~{\rm km~s^{-1}}$) and a hot, diffuse intracluster medium \citep[ICM, $\text{T}\sim10^7 - 10^8$~K, $n\sim10^{-4}~\text{cm}^{-3}$;][]{Kaiser1986,White1997}. 

Cluster members are subject to a diverse range of environmental effects that drive the observed increase in quenched (or quenching) galaxies with respect to the field \citep{Balogh1998, Balogh1999, Gomez2003, Boselli2006}. Dynamical interactions between two or more systems can cause tidal stripping of stars and gas \citep{Moore1999,Iono2005} or compression of the interstellar medium \citep[ISM;][]{Nehlig2016,Kaneko2018}. Harassment by frequent, high-velocity galaxy--galaxy encounters is capable of tidally heating the ISM, shutting off inflow and evaporating gas reservoirs \citep{Moore1996,Moore1998,Fujita1998}. Viscous stripping or thermal evaporation of the cold gas content of galaxies surrounded by a hot ICM can also suppress star formation \citep{Cowie1977,Livio1980,Nulsen1982}. 
Starvation occurs when there is a lack of gas cooling and/or a cut off in the external gas supply, either to the halo or to the galaxy itself, meaning reservoirs consumed in star formation are not replenished \citep{Larson1980,Balogh2000,Bekki2002}. For galaxies entering a cluster, the interaction between the ISM and ICM has been observed to be strong enough to rapidly remove gas from the disk via ram pressure stripping \citep{Gunn1972,Hester2006,Chung2009b,Fumagalli2014}. More recently, evidence has also been presented for  elevated gas densities caused by ram pressure, affecting the star formation efficiency (SFE) and the transition of atomic hydrogen gas to its molecular phase \citep{Ebeling2014,Lee2017,Mok2017,Vulcani2018,Moretti2020a,Moretti2020b}.

The common outcome of such processes is that they leave different imprints on galaxies' cold gas reservoirs. Thus, to understand the nature of these effects, we must establish their influence upon the constituent phases of the ISM. Pursuit of this understanding has demonstrated the widespread and systematic depletion of atomic hydrogen (\HInospace) gas reservoirs --- considered the gas supply for future star formation---in cluster galaxies with respect to the field \citep[e.g.,][]{Davies1973,Chamaraux1980,Giovanelli1985,Gavazzi2008, Chung2009b, Hughes2009,  Cortese2011,  Jaffe2016, Brown2017, Healy2020}.

Similarly, there have been many efforts to determine cluster galaxies' molecular hydrogen gas (H$_2$) content, the fuel for on-going star formation. The low temperatures found in giant molecular gas clouds and lack of a dipole moment emission mechanism make the \Htwo molecule very difficult to observe directly. Instead the bulk of molecular gas is usually traced via indirect means, the most common and reliable of which are the low-$J$ rotational transitions of carbon monoxide (CO), although even this method is not without its uncertainties \citep[see][for a thorough review of this topic]{Bolatto2013}.

Because of the more stringent instrumental and observational requirements relative to longer-wavelength radio observations, the findings from studies of molecular gas have necessarily focused on smaller samples and lacked the consistency found in \HI studies. A number of prominent early works using unresolved detections found no differences between the molecular gas content of field and cluster galaxies, concluding that molecular gas disks that are bound deep within the gravitational potential well of the system are largely immune to environmental effects \citep{Stark1986,Kenney1986,Kenney1989,Boselli1997}. However, more recent efforts to survey global molecular gas reservoirs in a large number of local cluster galaxies tend to favour a scenario whereby the star-forming gas is depleted in dense environments, albeit to a lesser extent than the \HI \citep{Boselli2002, Wilson2009, Corbelli2012, Boselli2014a, Mok2016, Chung2017, Koyama2017, Stevens2021}. 

Studies using resolved observations of molecular gas in a small number of systems generally support the picture of environmental mechanisms perturbing molecular gas reservoirs in clusters, and demonstrate that environmental influences often manifest themselves on the sub-kpc scale of bars, disks, streams, tails, and warps \citep[e.g.,][]{Nehlig2016, Mok2017, Lee2017, Lee2018, Moretti2018, Jachym2019, Zabel2019, Cramer2019, Cramer2020, Lizee2021}. Interestingly, some of these studies find that environmental mechanisms have contrasting effects on the gas content or star formation properties of galaxies. For example, elevated star formation has been found at the ISM--ICM interface in cluster galaxies, implying that ram-pressure stripping is able to either increase the amount of gas available for star formation or the SFE of the surviving gas \citep[or a combination of both; e.g.,][]{Ebeling2014, Nehlig2016, Zabel2020, Lizee2021}. On the other hand, \citet{Moretti2018} find that ram-pressure stripping decreases the SFE of stripped molecular gas, and \citet{Mok2017} report enhanced molecular gas masses in cluster galaxies relative to the field \citep[see also][]{Moretti2020b}, concluding that the environment is aiding the transition from atomic to molecular gas while lowering the SFE. Recent work attributes the triggering of active galactic nuclei (AGN) to central gas flows driven by ram pressure, suggesting a causal link between galaxy environment and energetic AGN feedback \citep{Poggianti2017b}.

Despite continuous progression in our understanding, the dichotomy in approaches taken by molecular gas studies---large statistical surveys of global properties versus resolved mapping of gas reservoirs in a small number of galaxies---has left two significant questions unanswered:

\begin{enumerate}
	\item What is the relationship between different environmental mechanisms and molecular gas density, morphology, kinematics, and chemistry?
	\item When, where, and how do environmental mechanisms alter the rate and efficiency of star formation?
\end{enumerate}

Answering these questions requires a merging of these two methodologies. In other words, we need resolved spectroscopic imaging of molecular gas disks across a large homogeneous sample of galaxies that are experiencing the full complement of environmental mechanisms. These observations must be combined with multi-wavelength data covering the full galactic ecosystem and accompanied by state-of-the-art models and simulations.

With this goal in mind, we present the Virgo Environment Traced in CO (VERTICO) survey, a Large Program with the Atacama Large Millimeter/submillimeter Array (ALMA) designed to map molecular gas in 51 Virgo Cluster galaxies on sub-kpc scales. The primary motivation of VERTICO is to understand the physical mechanisms that drive galaxy evolution in dense environments. We also aim to provide a homogeneous legacy dataset for studying galaxy evolution in the nearest massive cluster to the Milky Way.

This first paper contains an overview of the VERTICO survey design and sample selection, observations and data reduction procedures, global molecular gas properties, and derived data products. We present early science results on the radial distribution of molecular gas in VERTICO galaxies and compare the molecular gas disk sizes and masses to a control sample of field galaxies. We also highlight areas for future work. The paper is structured as follows: Section \ref{sec:Sample} presents the sample; Section \ref{sec:Observations} describes the observations, including our data reduction method and integrated CO properties; Section \ref{sec:Derived Data Products} gives the methodology and examples of key data products; Section \ref{sec:size-mass relation} is a brief comparative analysis of molecular gas disk sizes and masses between VERTICO and a sample of field galaxies; finally, Section \ref{sec:Summary} presents a summary of this paper and looks forward to the next steps.

Throughout this paper, we assume a common distance of 16.5 Mpc to all Virgo galaxies based upon the Virgo Cluster distance found by \citet{Mei2007}. Where relevant, all astrophysical quantities are derived using a Kroupa initial mass function \citep[IMF;][]{Kroupa2001} or rescaled from literature values using
\begin{equation}
	M_{\star, \, {\rm K}} = 1.06 ~ M_{\star,\,{\rm C}} = 0.62 ~ M_{\star,\,{\rm S}},
\end{equation}
where the subscripts K, C, and S denote the Kroupa, \citet{Chabrier2003}, and \citet{Salpeter1955} IMFs, respectively \citep{Salim2007,Elbaz2007,Zahid2012,Speagle2014}.

\section{The VERTICO Sample} \label{sec:Sample}

VERTICO targets 51 Virgo Cluster galaxies included in the Very Large Array Imaging of Virgo in Atomic gas (VIVA) survey \citep{Chung2009b}. The full VIVA survey contains 53 galaxies, however, we exclude two very low-mass systems (IC3355 and VCC2062) that were deemed unlikely to be detected. VIVA was selected by \citet{Chung2009b} to sample a range of star formation properties in the classification scheme published by \citet[][normal, enhanced, anemic, truncated]{Koopmann2004}, which is in turn based upon the spatial distribution of H$\alpha$ and R-band emission. The resulting sample of primarily late-type galaxies spans a broad range in stellar mass ($10^{8.3} \leq$ $M_\star/{\rm M_\odot} \leq 10^{11}$) and specific star formation rate (${\rm sSFR}={\rm SFR}/M_\star$; $10^{-11.5} \leq$ sSFR$/{\rm yr^{-1}} \leq 10^{-9.5}$). VERTICO targets have existing resolved multi-wavelength observations tracing their stellar component, star formation activity, and \HI gas content. Galaxy \HI gas reservoirs exhibit signatures of the full complement of environmental effects, including the gas tails and truncated disks typical of stripping, fading gas disks of starvation, and morphological asymmetries and kinematic misalignment from gravitational perturbations \citep{Chung2009b,Yoon2017}. Every galaxy has existing resolved multi-wavelength observations tracing the stellar component and star formation activity \citep[e.g.,][see Section \ref{sec:Virgo Cluster} for further details]{Wright2010, Alam2015, Martin2005}. Furthermore, 15 galaxies already have archival Atacama Compact Array (ACA) observations of the $^{12}$CO($2-1$) emission line,  hereafter CO($2-1$). The remaining 36 targets were observable in CO($2-1$) within a feasible amount of time for an ALMA Large Program using the ACA. The full VERTICO sample is listed in Table \ref{tab:VERTICO-sample}. Optical inclinations and East-of-North position angles are calculated from fits to the Sloan Digital Sky Survey \citep[SDSS;][]{York2000,Alam2015} $r$-band photometry described in Section \ref{sec:PVD}.

\startlongtable
\begin{deluxetable*}{lccccc}
\tablecaption{The VERTICO target sample.}
\tablehead{\colhead{Galaxy} & \colhead{R.A. (J2000)} & \colhead{Dec. (J2000)} & \colhead{$v_{\rm opt}$} & \colhead{$i$} & \colhead{P.A.}\\ \colhead{ } & \colhead{ } & \colhead{ } & \colhead{$\mathrm{km\,s^{-1}}$} & \colhead{$\mathrm{{}^{\circ}}$} & \colhead{$\mathrm{{}^{\circ}}$}}
\startdata
IC3392\tablenotemark{\text{\textasteriskcentered}} & $12^\mathrm{h}28^\mathrm{m}43.27^\mathrm{s}$ & $14^\circ59{}^\prime57.48{}^{\prime\prime}$ & 1678 & 68 & 219 \\
IC3418\tablenotemark{\text{\textasteriskcentered}} & $12^\mathrm{h}29^\mathrm{m}43.50^\mathrm{s}$ & $11^\circ24{}^\prime08.00{}^{\prime\prime}$ & 38 & 62 & 233 \\
NGC4064 & $12^\mathrm{h}04^\mathrm{m}11.26^\mathrm{s}$ & $18^\circ26{}^\prime39.12{}^{\prime\prime}$ & 1000 & 70 & 150 \\
NGC4189\tablenotemark{\text{\textasteriskcentered}} & $12^\mathrm{h}13^\mathrm{m}47.47^\mathrm{s}$ & $13^\circ25{}^\prime34.68{}^{\prime\prime}$ & 1995 & 42 & 70 \\
NGC4192\tablenotemark{\text{\textasteriskcentered}} & $12^\mathrm{h}13^\mathrm{m}48.58^\mathrm{s}$ & $14^\circ53{}^\prime57.12{}^{\prime\prime}$ & -118 & 83 & 333 \\
NGC4216\tablenotemark{\text{\textasteriskcentered}} & $12^\mathrm{h}15^\mathrm{m}54.19^\mathrm{s}$ & $13^\circ08{}^\prime54.96{}^{\prime\prime}$ & 30 & 90 & 20 \\
NGC4222 & $12^\mathrm{h}16^\mathrm{m}22.56^\mathrm{s}$ & $13^\circ18{}^\prime25.20{}^{\prime\prime}$ & 225 & 90 & 238 \\
NGC4254\tablenotemark{\text{\textasteriskcentered}}\tablenotemark{\text{\textdagger}} & $12^\mathrm{h}18^\mathrm{m}49.68^\mathrm{s}$ & $14^\circ25{}^\prime05.52{}^{\prime\prime}$ & 2453 & 39 & 243 \\
NGC4293\tablenotemark{\text{\textasteriskcentered}}\tablenotemark{\text{\textdagger}} & $12^\mathrm{h}21^\mathrm{m}13.47^\mathrm{s}$ & $18^\circ23{}^\prime03.12{}^{\prime\prime}$ & 717 & 67 & 239 \\
NGC4294 & $12^\mathrm{h}21^\mathrm{m}17.81^\mathrm{s}$ & $11^\circ30{}^\prime39.24{}^{\prime\prime}$ & 421 & 74 & 151 \\
NGC4298\tablenotemark{\text{\textasteriskcentered}}\tablenotemark{\text{\textdagger}} & $12^\mathrm{h}21^\mathrm{m}33.12^\mathrm{s}$ & $14^\circ36{}^\prime19.80{}^{\prime\prime}$ & 1122 & 52 & 132 \\
NGC4299\tablenotemark{\text{\textasteriskcentered}} & $12^\mathrm{h}21^\mathrm{m}40.71^\mathrm{s}$ & $11^\circ30{}^\prime06.12{}^{\prime\prime}$ & 209 & 14 & 128 \\
NGC4302\tablenotemark{\text{\textasteriskcentered}} & $12^\mathrm{h}21^\mathrm{m}42.24^\mathrm{s}$ & $14^\circ35{}^\prime57.12{}^{\prime\prime}$ & 1111 & 90 & 356 \\
NGC4321\tablenotemark{\text{\textasteriskcentered}}\tablenotemark{\text{\textdagger}} & $12^\mathrm{h}22^\mathrm{m}54.77^\mathrm{s}$ & $15^\circ49{}^\prime33.24{}^{\prime\prime}$ & 1579 & 32 & 280 \\
NGC4330 & $12^\mathrm{h}23^\mathrm{m}16.95^\mathrm{s}$ & $11^\circ22{}^\prime04.08{}^{\prime\prime}$ & 1567 & 90 & 238 \\
NGC4351\tablenotemark{\text{\textasteriskcentered}} & $12^\mathrm{h}24^\mathrm{m}01.30^\mathrm{s}$ & $12^\circ12{}^\prime15.12{}^{\prime\prime}$ & 2388 & 48 & 251 \\
NGC4380\tablenotemark{\text{\textasteriskcentered}} & $12^\mathrm{h}25^\mathrm{m}22.16^\mathrm{s}$ & $10^\circ01{}^\prime00.12{}^{\prime\prime}$ & 935 & 61 & 158 \\
NGC4383\tablenotemark{\text{\textasteriskcentered}} & $12^\mathrm{h}25^\mathrm{m}25.47^\mathrm{s}$ & $16^\circ28{}^\prime11.64{}^{\prime\prime}$ & 1663 & 56 & 17 \\
NGC4388 & $12^\mathrm{h}25^\mathrm{m}46.61^\mathrm{s}$ & $12^\circ39{}^\prime46.44{}^{\prime\prime}$ & 2538 & 83 & 271 \\
NGC4394 & $12^\mathrm{h}25^\mathrm{m}55.66^\mathrm{s}$ & $18^\circ12{}^\prime52.20{}^{\prime\prime}$ & 772 & 32 & 312 \\
NGC4396\tablenotemark{\text{\textasteriskcentered}} & $12^\mathrm{h}25^\mathrm{m}59.66^\mathrm{s}$ & $15^\circ40{}^\prime10.20{}^{\prime\prime}$ & -115 & 83 & 304 \\
NGC4402\tablenotemark{\text{\textdagger}} & $12^\mathrm{h}26^\mathrm{m}07.34^\mathrm{s}$ & $13^\circ06{}^\prime45.00{}^{\prime\prime}$ & 190 & 80 & 270 \\
NGC4405 & $12^\mathrm{h}26^\mathrm{m}07.11^\mathrm{s}$ & $16^\circ10{}^\prime51.60{}^{\prime\prime}$ & 1751 & 46 & 18 \\
NGC4419 & $12^\mathrm{h}26^\mathrm{m}56.35^\mathrm{s}$ & $15^\circ02{}^\prime51.36{}^{\prime\prime}$ & -228 & 74 & 131 \\
NGC4424\tablenotemark{\text{\textasteriskcentered}}\tablenotemark{\text{\textdagger}} & $12^\mathrm{h}27^\mathrm{m}11.69^\mathrm{s}$ & $09^\circ25{}^\prime14.16{}^{\prime\prime}$ & 447 & 61 & 274 \\
NGC4450\tablenotemark{\text{\textasteriskcentered}} & $12^\mathrm{h}28^\mathrm{m}29.23^\mathrm{s}$ & $17^\circ05{}^\prime04.56{}^{\prime\prime}$ & 2048 & 51 & 170 \\
NGC4457\tablenotemark{\text{\textasteriskcentered}}\tablenotemark{\text{\textdagger}} & $12^\mathrm{h}28^\mathrm{m}59.02^\mathrm{s}$ & $03^\circ34{}^\prime14.16{}^{\prime\prime}$ & 738 & 37 & 256 \\
NGC4501\tablenotemark{\text{\textasteriskcentered}} & $12^\mathrm{h}31^\mathrm{m}59.33^\mathrm{s}$ & $14^\circ25{}^\prime10.92{}^{\prime\prime}$ & 2120 & 65 & 320 \\
NGC4522 & $12^\mathrm{h}33^\mathrm{m}39.72^\mathrm{s}$ & $09^\circ10{}^\prime26.76{}^{\prime\prime}$ & 2332 & 82 & 35 \\
NGC4532\tablenotemark{\text{\textasteriskcentered}} & $12^\mathrm{h}34^\mathrm{m}19.35^\mathrm{s}$ & $06^\circ28{}^\prime05.52{}^{\prime\prime}$ & 2154 & 64 & 159 \\
NGC4533 & $12^\mathrm{h}34^\mathrm{m}22.03^\mathrm{s}$ & $02^\circ19{}^\prime33.24{}^{\prime\prime}$ & 1753 & 80 & 342 \\
NGC4535\tablenotemark{\text{\textasteriskcentered}}\tablenotemark{\text{\textdagger}} & $12^\mathrm{h}34^\mathrm{m}20.26^\mathrm{s}$ & $08^\circ11{}^\prime53.52{}^{\prime\prime}$ & 1973 & 48 & 12 \\
NGC4536\tablenotemark{\text{\textasteriskcentered}}\tablenotemark{\text{\textdagger}} & $12^\mathrm{h}34^\mathrm{m}27.12^\mathrm{s}$ & $02^\circ11{}^\prime16.08{}^{\prime\prime}$ & 1894 & 74 & 118 \\
NGC4548\tablenotemark{\text{\textasteriskcentered}}\tablenotemark{\text{\textdagger}} & $12^\mathrm{h}35^\mathrm{m}26.64^\mathrm{s}$ & $14^\circ29{}^\prime43.80{}^{\prime\prime}$ & 498 & 37 & 318 \\
NGC4561\tablenotemark{\text{\textasteriskcentered}} & $12^\mathrm{h}36^\mathrm{m}08.14^\mathrm{s}$ & $19^\circ19{}^\prime21.72{}^{\prime\prime}$ & 1441 & 28 & 60 \\
NGC4567 & $12^\mathrm{h}36^\mathrm{m}33.07^\mathrm{s}$ & $11^\circ15{}^\prime29.16{}^{\prime\prime}$ & 2213 & 49 & 251 \\
NGC4568\tablenotemark{\text{\textasteriskcentered}} & $12^\mathrm{h}36^\mathrm{m}34.34^\mathrm{s}$ & $11^\circ14{}^\prime21.84{}^{\prime\prime}$ & 2260 & 70 & 211 \\
NGC4569\tablenotemark{\text{\textasteriskcentered}}\tablenotemark{\text{\textdagger}} & $12^\mathrm{h}36^\mathrm{m}50.12^\mathrm{s}$ & $13^\circ09{}^\prime55.08{}^{\prime\prime}$ & -220 & 69 & 203 \\
NGC4579\tablenotemark{\text{\textasteriskcentered}}\tablenotemark{\text{\textdagger}} & $12^\mathrm{h}37^\mathrm{m}43.44^\mathrm{s}$ & $11^\circ49{}^\prime05.52{}^{\prime\prime}$ & 1627 & 40 & 273 \\
NGC4580\tablenotemark{\text{\textasteriskcentered}} & $12^\mathrm{h}37^\mathrm{m}48.38^\mathrm{s}$ & $05^\circ22{}^\prime06.24{}^{\prime\prime}$ & 1227 & 46 & 337 \\
NGC4606 & $12^\mathrm{h}40^\mathrm{m}57.62^\mathrm{s}$ & $11^\circ54{}^\prime43.56{}^{\prime\prime}$ & 1653 & 69 & 38 \\
NGC4607 & $12^\mathrm{h}41^\mathrm{m}12.39^\mathrm{s}$ & $11^\circ53{}^\prime09.60{}^{\prime\prime}$ & 2284 & 90 & 2 \\
NGC4651\tablenotemark{\text{\textasteriskcentered}} & $12^\mathrm{h}43^\mathrm{m}42.72^\mathrm{s}$ & $16^\circ23{}^\prime37.68{}^{\prime\prime}$ & 788 & 53 & 75 \\
NGC4654\tablenotemark{\text{\textasteriskcentered}}\tablenotemark{\text{\textdagger}} & $12^\mathrm{h}43^\mathrm{m}56.76^\mathrm{s}$ & $13^\circ07{}^\prime32.52{}^{\prime\prime}$ & 1035 & 61 & 300 \\
NGC4689\tablenotemark{\text{\textasteriskcentered}}\tablenotemark{\text{\textdagger}} & $12^\mathrm{h}47^\mathrm{m}45.68^\mathrm{s}$ & $13^\circ45{}^\prime42.12{}^{\prime\prime}$ & 1522 & 38 & 341 \\
NGC4694\tablenotemark{\text{\textasteriskcentered}}\tablenotemark{\text{\textdagger}} & $12^\mathrm{h}48^\mathrm{m}15.08^\mathrm{s}$ & $10^\circ59{}^\prime00.60{}^{\prime\prime}$ & 1211 & 62 & 323 \\
NGC4698\tablenotemark{\text{\textasteriskcentered}} & $12^\mathrm{h}48^\mathrm{m}22.99^\mathrm{s}$ & $08^\circ29{}^\prime15.00{}^{\prime\prime}$ & 1032 & 66 & 347 \\
NGC4713\tablenotemark{\text{\textasteriskcentered}} & $12^\mathrm{h}49^\mathrm{m}57.65^\mathrm{s}$ & $05^\circ18{}^\prime39.60{}^{\prime\prime}$ & 631 & 45 & 89 \\
NGC4772 & $12^\mathrm{h}53^\mathrm{m}29.12^\mathrm{s}$ & $02^\circ10{}^\prime06.24{}^{\prime\prime}$ & 1042 & 60 & 325 \\
NGC4808\tablenotemark{\text{\textasteriskcentered}} & $12^\mathrm{h}55^\mathrm{m}48.94^\mathrm{s}$ & $04^\circ18{}^\prime15.12{}^{\prime\prime}$ & 738 & 72 & 127 \\
VCC1581 & $12^\mathrm{h}34^\mathrm{m}45.05^\mathrm{s}$ & $06^\circ18{}^\prime03.24{}^{\prime\prime}$ & 2141 & 43 & 329
\enddata
\tablecomments{Columns are 
(1) galaxy name and unique identifier in this paper; 
(2) right ascension (J2000) of the galaxy optical center; 
(3) declination (J2000) of the galaxy optical center; 
(4) optical heliocentric recession velocity; 
(5) optical $r-$band inclination; 
(6) optical $r-$band position angle of the kinematically redshifted half of the galaxy, calculated East-of-North. 
Columns (2)---(4) are drawn from the NASA/IPAC Extragalactic Database (\url{https://ned.ipac.caltech.edu/}). 
Columns (5) and (6) are calculated using fits to SDSS photometry described in Section \ref{sec:PVD}.
This table is published in its entirety in machine-readable format.}
\tablenotetext{\text{\textasteriskcentered}}{Data are from 7m and Total Power arrays. Other observations are 7m array only.}
\tablenotetext{\text{\textdagger}}{Data are archival ALMA ACA (7m + total power) $^{12}$CO($2-1$) observations 
at comparable sensitivity to our Cycle 7 observations. The sources of these data are the PHANGS-ALMA program 
\citep[][14 galaxies]{Leroy2021a} and one regular PI program \citep[][NGC4402]{Cramer2019}. 
See Section \ref{sec:Observations} and the Acknowledgments section for further details.}
\label{tab:VERTICO-sample}
\end{deluxetable*}

\subsection{The Virgo Cluster} \label{sec:Virgo Cluster}

The Virgo Cluster resides at a distance of 16.5 Mpc and contains thousands of member galaxies, making it the closest massive galaxy cluster to the Milky Way. The main body of the cluster centered on M87 has an estimated mass $M_{200} = 10^{14 - 14.6} ~ {\rm M}_\odot$ and radius $r_{200} = 1.08 - 1.55~\text{Mpc}$ \citep{Bohringer1994,Nulsen1995,Girardi1998,Schindler1999,McLaughlin1999,Gavazzi1999b,Mei2007,Urban2011,Kim2014,Boselli2018}, where $M_{200}$ is the total mass within $r_{200}$, the radius at which the enclosed mean mass density is 200 times the critical cosmic mass density. A comprehensive list of mass, size, and distance estimates for Virgo is provided by \citet[][table 1 in that work]{Boselli2018}. The cluster is dynamically young, contains significant substructure and is actively accreting members. This ensures Virgo's membership includes both infalling and virialised systems on a wide range of orbits throughout the cluster \citep{Tully1984,Gavazzi1999b,Karachentsev2014, Sorce2016, Yoon2017, Morokuma-Matsui2021}.

For these reasons, the Virgo region has been remarkably well surveyed at almost every wavelength: X-ray (ROSAT -- \citealt{Bohringer1994}; ASCA -- \citealt{Shibata2001}; XMM-Newton -- \citealt{Urban2011}), ultra-violet (UV; the GALEX Ultra-violet Virgo Cluster Survey, GUViCS -- \citealt{Boselli2011}), optical (SDSS -- \citealt{York2000}; the Next Generation Virgo Survey, NGVS -- \citealt{Ferrarese2012}; the Virgo Environmental Survey Tracing Ionised Gas Emission, VESTIGE -- \citealt{Boselli2018}), infrared (the Wide-field Infrared Survey Explorer, WISE -- \citealt{Wright2010}; the Herschel Reference Survey, HRS -- \citealt{Boselli2010}; the Herschel Virgo Cluster Survey, HeViCS -- \citealt{Davies2010}) and $21~\text{cm}$ radio  \citep[VIVA;][]{Chung2009b}. These surveys map the hot ICM, and the stellar mass, dust, SFR, and atomic and ionized gas properties of Virgo members in extraordinary detail. Almost all these data have VERTICO's angular resolution or better, while the $21~\text{cm}$ observations are coarser ($\sim15\arcsec$). This combination of richness, proximity, wavelength coverage, and data quality is not available for any other galaxy cluster.

At mm wavelengths, \citet{Boselli2014a} provide a census of unresolved CO($1-0$) observations and global molecular gas properties for approximately 150 Virgo members, showing that Virgo galaxies are, on average, deficient in molecular gas in comparison to similar field galaxies. Resolved studies of CO($2-1$) and CO($3-2$) have mapped the distribution of molecular gas at 1--2 kpc scale in $\sim30$ late-type cluster members \citep{Pappalardo2012,Mok2017} and there are also published sub-kiloparsec resolution CO($2-1$) observations for a handful of Virgo spirals \citep{Lee2017,Cramer2020,Lizee2021}. The key contribution of VERTICO to this field is homogeneous observations of molecular gas at sub-kiloparsec resolution for a large sample of cluster galaxies.

\subsection{Sample Properties} \label{sec:Sample Properties}

\begin{figure*}
	\centering
	\includegraphics{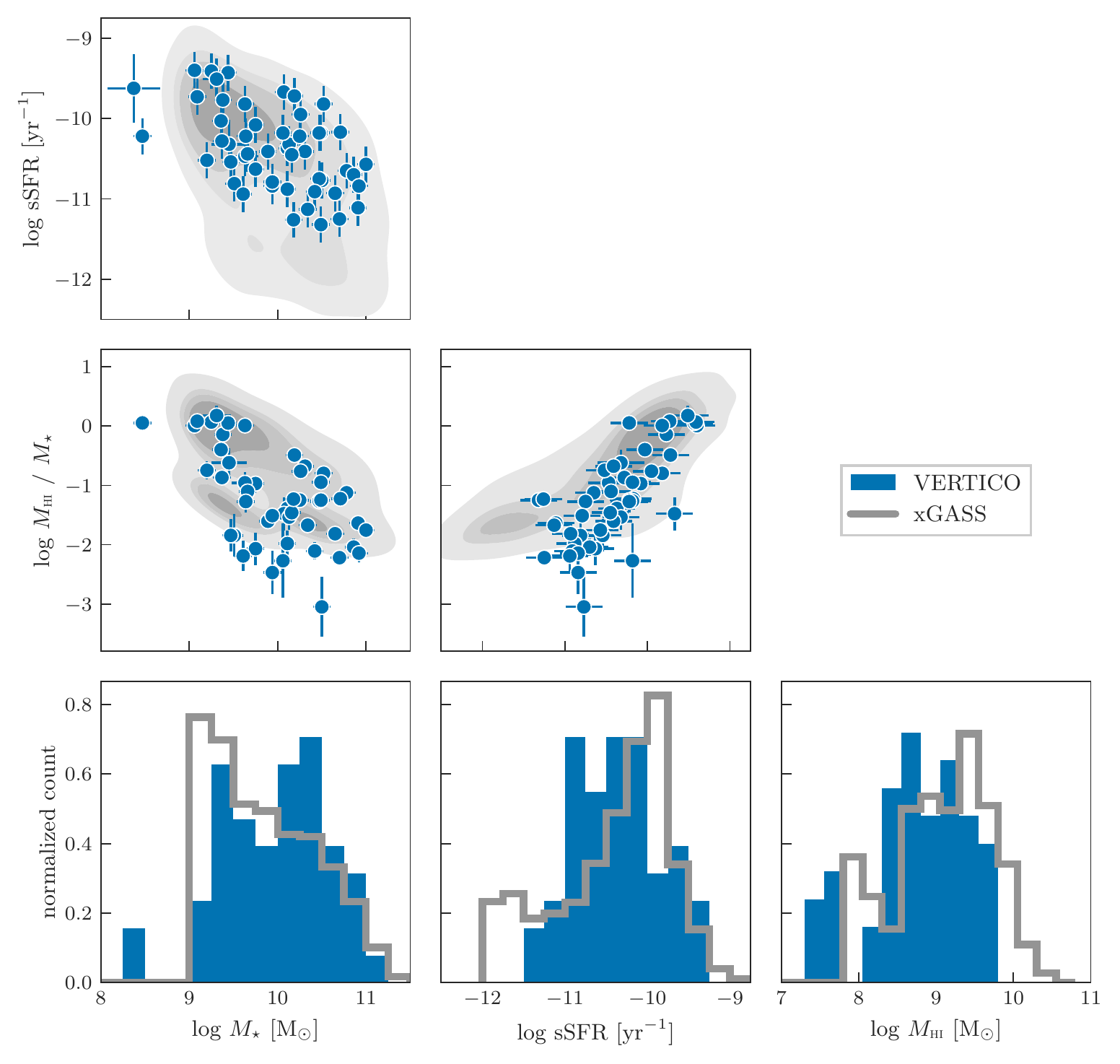}
	\caption{Global properties of the VERTICO sample (blue points and histograms) compared to the representative xGASS survey \citep[grey distributions, 1179 galaxies;][]{Catinella2018}. Left to right from the upper left; sSFR vs. stellar mass, \HI gas fraction vs. stellar mass, \HI gas fraction vs. sSFR, the normalized stellar mass, sSFR, and \HI mass distributions. Panels in the same row share the same $y$-axis range.}
	\label{fig:mass_sfr_mhi_corner}
\end{figure*}

The global stellar mass and star formation rate (SFR) estimates used in this paper are drawn from the $z=0$ Multi-wavelength Galaxy Database \citep[$z0$MGS; ][]{Leroy2019}. They were downloaded via the NASA/IPAC Infrared Science Archive\footnote{\url{https://irsa.ipac.caltech.edu/data/WISE/z0MGS/overview.html}}. The SFRs are derived from GALEX NUV and WISE band 4 (22 $\micron$) luminosities, while the stellar masses are based upon a variable WISE band 1 (3.4 $\micron$) mass-to-light ratio predicted using the specific SFR-like quantity, SFR-to-WISE band 1 luminosity. Global \HI mass estimates are taken from the VIVA survey. The median uncertainties are 0.1 dex for stellar and \HI mass, and 0.2 dex for SFR.  One galaxy, IC3418, is not included in the $z0$MGS, so we adopt the literature values of $M_\star = 10^{8.37}~ \text{M}_\odot$ and $\textrm{SFR}=0.1~\text{M}_\odot~\text{yr}^{-1}$ \citep{Fumagalli2011, Jachym2013}. These properties are derived using spectral energy distribution fits to optical and UV observations, and a far-UV luminosity--SFR relation modified for dwarf galaxies. Without formal uncertainties quoted in these works, we assign an uncertainty of $0.3~$dext to the logarithmic values of stellar mass and SFR for IC3418.

Figure \ref{fig:mass_sfr_mhi_corner} presents the scaling relations between and distributions of global stellar mass, sSFR, and \HI properties of the VERTICO sample (blue points and histograms). We compare the VERTICO sample with the 1,179 galaxies in the extended GALEX Arecibo SDSS Survey (xGASS, grey distributions) that have their `best' SFR measurement within that catalog and including \HI non-detections as $3\sigma$ upper limits \citep{Catinella2018}\footnote{\url{https://xgass.icrar.org/}}. Following \citet{Catinella2018}, the xGASS sample is weighted to correct for the flat stellar mass distribution and recover a volume-limited sample for $M_\star  \geq \ 10^9$ M$_\odot$ and $z\leq0.05$. The lower left panel shows the normalized distribution of stellar mass for both samples. Barring two galaxies (IC3418, VCC1581), VERTICO targets have $M_\star \geq 10^9$ M$_\odot$. Compared to the representative xGASS sample, VERTICO has an excess of galaxies above $M_\star \sim 10^{10}$ M$_\odot$. The stellar mass--sSFR relation is provided in the upper left and the sSFR distribution is shown in lower center. VERTICO contains galaxies that are either star-forming (i.e., blue cloud) or transitioning from star-forming towards quiescence (i.e., green valley) with no systems that are already quenched (sSFR $\leq 10^{-11.5}$ yr$^{-1}$). We note that this bias towards star-forming or transitioning galaxies is by selection, with the VIVA survey not including quenched or early-type galaxies. Atomic gas fractions ($M_{\text{\sc HI}} / M_\star$) as a function of stellar mass and sSFR are shown in the left and center panels of the second row, respectively. VERTICO galaxies exhibit a broad range in \HI gas fraction for their stellar and star formation properties, spanning $\sim 3$ dex in gas fraction at both fixed stellar mass and sSFR. The distribution of $M_{\text{\sc HI}}$ in the lower right panel shows that the VERTICO sample is marginally offset to lower \HI gas masses compared to the representative xGASS sample. \citet{Yoon2017} demonstrate that the majority of the VIVA (and therefore the VERTICO) sample entered the cluster relatively recently (i.e., are not yet virialized). Combined with the targeting of late-type galaxies that were likely to be detected in \HInospace, this naturally results in an absence of passive systems. Indeed, in the SFR--$M_\star$ parameter space, \citet{Mun2021} show that the Virgo Cluster has a well-populated quiescent sequence that is not sampled by VIVA. Given we are missing the bulk of the passive, gas-poor population, our sample thus cannot be considered representative of Virgo’s entire population. This is by construction with the sample selection designed to target galaxies that are either being, will be, or recently have been actively quenched. Despite the star-forming nature of the sample, VERTICO probes a large range in \HI gas fraction, covering the full gas-rich to gas-poor parameter space at both fixed stellar mass and sSFR.

\begin{figure*}
	\centering
	\includegraphics{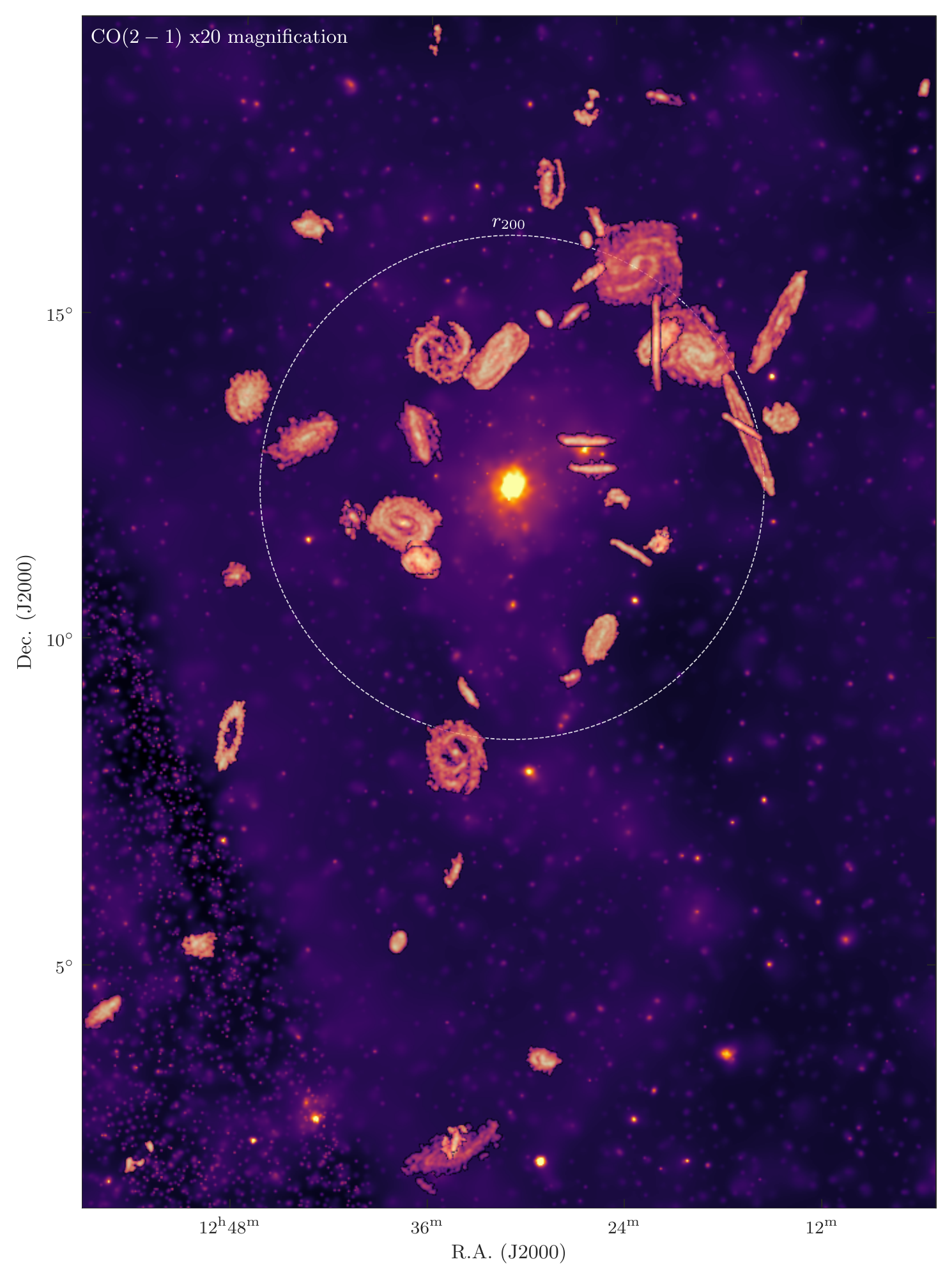}
	\caption{Background-subtracted, exposure-corrected ROSAT All Sky Survey mosaic of the Virgo Cluster (hard band: $0.4-2.4$ keV). We overlay the VERTICO CO($2-1$) peak temperature maps for the 49 detected galaxies, increased in angular size by a factor of 20 for illustration. CO($2-1$) observations of three galaxies (NGC4254, NGC4321, NGC4501) do not cover the full extent of the CO disk and, as such, their peak temperature maps are rectangular in shape. The white dashed circle denotes the radius of the Virgo cluster, $r_{200}$ = 3.9$^\circ$ \citep[$\sim1.08$ Mpc;][]{Urban2011}.}
	\label{fig:virgoXray}
\end{figure*}

Figure \ref{fig:virgoXray} shows the ROSAT All Sky Survey mosaic of the Virgo Cluster (hard band: 0.4--2.4 keV) with the VERTICO CO($2-1$) peak temperature maps overlaid. ROSAT images and exposure maps were obtained from NASA's High Energy Astrophysics Science Archive Research Center\footnote{\url{https://heasarc.gsfc.nasa.gov/}}.  An exposure-corrected X-ray mosaic of the Virgo cluster area was produced with the {\tt reproject\_image\_grid} function from the Chandra Interactive Analysis of Observations\footnote{\url{https://cxc.cfa.harvard.edu/ciao/}} software package \citep[CIAO;][]{Fruscione2006}.  This mosaic was then adaptively smoothed to emphasize emission on different scales with the CIAO function {\tt csmooth} \citep{Ebeling2006}. The figure shows VERTICO CO($2-1$) peak temperature maps at the locations of the target galaxies relative to the ROSAT image. VERTICO galaxies are distributed throughout the Virgo Cluster with a large range of cluster-centric radii ($\sim0.2-2\times~ r_{200}$). For illustration, the angular size of the VERTICO maps has been increased by a factor of twenty. See Section \ref{sec:Derived Data Products} for a full description of the CO maps.

\section{Observations} \label{sec:Observations}

We observed 36 targets in CO($2-1$), $^{13}$CO($2-1$), C$^{18}$O($2-1$), and ALMA Band 6 continuum in ALMA Cycle 7. We set a spectral bandwidth of $1875~{\rm MHz}$ to ensure we cover emission from the target itself and potential interactions. We average the basic Frequency Division Mode channel spacing by a factor of four to achieve a raw spectral resolution of $1.953~{\rm MHz} \approx 2.5 ~ {\rm km ~ s^{-1}}$. We bin by a further factor of four during the data processing to yield cubes with a final resolution of $\sim10 ~ {\rm km ~ s^{-1}}$. While all lines are observed with the same bandwidth and approximate channel width, tuning constraints on the spectral setup means that we do not require the line frequencies to be in the center of the spectral window. The target galaxies are combined with archival ACA CO($2-1$) data (7-m and Total Power arrays) for 14 massive Virgo Cluster spirals from the ALMA component of the Physics at High Angular resolution in Nearby Galaxies project \citep[PHANGS-ALMA;][]{Leroy2021a} and one from a regular program \citep[2016.1.00912.S;][]{Cramer2020} to make the final VERTICO sample of 51 galaxies.

Observations and theory suggest that gas transitions from predominantly atomic to molecular at $\sim 10~{\rm M}_\odot\, {\rm pc}^{-2}$ at solar metallicity \citep{Leroy2008,Krumholz2009}. CO($3-2$) James Clerk Maxwell Telescope (JCMT) $15\arcsec$ observations of 13 \HInospace-selected Virgo galaxies demonstrate \Htwo surface densities of $\sim 6 ~ {\rm M}_\odot~ {\rm pc}^{-2}$ at the outskirts of molecular disks \citep{Mok2017}. We choose a $5\sigma$ sensitivity limit of $8.5 ~{\rm M}_\odot ~{\rm pc}^{-2}$ per 10~km~s$^{-1}$ channel, ensuring we detect the diffuse gas that is below the atomic-to-molecular transition density and most susceptible to environmental influence. This mass surface density sensitivity corresponds to a root mean square (rms) of 10.6 mJy beam$^{-1}$ per $10 ~ {\rm km ~ s^{-1}}$ channel. The total integration time required to successfully reach this sensitivity across all 36 Cycle-7 targets was 186.5 hours. Each galaxy was observed with a mosaic (between three and 31 pointings with an average of 13) with Nyquist spacing. We obtained Total Power observations for 25 out of 36 Cycle-7 targets where galaxy CO disks were expected to extend more than 29\arcsec.

\subsection{Data Reduction} \label{sec:DataReduction}

For the galaxies in the VERTICO Cycle-7 sample, we used the calibrated {\em uv} data delivered by ALMA and we imaged all the available $J=2-1$ CO lines  [CO($2-1$), $^{13}$CO($2-1$), C$^{18}$O($2-1$)]. All galaxies were observed with ACA mosaics. The spatial extent of each mosaic was set to cover the CO($3-2$) emission from the JCMT--Next Generation Virgo Legacy Survey \citep[JCMT--NGLS;][]{Wilson2009} if available, or the Herschel Spectral and Photometric Imaging Receiver (SPIRE) $250\micron$ flux maps published by \citet{Ciesla2012} and downloaded from the Herschel Database in Marseille\footnote{\url{http://hedam.lam.fr}}. For the 14 VERTICO targets that are also part of PHANGS-ALMA [which have CO($2-1$), C$^{18}$O($2-1$), ACA 7m and Total Power data], as well as for the one archival target that was not part of PHANGS-ALMA [NGC4402; CO($2-1$) only, no Total Power data], we retrieved the raw ACA {\em uv} data from the ALMA archive and calibrated it using CASA version 5.6. Th
ree of these galaxies (NGC4254, NGC4321, NGC4535) remained in ALMA quality assurance (QA3) at the time of VERTICO imaging and the calibrated {\em uv} data for these galaxies were kindly provided to us in private communication by Adam Leroy on behalf of the PHANGS-ALMA team. The $^{13}$CO($2-1$) and C$^{18}$O($2-1$) observations have been processed in the same manner as the CO($2-1$) data. However, the presentation of those data, along with an expanded analysis using spectral stacking, will be the focus of future work.

To image the VERTICO ACA data, we used the PHANGS-ALMA Imaging Pipeline Version 1.0 \citep{Leroy2021b} with three modifications to adapt the pipeline for these ACA-only images (note that the PHANGS survey and pipeline papers, as well as delivered data products, use and describe Version 2.0 of this pipeline). First, we added a continuum subtraction step to the PHANGS-ALMA pipeline. For 26 galaxies where the delivered products from ALMA suggested the continuum was detected, we applied continuum subtraction in {\em uv} space to perform a first-order fit to the line-free channels across all spectral windows. Second, we removed the steps in building the single-scale clean mask where the mask was expanded in velocity space. This change was necessary to keep the mask from expanding out to include strong sidelobes present in some of the VERTICO data. Third, in addition to the maximum resolution cubes produced by the pipeline (median resolution $= 8\arcsec$), we also produced data cubes with a $9\arcsec$ beam [CO($2-1$) only, and excluding NGC4321 where the maximum resolution is $10\arcsec$] and a $15\arcsec$ beam (all lines). As a guide, $1\arcsec \approx 80~{\rm pc}$ at the distance of Virgo. For simplicity, these lower-resolution cubes were produced using the CASA task {\tt imsmooth} rather than applying a {\em uv} taper at the initial imaging stage.

For VERTICO imaging, we use Briggs weighting \citep[robust$=0.5$;][]{Briggs1995} and set the target velocity resolution to be 10~km~s$^{-1}$ with the local standard of rest (LSR) as our velocity reference frame, using the radio definition of velocity. We defined the reference phase center to be the centroid pixel determined from a map of the two-dimensional primary beam response in a single velocity channel. This step was necessary as the phase centers specified in the observing stage were not always at the precise center of the resulting image mosaic. Using the observed offset phase center could produce a mildly to strongly asymmetric shape and sidelobe response of the point spread function (PSF). We first carried out a multi-scale clean down to $\mathrm{S/N} = 4$, followed by a single-scale clean down to $\mathrm{S/N} = 1$ in masked regions. All cubes were visually inspected for obvious problems or imaging errors. ALMA Band 6 observations have a $5-10\%$ flux calibration uncertainty\footnote{See Chapter 10 in the ALMA Cycle 7 Technical Handbook\\\url{https://almascience.nrao.edu/documents-and-tools/cycle7/alma-technical-handbook/}}.

For the Total Power data for galaxies in the VERTICO Cycle-7 sample, we started with the raw data delivered by ALMA and used version 1.0 of the PHANGS-ALMA Total Power pipeline \citep{Herrera2020,Leroy2021b}. The only modification we made to the pipeline was to remove the initial velocity binning, so that the Total Power data were processed at their native velocity resolution of $\sim 3$~km~s$^{-1}$. We typically imaged a range of 1000~km~s$^{-1}$ around the mean velocity of the galaxy. We then fit and removed a first-order baseline using the highest and lowest 200~km~s$^{-1}$ of the cube. For five galaxies (IC3392, NGC4380, NGC4383, NGC4580, NGC4651), the baseline region was shifted to avoid an atmospheric ozone line. For two galaxies (NGC4302, NGC4698), the ozone line overlaps with the CO($2-1$) line at some velocities and so the Total Power fluxes for those galaxies are less reliable. All Total Power cubes were inspected to check for any problems in the data reduction. For the PHANGS-VERTICO galaxies, calibrated Total Power cubes were kindly provided to us by Adam Leroy on behalf of the PHANGS-ALMA team in private communication.

For all galaxies and lines for which Total Power data were available, the Total Power data were combined with the ACA data via feathering using the PHANGS-ALMA pipeline. This technique is described in full in section 6 of \citet{Leroy2021b}. The final high-resolution data cubes were binned by a factor of two to produce $\ge 3$ pixels across the beam. We also produced data cubes binned by factors of four (for the native and $9\arcsec$ resolution images) and eight (for the $15\arcsec$ resolution images).

\begin{figure*}
	\centering
	\includegraphics[width=\linewidth]{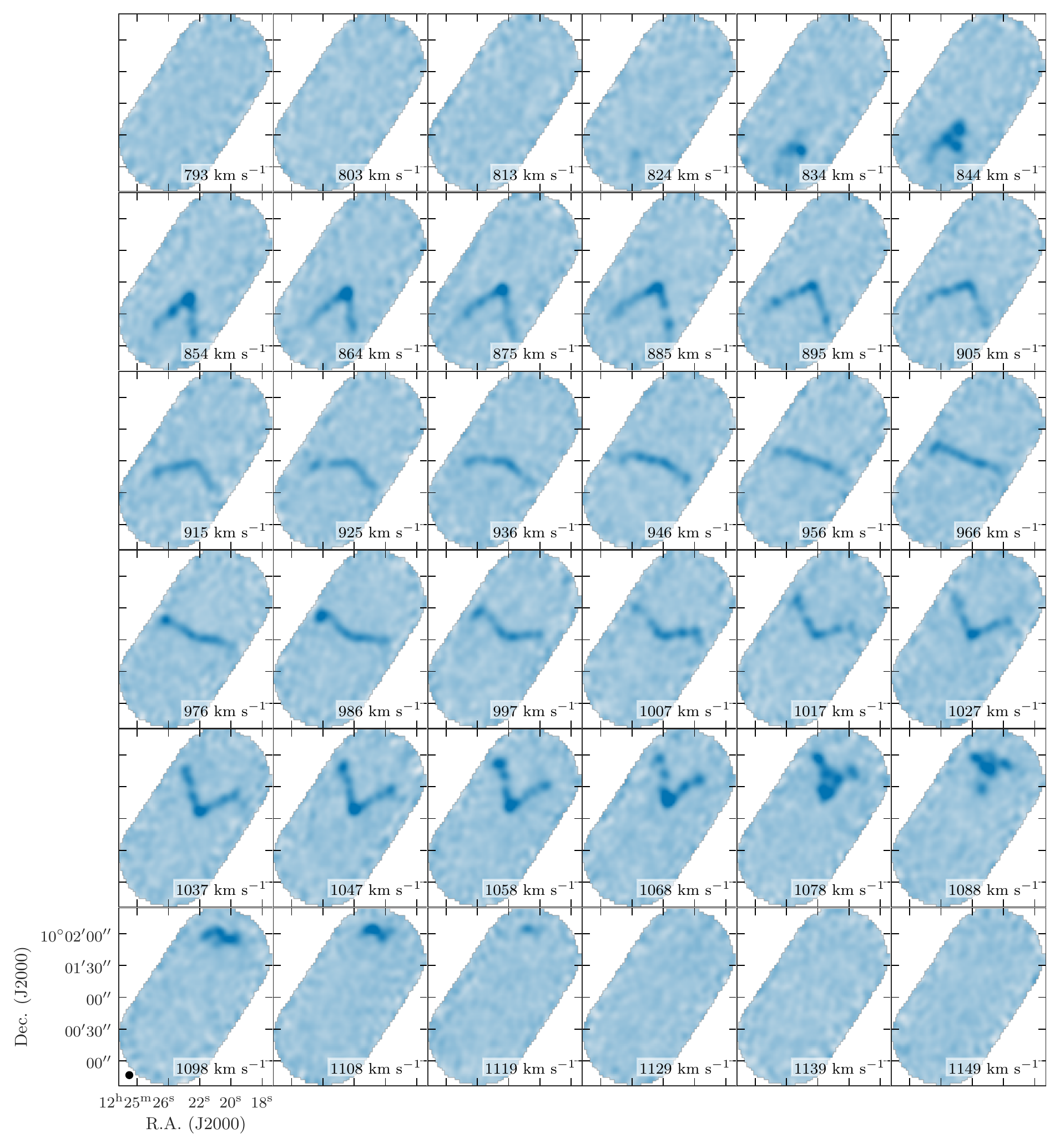}
	\caption{Channel maps for NGC4380. The circular beam size illustrated in the lower left panel is $7.5\arcsec$, the channel width is 10.6~km~s$^{-1}$, and the rms intensity in one channel is 3.5 mK. The systemic velocity of each channel is given in the lower right-hand corner. The color scale is fixed between channels.}
	\label{fig:NGC4380_channel_maps}
\end{figure*}

As an example, Figure \ref{fig:NGC4380_channel_maps} shows the integrated intensity channel maps for NGC4380. Each channel is 10.6~km~s$^{-1}$ wide with the channel systemic velocity shown in the bottom right corner of each panel. The maps span the velocity range over which we detect CO($2-1$) emission. The color scale is fixed from channel to channel.

\section{Derived Data Products} \label{sec:Derived Data Products}

We calculate moment maps, position--velocity diagrams, and radial profiles for each galaxy from a masked signal cube. The masking process follows a revised version of the signal identification scheme described in \citet{Sun2018}. This method uses a spatially and spectrally varying noise estimate (i.e., we estimate the noise at each pixel in every channel) computed from the signal cubes before primary beam correction and is described in detail in Section 7.2 of \citet{Leroy2021b}. The original code is publicly available%
\footnote{\url{https://github.com/astrojysun/Sun_Astro_Tools/tree/master/sun_astro_tools}} and the steps are as follows:
\begin{enumerate}
	\item Generate a core mask for spaxels with S/N $\geq 3.5$ in at least three consecutive channels.
	\item Generate a wing mask for spaxels with S/N $\geq 2$ in at least two consecutive channels.
	\item Combine the core mask with the wing mask to define a signal mask that encapsulates all detected spaxels.
	\item Prune spaxels from the signal mask if the projected area of connected neighbors on the sky is smaller than one beam.
	\item Expand the signal mask along the spatial dimensions by a given number of pixels or a fraction of the beam size.
	\item Expand the signal mask by two velocity channels.
	\item Apply the final signal mask to produce a signal cube from which all moments are calculated.
\end{enumerate}

We now describe the science-ready data products that are derived from these masked cubes.

\subsection{Moment Maps} \label{sec:MomentMaps}

\begin{figure*}
	\centering
	\includegraphics[width=\linewidth]{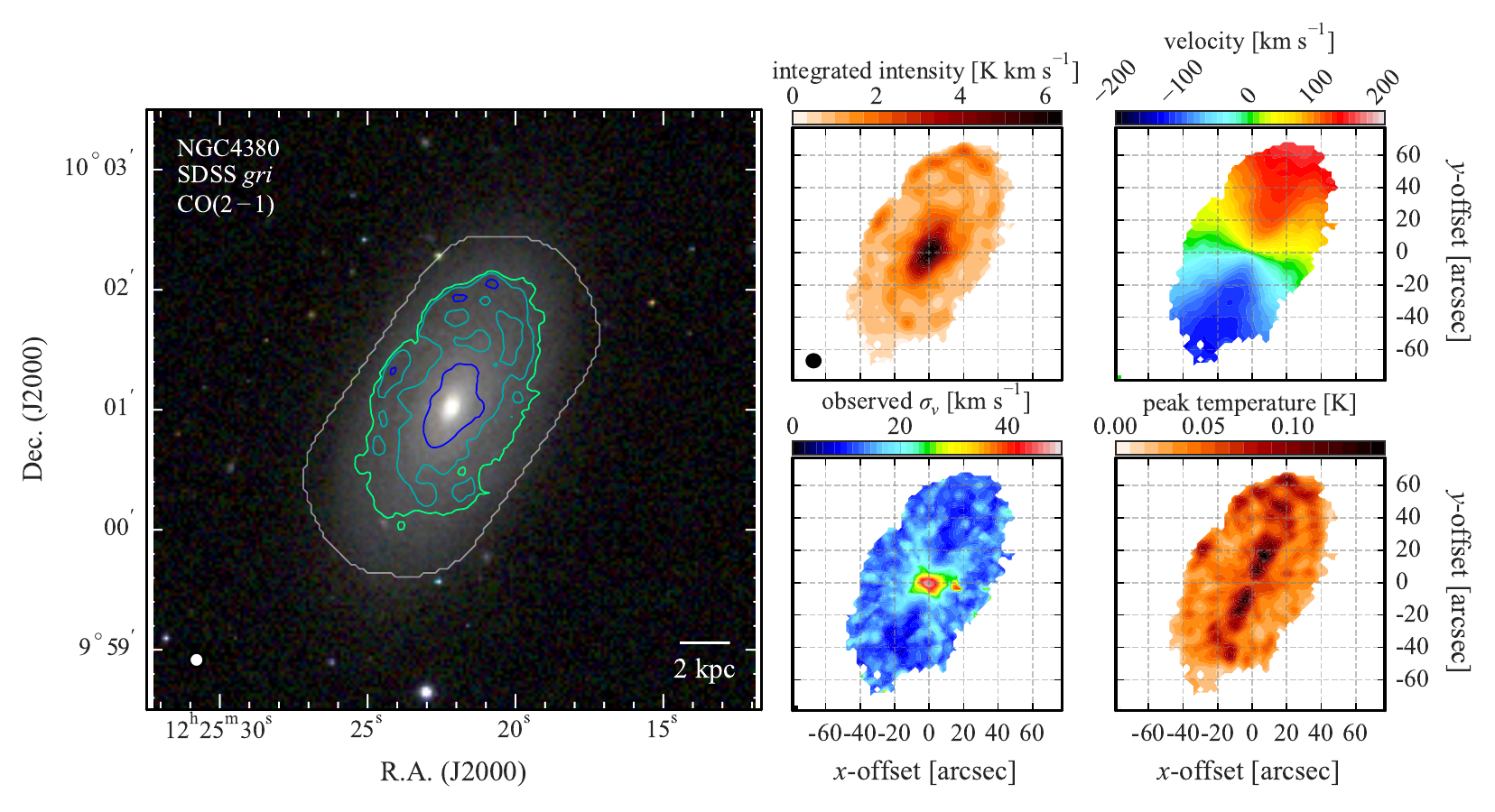}
	\caption{An example of the CO($2-1$) data products available for each galaxy in the VERTICO survey. The left panel shows the SDSS $gri$ composite image for NGC4380 with molecular gas surface brightness contours at the 10th, 50th, and 90th percentiles of the distribution. The field of view of the ACA observations is defined where the primary beam response drops to 50\% and is illustrated by the gray line. The rounded synthesized beam is $7.5\arcsec$ in diameter and illustrated in the bottom left corner. This beam corresponds to $\sim$600 pc at the distance of Virgo (16.5 Mpc). The VERTICO CO($2-1$) data products available for each galaxy include maps of integrated intensity (upper center panel), the velocity field (upper right), observed line width, $\sigma_v$ (lower center), and peak temperature (lower left). The $x$- and $y$-axes of each moment map shows the angular offset from the optical center listed in Table \ref{tab:VERTICO-sample}. {\it The complete Figure Set containing 49 panel plots for all VERTICO detected galaxies is available in the online journal.}}.
	\label{fig:NGC4380_panel_plot}
\end{figure*}

We compute the zeroth, first, second, and eighth-order two-dimensional moment maps from the CO($2-1$) spectral line cubes. In order, these are: 
\begin{itemize}
	\item Integrated intensity of the spectrum along the spectral axis in K~km~s$^{-1}$.
	\item Intensity weighted spectral coordinate in km~s$^{-1}$, often referred to as the velocity field.
	\item Observed line width ($\sigma_v$, i.e., not corrected for broadening) along the spectral axis in km~s$^{-1}$.
	\item Peak brightness temperature value of the spectrum in K.
\end{itemize}

Although not shown in this paper, we calculate statistical uncertainty maps for the pixel-by-pixel integrated intensity, velocity field, and observed line width maps. The uncertainty on the integrated intensity is
\begin{equation}
	u_{I}= \sqrt{N} \sigma_{T} \Delta v \,,
\end{equation}
where $N$ is the number of channels included in the mask, $\sigma_{T}$ the rms uncertainty across the integrated intensity map, and $\Delta v$ is the velocity channel width. We then compute the uncertainty on the velocity field,
\begin{equation}
	u_{\rm vel} = \frac{\Delta v_{\rm{line}}}{2 \sqrt{3}} \frac{u_{I}}{I} \,,
\end{equation}
where $I$ is the integrated intensity and $\Delta v_{\rm{line}}$ the spectral line width over which $u_{\rm vel}$ is calculated. The uncertainty on the observed line width is given by
\begin{equation}
	u_{\sigma_v} = \frac{u_{I}}{I} \frac{\left(\Delta v_{\rm{line}}\right)^{2}}{8 \sqrt{5}} \frac{1}{\sigma_v} \,,
\end{equation}
where $\sigma_v$ is the observed line width map. The derivation for these equations is provided in Wilson et al. (2021, in prep.).

Figure \ref{fig:NGC4380_panel_plot} illustrates the high quality of the VERTICO CO($2-1$) moment maps for NGC4380. An archetypal unbarred-spiral, we choose this galaxy to showcase the VERTICO as it appears to be relatively unperturbed by its environment and has a bright, extended CO($2-1$)  gas disk. The left panel shows the SDSS $gri$ composite image with molecular gas surface brightness contours at the 10th, 50th, and 90th percentiles. The integrated intensity, velocity field, peak temperature, and observed CO($2-1$) line width maps are provided clockwise from the top middle panel. We calculate NGC4380's $r$-band inclination to be $61\degr$. Equivalent panel plots of the CO data products for the 49 detected VERTICO galaxies are \textbf{in the online version of Figure \ref{fig:NGC4380_panel_plot}}.

The synthesized $7.5\arcsec$ beam of the NGC4380 data corresponds to approximately 600 pc at the distance of Virgo (16.5 Mpc). On this scale, the VERTICO observations reveal the imprint of stellar structure in the molecular gas distribution of NGC4380 in glorious detail. The Northern and North-Eastern outskirts of the gas disk have a ridge of emission that is not found on the opposing side of the galaxy. We can also see a strong CO feature straddling the nucleus from the South East to the North West. Examining the central region in the peak temperature map shows there are two symmetrical peaks of CO emission such as those commonly found in the centers of barred galaxies \citep[e.g.,][]{Kenney1992, Muraoka2016}.

\subsection{Position--Velocity Diagrams}
\label{sec:PVD}

\begin{figure}
	\centering
	\includegraphics[width=\linewidth]{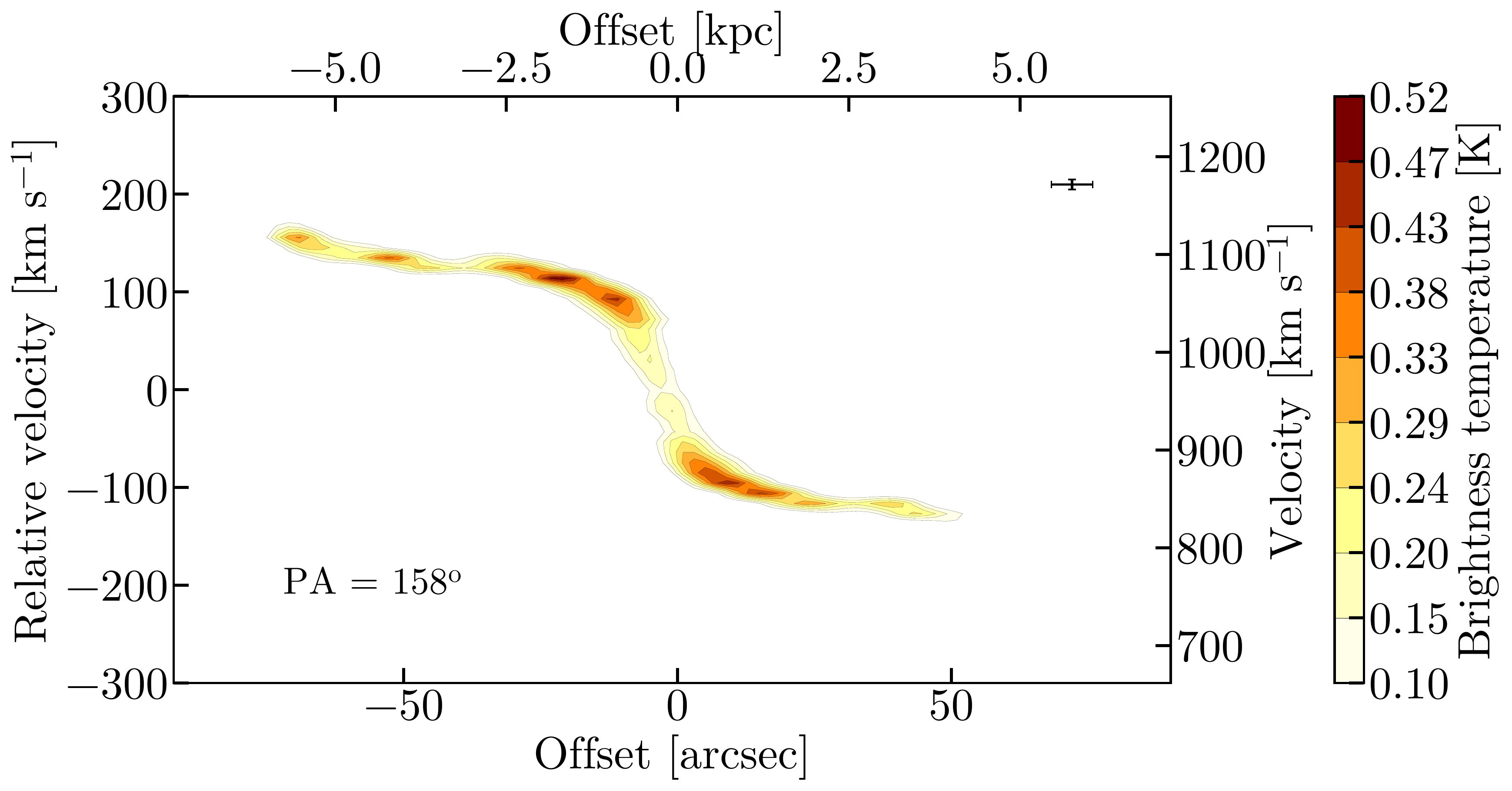}
	\includegraphics[width=\linewidth]{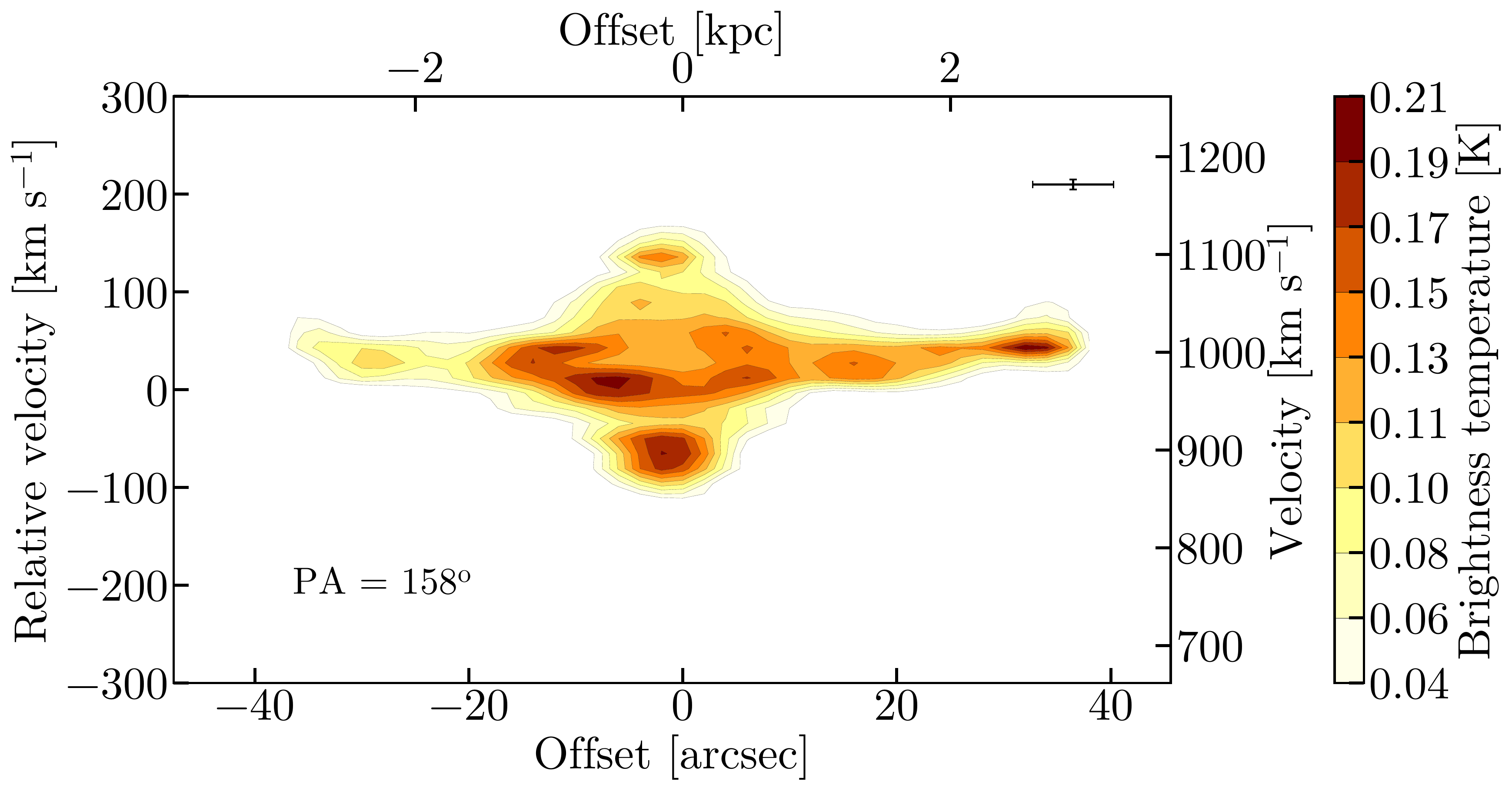}
	\caption{Position--velocity diagrams along the major (upper panel) and minor (lower panel) axes of NGC4380, extracted along the respective axes with a slit one beam in width. The bottom and top $x$-axes show the angular and physical offset from the galaxy center, respectively. The $y$-axis shows the relative offset from the systemic velocity listed in Table \ref{tab:VERTICO-sample}. The position angle used is $158\degr$ and the average uncertainty in each axis is illustrated in the top right corner. The channel width is $10~{\rm km~s^{-1}}$.}
	\label{fig:NGC4380_PVD}
\end{figure}

We create major- and minor-axis position--velocity diagrams (PVD) from the CO($2-1$) signal cubes. The major-axis position angles are provided in Table \ref{tab:VERTICO-sample} and were derived by fitting Kron ellipses to the background-subtracted, masked SDSS $r$-band images using the {\sc Photutils} {\sc Python} package \citep{larry_bradley_2020_4044744}. All Kron ellipses pass visual inspection to ensure a robust fit before being used. The position--velocity slit width is set to one beam width for each galaxy.

Figure \ref{fig:NGC4380_PVD} shows both PVDs for NGC4380. The two $x$-axes denote the positional offset about the optical center of the galaxy (provided in Table \ref{tab:VERTICO-sample}) in physical and angular units. The major-axis PVD is shown in the upper panel and reveals the increase in high velocity emission at negative offsets compared to positive offsets. This asymmetry is also apparent in the Northern regions of the peak temperature map in Figure \ref{fig:NGC4380_panel_plot} and is discussed in Section \ref{sec:MomentMaps}. The asymmetry that shows higher brightness temperatures at positive relative velocity offsets in the minor-axis PVD (lower panel) highlights the concentration of emission along the North-East edge of the gas disk. More work is needed to determine whether the source of this asymmetry is secular (e.g., spiral arms, gas streaming motions) or environment-driven (e.g., starvation, gravitational and/or ICM interaction). Channel maps for NGC4380's unmasked CO($2-1$) cube are shown in Figure \ref{fig:NGC4380_channel_maps}.

\subsection{Global CO Fluxes}\label{sec:COFluxes}
The global CO line luminosity, $L_{\rm CO}$, and integrated flux, $S_{\rm CO}$, are calculated from global spectra derived from masked data cubes. In an effort to include fainter emission at the edges of each disk in the global estimate, we adopt a 2D masking process to calculate these global properties (adapted from the 3D method used to produce the moment maps described in Section \ref{sec:MomentMaps}, which we refer to as the \citealt{Sun2018} method). The 2D masking process for making the spectra is as follows:
\begin{enumerate}
	\item Create a 2D mask where pixels are masked if the conditions for masking described in Section \ref{sec:MomentMaps} are met at that {\em xy} position in every channel.
	\item Dilate this mask in the {\em xy} plane by the rounded beam size listed in Table \ref{tab:co_props}.
	\item Replicate the dilated mask in every channel and apply to the primary beam corrected, continuum subtracted CO data cubes.
\end{enumerate}

The full Figure Set for spectra derived from these masked cubes for every galaxy are shown after the main text of the article. The integrated CO line properties and velocity widths at zero intensity over which the line fluxes are measured are provided in Table \ref{tab:co_props}. We provide measurement uncertainties on the global CO line luminosities and fluxes. True uncertainties should also include the $5-10\%$ ALMA band-6 calibration accuracy and distance uncertainties, added in quadrature, that are not accounted for in the quoted values. For the latter, the determination of reliable distance for cluster galaxies is non-trivial. However, the standard deviation of VERTICO galaxy distances taken from the $z0$MGS database \citep{Leroy2019} is 1.65 Mpc, which consistent with the range of Virgo size estimates \citep[$r_{200} \approx 1.05 - 1.55$ Mpc;][]{Boselli2018}. There are four marginal detections with S/N$<4$ (NGC4533, NGC4698, NGC4561 and NGC4772) and two non-detections (IC3418 and VCC1581) for which we provide $3\sigma$ upper limits.

\begin{longrotatetable}
	\begin{deluxetable*}{lccccccccc}
		\tablecaption{Global CO($2-1$) line properties for the 51 VERTICO galaxies.}
		\tablehead{\colhead{Galaxy} & \colhead{S/N} & \colhead{$\theta_{\rm b}$} & \colhead{rms} & \colhead{$v_{\rm lsr}$} & \colhead{$\Delta v_{\rm{line}}$} & \colhead{$S_{\rm CO}$} & \colhead{log $L_{\rm CO}$} & \colhead{$\overline{I}_{\rm CO}$} & \colhead{log $M_{\rm mol}$}\\ \colhead{ } & \colhead{ } & \colhead{$\mathrm{{}^{\prime\prime}}$} & \colhead{$\mathrm{mJy}$} & \colhead{$\mathrm{km\,s^{-1}}$} & \colhead{$\mathrm{km\,s^{-1}}$} & \colhead{$\mathrm{Jy\,km\,s^{-1}}$} & \colhead{$\mathrm{K\,km\,s^{-1}\,pc^{2}}$} & \colhead{$\mathrm{K\,km\,s^{-1}}$} & \colhead{$\mathrm{M_{\odot}}$} \\ \colhead{(1)} &  \colhead{(2)} &  \colhead{(3)} &  \colhead{(4)} &  \colhead{(5)} &  \colhead{(6)} &  \colhead{(7)} &  \colhead{(8)} &  \colhead{(9)} &  \colhead{(10)}}
		\startdata
		IC3392 & 28.3 & 8.4 & 134 & 1690 & 206 & 259 $\pm$ 9 & 7.63 $\pm$ 0.02 & 1.87 $\pm$ 0.07 & 8.37 $\pm$ 0.02 \\
		IC3418\tablenotemark{a} &  & 8.1 & 101 &  &  & $<$30.29 & $<$6.70 & $<$11.44 & $<$7.44 \\
		NGC4064 & 23.3 & 9.0 & 32 & 954 & 225 & 285 $\pm$ 12 & 7.67 $\pm$ 0.02 & 2.48 $\pm$ 0.11 & 8.41 $\pm$ 0.02 \\
		NGC4189 & 38.9 & 7.5 & 202 & 2133 & 289 & 545 $\pm$ 14 & 7.95 $\pm$ 0.01 & 1.59 $\pm$ 0.04 & 8.69 $\pm$ 0.01 \\
		NGC4192 & 74.4 & 9.2 & 358 & -143 & 518 & 1997 $\pm$ 27 & 8.52 $\pm$ 0.01 & 2.78 $\pm$ 0.04 & 9.26 $\pm$ 0.01 \\
		NGC4216 & 40.9 & 7.2 & 304 & 160 & 621 & 1009 $\pm$ 25 & 8.22 $\pm$ 0.01 & 1.37 $\pm$ 0.03 & 8.96 $\pm$ 0.01 \\
		NGC4222 & 12.5 & 8.4 & 26 & 230 & 255 & 128 $\pm$ 10 & 7.33 $\pm$ 0.03 & 0.58 $\pm$ 0.05 & 8.06 $\pm$ 0.03 \\
		NGC4254 & 185.0 & 8.6 & 331 & 2447 & 320 & 8369 $\pm$ 45 & 9.14 $\pm$ 0.00 & 4.64 $\pm$ 0.03 & 9.88 $\pm$ 0.00 \\
		NGC4293 & 48.8 & 7.6 & 328 & 936 & 297 & 810 $\pm$ 17 & 8.13 $\pm$ 0.01 & 3.80 $\pm$ 0.08 & 8.86 $\pm$ 0.01 \\
		NGC4294 & 6.4 & 8.2 & 29 & 376 & 173 & 62 $\pm$ 10 & 7.01 $\pm$ 0.07 & 0.37 $\pm$ 0.06 & 7.75 $\pm$ 0.07 \\
		NGC4298 & 82.7 & 7.5 & 191 & 1140 & 278 & 1336 $\pm$ 16 & 8.35 $\pm$ 0.01 & 2.50 $\pm$ 0.03 & 9.08 $\pm$ 0.01 \\
		NGC4299 & 5.2 & 8.3 & 117 & 239 & 112 & 42 $\pm$ 8 & 6.84 $\pm$ 0.08 & 0.19 $\pm$ 0.04 & 7.58 $\pm$ 0.08 \\
		NGC4302 & 67.2 & 7.7 & 313 & 1158 & 410 & 1375 $\pm$ 20 & 8.36 $\pm$ 0.01 & 3.39 $\pm$ 0.05 & 9.09 $\pm$ 0.01 \\
		NGC4321 & 187.9 & 10.2 & 433 & 1590 & 279 & 7635 $\pm$ 41 & 9.10 $\pm$ 0.00 & 3.45 $\pm$ 0.02 & 9.84 $\pm$ 0.00 \\
		NGC4330 & 16.4 & 7.7 & 32 & 1566 & 298 & 241 $\pm$ 15 & 7.60 $\pm$ 0.03 & 1.19 $\pm$ 0.07 & 8.34 $\pm$ 0.03 \\
		NGC4351 & 11.8 & 8.5 & 149 & 2324 & 124 & 79 $\pm$ 7 & 7.11 $\pm$ 0.04 & 0.37 $\pm$ 0.03 & 7.85 $\pm$ 0.04 \\
		NGC4380 & 35.3 & 7.5 & 249 & 990 & 307 & 432 $\pm$ 12 & 7.86 $\pm$ 0.01 & 0.90 $\pm$ 0.03 & 8.59 $\pm$ 0.01 \\
		NGC4383 & 22.9 & 8.1 & 184 & 1715 & 206 & 263 $\pm$ 11 & 7.64 $\pm$ 0.02 & 1.26 $\pm$ 0.05 & 8.37 $\pm$ 0.02 \\
		NGC4388 & 49.3 & 8.1 & 25 & 2501 & 486 & 859 $\pm$ 17 & 8.15 $\pm$ 0.01 & 2.86 $\pm$ 0.06 & 8.89 $\pm$ 0.01 \\
		NGC4394 & 12.0 & 8.7 & 24 & 929 & 215 & 105 $\pm$ 9 & 7.24 $\pm$ 0.04 & 0.44 $\pm$ 0.04 & 7.98 $\pm$ 0.04 \\
		NGC4396 & 10.0 & 7.8 & 217 & -119 & 193 & 124 $\pm$ 12 & 7.32 $\pm$ 0.04 & 0.61 $\pm$ 0.06 & 8.05 $\pm$ 0.04 \\
		NGC4402 & 83.1 & 7.6 & 30 & 248 & 357 & 1402 $\pm$ 17 & 8.37 $\pm$ 0.01 & 5.00 $\pm$ 0.06 & 9.10 $\pm$ 0.01 \\
		NGC4405 & 18.3 & 7.7 & 33 & 1752 & 175 & 206 $\pm$ 11 & 7.53 $\pm$ 0.02 & 1.99 $\pm$ 0.11 & 8.27 $\pm$ 0.02 \\
		NGC4419 & 121.2 & 8.1 & 32 & -172 & 447 & 1399 $\pm$ 12 & 8.37 $\pm$ 0.00 & 5.71 $\pm$ 0.05 & 9.10 $\pm$ 0.00 \\
		NGC4424 & 34.1 & 7.9 & 101 & 448 & 158 & 228 $\pm$ 7 & 7.58 $\pm$ 0.01 & 1.63 $\pm$ 0.05 & 8.31 $\pm$ 0.01 \\
		NGC4450 & 31.8 & 7.6 & 233 & 1960 & 340 & 523 $\pm$ 16 & 7.94 $\pm$ 0.01 & 1.14 $\pm$ 0.04 & 8.67 $\pm$ 0.01 \\
		NGC4457 & 63.6 & 7.8 & 238 & 905 & 297 & 1154 $\pm$ 18 & 8.28 $\pm$ 0.01 & 4.16 $\pm$ 0.07 & 9.02 $\pm$ 0.01 \\
		NGC4501 & 219.6 & 8.0 & 302 & 2277 & 589 & 5499 $\pm$ 25 & 8.96 $\pm$ 0.00 & 7.61 $\pm$ 0.03 & 9.69 $\pm$ 0.00 \\
		NGC4522 & 20.8 & 8.5 & 32 & 2351 & 217 & 223 $\pm$ 11 & 7.57 $\pm$ 0.02 & 1.13 $\pm$ 0.05 & 8.30 $\pm$ 0.02 \\
		NGC4532 & 18.3 & 7.1 & 151 & 2037 & 227 & 198 $\pm$ 11 & 7.52 $\pm$ 0.02 & 1.03 $\pm$ 0.06 & 8.25 $\pm$ 0.02 \\
		NGC4533 & 1.7 & 7.8 & 35 & 1698 & 103 & 10 $\pm$ 6 & 6.20 $\pm$ 0.26 & 0.04 $\pm$ 0.03 & 6.94 $\pm$ 0.26 \\
		NGC4535 & 85.4 & 8.2 & 680 & 1978 & 270 & 2862 $\pm$ 34 & 8.68 $\pm$ 0.01 & 2.27 $\pm$ 0.03 & 9.41 $\pm$ 0.01 \\
		NGC4536 & 88.4 & 8.6 & 452 & 1817 & 389 & 2465 $\pm$ 28 & 8.61 $\pm$ 0.01 & 1.90 $\pm$ 0.02 & 9.35 $\pm$ 0.01 \\
		NGC4548 & 58.7 & 7.6 & 354 & 509 & 336 & 1085 $\pm$ 18 & 8.26 $\pm$ 0.01 & 1.01 $\pm$ 0.02 & 8.99 $\pm$ 0.01 \\
		NGC4561 & 2.3 & 7.9 & 253 & 1433 & 92 & 23 $\pm$ 10 & 6.58 $\pm$ 0.19 & 0.14 $\pm$ 0.06 & 7.31 $\pm$ 0.19 \\
		NGC4567 & 91.7 & 7.5 & 21 & 2295 & 258 & 766 $\pm$ 8 & 8.10 $\pm$ 0.00 & 2.60 $\pm$ 0.03 & 8.84 $\pm$ 0.00 \\
		NGC4568 & 232.6 & 8.6 & 168 & 2279 & 382 & 2865 $\pm$ 12 & 8.68 $\pm$ 0.00 & 9.58 $\pm$ 0.04 & 9.41 $\pm$ 0.00 \\
		NGC4569 & 158.6 & 7.5 & 433 & -222 & 492 & 4204 $\pm$ 27 & 8.85 $\pm$ 0.00 & 6.00 $\pm$ 0.04 & 9.58 $\pm$ 0.00 \\
		NGC4579 & 109.1 & 8.0 & 441 & 1517 & 458 & 2277 $\pm$ 21 & 8.58 $\pm$ 0.00 & 1.98 $\pm$ 0.02 & 9.31 $\pm$ 0.00 \\
		NGC4580 & 40.1 & 8.4 & 179 & 1041 & 215 & 396 $\pm$ 10 & 7.82 $\pm$ 0.01 & 2.13 $\pm$ 0.05 & 8.55 $\pm$ 0.01 \\
		NGC4606 & 25.0 & 7.6 & 21 & 1653 & 165 & 154 $\pm$ 6 & 7.41 $\pm$ 0.02 & 0.84 $\pm$ 0.03 & 8.14 $\pm$ 0.02 \\
		NGC4607 & 33.9 & 7.5 & 26 & 2283 & 299 & 420 $\pm$ 12 & 7.84 $\pm$ 0.01 & 2.11 $\pm$ 0.06 & 8.58 $\pm$ 0.01 \\
		NGC4651 & 51.0 & 7.9 & 165 & 805 & 368 & 640 $\pm$ 13 & 8.03 $\pm$ 0.01 & 1.94 $\pm$ 0.04 & 8.76 $\pm$ 0.01 \\
		NGC4654 & 90.3 & 7.5 & 468 & 1042 & 377 & 2348 $\pm$ 26 & 8.59 $\pm$ 0.00 & 2.54 $\pm$ 0.03 & 9.33 $\pm$ 0.00 \\
		NGC4689 & 62.2 & 7.6 & 293 & 1645 & 239 & 1175 $\pm$ 19 & 8.29 $\pm$ 0.01 & 1.65 $\pm$ 0.03 & 9.02 $\pm$ 0.01 \\
		NGC4694 & 20.8 & 7.1 & 105 & 1234 & 278 & 188 $\pm$ 9 & 7.49 $\pm$ 0.02 & 1.05 $\pm$ 0.05 & 8.23 $\pm$ 0.02 \\
		NGC4698 & 4.6 & 8.1 & 253 & 1023 & 451 & 97 $\pm$ 21 & 7.21 $\pm$ 0.10 & 0.29 $\pm$ 0.06 & 7.94 $\pm$ 0.10 \\
		NGC4713 & 23.2 & 8.4 & 184 & 652 & 204 & 265 $\pm$ 11 & 7.64 $\pm$ 0.02 & 0.80 $\pm$ 0.03 & 8.38 $\pm$ 0.02 \\
		NGC4772 & 1.6 & 7.5 & 31 & 1055 & 410 & 28 $\pm$ 17 & 6.66 $\pm$ 0.27 & 0.19 $\pm$ 0.11 & 7.40 $\pm$ 0.27 \\
		NGC4808 & 44.9 & 8.2 & 91 & 779 & 307 & 607 $\pm$ 14 & 8.00 $\pm$ 0.01 & 1.80 $\pm$ 0.04 & 8.74 $\pm$ 0.01 \\
		VCC1581\tablenotemark{a} &  & 8.7 & 25 &  &  & $<$7.55 & $<$6.11 & $<$3.25 & $<$6.84
		\enddata
		\tablecomments{Columns are 
			(1) galaxy identifier; 
			(2) signal-to-noise of the integrated spectrum; 
			(3) diameter of the circularized synthesized beam; 
			(4) integrated spectrum rms in 10.6 km~s$^{-1}$ channels; 
			(5) local standard of rest recession velocity of the line center, calculated as the midpoint of $\Delta v_{\rm{line}}$; 
			(6) velocity width of the line at zero intensity; 
			(7) velocity integrated flux; 
			(8) CO line luminosity given by Equation \ref{eq:Lco}; 
			(9) mean velocity integrated CO intensity; 
			(10) logarithm of molecular gas mass given by Equation \ref{eq:Mmol}. 
		This table is published in its entirety in machine-readable format. We do not include the typical band-6 calibration uncertainty of $5-10\%$ (0.02 -- 0.04 dex) in the flux measurement uncertainties. One may add this in quadrature to account for this.}
		\tablenotetext{a}{Values are $3\sigma$ upper limits calculated over an on-sky circle with 
			radius=$30\arcsec$ (projected radius $\approx 2.4~\text{kpc}$) and a line width of 100 km~s$^{-1}$. 
			The positions and systemic velocities used are listed in Table \ref{tab:VERTICO-sample}}
		\label{tab:co_props}
	\end{deluxetable*}
\end{longrotatetable}

Following \citet{Solomon2005}, we calculate the CO line luminosity, $L_{\rm CO}$ in K~km~s$^{-1}$~pc$^2$, expressed as the product of the velocity integrated source brightness temperature and the source area
\begin{equation} \label{eq:Lco}
	L_{\rm CO} = 3.25 \times 10^7 \, S_{\rm CO} \, \nu_{\rm obs}^{-2} \, D_L^2 \,,
\end{equation} %
\noindent where $S_{\rm CO}$ is the velocity integrated flux in Jy~km~s$^{-1}$, $\nu_{\rm obs}$ is the observed frequency in GHz, and $D_L$ is the luminosity distance to the source in Mpc. For clarity, $L_{\rm CO}$ is $L^{\prime}_{\rm CO}$ in equation 3 of \citet{Solomon2005}.

\begin{figure}
	\centering
	\includegraphics{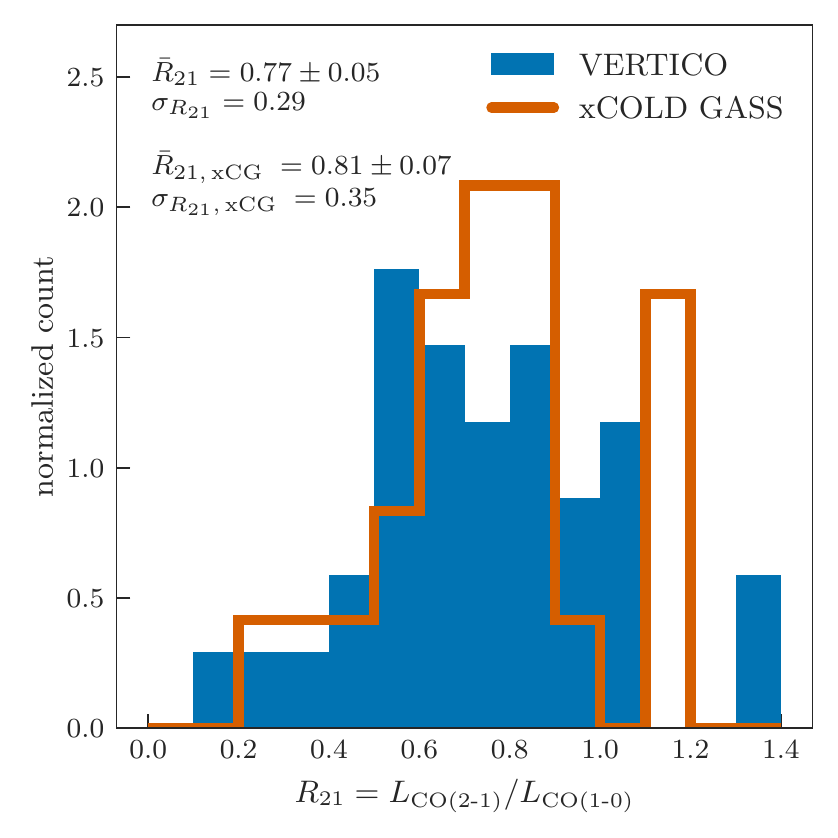}
	\caption{Normalised distribution of CO($2-1$)/CO($1-0$) global line ratios for the 35 VERTICO galaxies (blue solid histogram) that have CO($1-0$) data presented in \citet[][table 11]{Boselli2014a}. The orange step histogram shows the 25 galaxies in xCOLD GASS with high-quality aperture-corrected IRAM 30-m CO($1-0$) and APEX CO($2-1$) data. The mean and standard deviation of each distribution are provided in the upper left corner.}
	\label{fig:vertico_hrs_r21}
\end{figure}

\begin{figure}
	\centering
	\includegraphics{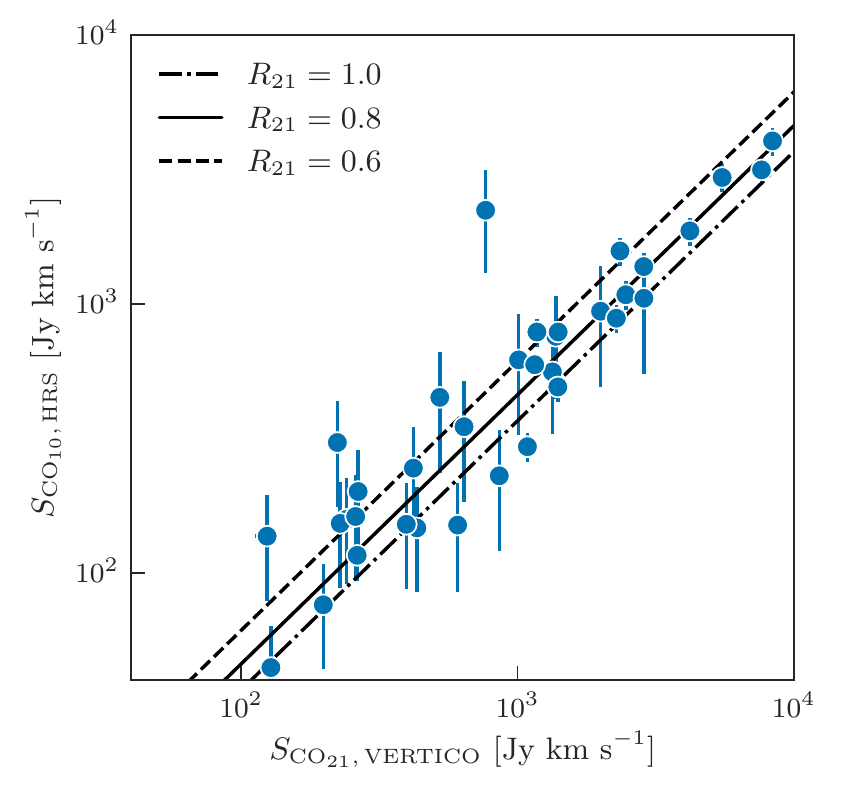}
	\caption{Correlation between observed CO($1-0$) and CO($2-1$) line flux for the 35 VERTICO galaxies that have high-quality CO($1-0$) data presented in \citet[][table 11;]{Boselli2014a}. The dashed, solid, and dot-dashed lines represent $R_{\rm 21}= 0.6,\, 0.8,\,\text{and}\,1$ respectively. For VERTICO, the average $R_{21}$ is $0.77 \pm 0.05$ while the median uncertainty on the CO($2-1$) line flux is 16 Jy~km~s$^{-1}$ and thus within the marker size on the $x-$axis scale.}
	\label{fig:Sco_B14_comparison}
\end{figure}

Figures \ref{fig:vertico_hrs_r21} and \ref{fig:Sco_B14_comparison} compare the CO($1-0$) and CO($2-1$) line luminosity ratio ($R_{21}$) and integrated flux densities for the 35 VERTICO galaxies that have CO($1-0$) data compiled and presented by \citet[][blue solid histogram]{Boselli2014a} for the HRS. Establishing $R_{21}$ across the sample is important for deriving molecular gas masses (e.g., Equation \ref{eq:Mmol}) and interpreting our results in the context of other studies. For this comparison, \citet{Boselli2014a} fluxes are converted from $T_R^*$ (observed antenna temperature corrected for atmospheric attenuation, radiative loss, and forward and rearward scattering and spillover efficiency) to $T_A^*$ (observed antenna temperature corrected for atmospheric attenuation, radiative loss, and rearward scattering and spillover efficiency) temperature scales via the expression
\begin{equation}
	T_A^*  =  \eta_{fss} \, T_R^* \,,
\end{equation}
where $\eta_{fss}$ is the forward scattering and spillover efficiency of the telescope \citep{Kutner1981}. Following \citet{Boselli2014a}, we adopt $\eta_{fss}  =  0.68$ for the National Radio Astronomy Observatory Kitt Peak 12-m telescope.

Figure \ref{fig:vertico_hrs_r21} shows the distribution of observed $R_{21} = $ CO($2-1$)/CO($1-0$) line luminosity ratios for the VERTICO-HRS sample (blue histogram).  We also show the observed $R_{21}$ ratio for the 25 xCOLD GASS galaxies with APEX CO($2-1$) and high-quality IRAM-30m CO($1-0$) detections as presented in \citet[][WCO\_FLAG and FLAG\_APEX = 1 in their catalog]{Saintonge2017}. We measure $\bar{R}_{21} = 0.77 \pm 0.05$ in VERTICO galaxies with a standard deviation of $\sigma = 0.3$, in agreement with the value of $\bar{R}_{21} = 0.79 \pm 0.03$ reported by \citet{Saintonge2017} and measured from their published data release, $\bar{R}_{21} = 0.81 \pm 0.07$, where their assumed metallicity-dependent $\alpha_{\rm CO} > 1 ~ {\rm M_\odot ~ pc^{-2} ~ (K~km~s^{-1})^{-1}}$. The minor difference between the reported and measured xCOLD GASS $R_{21}$ values is due to the exclusion of galaxies based on the quality of the APEX data in the published value (Saintonge, private communication). Based on our measured value and its agreement with the literature, we adopt a constant value of $R_{21} = 0.8$ throughout this work. We note that this is comparable to but slightly higher than reported values in other works \citep[][]{Leroy2013,denBrok2021,Yajima2021}. Further characterization of $R_{21}$ and its variation across the VERTICO sample will be the subject of future work. 

Figure \ref{fig:Sco_B14_comparison} plots the relation between the CO($1-0$) and CO($2-1$) integrated flux densities for the 35 VERTICO-HRS galaxies. For illustration, $R_{21} = 1,\, 0.8$, and $0.5$, are denoted by the dot-dashed, solid, and dashed lines respectively. The outlier with the lowest $R_{21}$ (above the main correlation) is NGC4567 which is in very close proximity to NGC4568. The original CO($1-0$) data for this galaxy are drawn from Five College Radio Astronomy Observatory observations ($45\arcsec$ beam size) published by \citet{Chung2009a}. Determining whether physical or observational effects (e.g., source confusion) are responsible for NGC4567's low $R_{21}$ value is beyond the scope of this paper.

\subsection{Global Molecular Gas Masses}\label{sec:GasMasses}
We convert CO luminosity to molecular gas mass, including the contribution from heavy elements, in units of solar mass using the relation
\begin{equation} \label{eq:Mmol}
	M_{\rm mol} = \frac{\alpha_{\rm CO}}{R_{21}} \, L_{\rm CO} \,,
\end{equation}
where $\alpha_{\rm CO} = 4.35 ~ {\rm M_\odot ~ pc^{-2} ~ (K~km~s^{-1})^{-1}}$, the molecular gas mass-to-CO$(1-0)$ luminosity ratio calculated for the Milky Way disk by \citet{Bolatto2013}, and ${\rm R_{21}} = 0.8$. Our $\alpha_{{\rm CO}}$ corresponds to a CO($1-0$)-to-\Htwo conversion factor of $X_{\rm CO} = 2 \times 10^{20}$ cm$^{-2}$ (K~km~s$^{-1}$)$^{-1}$ that is consistent with other surveys of CO emission in nearby galaxies (e.g., HERACLES -- \citealt{Leroy2009};  JCMT-NGLS -- \citealt{Wilson2009}; xCOLD GASS -- \citealt{Saintonge2011b,Saintonge2017}; EDGE-CALIFA -- \citealt{Bolatto2017}).

All molecular gas masses quoted in this paper include the 36\% contribution of helium \citep{Kennicutt2012,Bolatto2013}. The true value of $\alpha_{{\rm CO}}$ can be up to a factor 5 lower in regions of increased average gas volume density such as in mergers and galaxy centers as has been shown in observations \citep[e.g.,][]{leroy2011,smith12,sandstrom2013} and demonstrated with modeling \citep[e.g.,][]{shetty11,narayanan11, narayanan12,olsen16}. We leave a more detailed modeling and application of $\alpha_{{\rm CO}}$ to future work.

\begin{figure*}
	\centering
	\includegraphics{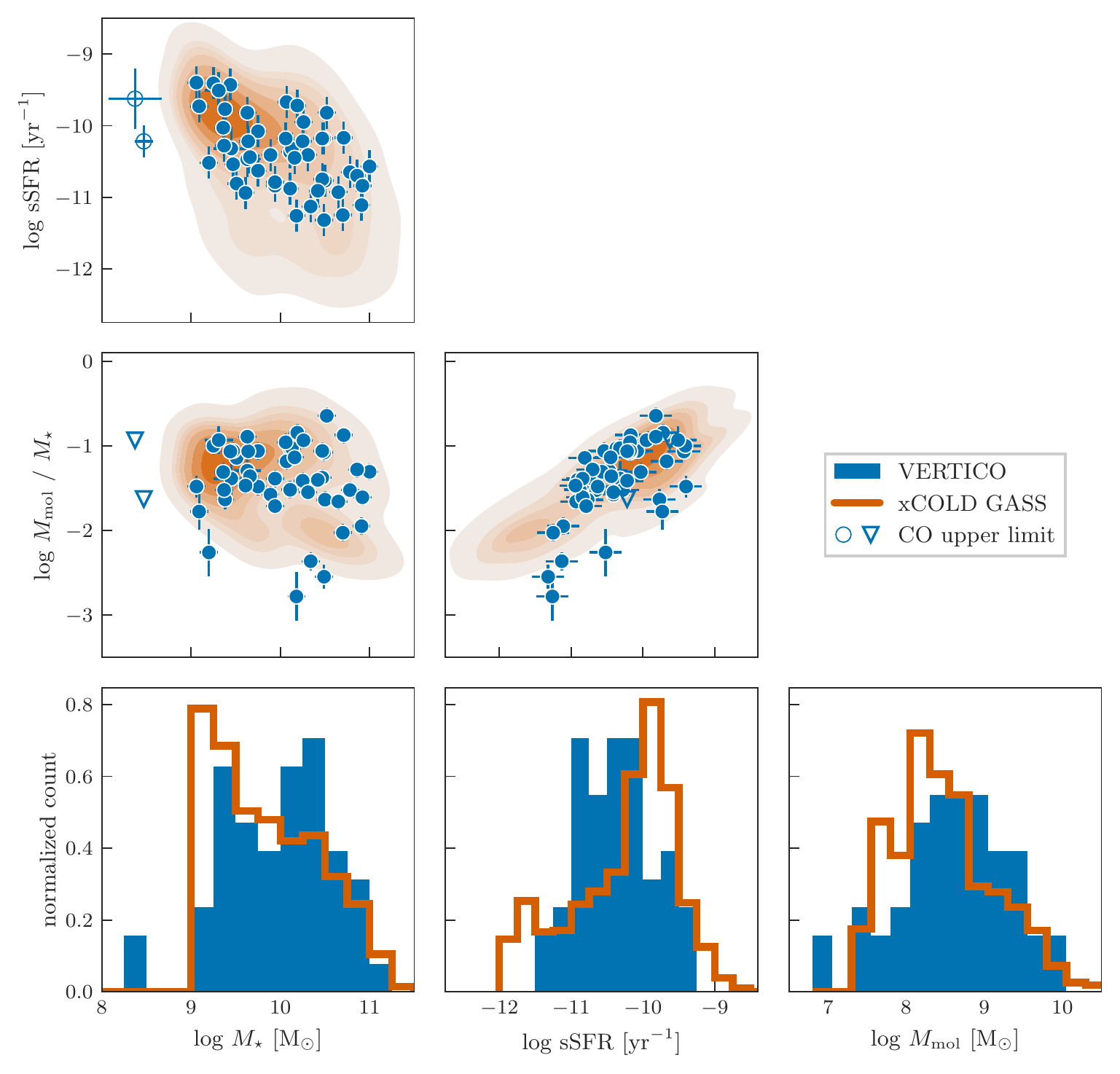}
	\caption{Global properties of the VERTICO sample (blue points and histograms) compared to the weighted xCOLD GASS catalog published by \citet[][orange distributions]{Saintonge2017}. We include the 528 xCOLD GASS galaxies that have their recommended (`best') SFR estimate. We adjust all molecular gas estimates to our constant conversion factor and the 199 CO non-detections in xCOLD GASS are included as $3\sigma$ upper limits. Left to right starting at the upper left; sSFR vs. stellar mass, molecular gas fraction vs. stellar mass, molecular gas fraction vs. sSFR, the normalized stellar mass, sSFR, and molecular gas mass distributions. The three histogram panels share the same $y$-axis range. The two galaxies that are undetected in CO are shown with open markers and inverted triangles denote their upper limits.}
	\label{fig:mass_sfr_mmol_corner}
\end{figure*}

Figure \ref{fig:mass_sfr_mmol_corner} compares the stellar mass, sSFR, and molecular gas mass properties of VERTICO galaxies (blue) with the volume-limited xCOLD GASS sample \citep[orange;][]{Saintonge2011b, Saintonge2017}\footnote{\url{http://www.star.ucl.ac.uk/xCOLD GASS/}}. As with xGASS, we apply the recommended weights published by \citet{Saintonge2017} to the xCOLD GASS data to achieve a volume-limited sample for $M_\star  \geq 10^9 ~ \text{M}_\odot$. We adjust the published xCOLD GASS molecular gas estimates to our assumed $\alpha_{\rm CO}$ in Equation \ref{eq:Mmol}. The upper left panel compares the distribution of xCOLD GASS and VERTICO in the stellar mass--sSFR plane, again highlighting VERTICO's selection of star-forming and quenching galaxies, rather than quiescent galaxies ({\em cf.} Figure \ref{fig:mass_sfr_mhi_corner}). The left and right panels in the middle row show molecular gas fraction ($M_{\rm mol} / M_\star$) as a function of stellar mass and sSFR, respectively. While there are a small number of VERTICO galaxies with low molecular gas fractions, the majority of the sample is either normal or rich in molecular gas at fixed stellar mass and sSFR. The lower panels show the volume-limited stellar mass, sSFR, and molecular gas mass distributions. Interestingly, there is a significant fraction of xCOLD GASS galaxies that are more star-forming than VERTICO, however, this does not translate into an excess in the molecular gas mass distribution. Although more investigation is needed, this may at least be partially explained by the increased fraction of massive VERTICO galaxies in comparison to the xCOLD GASS sample (lower left panel).

\subsection{CO Radial Profiles} \label{sec:Radial Profiles}

\begin{figure*}
	\centering
	\includegraphics{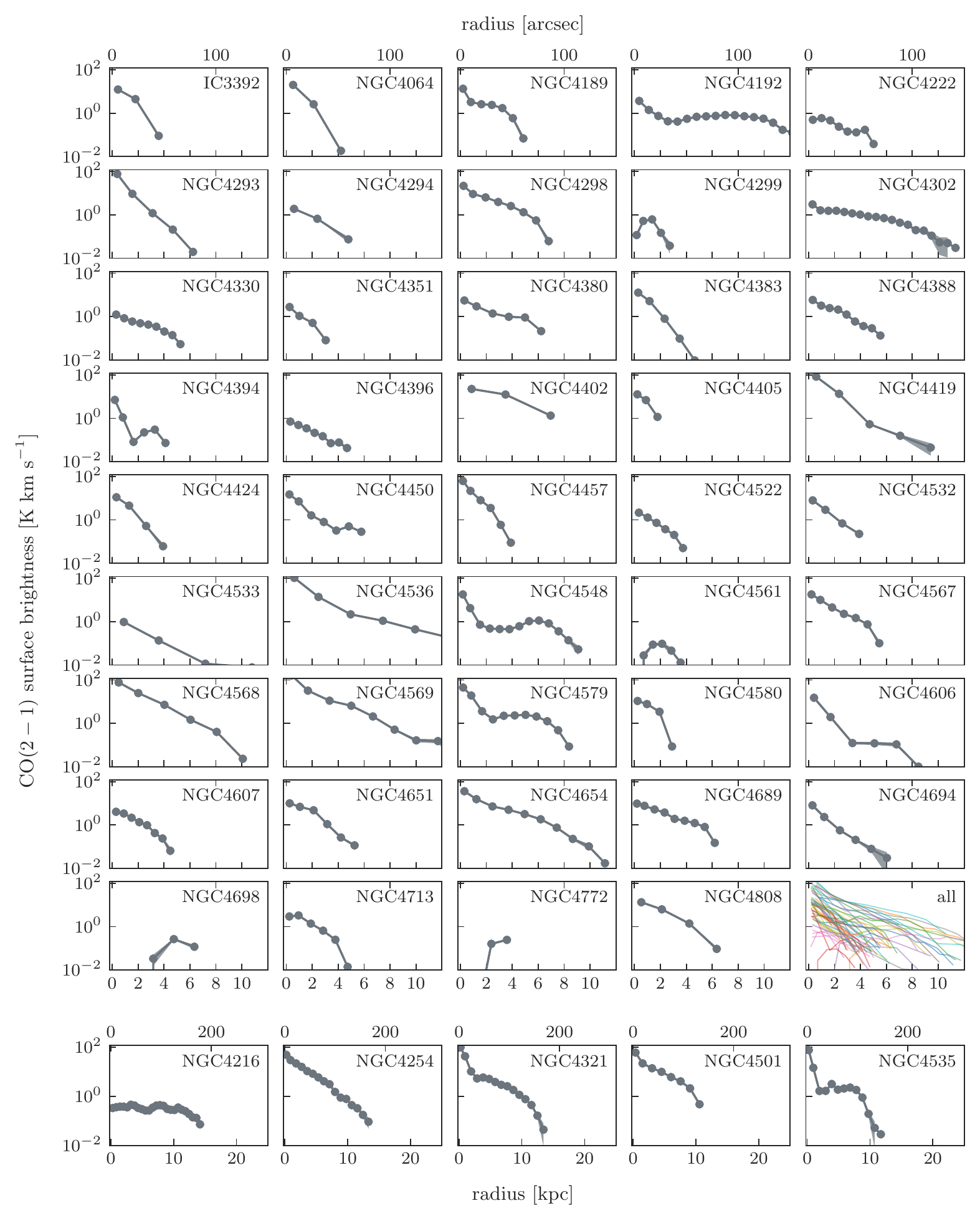}
	\caption{Azimuthally averaged surface brightness profiles as a function of galactocentric radius in kpc (bottom $x$-axis) and arcseconds (top). Radial profiles are denoted by the grey points while their uncertainties are illustrated by the shaded region. The five largest galaxies are shown in the last row with a different $x$-axis range. We show all the profiles in the last panel of the ninth row, highlighting the diversity of profile shapes. The surface brightness radial profiles are not corrected for inclination. Table \ref{tab:co_radii} contains the 50\% and 90\% flux radii ($r_{50,\, {\rm mol}}$ and $r_{90,\,{\rm mol}}$, respectively) and inclination-corrected iso-density radii ($r_{5,\,{\rm mol}}$) for each galaxy, where possible. }
	\label{fig:radialprofiles}
\end{figure*}

We measure azimuthally averaged CO intensity radial profiles in elliptical annuli overlaid on the integrated intensity maps described in Section \ref{sec:MomentMaps}. Annuli are centered on the optical position and aligned with the major-axis position angle. Eccentricity is derived from the optical inclination of the galaxy provided in Table \ref{tab:VERTICO-sample}. To ensure the independence of each measurement, the annulus width along the minor axis is set to the synthesized beam size. The corresponding radii are defined as the mean galacto-centric radius of pixels in each annuli. The average intensity is calculated as the summed emission divided by the total area. This means that non-detections (i.e., masked pixels) within each annulus are included in this calculation as zero intensity. This approach is conservative, especially where galaxies have asymmetric or fragmented gas disks, and results in radial profiles that can be considered lower limits on the true average CO surface brightness, particularly in the outskirts. Future work will explore the impact of this method on the measured radial profiles and iso-density radii.

For highly inclined galaxies this method does not work well as the axial ratio becomes too small causing highly eccentric annuli with the major axes extending beyond the galactic disk. Therefore, for galaxies with inclination $\geq80\degr$, we use a slice that is one beam thick aligned with the major axis. The integrated intensity as a function of radius is then calculated by averaging the emission in each slice.

Uncertainties on all the intensity profiles are calculated as,
\begin{equation}
	u_{\rm rp} = \sigma(I_{\rm CO, \, pix}) \; \sqrt{N_{\rm beam}},
\end{equation}
where $\sigma(I_{\rm CO, \, pix})$ is the standard deviation of integrated intensity in each pixel and $N_{\rm beam}$ is the number of beams within each annulus.

Figure \ref{fig:radialprofiles} shows the mean surface brightness radial profiles of 49 VERTICO galaxies (IC3418 and VCC1581 are non-detections). The profiles are denoted by gray points and their uncertainties are shown by the gray shaded regions. Radial profiles in surface brightness units are not corrected for inclination.

Although quantitative analyses of the molecular gas distribution in VERTICO galaxies will be the subject of upcoming work, here we comment briefly on the large range of gas disk morphologies exhibited by the CO surface brightness profiles. A significant number of  profiles do not decrease steadily as a function of radius, showing bumps in the CO distribution at larger radii (e.g., NGC4189, NGC4450, NGC4535, NGC4548, NGC4579, NGC4808) or, perhaps most interestingly, signs of truncation at the outer edge of the disk (e.g., NGC4064, NGC4299, NGC4402, NGC4457, NGC4532, NGC4535, NGC4580, NGC4607). This variation contrasts with the findings from studies of nearby field galaxies where the CO radial profiles are typically exponential with a comparable scale length as the stellar disk \citep{Regan2001, Helfer2003, Leroy2009, Schruba2011, Bigiel2012, Leroy2013, Bolatto2017} but is qualitatively closer to the observed surface brightness profiles of early-type galaxies, which do often show such enhancements and truncation \citep[e.g.,][]{Davis2013}. Given the observed diversity in the VERTICO galaxies' stellar morphologies and the efficacy of CO as a dynamical tracer, the variety of shapes seen in the inner regions of VERTICO radial profiles could be driven by stellar bars or other dynamical features such as bulges or arms. \citet{Chown2019} clearly show that barred spiral galaxies exhibit a large range of molecular gas radial profile shapes, molecular gas concentrations, and star formation histories in their inner regions. Furthermore, those authors find that the level of central star formation enhancement is correlated with gas concentration in barred galaxies. Related works show similar diversity in central molecular gas concentration \citep[e.g.,][]{Sheth2005}, star formation properties \citep[e.g.,][]{Lin2017,Lin2020}, and HI surface density \citep[e.g.,][]{Wang2014}. It is clear that the combination of gas morphology with the high angular resolution and sensitivity of these observations results in profiles that frequently depart from smoothly decreasing decline. We also note that we do not see a central hole in the CO surface brightness distribution for many galaxies, as is commonly found in other surveys \cite[e.g.,][]{Bigiel2012}. In contrast, the observed CO surface brightness distribution of VERTICO galaxies is closer to the typical shapes of H$\alpha$ emission in Virgo Cluster galaxies found by \citet{Koopmann2001}. In this work, the H$\alpha$ disks -- as a close tracer of star formation -- exhibit a similar range concentration and morphology to the CO, including systems that show elevated emission in the circumnuclear regions and truncation of star formation in the outer optical disk. A large range in concentration is also reported by studies of UV disk morphology in nearby galaxies \citep[e.g.,][]{MunozMateos2009}. Improved characterization and modeling of the CO radial profiles is required to understand the observed trends and differences between VERTICO and field galaxy samples.

\startlongtable
\begin{deluxetable}{lccc}
	\tablecaption{Molecular gas disk radii estimates.}
	\tablehead{\colhead{Galaxy} & \colhead{$r_{50,\, {\rm mol}}$} & \colhead{$r_{90,\, {\rm mol}}$} & \colhead{$r_{5,\, {\rm mol}}$}\\ \colhead{ } & \colhead{$\mathrm{kpc}$} & \colhead{$\mathrm{kpc}$} & \colhead{$\mathrm{kpc}$}}
	\startdata
IC3392 & 1.16 & 2.32 & 2.41 \\
NGC4064 & 1.06 & 2.18 & 2.08 \\
NGC4189 & 1.27 & 3.23 & 3.57 \\
NGC4192 & 5.10 & 9.70 &  \\
NGC4216 & 6.51 & 11.58 &  \\
NGC4222 & 1.55 & 4.01 &  \\
NGC4254 & 1.94 & 5.72 & 8.33 \\
NGC4293 & 0.79 & 1.86 & 2.60 \\
NGC4294 & 1.55 & 3.20 &  \\
NGC4298 & 1.45 & 3.94 & 4.69 \\
NGC4299 & 1.15 & 1.85 &  \\
NGC4302 & 2.86 & 6.74 &  \\
NGC4321 & 1.03 & 5.95 & 9.74 \\
NGC4330 & 1.65 & 3.76 &  \\
NGC4351 & 0.88 & 2.12 & 0.80 \\
NGC4380 & 1.63 & 4.77 & 1.95 \\
NGC4383 & 0.84 & 1.84 & 1.94 \\
NGC4388 & 1.46 & 3.30 & 0.42 \\
NGC4394 & 0.50 & 2.85 & 0.82 \\
NGC4396 & 1.30 & 3.18 &  \\
NGC4402 & 2.69 & 5.21 & 5.63 \\
NGC4405 & 0.69 & 1.35 & 1.74 \\
NGC4419 & 1.24 & 2.70 & 3.69 \\
NGC4424 & 0.92 & 1.94 & 1.97 \\
NGC4450 & 0.85 & 2.78 & 2.04 \\
NGC4457 & 0.65 & 1.84 & 2.87 \\
NGC4501 & 2.20 & 6.54 & 8.97 \\
NGC4522 & 1.08 & 2.46 &  \\
NGC4532 & 0.95 & 2.30 & 1.65 \\
NGC4533 & 1.90 & 4.58 &  \\
NGC4535 & 0.74 & 6.39 & 8.29 \\
NGC4536 & 1.33 & 3.62 & 4.31 \\
NGC4548 & 0.78 & 6.31 & 1.29 \\
NGC4561 & 1.91 & 2.87 &  \\
NGC4567 & 1.06 & 3.16 & 3.71 \\
NGC4568 & 1.47 & 3.84 & 5.31 \\
NGC4569 & 1.12 & 4.12 & 6.42 \\
NGC4579 & 0.85 & 5.15 & 6.75 \\
NGC4580 & 0.98 & 2.01 & 2.50 \\
NGC4606 & 0.87 & 2.13 & 1.53 \\
NGC4607 & 1.19 & 2.75 &  \\
NGC4651 & 1.30 & 2.67 & 2.96 \\
NGC4654 & 1.52 & 4.97 & 6.15 \\
NGC4689 & 1.47 & 4.10 & 4.76 \\
NGC4694 & 0.85 & 2.41 & 1.38 \\
NGC4698 & 4.91 & 5.94 &  \\
NGC4713 & 1.17 & 2.72 & 1.98 \\
NGC4772 & 2.84 & 3.45 &  \\
NGC4808 & 1.65 & 3.59 & 3.32
	\enddata
	\tablecomments{Columns are (1) galaxy identifier; (2) radius containing 
		50\% of the global CO($2-1$) flux; (3) radius containing 90\% of the global CO($2-1$) flux; 
		(4) inclination-corrected molecular gas iso-density radius at 
		$\Sigma_{\rm mol} (r) = 5 ~ {\rm M}_\odot ~ {\rm pc}^{-2}$; At the distance of Virgo 
		$1\arcsec = 80~\text{pc}$. 
	This table is published in its entirety in machine-readable format.}
	\label{tab:co_radii}
\end{deluxetable}

\section{The Molecular Gas Size--Mass Relation} \label{sec:size-mass relation}

\begin{figure}
	\centering
	\includegraphics{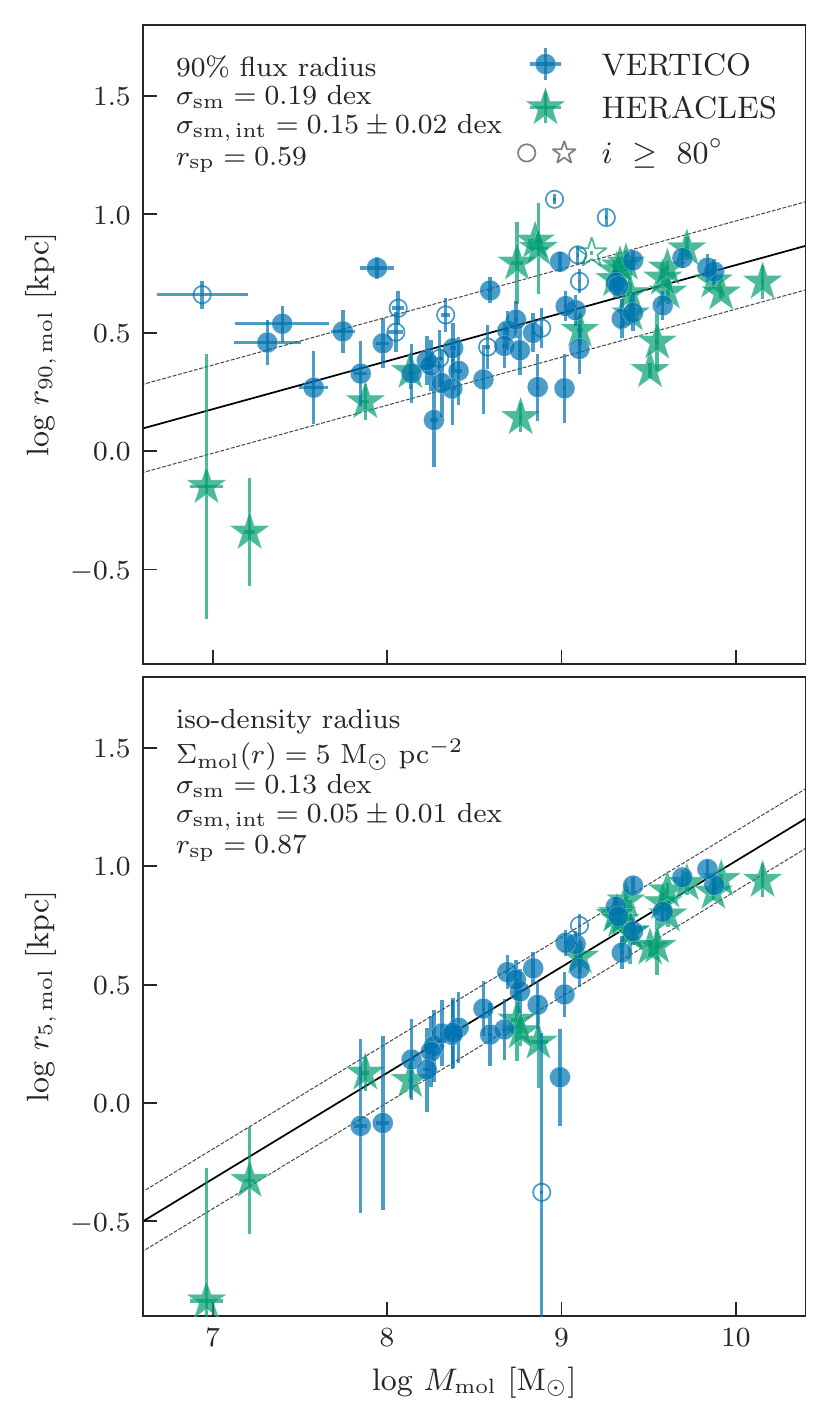}
	\caption{The molecular gas size--mass relation for 90\% flux radii ($r_{90,\, {\rm mol}}$, upper panel) and inclination-corrected iso-density radii ($r_{5,\,{\rm mol}}$, lower panel). The former enclose 90\% of the molecular gas flux (and thus mass for a fixed $\alpha_{\rm CO}$), while the latter are defined as the radius where $\Sigma_{\rm mol} (r) = 5 ~ {\rm M}_\odot ~ {\rm pc}^{-2}$. Galaxies from the VERTICO and HERACLES surveys are denoted by blue and green points, respectively. The rms scatter ($\sigma_{\rm sm}$), intrinsic scatter accounting for errors ($\sigma_{\rm sm, \, int}$), and Spearman rank coefficient ($r_{sp}$) for the combined VERTICO-HERACLES relation are given in the upper left of each panel. The best-fitting relation for all galaxies is shown by the black solid line with $\pm 1 \sigma_{\rm sm}$ denoted by the black dashed lines (see text for fit parameters). Galaxies with inclination $\geq80\degr$ are denoted by open markers and not included in the fits 
to the data. }
	\label{fig:size_mass_relation}
\end{figure}

The richness and quality of the VERTICO data enables many research avenues. As an initial demonstration of the survey's scientific power, we now focus on establishing the relationship between the size and mass of molecular gas disks in the VERTICO sample, and consider the effect of environment on this relation. The size--mass relation has been explored in detail for stellar and \HI distributions in both theory and observation \citep[e.g.,][and references therein]{Wang2016, Stevens2019, Trujillo2020, SanchezAlmeida2020}. \citet{Stevens2019} and \citet{SanchezAlmeida2020} respectively demonstrate for \HI and stellar content that the tight correlation between galaxy radius---defined at a fixed surface density---and global mass is mathematically inevitable, given the limited range of physically plausible surface density profile shapes these components can have (e.g., saturated exponentials for \HInospace, S\'{e}rsic stellar profiles). Indeed, \citet{Stevens2019} find that only a drastic change in \HI disk morphology is able to cause significant deviation from the \HI size--mass relation. Hereafter, the term `size--mass relation' refers to the molecular gas size--mass relation unless explicitly stated otherwise.

We use galaxies from the Heterodyne Receiver Array CO Line Extragalactic Survey \citep[HERACLES;][]{Leroy2009} as a convenient nearby field control sample ( $2 < D \, / \, {\rm Mpc} < 25$), which spans a comparable range in galaxy stellar mass and sSFR ($10^{8.5} < {\rm M_\star \, / \, M_\odot} < 10^{11}$ and $10^{-11.5} < {\rm sSFR \, / \, yr^{-1}} < 10^{-9.2}$, respectively). As with VERTICO, the stellar mass and star formation rate estimates are drawn from the $z0$MGS database \citep{Leroy2019}. There are 48 HERACLES galaxies with public CO($2-1$) data cubes that have an angular resolution of $13\arcsec$ in 5~km~s$^{-1}$ wide channels\footnote{\url{https://www.iram-institute.org/EN/content-page-242-7-158-240-242-0.html}}. Excluding 18 non-detections and five galaxies that overlap with the VERTICO sample (NGC4579, NGC4569, NGC4536, NGC4321, NGC4254) leaves 25 galaxies for this comparison. Starting from the public cubes at their native $13\arcsec$ angular resolution and 10~km~s$^{-1}$ channel width, we apply the same methodology used for the VERTICO data and described in Sections \ref{sec:COFluxes} and \ref{sec:GasMasses} to derive the molecular gas masses. The VERTICO masses and radii used in this section are presented in Table \ref{tab:co_props} and Table \ref{tab:co_radii}, respectively.

Figure \ref{fig:size_mass_relation} shows the combined VERTICO and HERACLES molecular gas size--mass relation for two measurements of galaxy size as measured from the radial profiles. The first, shown in the right panel, is the radius enclosing 90\% of the CO flux, $r_{90, \, {\rm mol}}$. From here on, we use the term `flux-percentage radius' to refer generally to radii calculated at a given percentage of flux. Since, in this work, we simply scale the CO emission by $\alpha_{\rm CO}$ to get molecular gas surface density, this 90\%-light radius is also the 90\%-mass radius. The second is the iso-density radius, $r_{5, \, {\rm mol}}$ shown in the lower panel, defined at fixed molecular gas surface density, $\Sigma_{\rm mol}(r) = 5 ~{\rm M}_\odot ~ {\rm pc}^{-2}$, and calculated from the inclination-corrected surface density radial profiles. The uncertainties shown are the observation beam size listed in Table \ref{tab:co_props}, converted into physical units. For highly inclined galaxies ($i\geq80\degr$), we assume an inclination of $80\degr$ when calculating the inclination correction and exclude these from the fits described below. Unlike the flux-percentage radii, estimates of iso-density radii are not possible for every galaxy so only the 62 galaxies (38 VERTICO, 24 HERACLES) where surface density profiles reach $\Sigma_{\rm mol} (r) = 5 ~{\rm M}_\odot ~ {\rm pc}^{-2}$ are shown in the lower panel. The severity of the inclination correction for the highly inclined galaxies ($cos(80\degr)=0.17$) means an $r_{5, \, {\rm mol}}$ measurement is not available.

Throughout this analysis, VERTICO and HERACLES galaxies are denoted by blue circles and green stars, respectively. We use the {\sc LtsFit Python} package to fit each relation. The method, described in detail by \citet[][Section 3.2]{Cappellari2013}, accounts for measurement uncertainties to determine the best-fit parameters and scatter. We turn off outlier detection for this work. Using this approach, we derive the best fit to the 90\% flux-percentage radius size--mass relation (black solid lines) as, 
\begin{equation}
	\log \left( \frac{r_{90, \, {\rm mol}}}{\rm kpc} \right) = 0.54 + 0.20 \left[\log \left( \frac{\rm M_{mol}}{\rm M_\odot} \right)  - 8.83\right] \,.
\end{equation}
Similarly, we fit the following relationship to the iso-density size--mass relation, 
\begin{equation}
	\log \left( \frac{r_{5, \, {\rm mol}}}{\rm kpc}\right) = 0.54 + 0.45 \left[\log \left( \frac{\rm M_{mol}}{\rm M_\odot} \right)  - 8.94\right]\,.
\end{equation}
Uncertainties on the fit parameters are provided in Table \ref{tab:mass-size_bestfit}. The rms scatter, $\sigma_{\rm sm}$, intrinsic scatter accounting for errors, $\sigma_{\rm sm, \, int}$, and Spearman rank coefficent, $r_{sp}$, are printed in the upper left of each panel. The former is denoted by the black dashed lines.

We find that the flux-percentage size--mass relation has significantly larger rms and intrinsic scatter ($\sigma_{\rm sm} =0.19$ dex and $\sigma_{\rm sm, \, int} =0.16 \pm 0.02$ dex) than the iso-density size--mass relation ($\sigma_{\rm sm} =0.13$ dex and $\sigma_{\rm sm, \, int} =0.05 \pm 0.01$ dex). This result agrees both qualitatively and quantitatively with studies of stellar and \HI disks that find iso-density radii reduce the scatter to $\lesssim0.1$ dex in the respective size--mass relations compared with flux-percentage radii \citep[][]{Saintonge2011a, Cortese2012, Wang2016, Stevens2019, Trujillo2020, SanchezAlmeida2020}. The larger Spearman rank coefficient ($r_{sp} = 0.87$) highlights the stronger statistical connection between global molecular gas mass and iso-density radii rather than flux-percentage radii ($r_{sp} = 0.59$).

\begin{figure}
	\centering
	\includegraphics{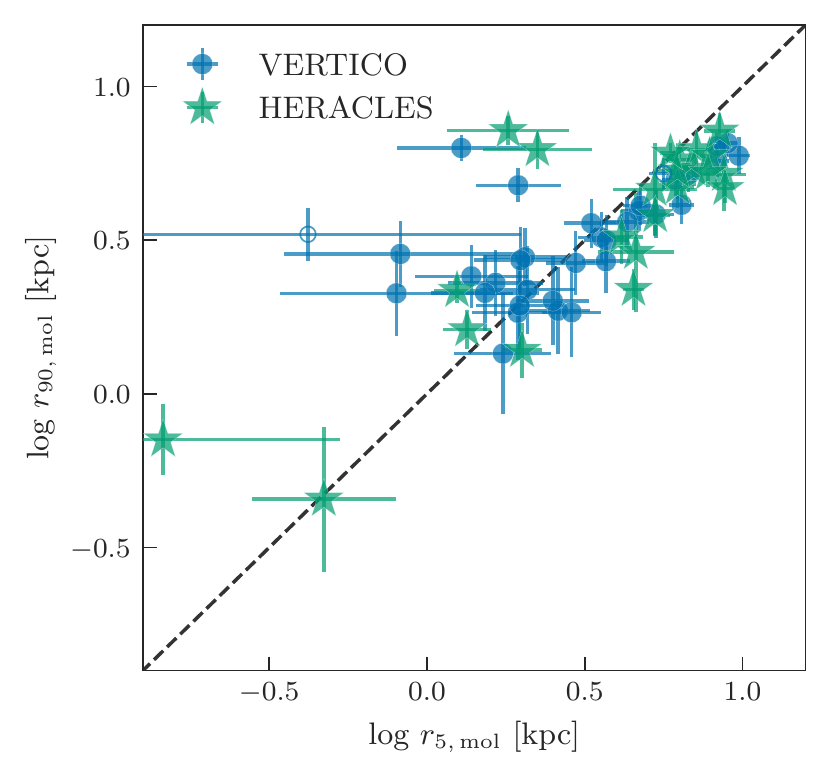}
	\caption{Iso-density radii ($r_{5, \, {\rm mol}}$) versus the 90\% flux radii ($r_{90,\, {\rm mol}}$) for VERTICO (blue) and HERACLES (green) galaxies. The iso-density radii are determined at $\Sigma (r_{5, \, {\rm mol}}) = 5 ~ {\rm M_\odot ~ pc}^{-2}$ and have been corrected for inclination. Galaxies to the bottom right tend to have CO distributions that are less concentrated than those in the upper left.}
	\label{fig:reff_riso}
\end{figure}

Figure \ref{fig:reff_riso} shows the comparison between these two size estimates for VERTICO and HERACLES galaxies. At fixed $r_{5, \, {\rm mol}}$, both VERTICO and HERACLES galaxies exhibit large range in $r_{90, \, {\rm mol}}$, and, while most galaxies tend to have consistently larger $r_{5, \, {\rm mol}}$ than $r_{90, \, {\rm mol}}$, this not the case for approximately 25\% of the combined sample. The scatter in $r_{90, \, {\rm mol}}$ at fixed $r_{5, \, {\rm mol}}$ reflects the different degrees of concentration in CO disks across both samples.

\begin{deluxetable*}{lccccccc}
	\tablecaption{Molecular gas size--mass relation best fit parameters for the VERTICO, HERACLES, and combined samples.}
	\tablehead{\colhead{Sample} & \colhead{$y$} & \colhead{$a$} & \colhead{$b$} & \colhead{pivot} & \colhead{$\sigma_{\rm sm}$} & \colhead{$\sigma_{\rm sm, \, int}$} & \colhead{$N$}}
	\startdata
	Combined &  & 0.54$\pm$0.02 & 0.20$\pm$0.03 & 8.83 & 0.19 & 0.16$\pm$0.02 & 62 \\
	VERTICO & $\log \left( \frac{r_{90, \, {\rm mol}}}{\rm kpc} \right) $ & 0.54$\pm$0.02 & 0.15$\pm$0.04 & 8.66 & 0.16 & 0.12$\pm$0.03 & 38 \\
    HERACLES &  & 0.61$\pm$0.03 & 0.27$\pm$0.05 & 9.18 & 0.19 & 0.16$\pm$0.03 & 29 \\
    Combined &  & 0.55$\pm$0.01 & 0.45$\pm$0.02 & 8.94 & 0.13 & 0.05$\pm$0.01 & 56 \\
    VERTICO & $\log \left( \frac{r_{5, \, {\rm mol}}}{\rm kpc} \right) $ & 0.52$\pm$0.02 & 0.46$\pm$0.03 & 8.83 & 0.12 & 0.04$\pm$0.02 & 33 \\
    HERACLES &  & 0.64$\pm$0.02 & 0.45$\pm$0.03 & 9.20 & 0.13 & 0.05$\pm$0.02 & 28
	\enddata
	\tablecomments{Columns are (1) the sample used to fit the molecular gas size--mass relation; (2) the definition of radius used; 
		(3) -- (5) the best fit parameters for $y = a + b (x - {\rm pivot})$ where $x$ is log $M_\star / {\rm M}_\odot$; 
		(6) the scatter about the fit, $\sigma_{\rm sm} = \sigma({\rm fit - radius})$; (7) intrinsic scatter around the linear relation 
		accounting for uncertainties \citep[$\sigma_{\rm sm, \, int}$ is called $\epsilon_y$ in equation 6 of][]{Cappellari2013}; and (8) the number
	of galaxies used in each fit.}
	\label{tab:mass-size_bestfit}
\end{deluxetable*}

We check the consistency of the size--mass relations between the VERTICO and HERACLES samples and, although we do not plot the fits, provide the best-fit parameters in Table \ref{tab:mass-size_bestfit}. Where only the HERACLES sample is fit, we include the galaxies that overlap with VERTICO with their radii and masses measured from the public HERACLES data. This comparison demonstrates that the flux-percentage size--mass relation varies considerably between the VERTICO and HERACLES samples. On the other hand, the slope and scatter of the iso-density size--mass relation are remarkably consistent for both VERTICO and HERACLES. Given that many of the VERTICO gas disks are clearly perturbed by environment, this suggest that the observed iso-density size--mass relation does not have an environmental dependence, a result that is consistent with the findings of the \HI size--mass relation studies \citep{Wang2016,Stevens2019}. Indeed, this invariance implies that external mechanisms act to suppress or remove galaxy gas content, rather than simply rearranging the distribution of gas within the disk. If environmental processes were to alter how gas is distributed throughout the galaxy without depleting the total amount of gas within the disk, one would expect such effects to drive galaxies away from the size--mass relation (above or below) as gas mass is conserved and iso-density radius is altered. It is reasonable to expect this scenario to result in a correlation of the form and scatter of size--mass relation with environment which we do not see in the comparison between VERTICO and HERACLES. This paints a picture whereby, for VERTICO at least, any perturbation of the molecular gas distribution is also coupled with a change in total mass of the gas disk. Although we caution that a much more detailed exploration is required, the consistency and tightness in the size--mass relation defined using iso-density radii highlights the interesting possibility of using molecular gas mass as a predictor of gas disk size.

\begin{figure*}
	\centering
	\includegraphics{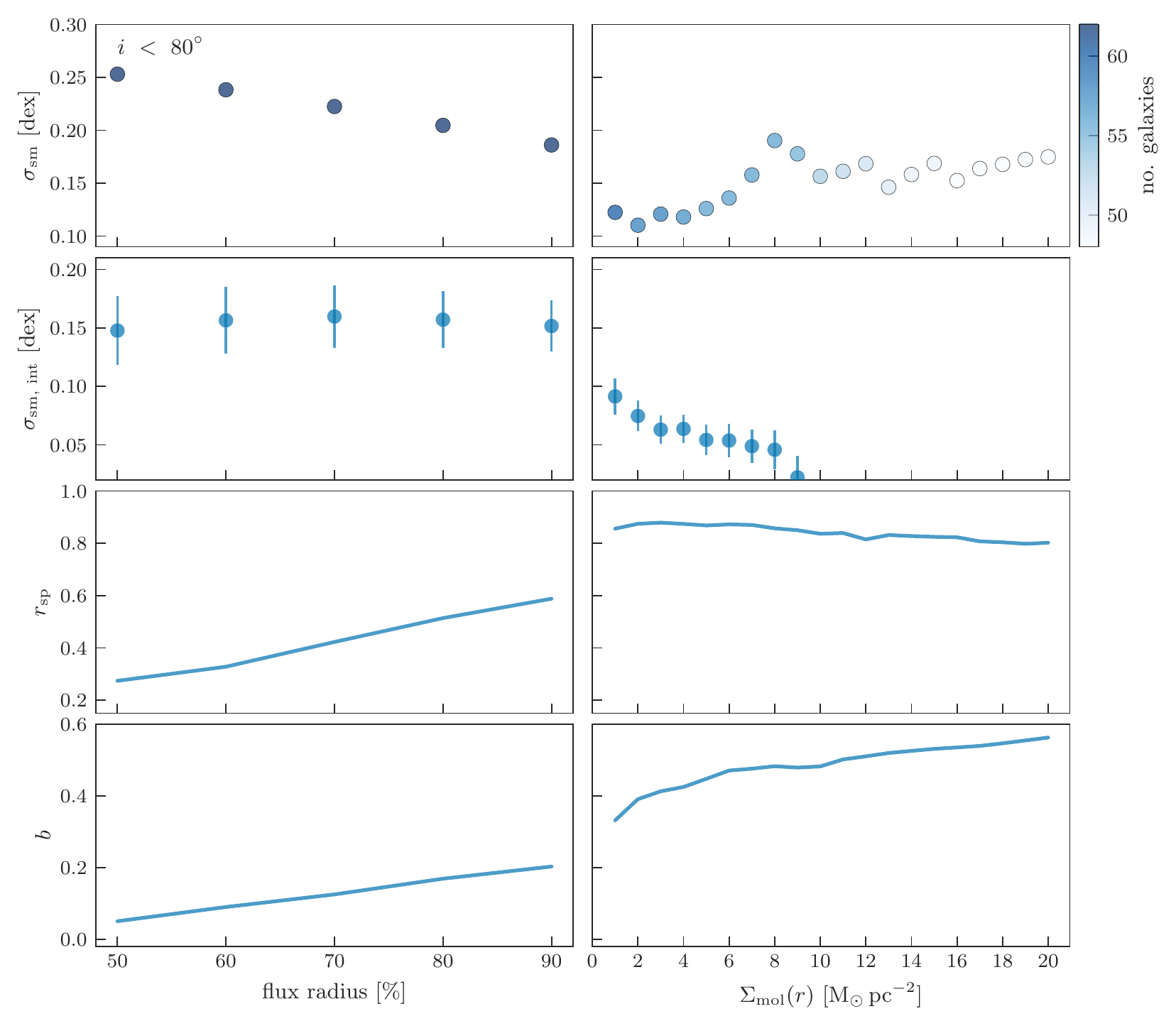}
	\caption{Scatter (top row), intrinsic scatter accounting for uncertainties (second row), Spearman rank coefficient (third row), and slope (bottom row) of the molecular gas size--mass relation as a function of flux-percentage radius percentage (left; i.e., radii are defined as enclosing a particular percentage of CO flux) and iso-density radius definition [right; i.e., radii are defined at mean molecular gas surface density, $\Sigma_{\rm mol} (r)$]. As with Figure \ref{fig:size_mass_relation}, we exclude galaxies with inclinations $\geq80\degr$ from the fits.}
	\label{fig:size--mass_scatter_rsp_slope}
\end{figure*}

We now investigate the definitions of flux-percentage radius and iso-density radius that produce the least scatter and strongest correlation in the two size--mass relations. To this end, we repeat the analysis above for flux-percentage radii enclosing 50\%, 60\%, 70\%, 80\%, and 90\% of the light, and iso-density radii in the range $\Sigma_{\rm mol} (r) = 1 - 20~{\rm M_\odot ~ pc^{-2}}$, in increments of $1 ~ {\rm M_\odot ~ pc^{-2}}$. The left- and right-hand columns in Figure \ref{fig:size--mass_scatter_rsp_slope} show the resulting rms scatter ($\sigma_{\rm sm} = \sigma({\rm fit - radius})$, top row), intrinsic scatter ($\sigma_{\rm sm,\, int}$, second row), Spearman rank coefficient (third row), and slope (bottom row) of the VERTICO-HERACLES size--mass relation versus the flux percentage at which radii are defined and iso-density $\Sigma_{\rm mol} (r)$, respectively. The colors of the points in the upper panels denote the number of galaxies in the given size--mass relation. 

The scatter in the flux-percentage radius size--mass relation decreases gradually as the percentage of flux enclosed by the radius increases. By contrast, the iso-density size--mass relation scatter increases with $\Sigma_{\rm mol} (r)$ above $\Sigma_{\rm mol} (r) \sim 10 ~ {\rm M_\odot ~ pc^{-2}}$. We find the rms scatter is minimized at $\Sigma_{\rm mol} (r) \sim 5 ~ {\rm M_\odot ~ pc^{-2}}$ while the intrinsic scatter is lowest in the range $\Sigma_{\rm mol} (r) \sim 5 - 10 ~ {\rm M_\odot ~ pc^{-2}}$. Using iso-density radii results in a stronger correlation, as quantified by the Spearman rank coeffecient, than flux-percentage radii regardless of the exact definition. However, the Spearman rank coefficient of the flux-percentage radius size--mass relation increases with the percentage of enclosed flux. We also see a slight decrease in the correlation strength for iso-densities of $\Sigma_{\rm mol} (r) \gtrsim 10 ~ {\rm M_\odot ~ pc^{-2}}$. The flux-percentage radius size--mass relation steadily steepens as the enclosed percentage of flux increases and, while the iso-density relation is always steeper, there is some variation its slope as a function of $\Sigma_{\rm mol}(r)$. The radius definitions where scatter and Spearman rank coefficient are minimized and maximized, respectively, motivate the choices of $r_{90,\, {\rm mol}}$ and $\Sigma_{\rm mol} (r) \sim 5 ~ {\rm M_\odot ~ pc^{-2}}$ used in Figures \ref{fig:size_mass_relation} and \ref{fig:reff_riso}.

Following the logic outlined in \citet{Stevens2019} and \citet{SanchezAlmeida2020}, it is expected that the iso-density size--mass relation should be consistently tighter than the flux-percentage radius relation because, broadly speaking, the surface density profiles decrease with radius. This means that for any two gas disks with the same total mass, their radial surface density profiles must intersect at a particular radius. Choosing a $\Sigma_{\rm mol} (r)$ that is close to the surface density at this crossing radius necessarily reduces the scatter.

The natural next step is to consider why the scatter in the molecular gas iso-density relation falls to the levels found in the stellar and \HI size--mass relations \citep[$\sim 0.1$ dex; cf.][]{Wang2016,Trujillo2020} at approximately $\Sigma_{\rm mol} (r) \sim 5 ~ {\rm M_\odot ~ pc^{-2}}$, increasing both above and below this value. While a full exploration of this issue is beyond the scope of this work, we note that the \HI size--mass relation is both tight and consistent across samples because of the physical threshold in central \HI surface densities of galaxies, above which gas becomes molecular \citep{Stevens2019}.  No equivalent restriction on central surface density applies to molecular gas (or stellar densities). Not only is the ceiling for central molecular surface densities much higher than \HInospace, but Figure \ref{fig:radialprofiles} shows variable and non-monotonic surface brightness (and therefore surface density) profiles are common.  Na\"{i}vely, one would 
expect that the higher the threshold surface density used to define the iso-density radius, the more susceptible to variation in the surface density profiles the relationship is, thus increasing the scatter.

Molecular gas density profiles tend to be less well described by monotonically decreasing functions of radius than \HI or stellar profiles, particularly at high resolution and sensitivity (e.g., Figure \ref{fig:radialprofiles}), and exhibit a wide range in concentration (e.g., Figure \ref{fig:reff_riso}). Despite this, the small scatter in the measured gas disk size at fixed total mass suggests that galaxies tend to move along the size--mass relation, rather than deviating from it. Of course, this poses the altogether more interesting question, what is driving the observed diversity in molecular gas radial profile shapes? Addressing this will be the subject of future work aimed at establishing a physically motivated definition for the size of molecular gas disks. However, it interesting that the optimal iso-density value of $\Sigma_{\rm mol} (r) = 5 ~ {\rm M_\odot ~ pc^{-2}}$ is consistent with the total gas surface densities of $\sim 10 ~{\rm M_\odot ~ pc^{-2}}$ where gas disks are theoretically predicted to transition from being \HInospace- to H$_2$-dominated with an atomic-to-molecular ratio of $\sim 0.5$ at solar metallicity \citep{Krumholz2009}. 

This work assumes a constant CO conversion factor across the sample. While this is likely a reasonable approximation given the mild gradients in $\alpha_{\rm CO}$ as a function of radius observed in most galaxy disks \citep{sandstrom2013}, it is important to consider how the prevalence of lower $\alpha_{\rm CO}$ values in the inner parts of galaxies as well as other variations in the conversion factor (e.g., metallicity-driven) could impact the form of our observed iso-density size--mass relation \citep{Wolfire2010, sandstrom2013, Heyer2015, Accurso2017}. Since $\alpha_{\rm CO}$ is accounted for in both axes, a global variation in the conversion factor -- without a change in shape between the CO and molecular gas surface density profiles -- would simply shift galaxies along the relation. However, radially varying $\alpha_{\rm CO}$, as seen in some galaxies by \citet{sandstrom2013} is more difficult to properly understand without further work. That said, an interesting quality of the iso-density size--mass relation is its insensitivity to the significant variations in the gas surface density profiles seen in the VERTICO sample. In other words, despite the large differences in CO distribution throughout the VERTICO sample, almost all galaxies fall on the combined sample size--mass relation. This suggests that changes in the conversion factor as a function of radius would have to dramatically alter the gas surface density profiles for the effect to be noticeable in the size--mass relation.

Lastly, it is notable that the optimal $\Sigma_{\rm mol} (r)$ should be in such good agreement with the predicted transition density, given the distance, $\alpha_{\rm CO}$, and flux calibration uncertainties on our observations, in addition to less well quantified physical effects such as metallicity variations and environment.

\section{Summary} \label{sec:Summary}

The VERTICO survey has mapped CO($2-1$) in 51 Virgo Cluster galaxies on sub-kpc scales. The primary motivation of this project is to understand the physical mechanisms that drive galaxy evolution in dense environments, and provide a diverse, homogeneous legacy dataset for studying galaxy evolution in our closest galaxy cluster.

The 36 targets observed in ALMA Cycle 7 are combined with archival CO($2-1$) data for 15 Virgo Cluster spirals to make the final VERTICO sample of 51 galaxies. Our final data cubes have a resolving beam of $\sim7-10\arcsec$, corresponding to $\sim0.6-0.8~\text{kpc}$ at the distance of Virgo, and $10~{\rm km~s^{-1}}$ velocity resolution. We provide global CO line luminosities and convert these into total molecular gas masses. We calculate R$_{21} = 0.8$ for the 35 galaxies with existing CO($1-0$) data, in general agreement with other surveys of nearby galaxies. 

We present the integrated intensity, velocity field, observed line width, and peak temperature maps for each galaxy. VERTICO's sensitivity and depth ensures that these maps reveal the imprint of stellar structure (e.g., spiral arms, bars, bulges) and environmental processes (e.g., warps, tails, depletion) in the gas morphology and kinematics in great detail. We measure integrated CO intensity radial profiles, which show a large range in gas disk morphologies across the VERTICO sample. A significant number of the profiles do not decrease steadily as a function of radius, showing bumps in the CO distribution at larger radii or signs of truncation at the outer edge of the disk.

We investigate the scaling relation between the size and mass of the molecular gas distribution in VERTICO galaxies. This is compared to the same relation for the HERACLES survey of field galaxies. We find the iso-density size--mass relation has less scatter and a stronger correlation than the flux-percentage radius size--mass relation. In agreement with studies of \HI disks, we suggest that the observed consistency of the iso-density size--mass relation between field and cluster galaxies suggests environment is not a driving factor in this relationship. We interpret this as evidence that the environmental processes which perturb the distribution of molecular gas in galaxies also affect the global gas mass. In this way, galaxies undergoing environmental transformation move along the size--mass relation rather than deviating from it. Finally, we investigate the effect that radius definition has on this correlation and determine the optimal molecular gas iso-density ($\Sigma_{\rm mol} (r) \sim 5 {\rm M_\odot ~ pc^{-2}}$) and flux-percentage ($r_{90,\, {\rm mol}}$) radius definitions that produce the least scatter and strongest correlation.

Our intent with this work is to provide an overview of the VERTICO survey and highlight its potential as a resource for revealing the role environment plays in galaxy evolution. To this end, VERTICO will be used to study the fate of molecular gas in cluster galaxies and the physics of environment-driven processes that perturb the star formation cycle. It is our hope that VERTICO advances our understanding and provides a valuable legacy resource that serves the community for years to come.

\acknowledgments \label{sec:acknowledgments}

The majority of this work was conducted on the traditional territory of the Lekwungen peoples. We acknowledge and respect the Songhees, Esquimalt and WS\'{A}NE\'{C} Nations whose historical relationships with the land continue to this day.

We thank the anonymous referee for a considered and constructive review that improved the manuscript.

This work was carried out as part of the VERTICO collaboration.

The authors wish to thank the members of the PHANGS-ALMA collaboration for their support and advice in reducing the VERTICO data. In particular, we thank Adam Leroy for graciously providing us with the PHANGS-ALMA ACA data and imaging pipeline in advance of publication.

TB acknowledges support from the National Research Council of Canada via the Plaskett Fellowship of the Dominion Astrophysical Observatory. CDW acknowledges support from the Natural Sciences and Engineering Research Council of Canada and the Canada Research Chairs program. TAD acknowledges support from STFC grant ST/S00033X/1. LP, JW, and KS acknowledge support from the Natural Science and Engineering Council of Canada. LC acknowledges support from the Australian Research Council’s Discovery Project and Future Fellowship funding schemes (DP210100337, FT180100066). ARHS acknowledges receipt of the Jim Buckee Fellowship at ICRAR-UWA. IDR acknowledges support from the ERC Starting Grant Cluster Web 804208. KPO is funded by NASA under award No 80NSSC19K1651. V. Villanueva acknowledges support from the scholarship CONICYT PFCHA/ CONICYT-FULBRIGHT BIO 2016 - 56160020 and funding from NRAO Student Observing Support (SOS) - SOSPA7-014. Support for this work was also provided by the National Research Foundation of Korea (NRF) grant No. 2018R1D1A1B07048314. BL acknowledges support from the National Science Foundation of China (12073002, 11721303). YMB gratefully acknowledges funding from the Netherlands Organization for Scientific Research (NWO) through Veni grant number 639.041.751. CW acknowledges the  support  of  the  National  Science  Foundation  award 1815251  (United  States)  held  by  Dr.  Susan  Kassin. Parts of this research were conducted by the Australian Research Council Centre of Excellence for All Sky Astrophysics in 3 Dimensions (ASTRO 3D), through project number CE170100013. MHH acknowledges support from the William and Caroline Herschel Postdoctoral Fellowship fund. CDPL has received funding from ASTRO 3D through project number CE170100013. PJE works on Whadjuk country and pays respect to the elders past, present and emerging of the Noongar people. CRC acknowledges support from STFC grant ST/R000840/1.

This paper makes use of the following ALMA data: ADS/JAO.ALMA\href{https://almascience.nrao.edu/asax/?result_view=observation&projectCode=\%222019.1.00763.L\%22}{\#2019.1.00763.L},\\ ADS/JAO.ALMA\href{https://almascience.nrao.edu/asax/?result_view=observation&projectCode=\%222017.1.00886.L\%22}{\#2017.1.00886.L},\\ ADS/JAO.ALMA\href{https://almascience.nrao.edu/asax/?result_view=observation&projectCode=\%222016.1.00912.S\%22}{\#2016.1.00912.S},\\ ADS/JAO.ALMA\href{https://almascience.nrao.edu/asax/?result_view=observation&projectCode=\%222015.1.00956.S\%22}{\#2015.1.00956.S}. \\
ALMA is a partnership of ESO (representing its member states), NSF (USA) and NINS (Japan), together with NRC (Canada), MOST and ASIAA (Taiwan), and KASI (Republic of Korea), in cooperation with the Republic of Chile. The Joint ALMA Observatory is operated by ESO, AUI/NRAO and NAOJ. The National Radio Astronomy Observatory is a facility of the National Science Foundation operated under cooperative agreement by Associated Universities, Inc.

In addition to the ALMA Science Archive, research made use of data and/or software provided by the following archives: 
\begin{itemize}
	\item NASA/IPAC Infrared Science Archive, which is funded by the National Aeronautics and Space Administration and operated by the California Institute of Technology
	\item NASA/IPAC Extragalactic Database (NED), which is operated by the Jet Propulsion Laboratory, California Institute of Technology, under contract with the National Aeronautics and Space Administration. 
	\item The High Energy Astrophysics Science Archive Research Center (HEASARC), which is a service of the Astrophysics Science Division at NASA/GSFC.
	\item The HRS data was accessed through the Herschel Database in Marseille (HeDaM - http://hedam.lam.fr) operated by CeSAM and hosted by the Laboratoire d'Astrophysique de Marseille.
\end{itemize}

We acknowledge the use of the ARCADE (ALMA Reduction in the CANFAR Data Environment) science platform. ARCADE is a ALMA Cycle 7 development study with support from the National Radio Astronomy Observatory, the North American ALMA Science center, and the National Research center of Canada. Our work used the facilities of the Canadian Astronomy Data center, operated by the National Research Council of Canada with the support of the Canadian Space Agency, and the Canadian Advanced Network for Astronomy Research, a consortium that serves data-intensive storage, access, and processing needs of university groups and centers engaged in astronomy research \citep{Gaudet2010}.

This research has made use of data from the Herschel Reference Survey (HRS) project. HRS is a Herschel Key Programme utilizing Guaranteed Time from the SPIRE instrument team, ESAC scientists and a mission scientist.

Here we acknowledge the key software used in this work:
\begin{itemize}
	\item \href{http://www.astropy.org}{Astropy}, a community-developed core {\sc Python} package for Astronomy and affiliated packages \citep[{\sc Spectral-cube, Reproject, Photutils};][]{astropy:2013, astropy:2018,larry_bradley_2020_4044744}.
	\item \href{https://casa.nrao.edu/}{The CASA (Common Astronomy Software Applications) package}, a suite of applications for the reduction and analysis of radio astronomical data with a {\sc Python} interface \citep{CASA2007}.
	\item PHANGS-ALMA Imaging Pipeline, a library of programs that use CASA, Astropy, and affiliated packages ({\sc Analysisutils, Spectral-cube, Reproject}) to process data from the calibrated visibilities to science-ready data cubes \citep{Leroy2021b}.
	\item \href{https://matplotlib.org/}{\sc Matplotlib}, a plotting library for the {\sc Python} programming language and its numerical mathematics extension {\sc NumPy} \citep{Hunter2007, 2020SciPy-NMeth}.
	\item \href{https://pandas.pydata.org/}{\sc Pandas}, an open source, BSD-licensed library providing high-performance, easy-to-use data structures and data analysis tools for the {\sc Python} programming language \citep{mckinney-proc-scipy-2010}.
	\item \href{https://cartavis.org/}{{\sc Carta}: The Cube Analysis and Rendering Tool for Astronomy}, an image visualization and analysis tool for radio astronomy data \citep{angus_comrie_2020_3746095}.
	\item \href{https://github.com/Stargrazer82301/AncillaryDataButton}{The Ancillary Data Button}, a selection of {\sc Python} scripts for acquiring and standardising imaging data from a wide range of telescopes \citep{Clark2018}.
	\item \href{https://github.com/johannesjmeyer/rsmf}{\sc Rsmf} (right-size my figures), a {\sc Python} library that provides a means to automatically adjust figure sizes and font sizes in {\sc Matplotlib} to fit those commonly used in scientific journals.
	\item \href{https://cxc.cfa.harvard.edu/ciao/}{The Chandra Interactive Analysis of Observations} application package developed by the Chandra X-ray Center for analyzing X-ray data \citep{Fruscione2006}.
\end{itemize}


\appendix 
\section{VERTICO CO(\texorpdfstring{$2-1$}{2--1}) Panel Plots}
\label{app:panel_plots}
   \figsetstart
   \figsetnum{4}
   \figsettitle{CO($2-1$) data products available for each galaxy in the VERTICO survey}

   \figsetgrpstart
   \figsetgrpnum{4.1}
   \figsetgrptitle{VERTICO CO($2-1$) data products for NGC4380}
   \figsetplot{NGC4380.pdf}
   \figsetgrpnote{An example of the CO($2-1$) data products available for each galaxy in the VERTICO survey. The left panel shows the SDSS $gri$ composite image for NGC4380 with molecular gas surface brightness contours at the 10th, 50th, and 90th percentiles of the distribution. The field of view of the ACA observations is defined where the primary beam response drops to 50\% and is illustrated by the gray line. The rounded synthesized beam is $7.5\arcsec$ in diameter and illustrated in the bottom left corner. This beam corresponds to $\sim$600 pc at the distance of Virgo (16.5 Mpc). The VERTICO CO($2-1$) data products available for each galaxy include maps of integrated intensity (upper center panel), the velocity field (upper right), observed line width, $\sigma_v$ (lower center), and peak temperature (lower left). The $x$- and $y$-axes of each moment map shows the angular offset from the optical center listed in Table \ref{tab:VERTICO-sample}. The galaxy name is provided in the upper left.}
   \figsetgrpend

   \figsetgrpstart
   \figsetgrpnum{4.2}
   \figsetgrptitle{VERTICO CO($2-1$) data products for IC3392}
   \figsetplot{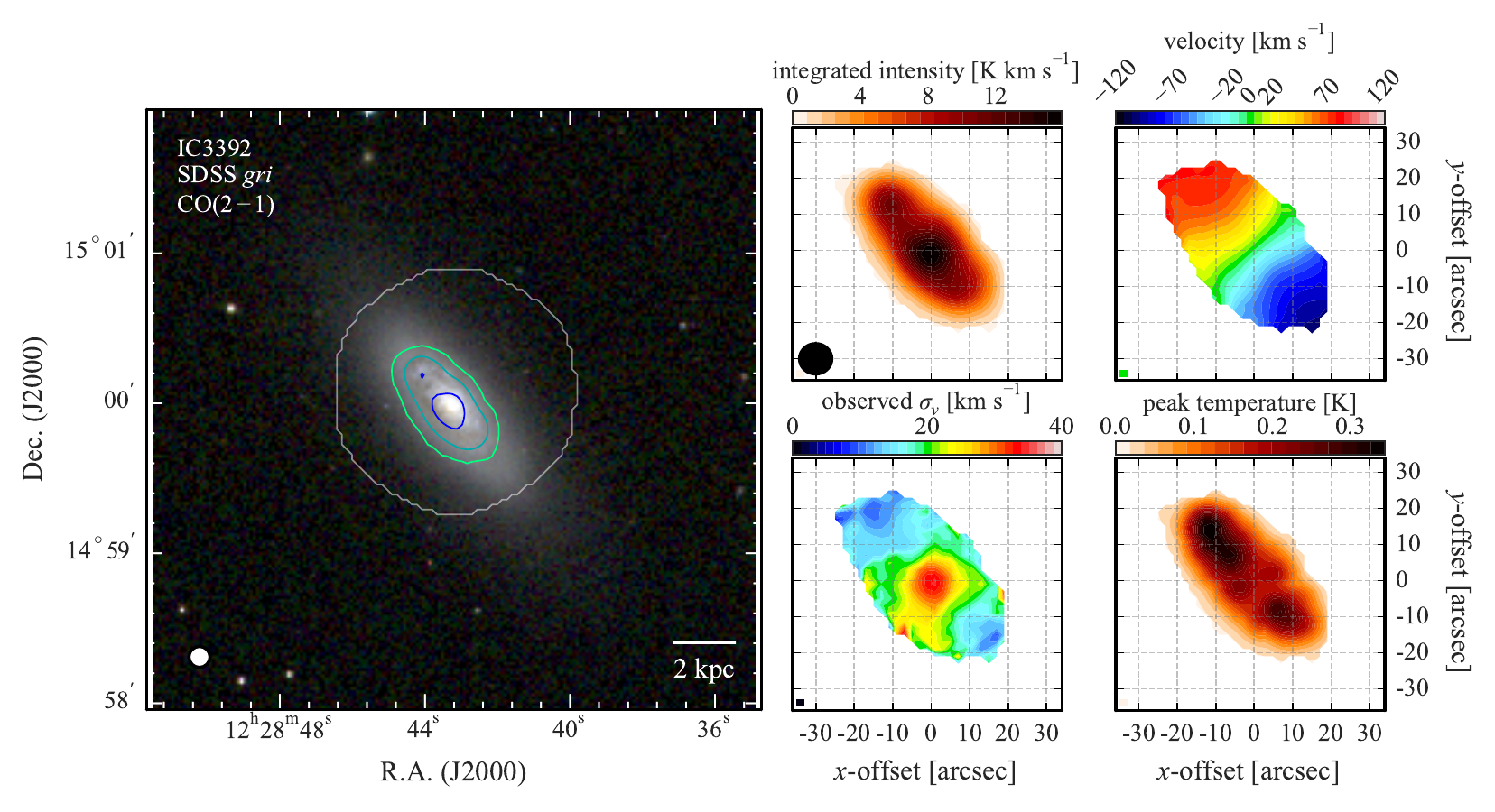}
   \figsetgrpnote{As in Figure 4.1.}
   \figsetgrpend

   \figsetgrpstart
   \figsetgrpnum{4.3}
   \figsetgrptitle{VERTICO CO($2-1$) data products for NGC4064}
   \figsetplot{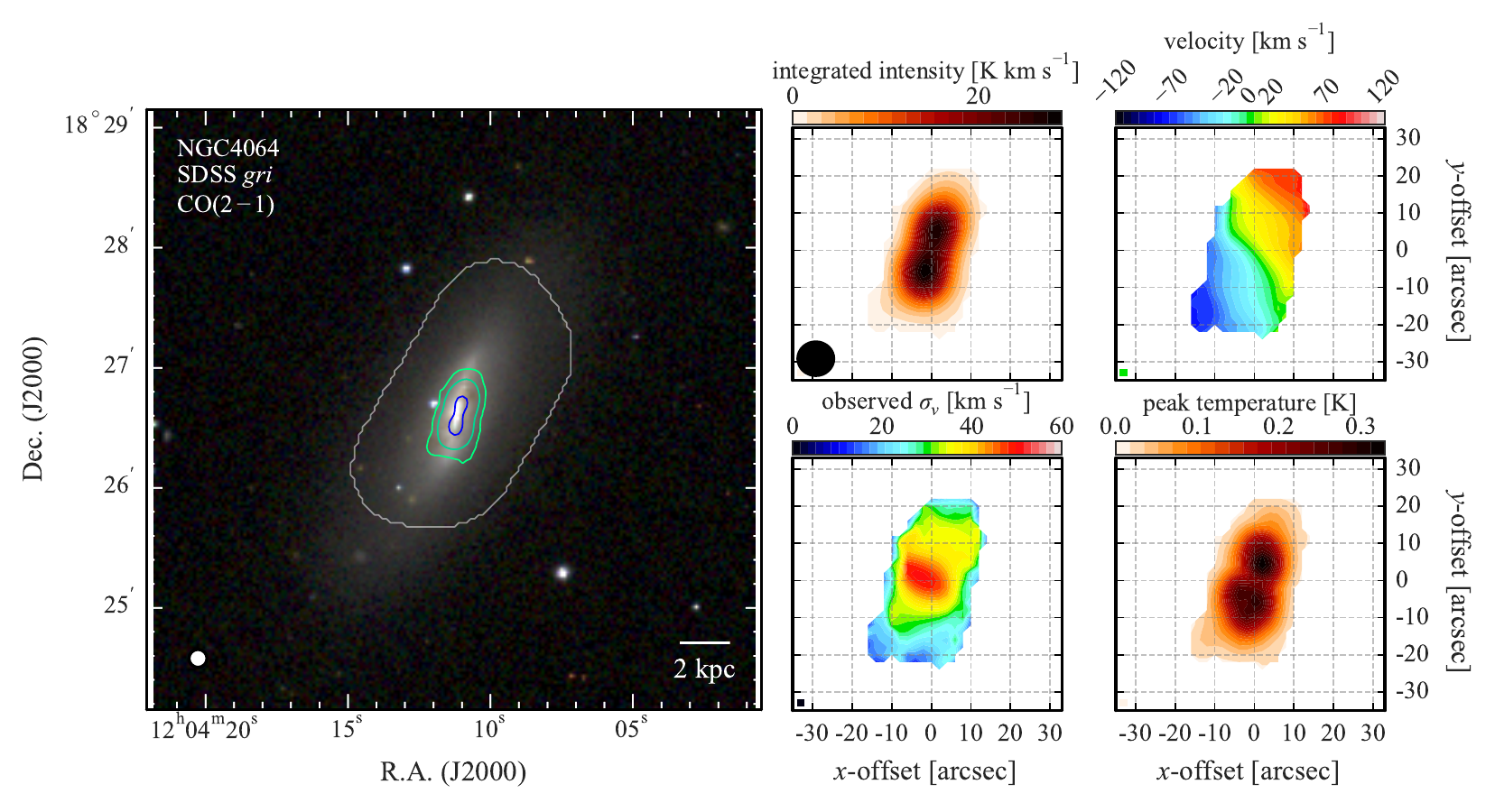}
   \figsetgrpnote{As in Figure 4.1.}
   \figsetgrpend

   \figsetgrpstart
   \figsetgrpnum{4.4}
   \figsetgrptitle{VERTICO CO($2-1$) data products for NGC4189}
   \figsetplot{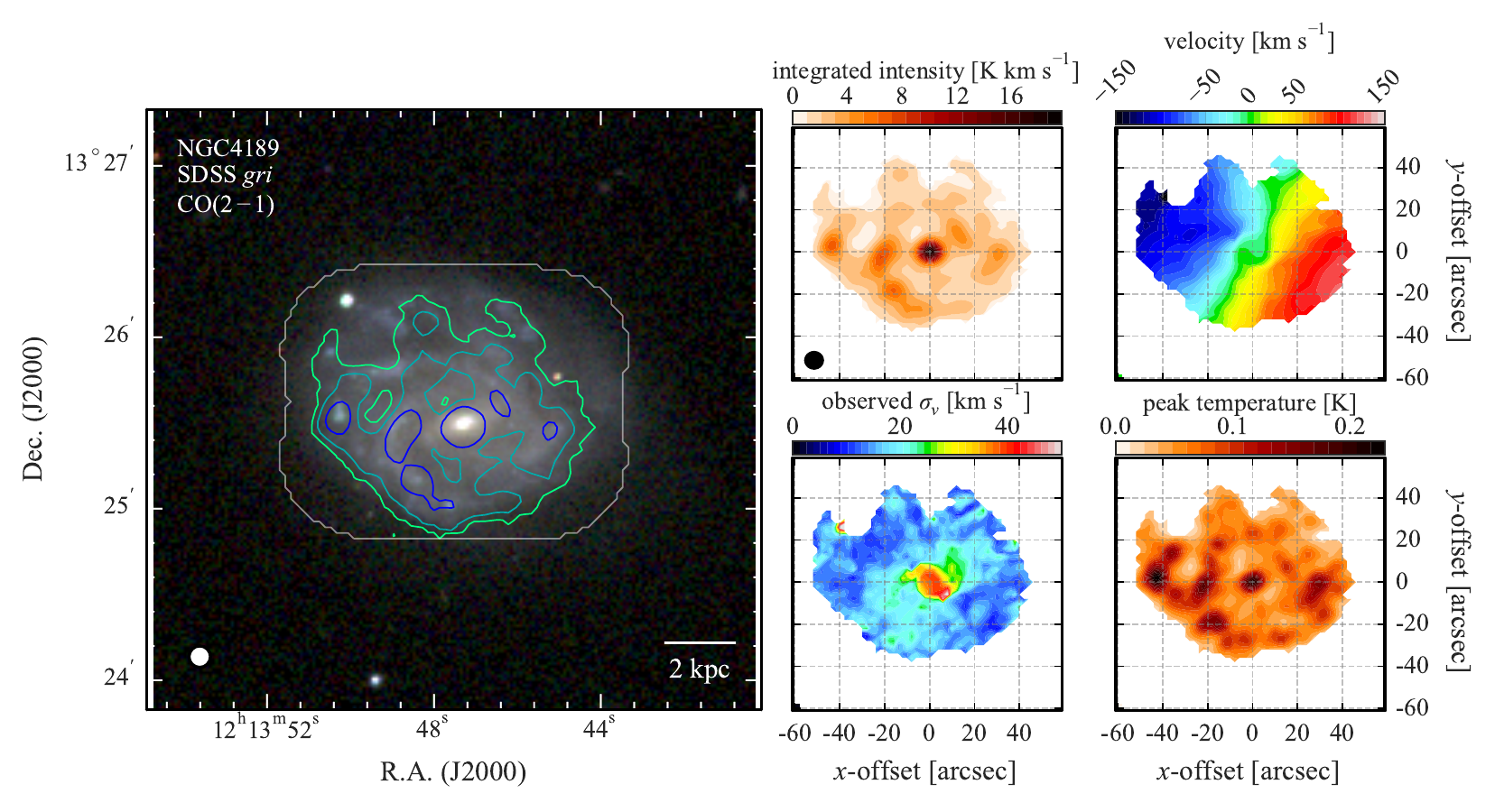}
   \figsetgrpnote{As in Figure 4.1.}
   \figsetgrpend

   \figsetgrpstart
   \figsetgrpnum{4.5}
   \figsetgrptitle{VERTICO CO($2-1$) data products for NGC4192}
   \figsetplot{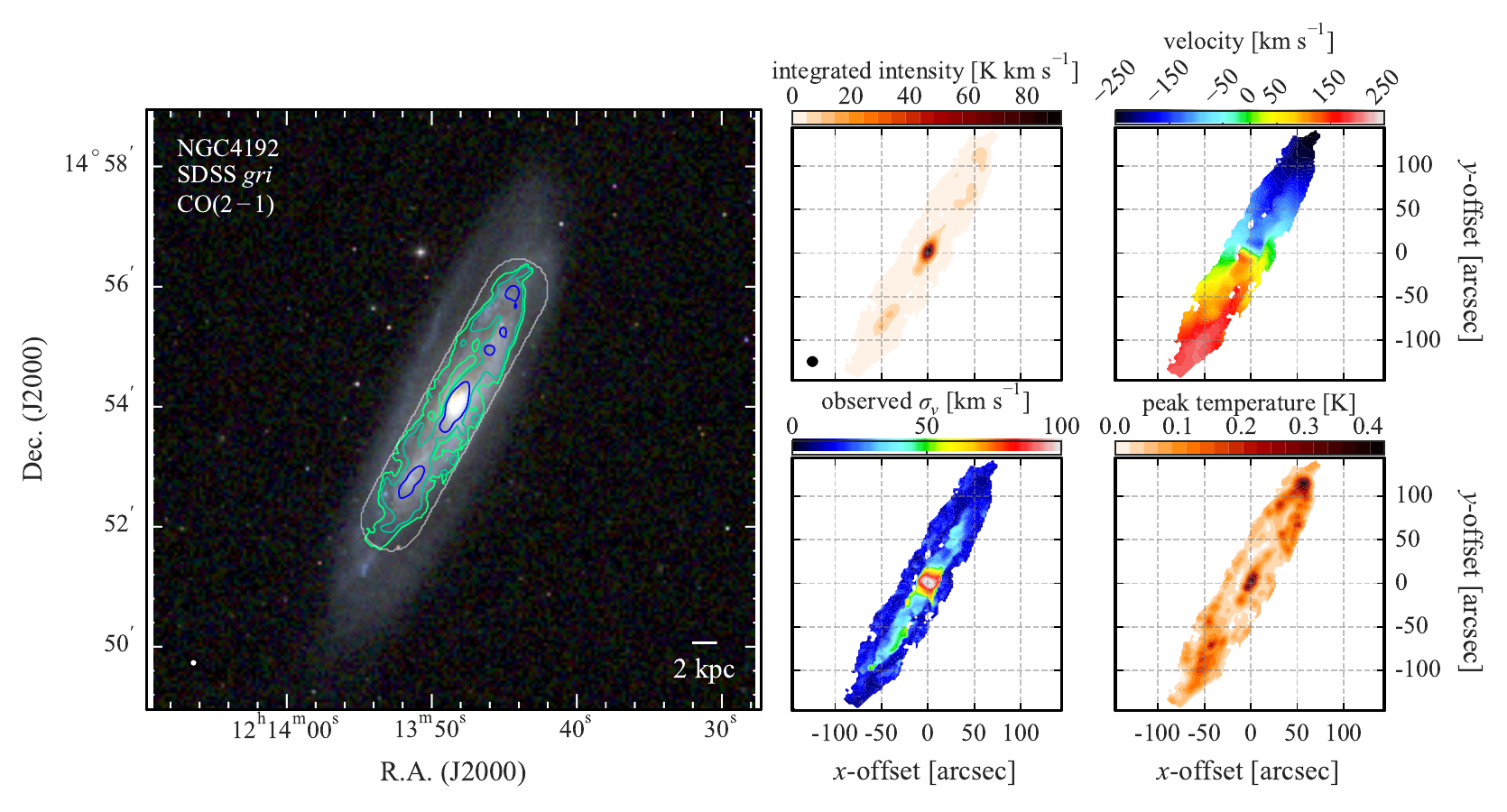}
   \figsetgrpnote{As in Figure 4.1.}
   \figsetgrpend

   \figsetgrpstart
   \figsetgrpnum{4.6}
   \figsetgrptitle{VERTICO CO($2-1$) data products for NGC4216}
   \figsetplot{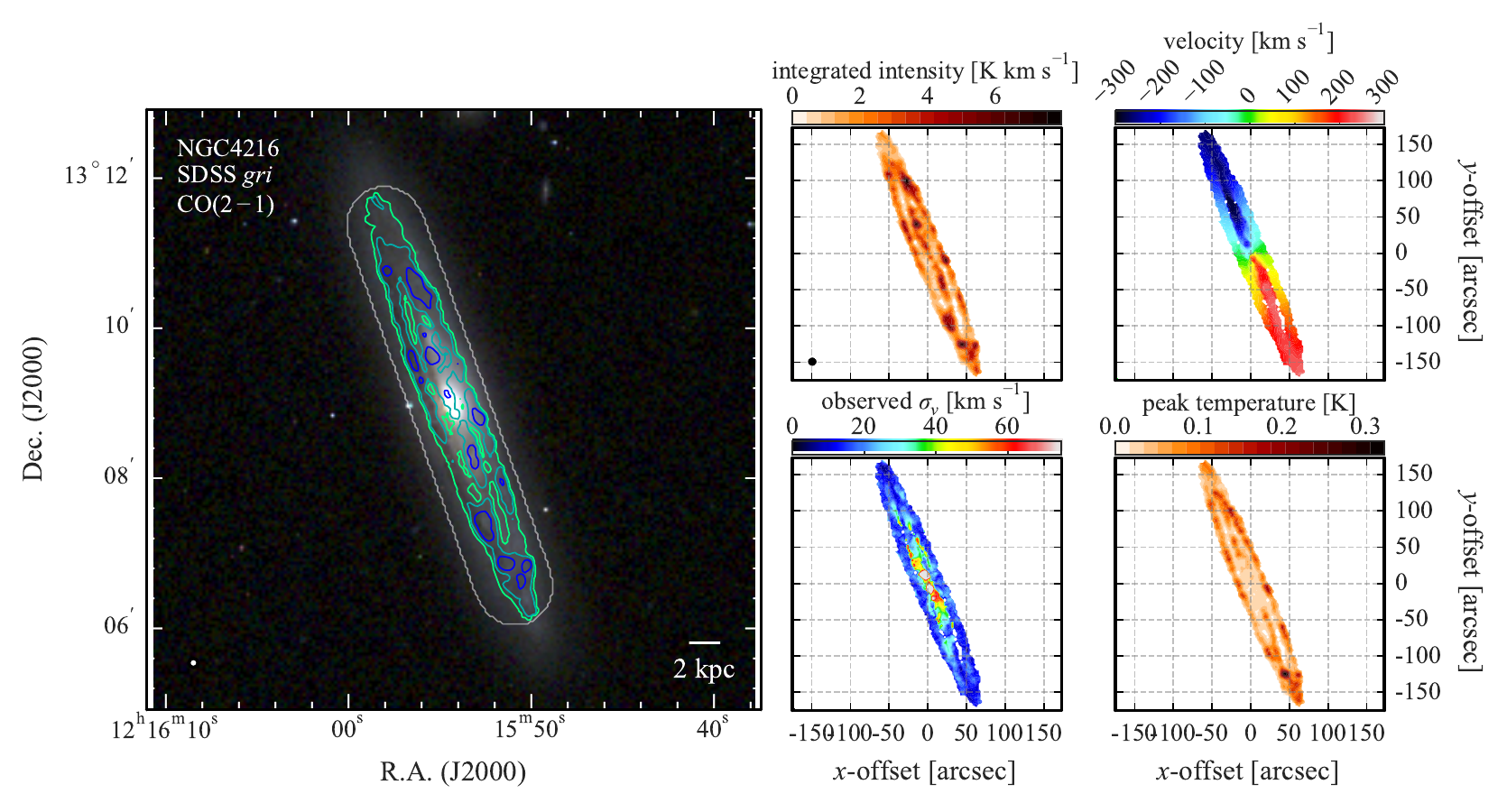}
   \figsetgrpnote{As in Figure 4.1.}
   \figsetgrpend

   \figsetgrpstart
   \figsetgrpnum{4.7}
   \figsetgrptitle{VERTICO CO($2-1$) data products for NGC4222}
   \figsetplot{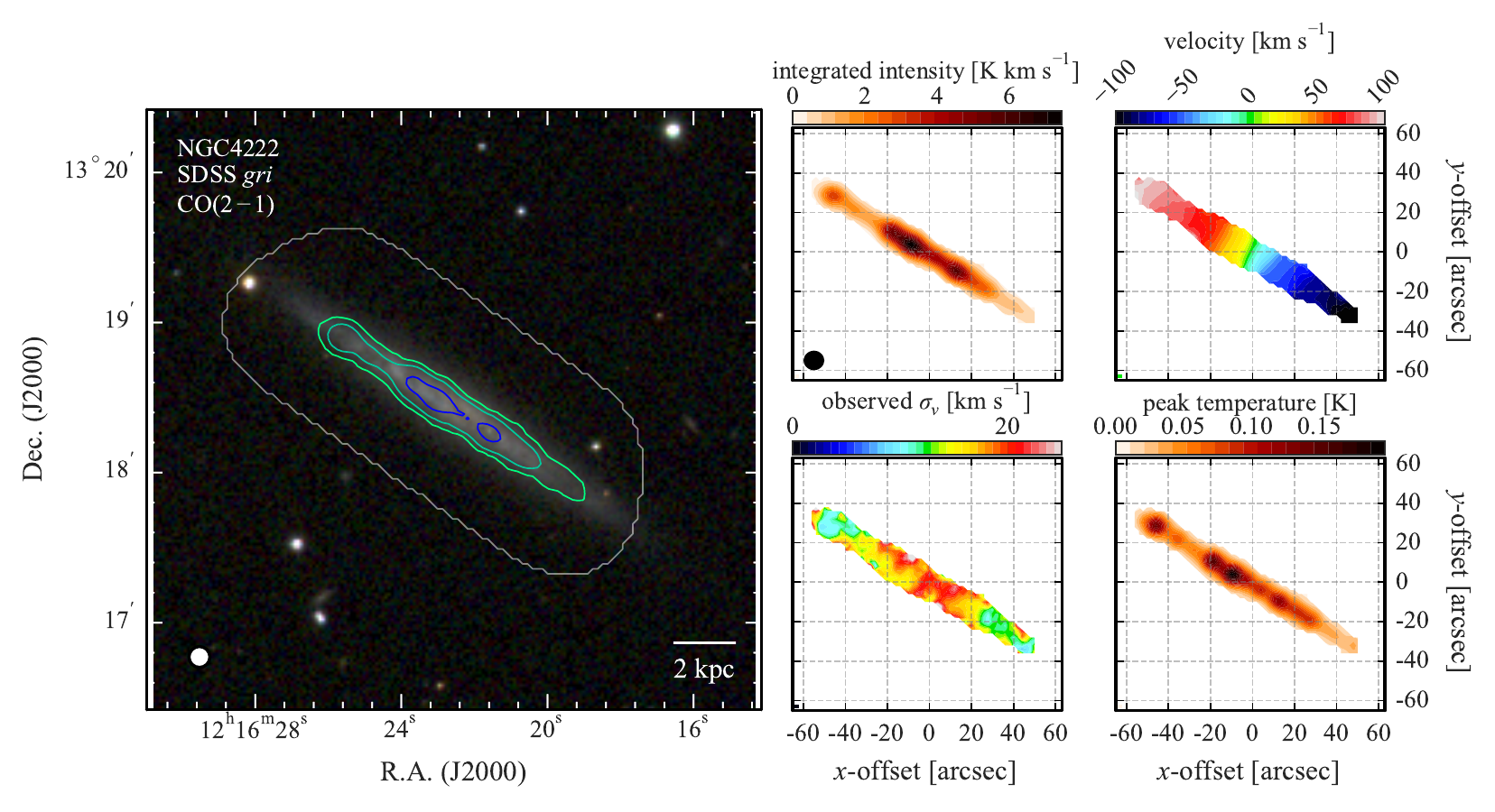}
   \figsetgrpnote{As in Figure 4.1.}
   \figsetgrpend

   \figsetgrpstart
   \figsetgrpnum{4.8}
   \figsetgrptitle{VERTICO CO($2-1$) data products for NGC4254}
   \figsetplot{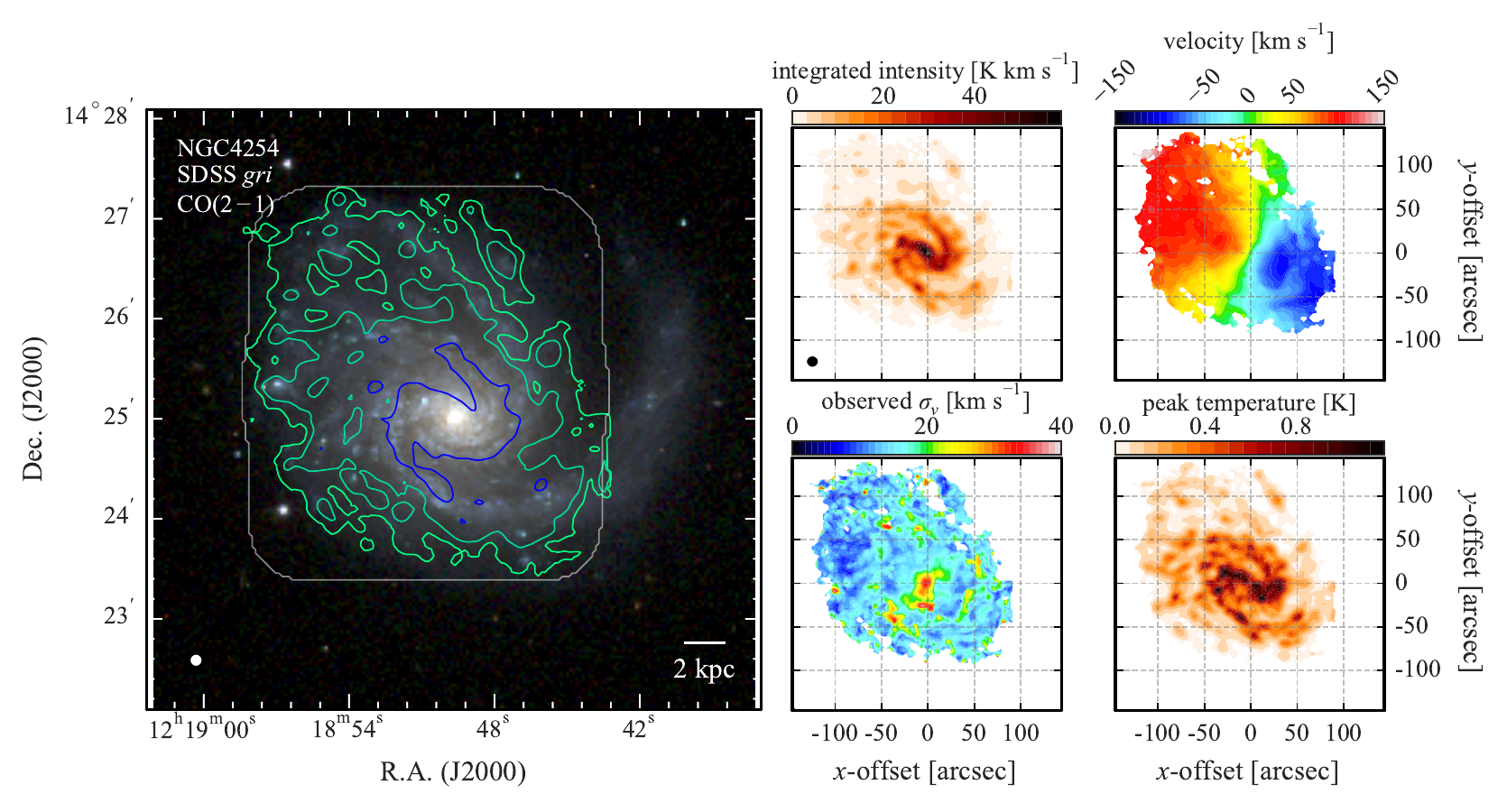}
   \figsetgrpnote{As in Figure 4.1.}
   \figsetgrpend

   \figsetgrpstart
   \figsetgrpnum{4.9}
   \figsetgrptitle{VERTICO CO($2-1$) data products for NGC4293}
   \figsetplot{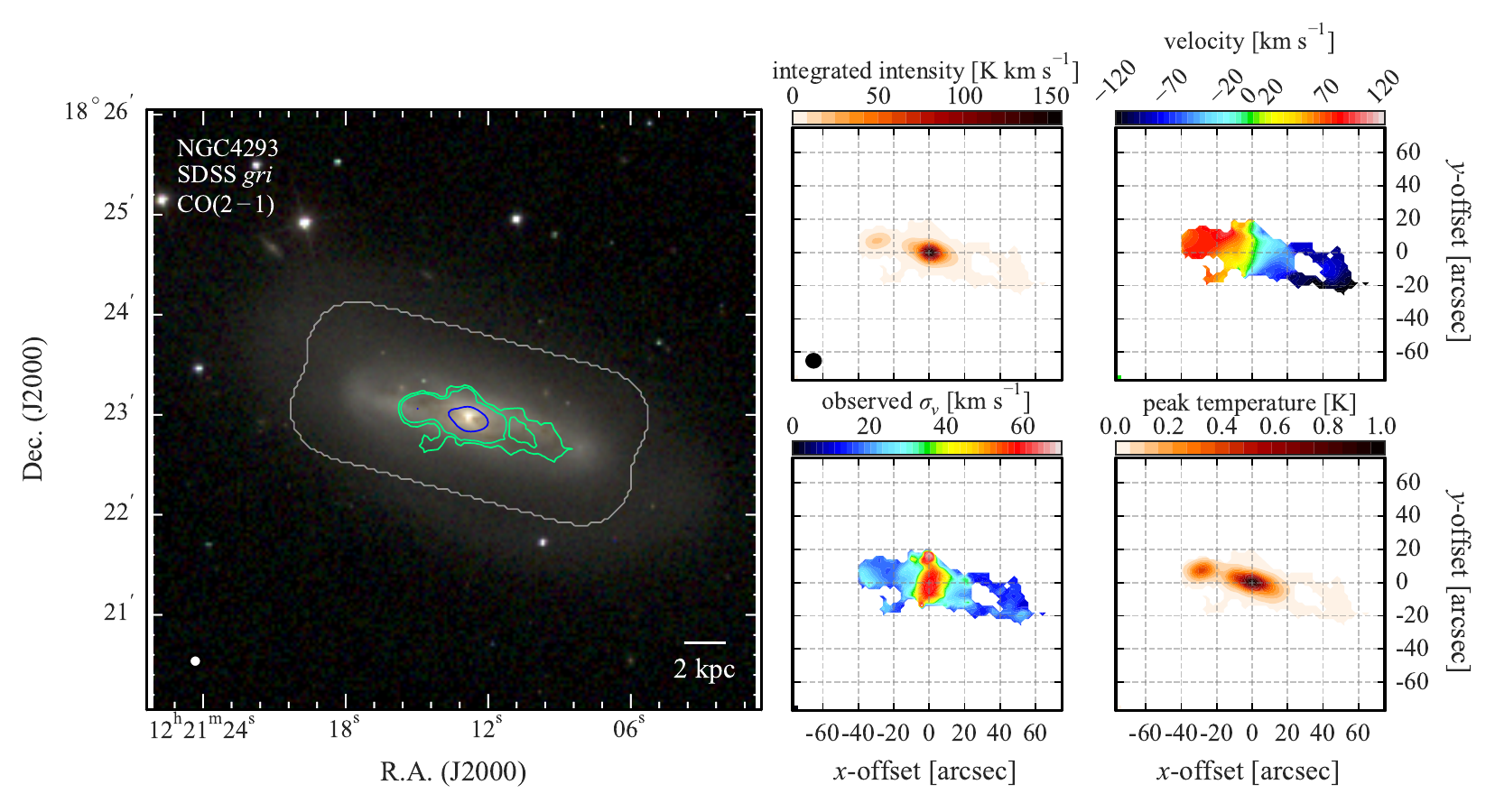}
   \figsetgrpnote{As in Figure 4.1. In the case of NGC4293, we caution against over-interpreting the skinny feature to the South-East of the main disk as this is likely related to the PSF pattern of the observations.}
   \figsetgrpend

   \figsetgrpstart
   \figsetgrpnum{4.10}
   \figsetgrptitle{VERTICO CO($2-1$) data products for NGC4294}
   \figsetplot{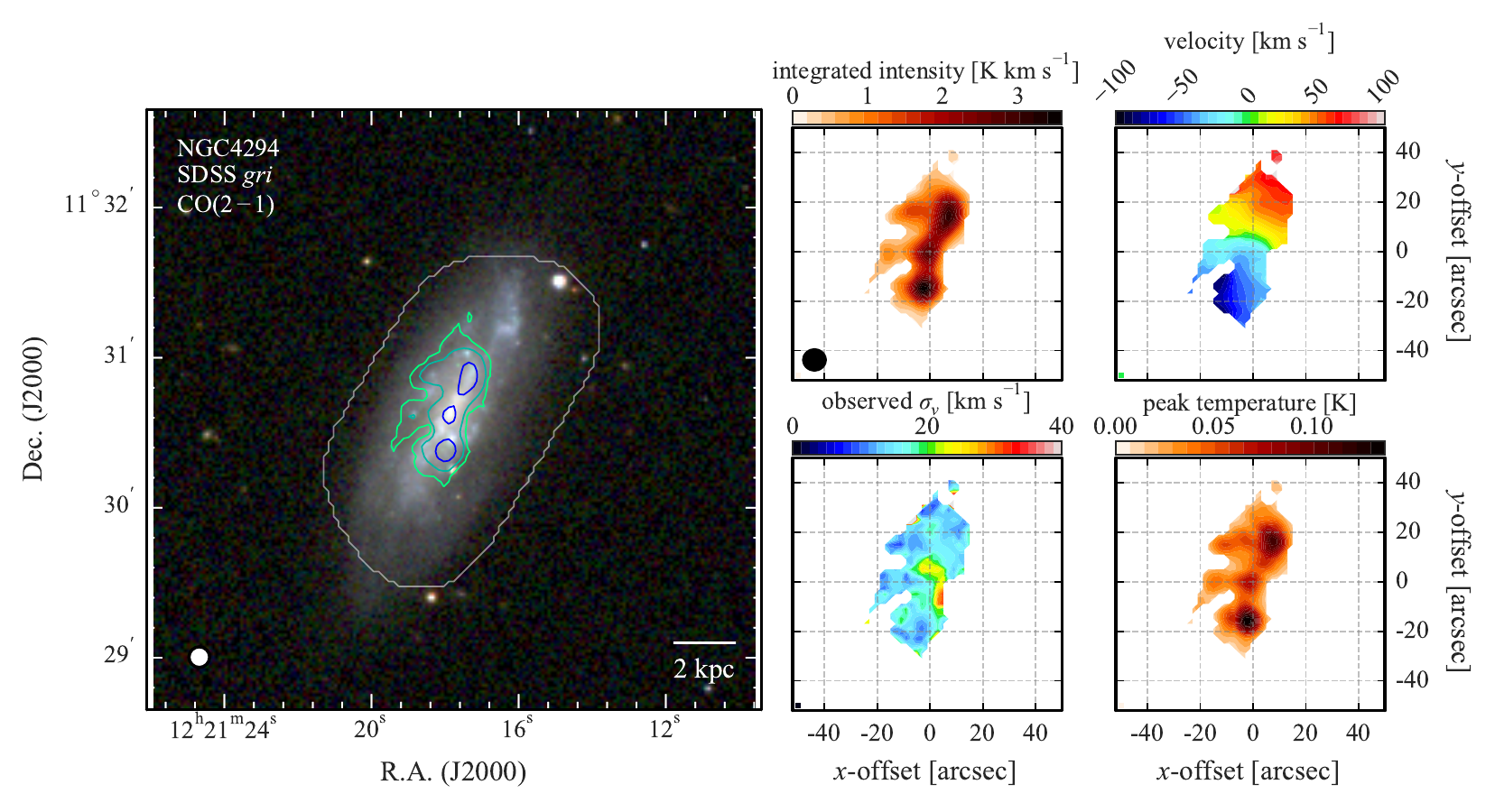}
   \figsetgrpnote{As in Figure 4.1.}
   \figsetgrpend

   \figsetgrpstart
   \figsetgrpnum{4.11}
   \figsetgrptitle{VERTICO CO($2-1$) data products for NGC4298}
   \figsetplot{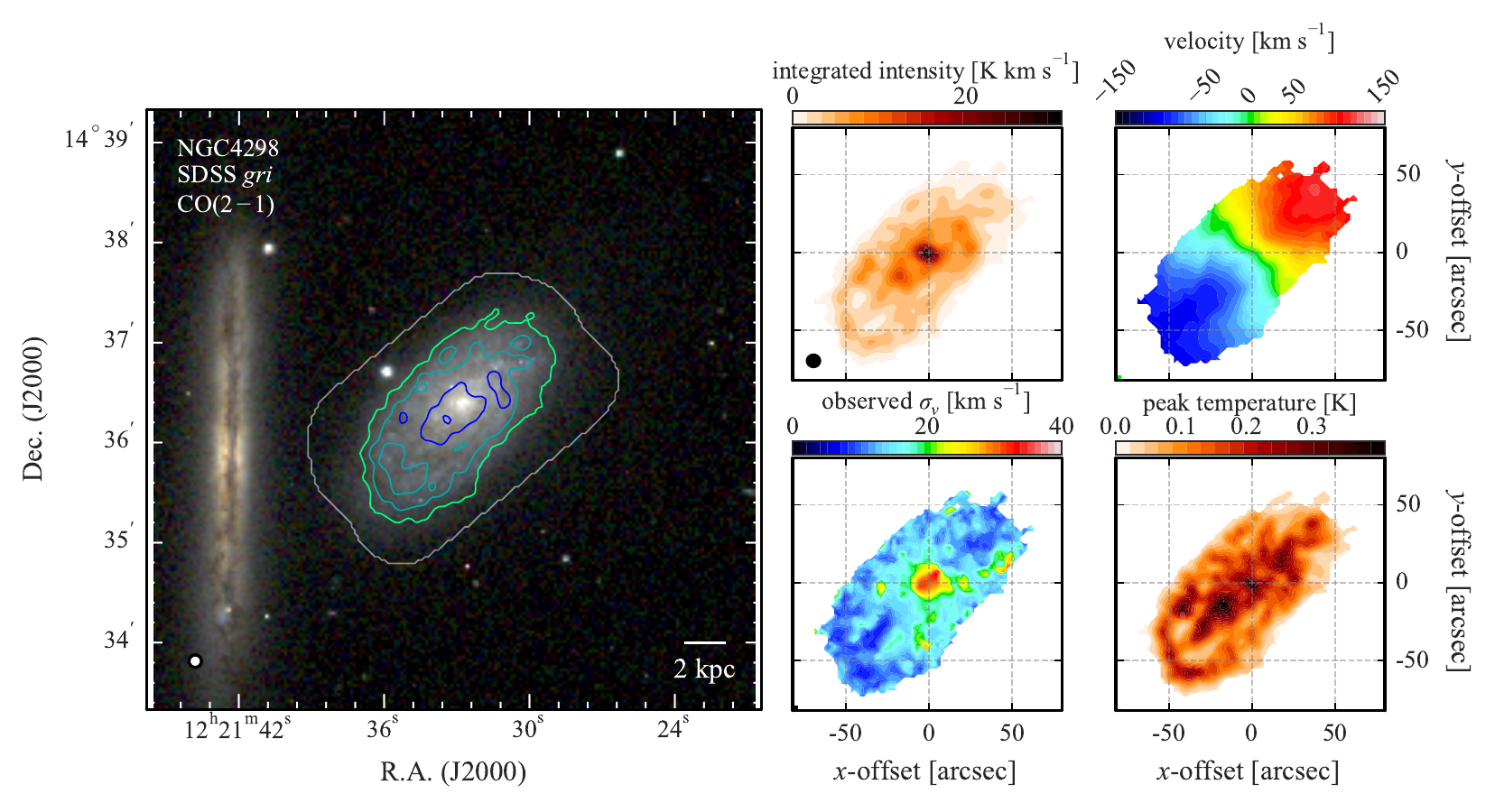}
   \figsetgrpnote{As in Figure 4.1.}
   \figsetgrpend

   \figsetgrpstart
   \figsetgrpnum{4.12}
   \figsetgrptitle{VERTICO CO($2-1$) data products for NGC4299}
   \figsetplot{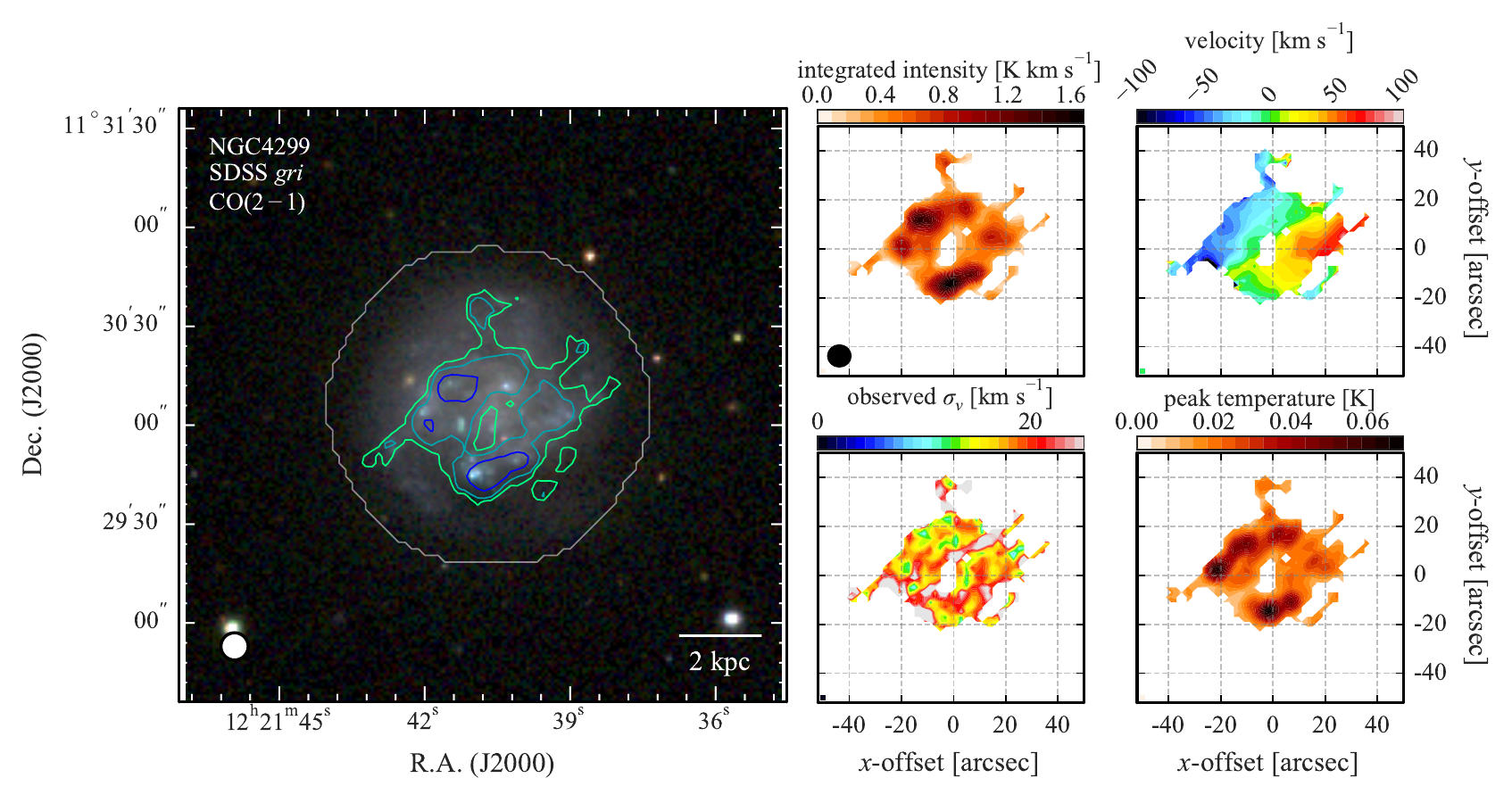}
   \figsetgrpnote{As in Figure 4.1.}
   \figsetgrpend

   \figsetgrpstart
   \figsetgrpnum{4.13}
   \figsetgrptitle{VERTICO CO($2-1$) data products for NGC4302}
   \figsetplot{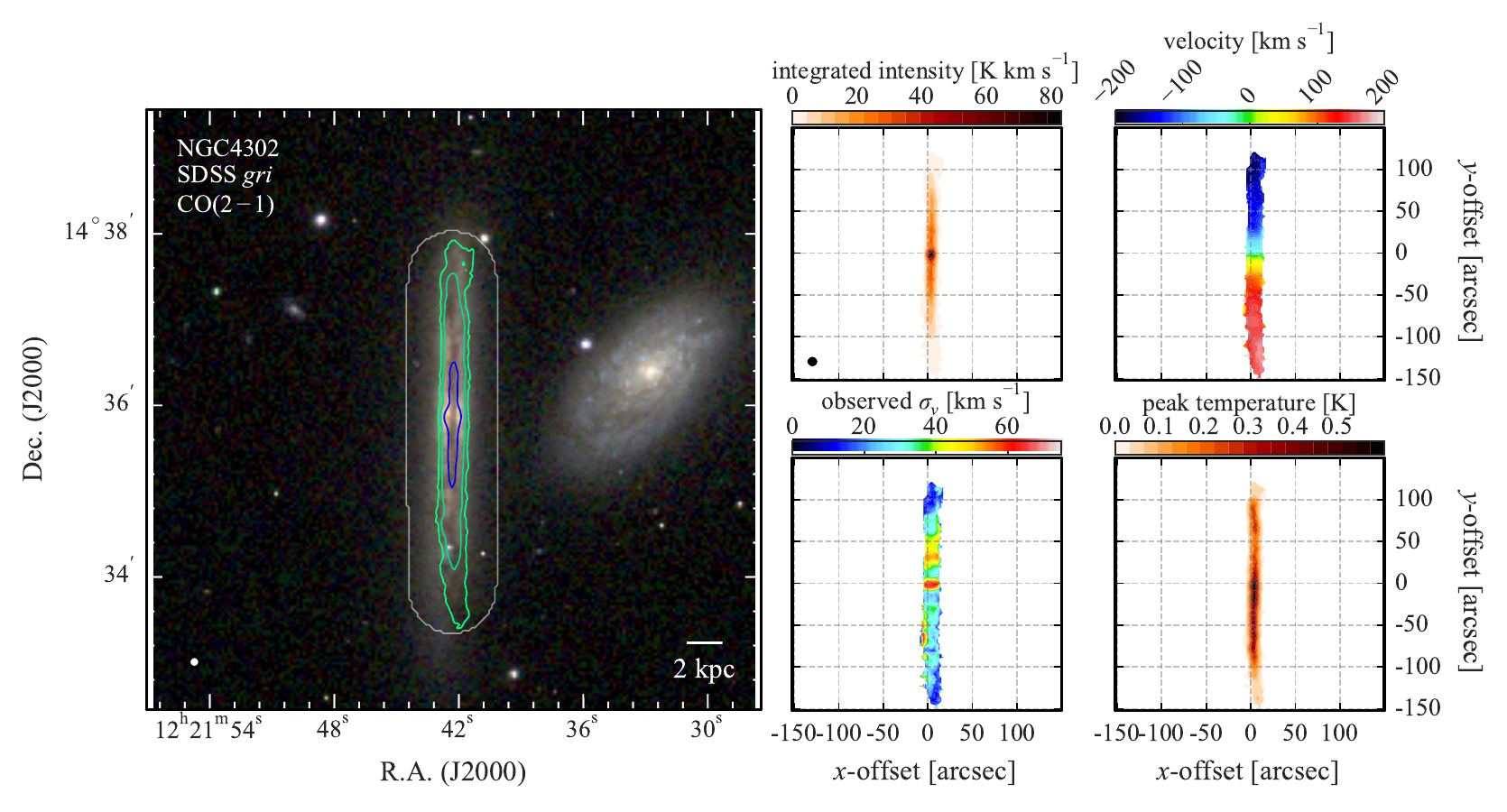}
   \figsetgrpnote{As in Figure 4.1.}
   \figsetgrpend

   \figsetgrpstart
   \figsetgrpnum{4.14}
   \figsetgrptitle{VERTICO CO($2-1$) data products for NGC4321}
   \figsetplot{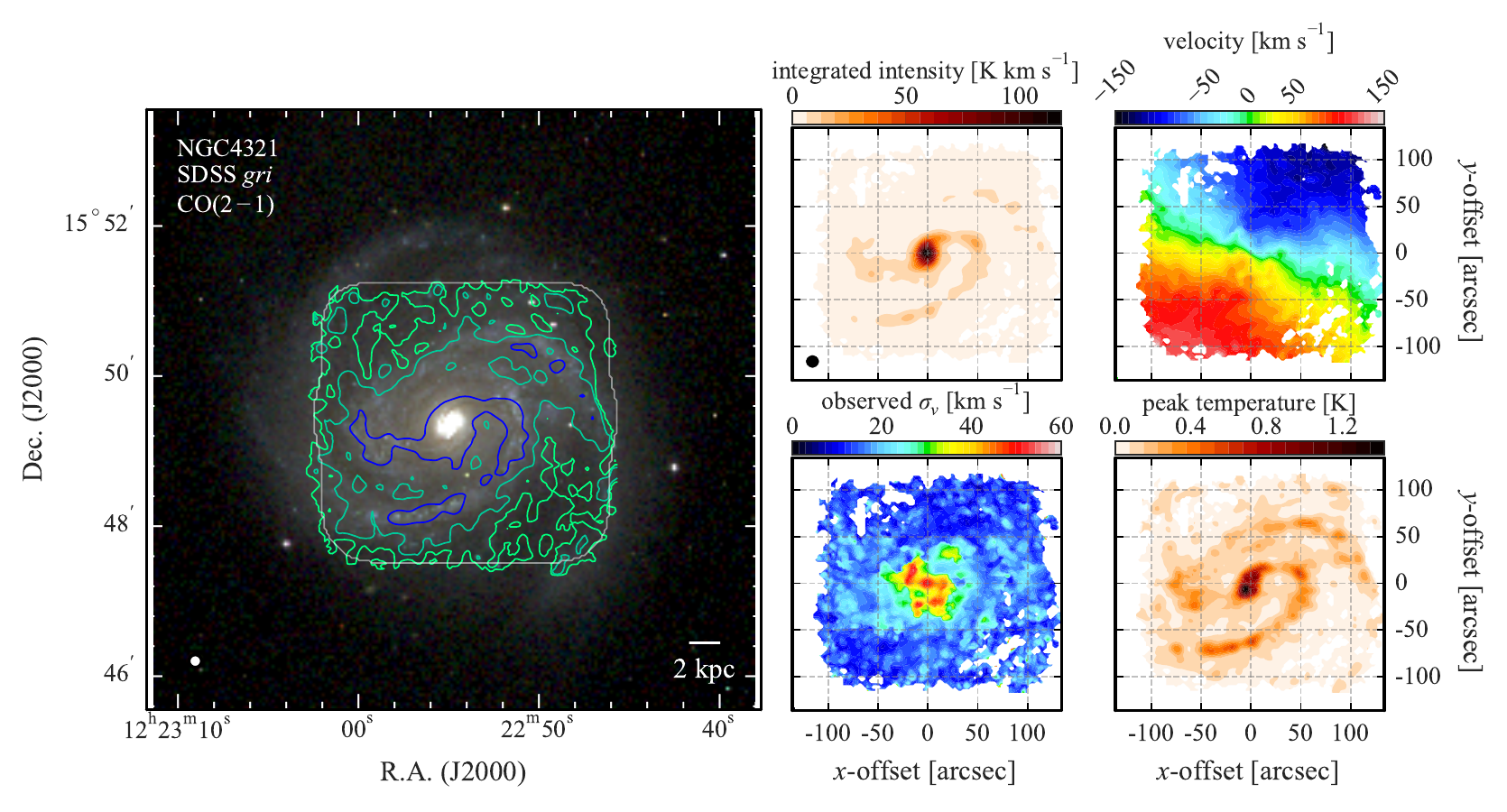}
   \figsetgrpnote{As in Figure 4.1.}
   \figsetgrpend

   \figsetgrpstart
   \figsetgrpnum{4.15}
   \figsetgrptitle{VERTICO CO($2-1$) data products for NGC4330}
   \figsetplot{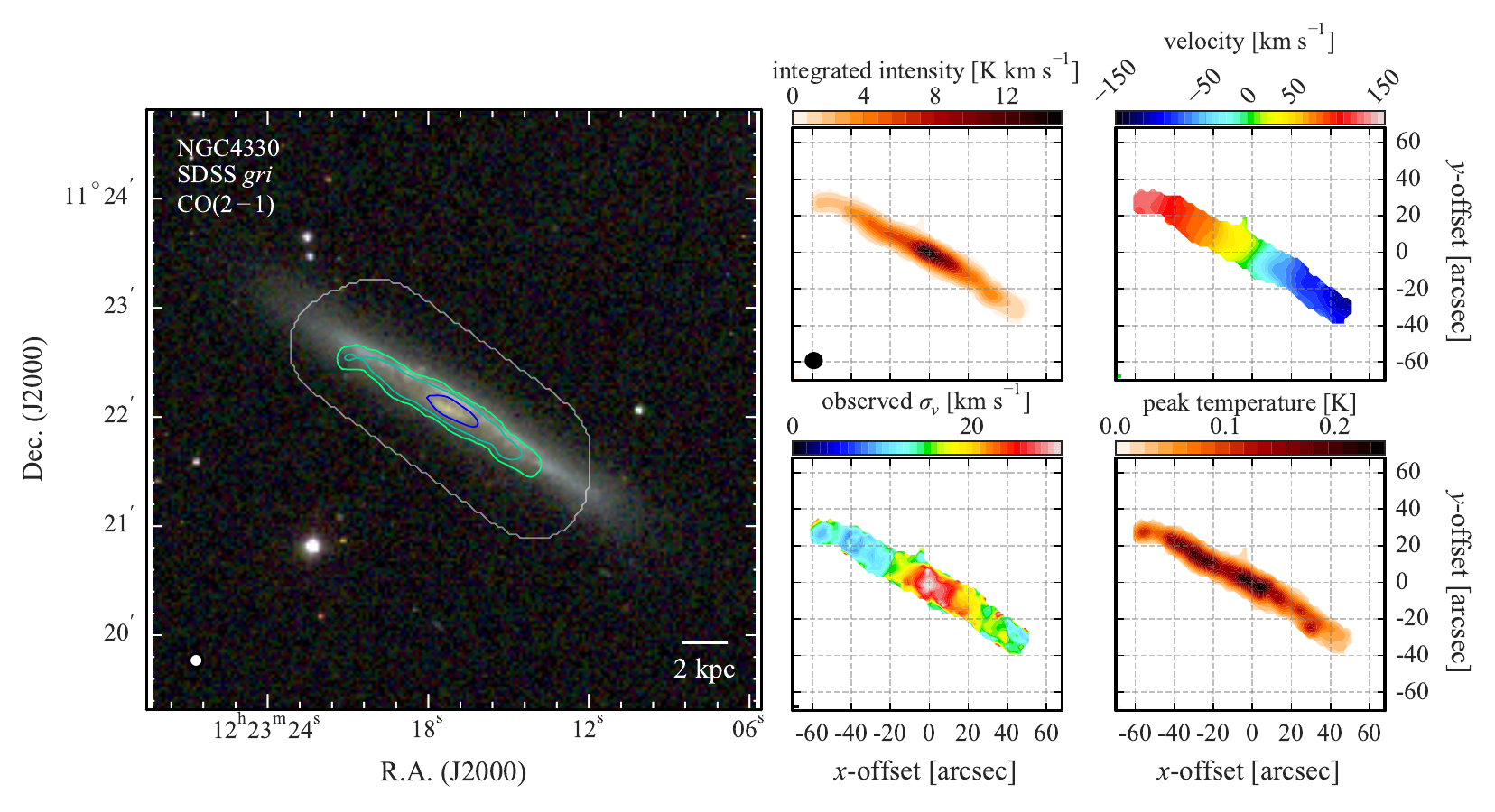}
   \figsetgrpnote{As in Figure 4.1.}
   \figsetgrpend

   \figsetgrpstart
   \figsetgrpnum{4.16}
   \figsetgrptitle{VERTICO CO($2-1$) data products for NGC4351}
   \figsetplot{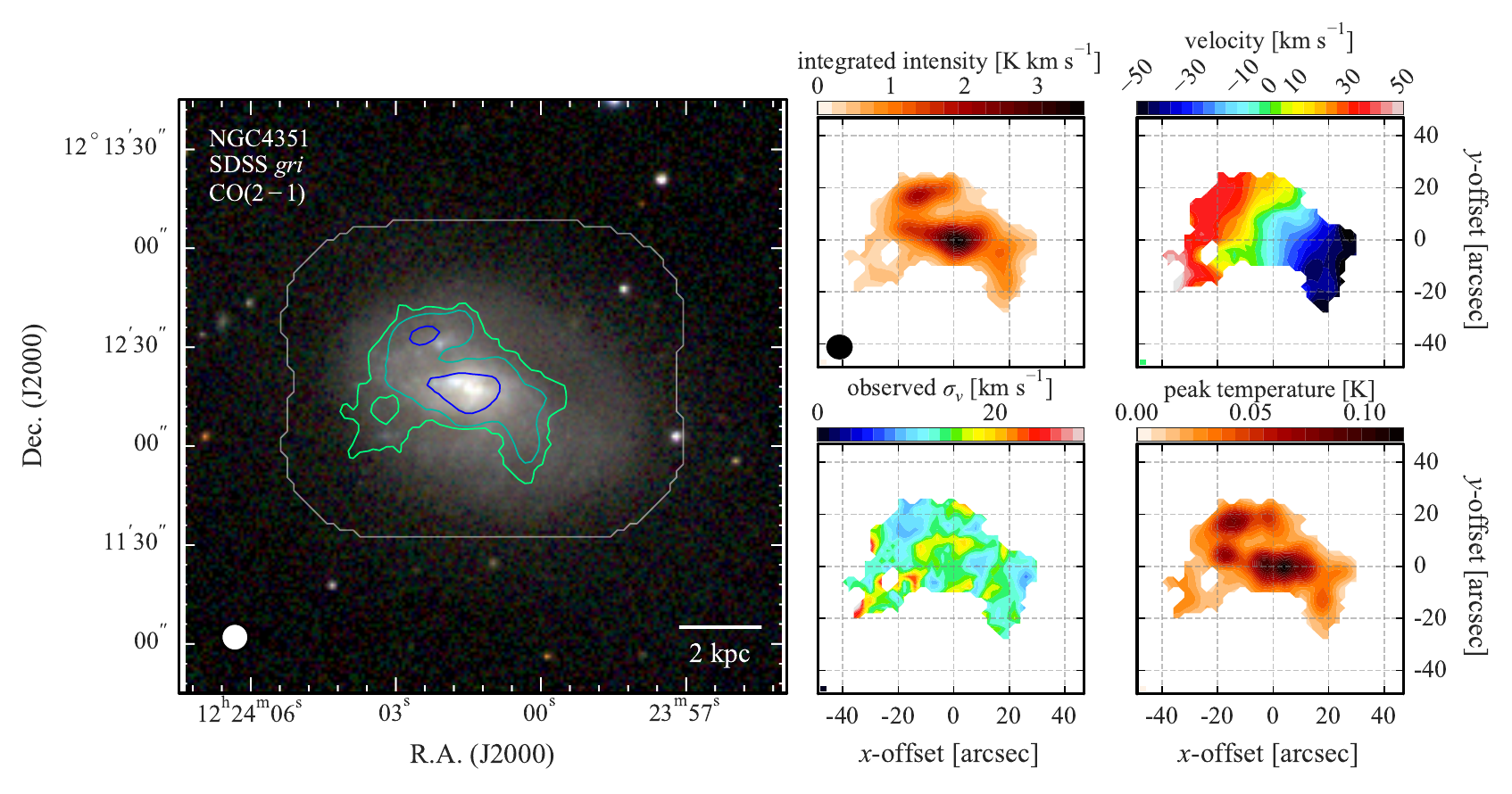}
   \figsetgrpnote{As in Figure 4.1.}
   \figsetgrpend

   \figsetgrpstart
   \figsetgrpnum{4.17}
   \figsetgrptitle{VERTICO CO($2-1$) data products for NGC4383}
   \figsetplot{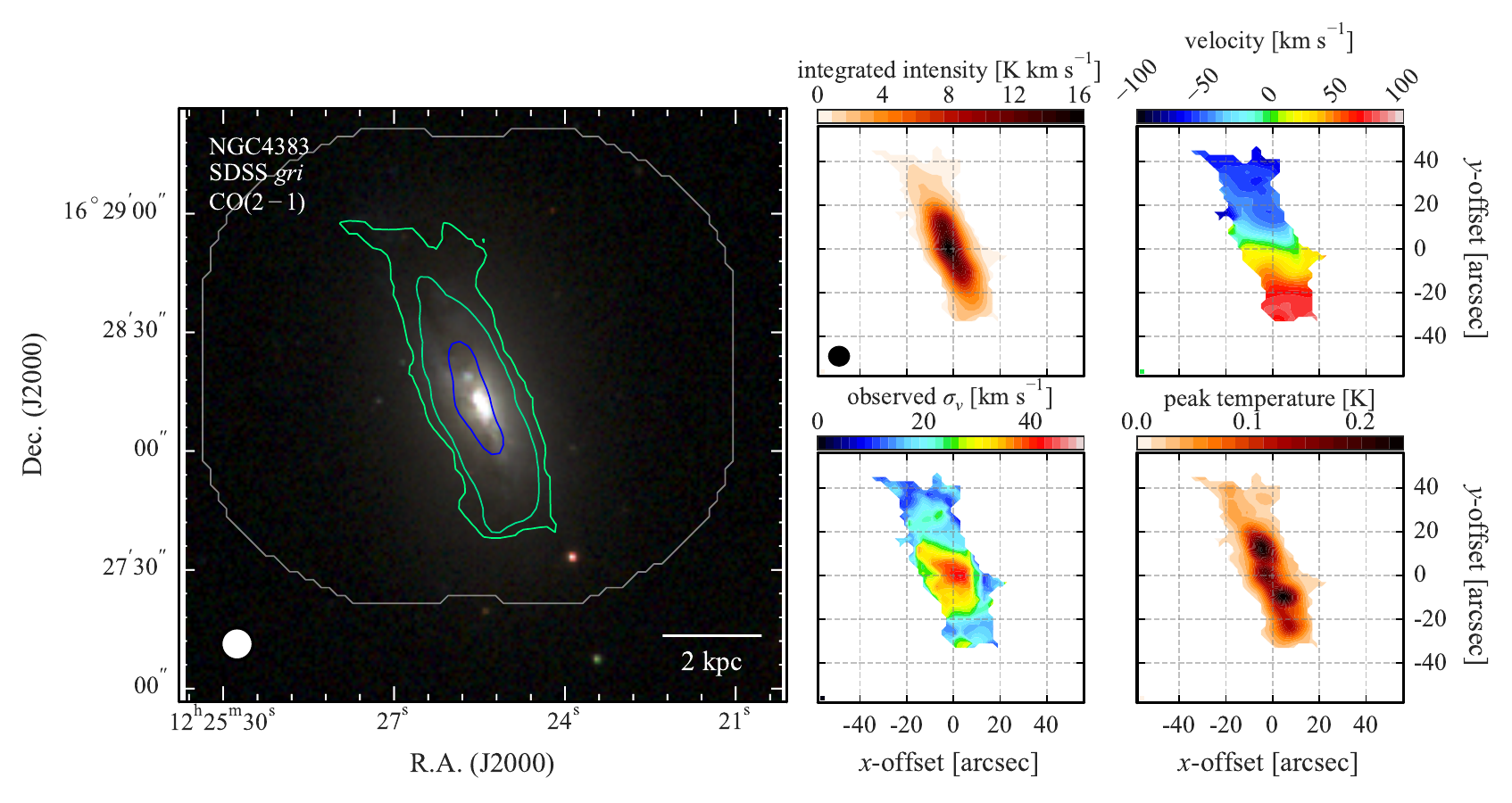}
   \figsetgrpnote{As in Figure 4.1.}
   \figsetgrpend

   \figsetgrpstart
   \figsetgrpnum{4.18}
   \figsetgrptitle{VERTICO CO($2-1$) data products for NGC4388}
   \figsetplot{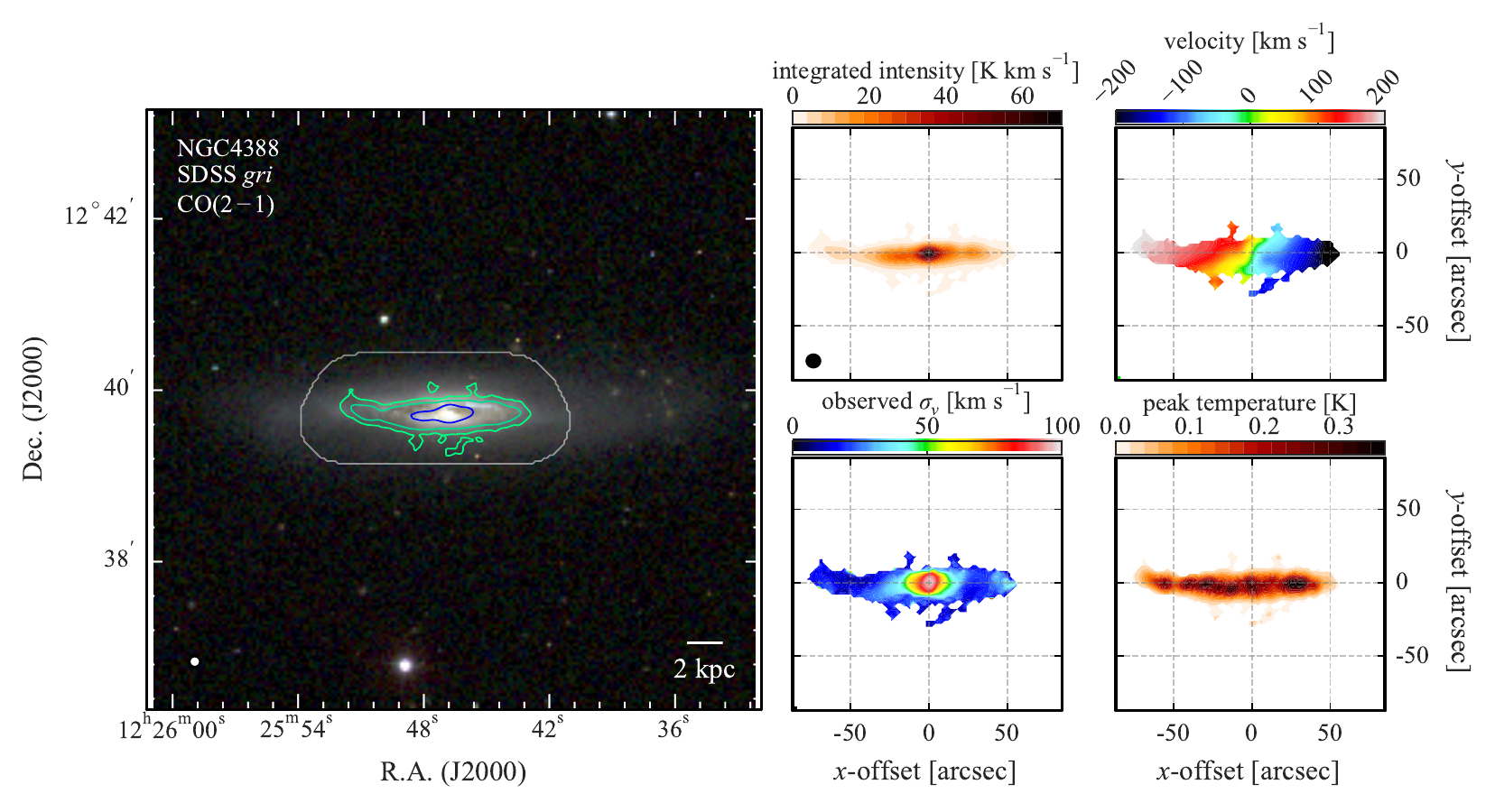}
   \figsetgrpnote{As in Figure 4.1.}
   \figsetgrpend

   \figsetgrpstart
   \figsetgrpnum{4.19}
   \figsetgrptitle{VERTICO CO($2-1$) data products for NGC4394}
   \figsetplot{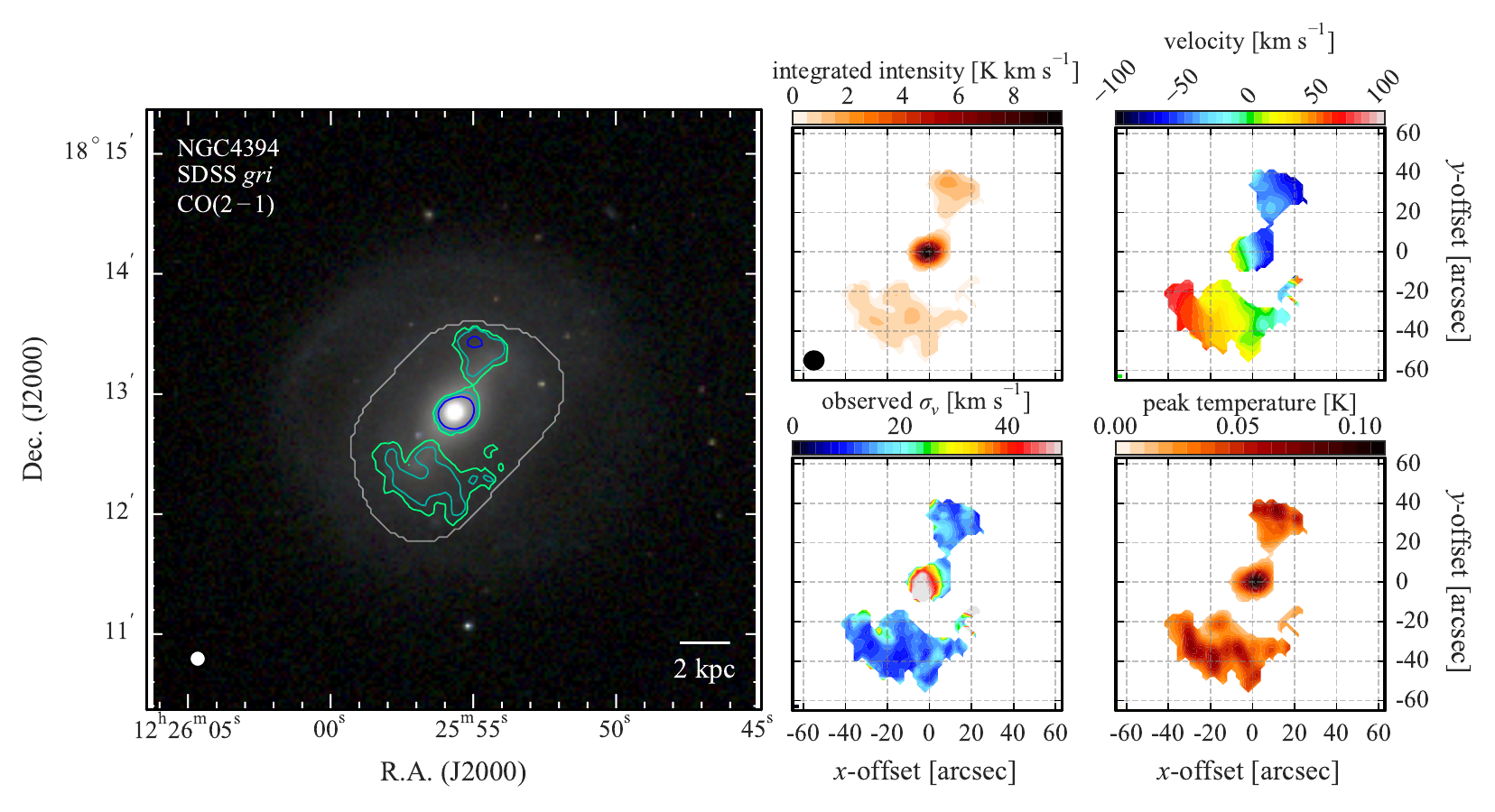}
   \figsetgrpnote{As in Figure 4.1.}
   \figsetgrpend

   \figsetgrpstart
   \figsetgrpnum{4.20}
   \figsetgrptitle{VERTICO CO($2-1$) data products for NGC4396}
   \figsetplot{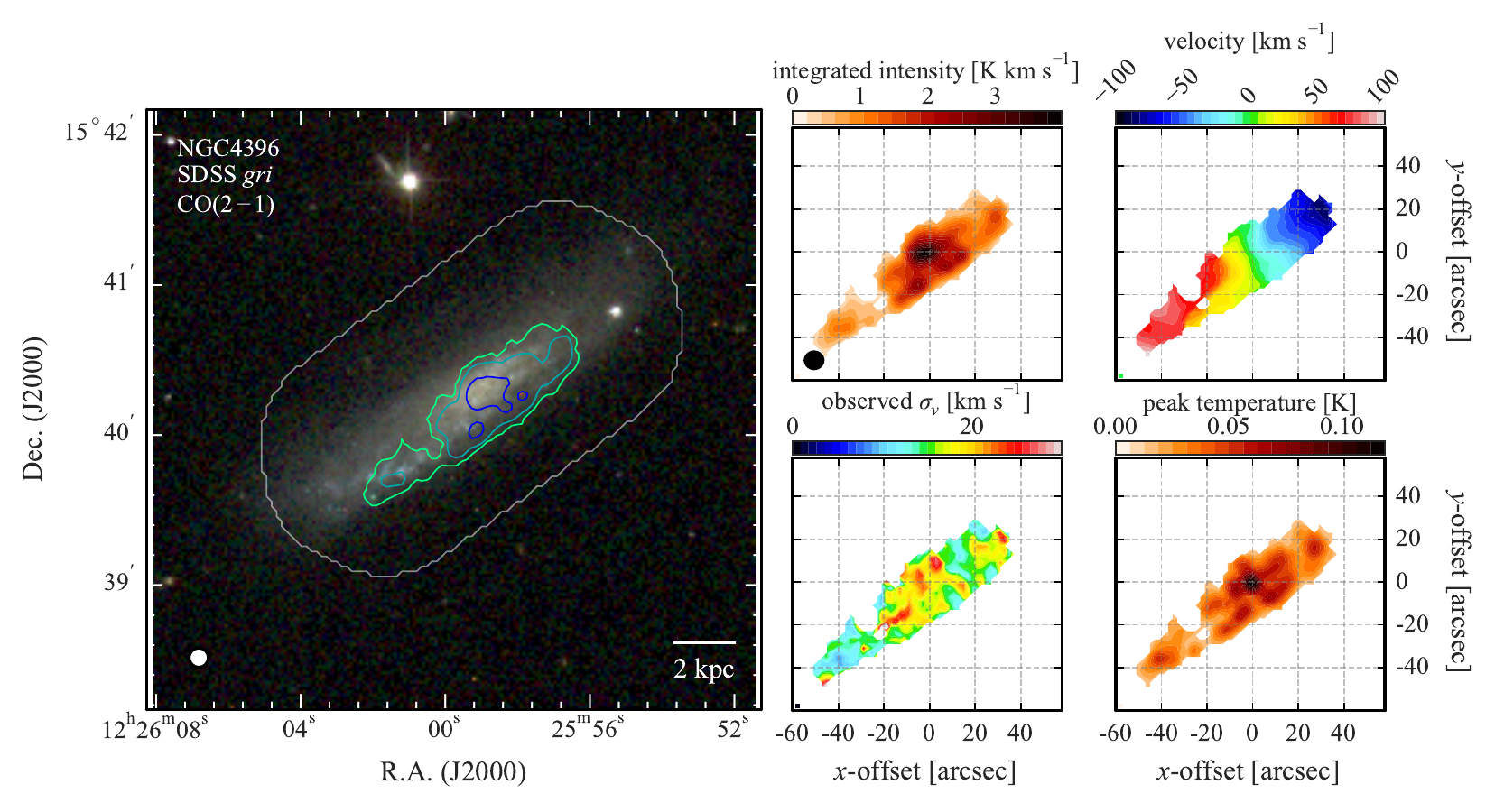}
   \figsetgrpnote{As in Figure 4.1.}
   \figsetgrpend

   \figsetgrpstart
   \figsetgrpnum{4.21}
   \figsetgrptitle{VERTICO CO($2-1$) data products for NGC4402}
   \figsetplot{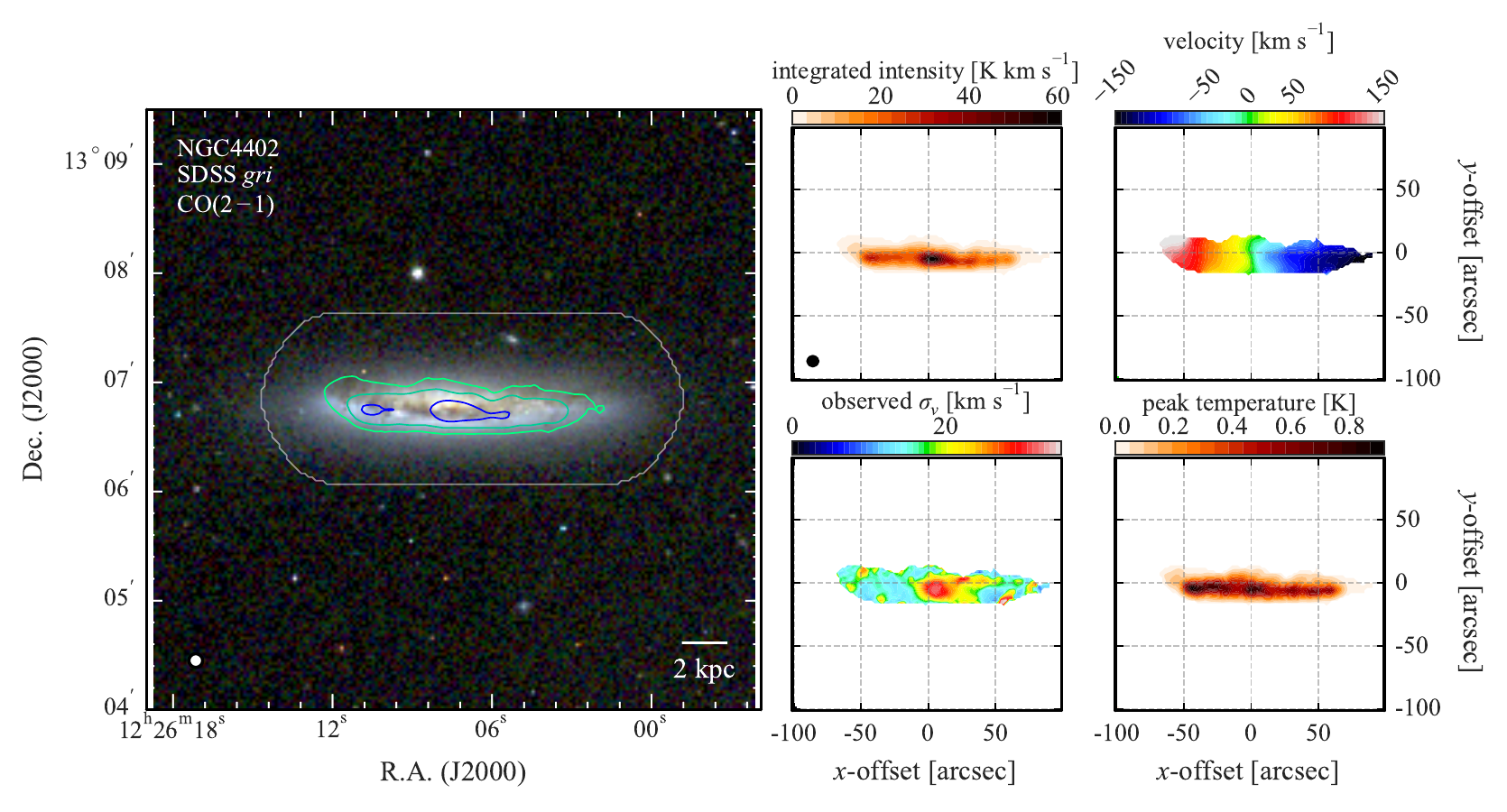}
   \figsetgrpnote{As in Figure 4.1.}
   \figsetgrpend

   \figsetgrpstart
   \figsetgrpnum{4.22}
   \figsetgrptitle{VERTICO CO($2-1$) data products for NGC4405}
   \figsetplot{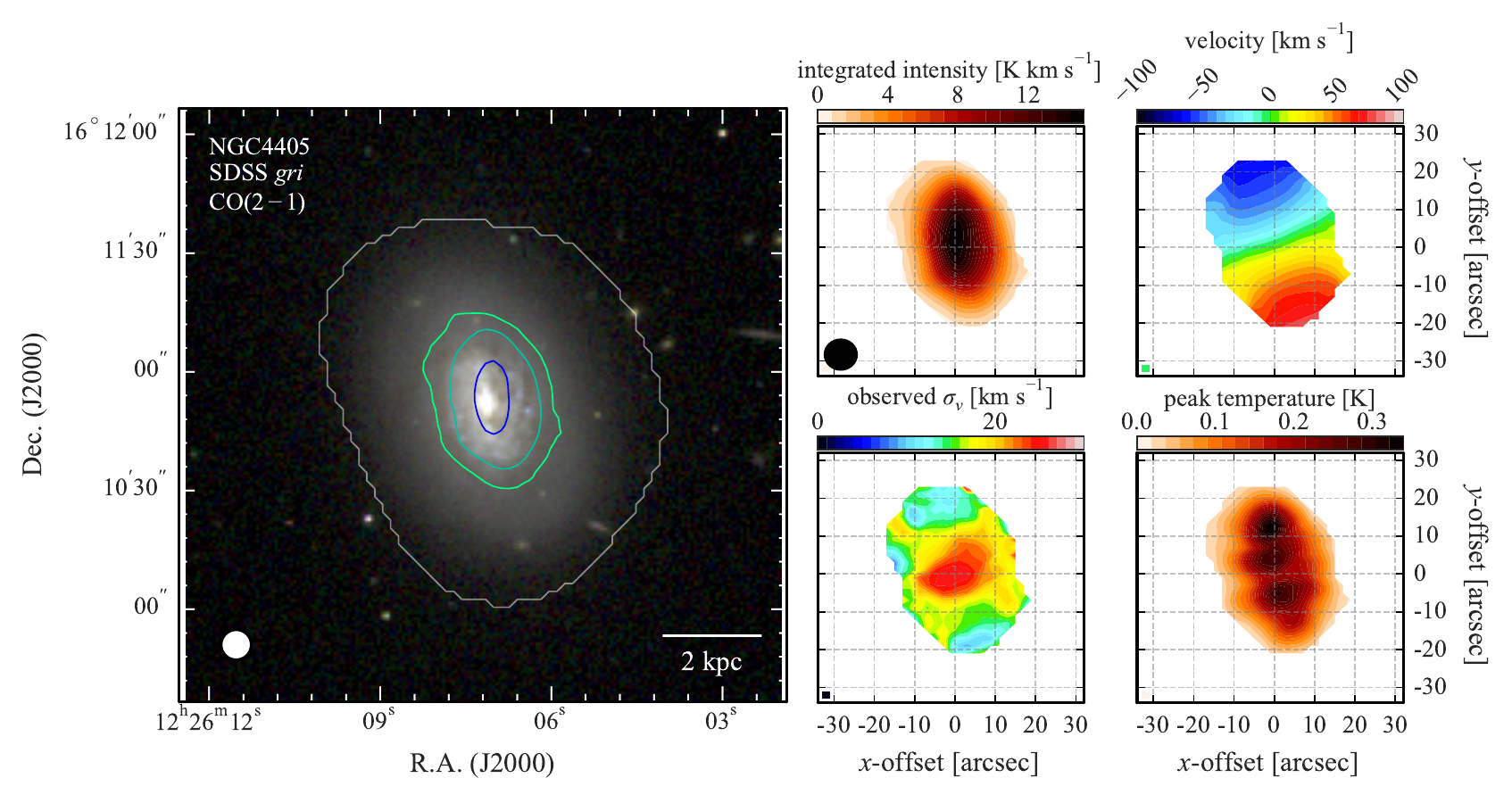}
   \figsetgrpnote{As in Figure 4.1.}
   \figsetgrpend

   \figsetgrpstart
   \figsetgrpnum{4.23}
   \figsetgrptitle{VERTICO CO($2-1$) data products for NGC4419}
   \figsetplot{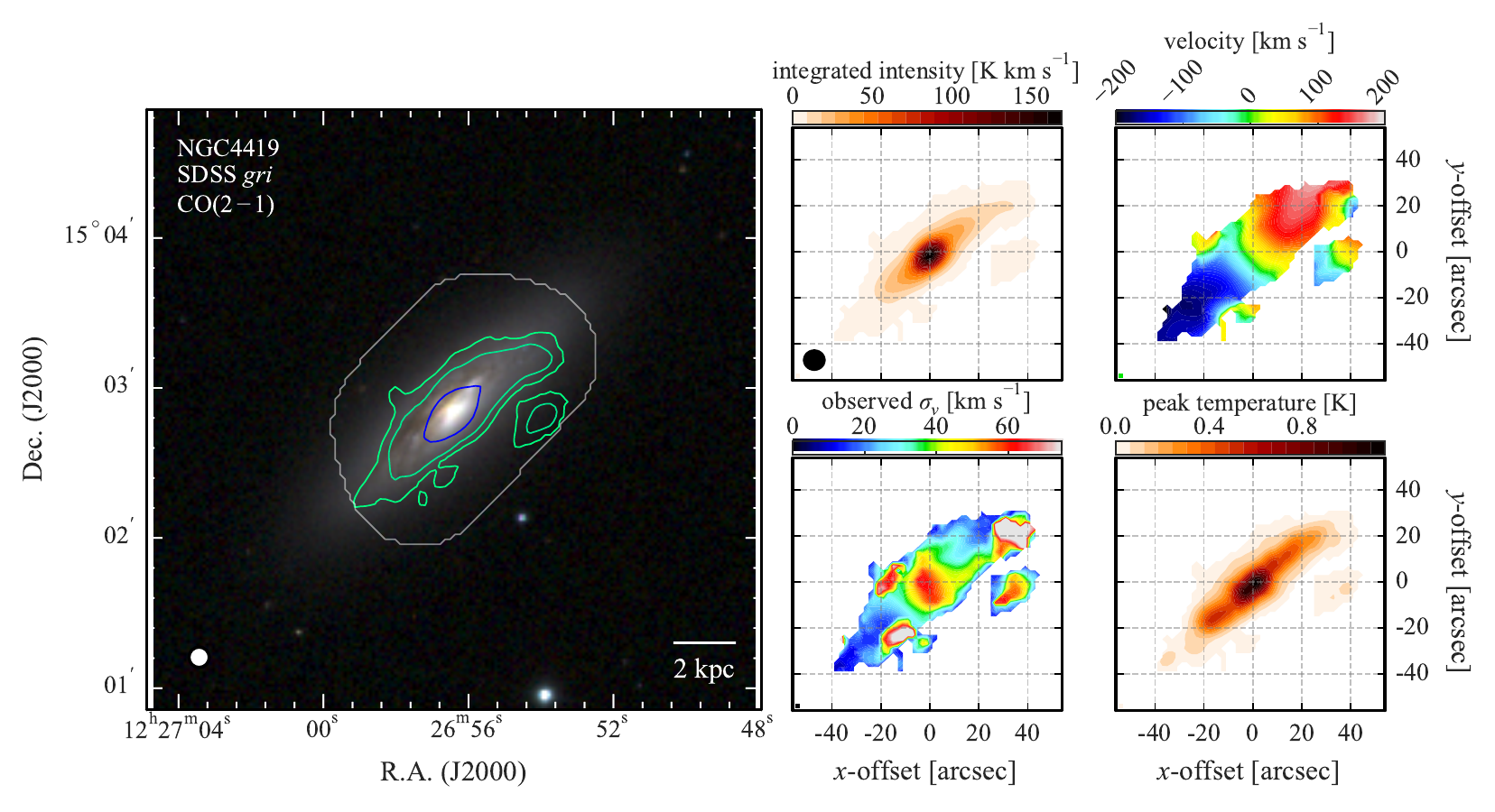}
   \figsetgrpnote{As in Figure 4.1.}
   \figsetgrpend

   \figsetgrpstart
   \figsetgrpnum{4.24}
   \figsetgrptitle{VERTICO CO($2-1$) data products for NGC4424}
   \figsetplot{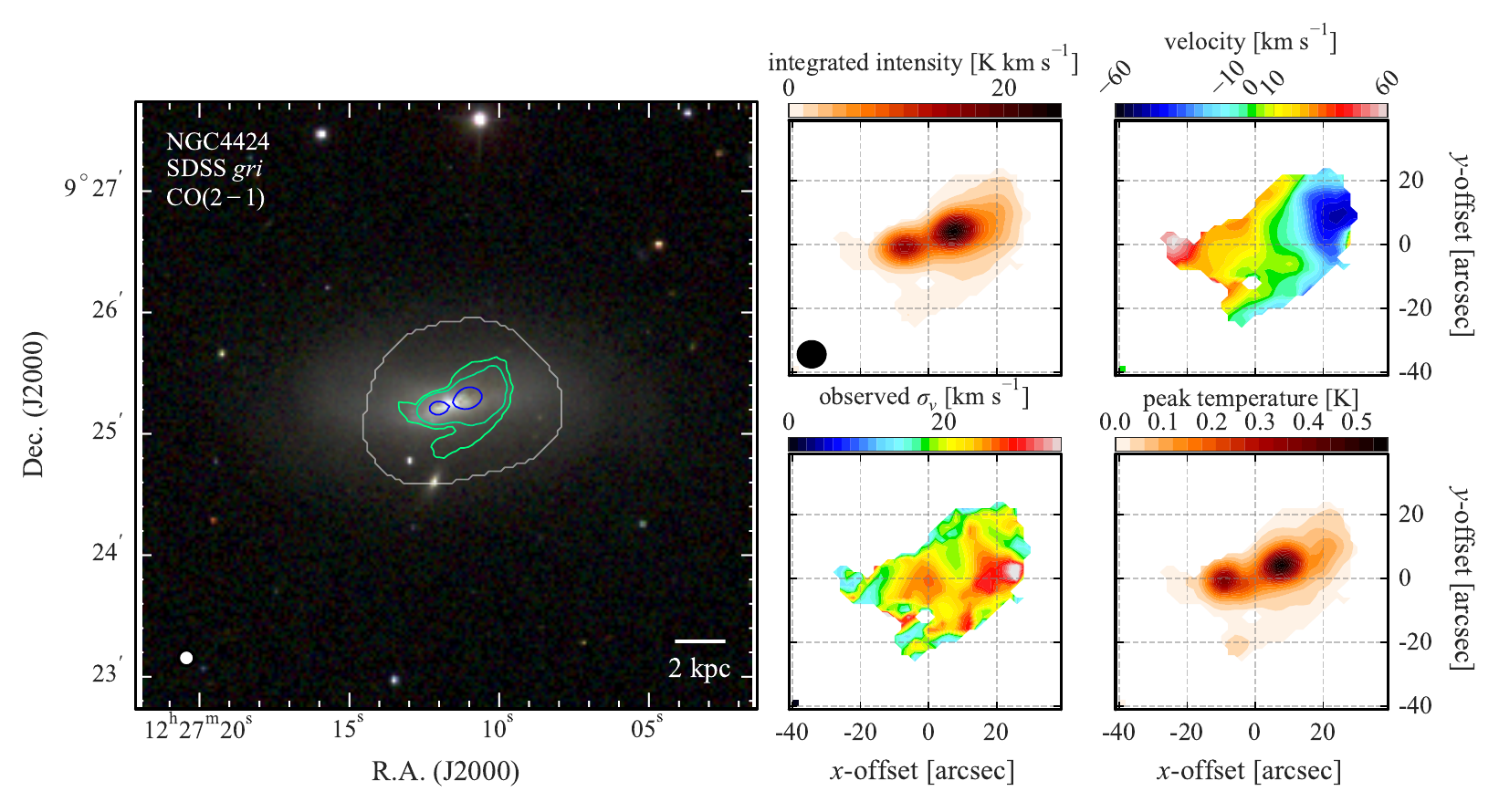}
   \figsetgrpnote{As in Figure 4.1.}
   \figsetgrpend

   \figsetgrpstart
   \figsetgrpnum{4.25}
   \figsetgrptitle{VERTICO CO($2-1$) data products for NGC4450}
   \figsetplot{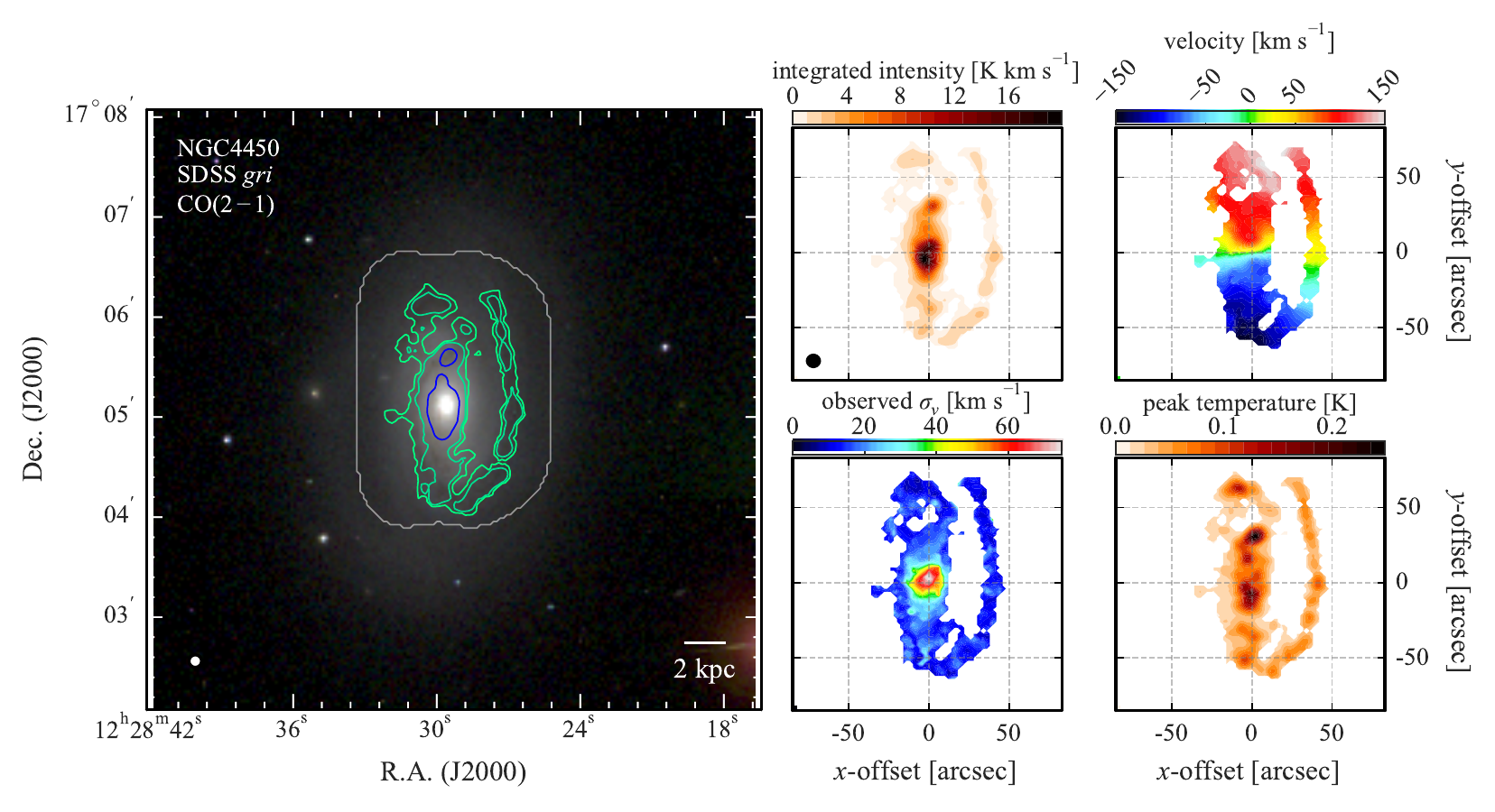}
   \figsetgrpnote{As in Figure 4.1.}
   \figsetgrpend

   \figsetgrpstart
   \figsetgrpnum{4.26}
   \figsetgrptitle{VERTICO CO($2-1$) data products for NGC4457}
   \figsetplot{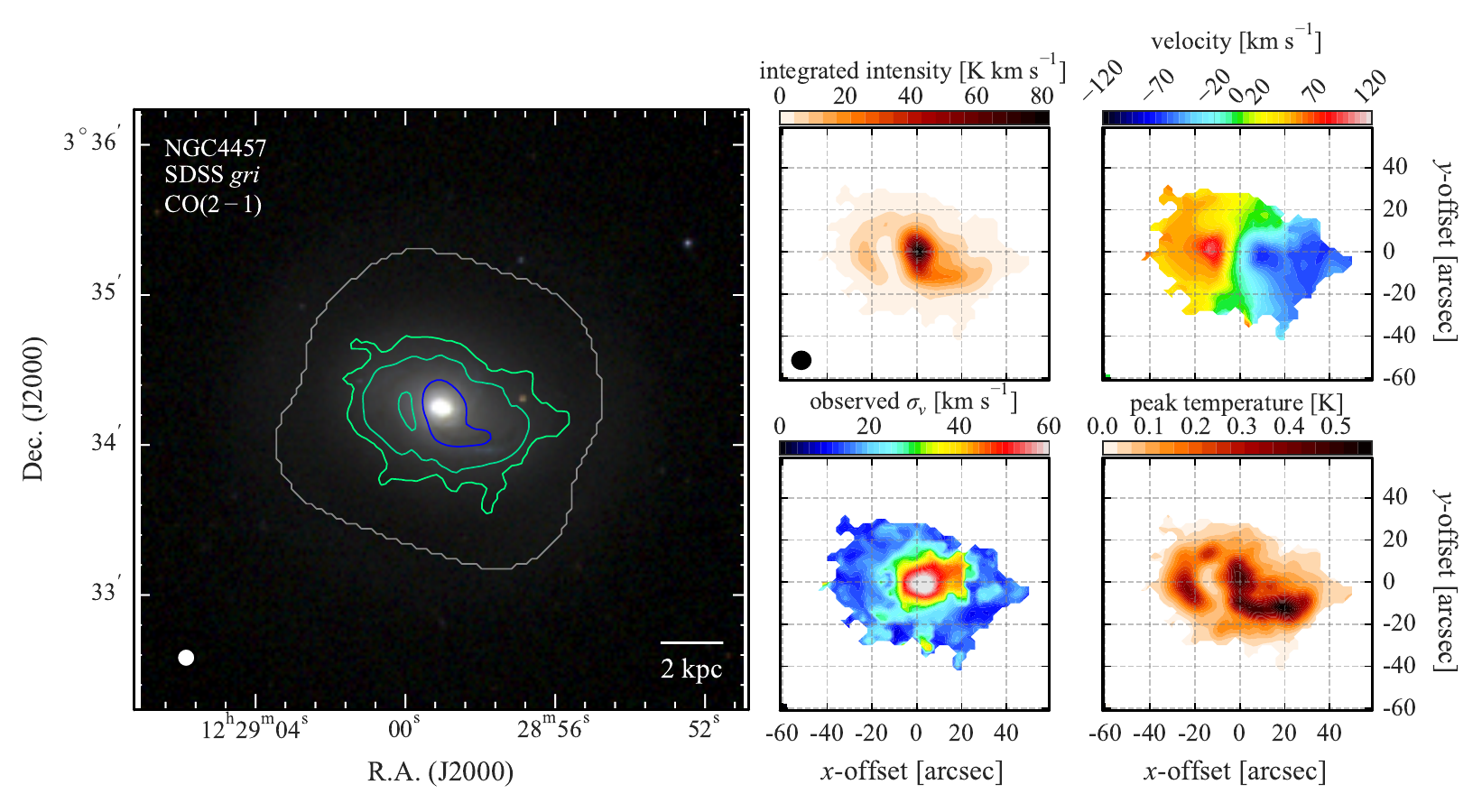}
   \figsetgrpnote{As in Figure 4.1.}
   \figsetgrpend

   \figsetgrpstart
   \figsetgrpnum{4.27}
   \figsetgrptitle{VERTICO CO($2-1$) data products for NGC4501}
   \figsetplot{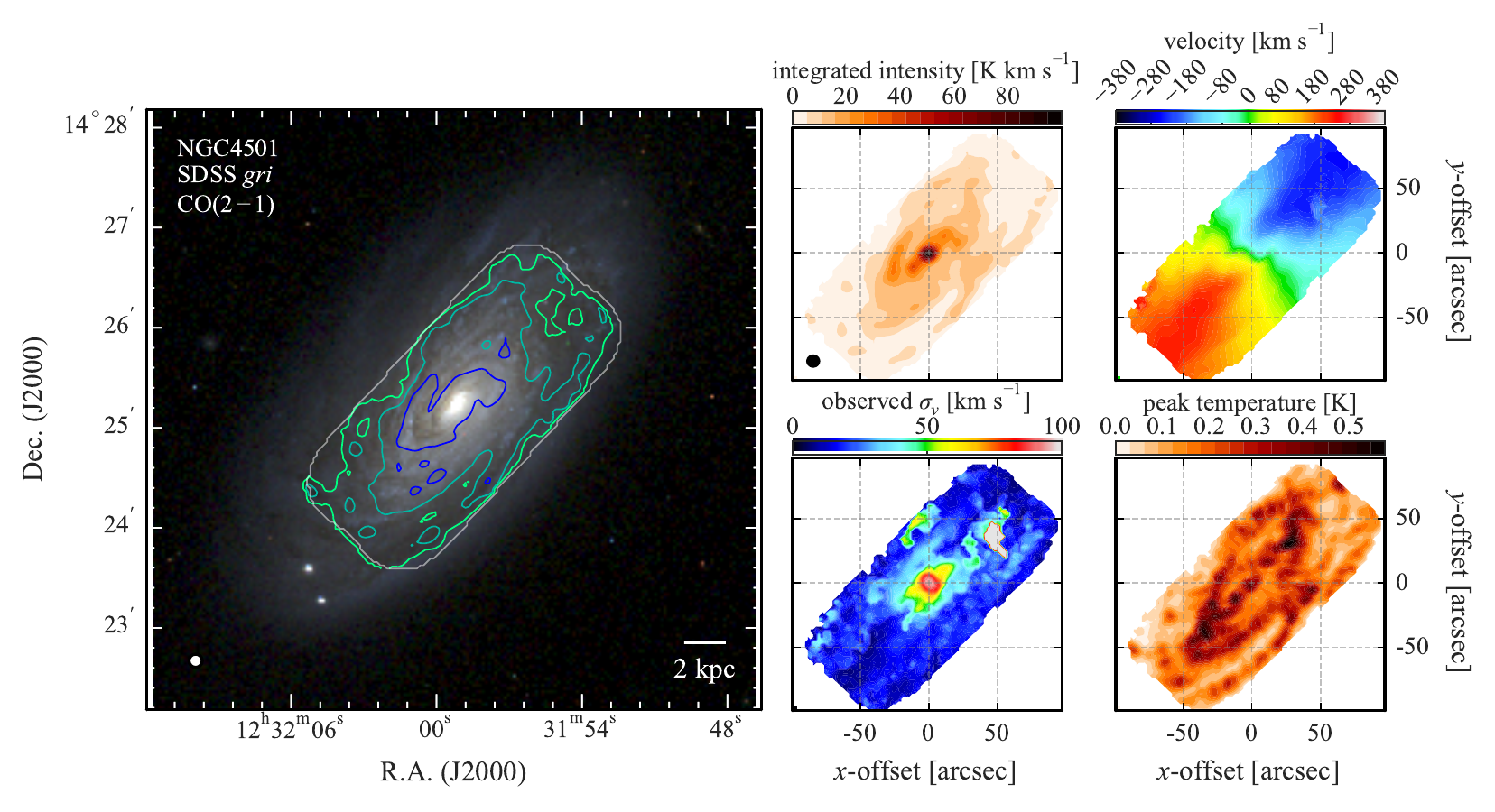}
   \figsetgrpnote{As in Figure 4.1.}
   \figsetgrpend

   \figsetgrpstart
   \figsetgrpnum{4.28}
   \figsetgrptitle{VERTICO CO($2-1$) data products for NGC4522}
   \figsetplot{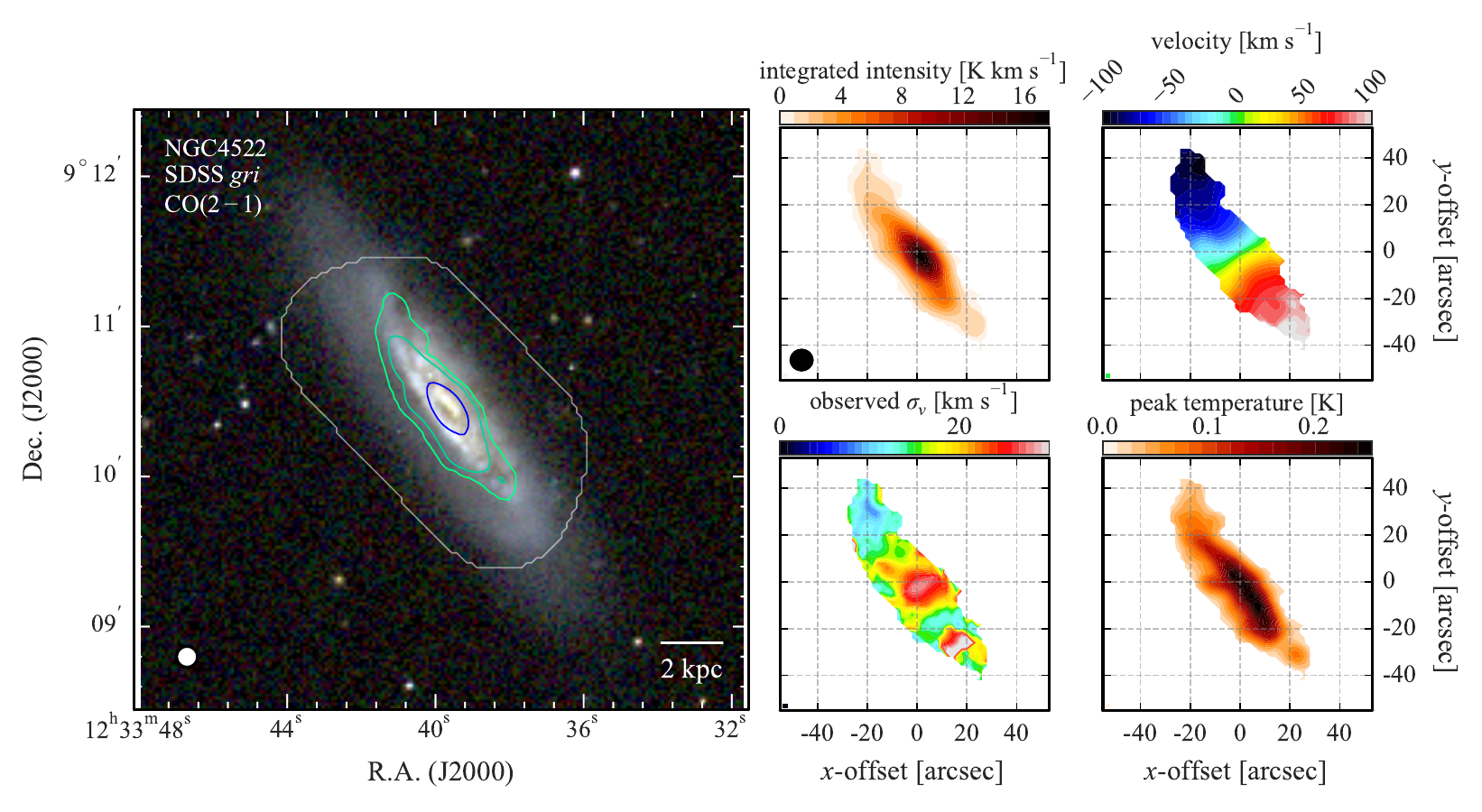}
   \figsetgrpnote{As in Figure 4.1.}
   \figsetgrpend

   \figsetgrpstart
   \figsetgrpnum{4.29}
   \figsetgrptitle{VERTICO CO($2-1$) data products for NGC4532}
   \figsetplot{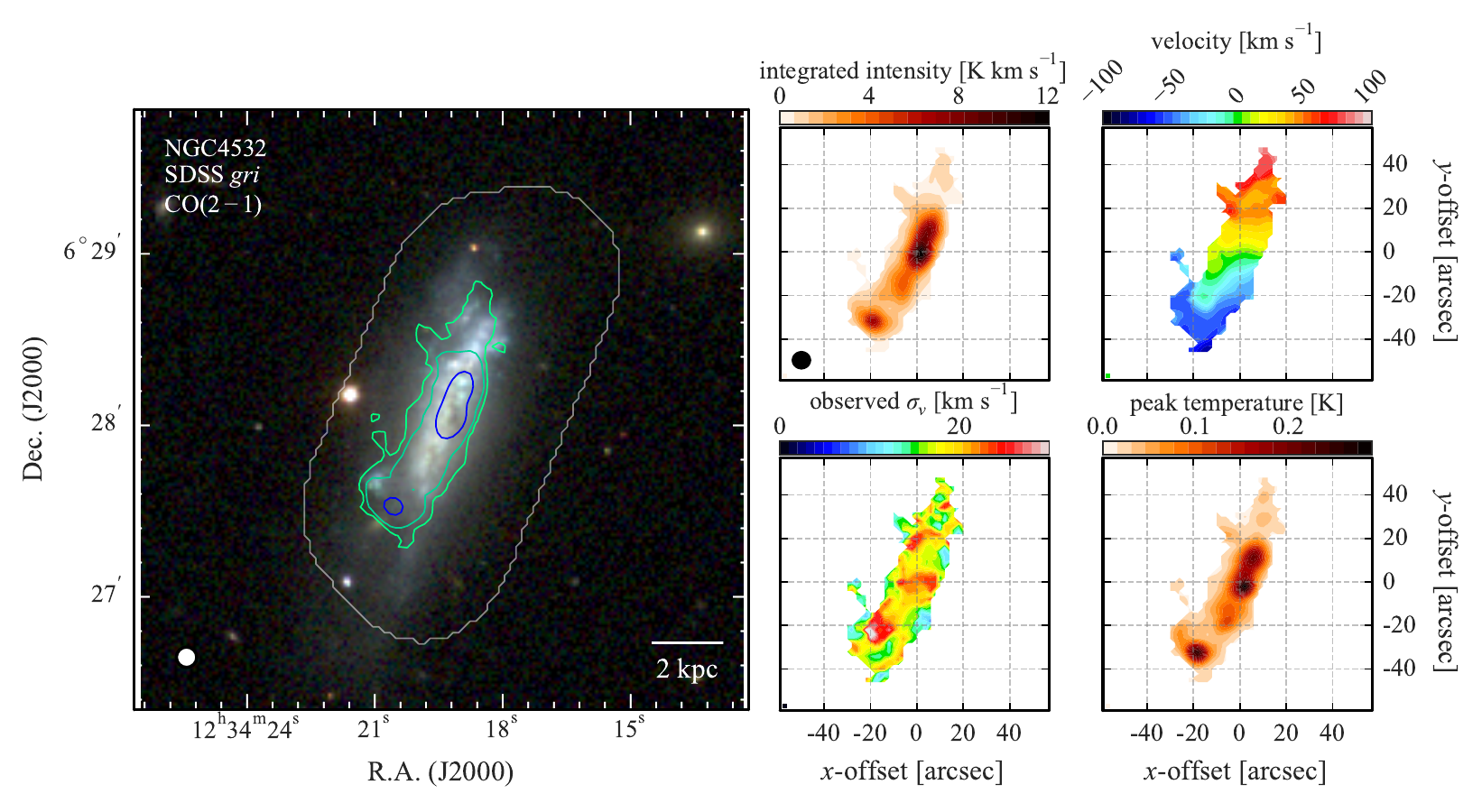}
   \figsetgrpnote{As in Figure 4.1.}
   \figsetgrpend

   \figsetgrpstart
   \figsetgrpnum{4.30}
   \figsetgrptitle{VERTICO CO($2-1$) data products for NGC4533}
   \figsetplot{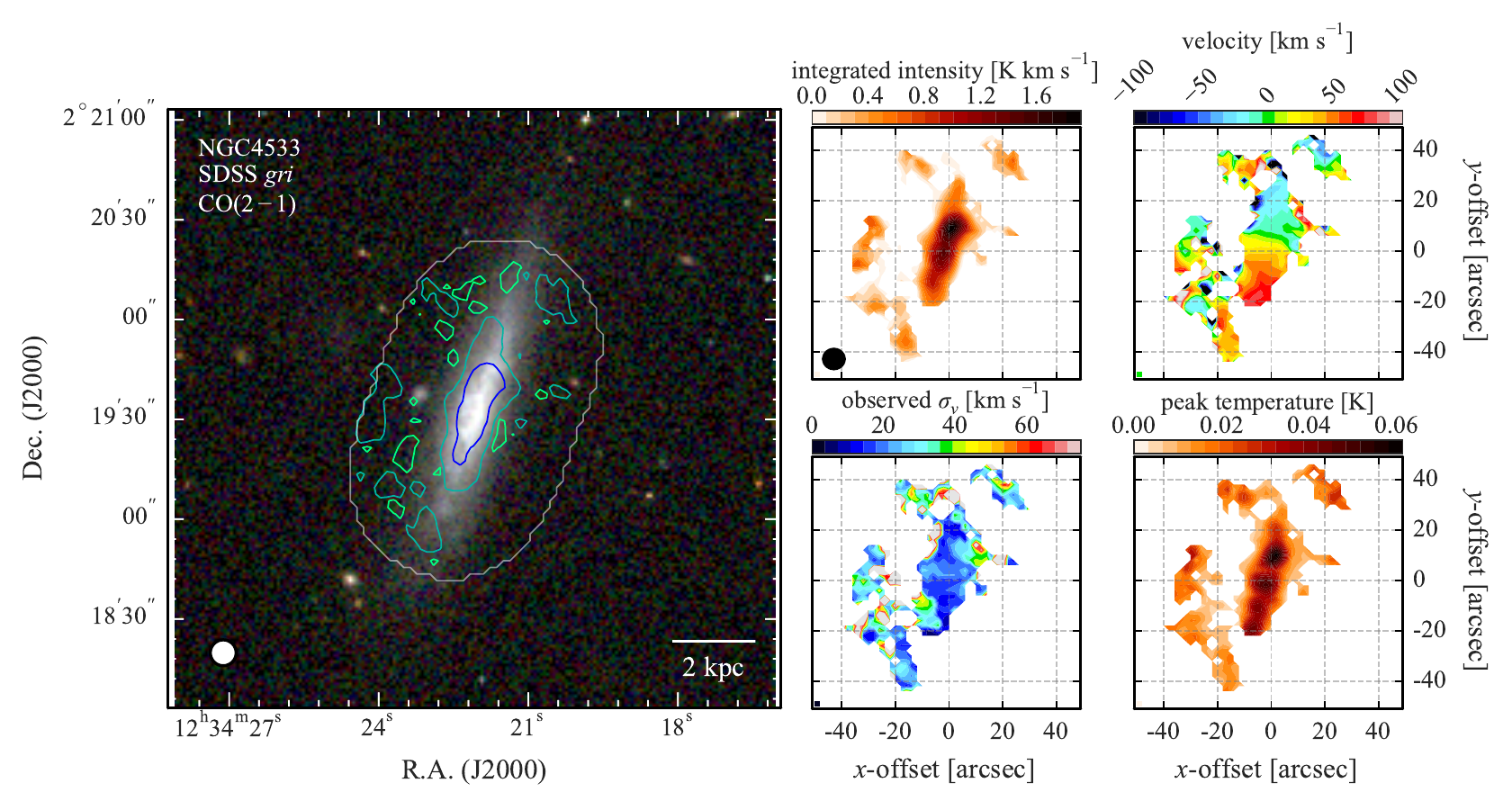}
   \figsetgrpnote{As in Figure 4.1.}
   \figsetgrpend

   \figsetgrpstart
   \figsetgrpnum{4.31}
   \figsetgrptitle{VERTICO CO($2-1$) data products for NGC4535}
   \figsetplot{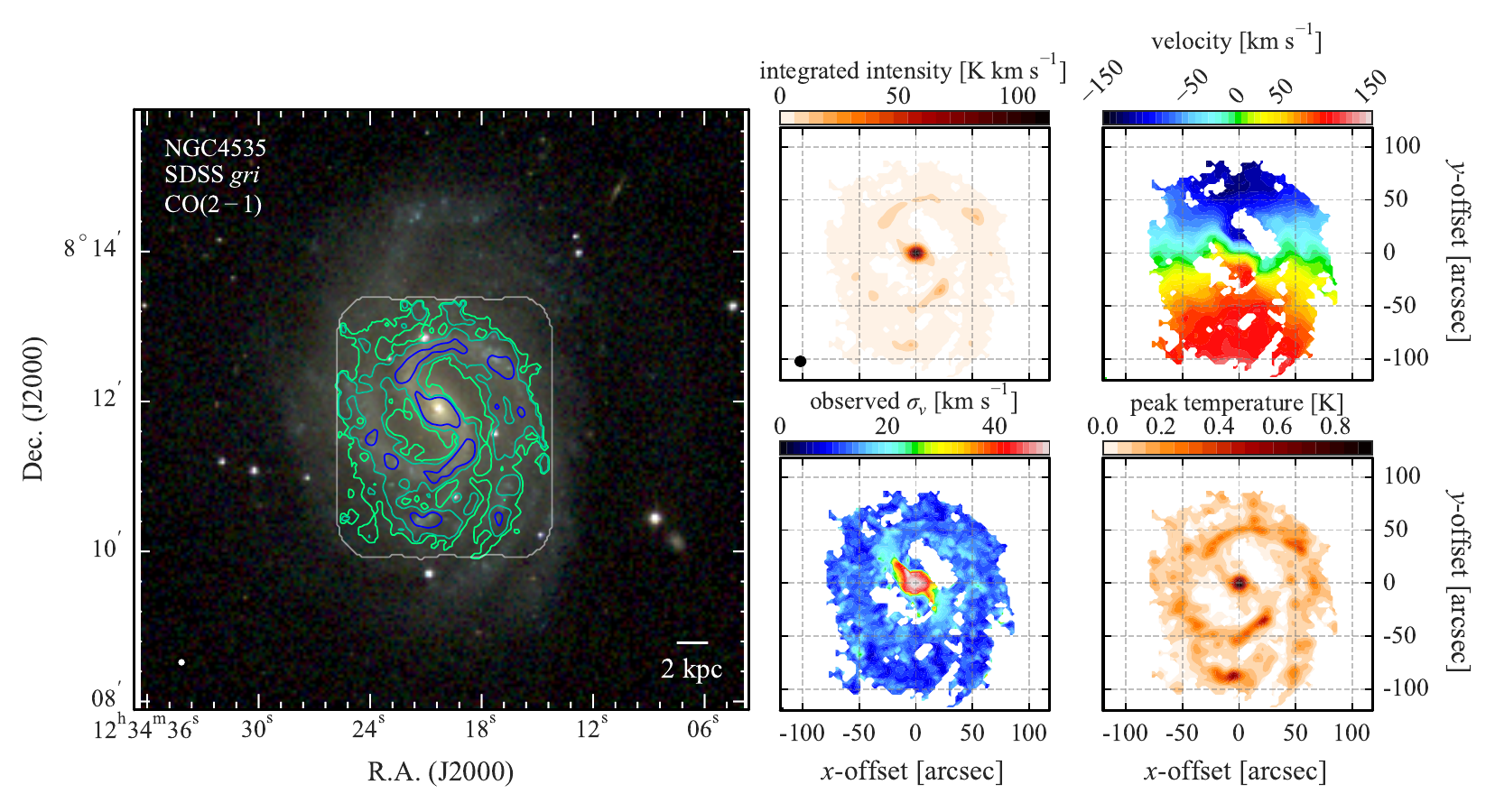}
   \figsetgrpnote{As in Figure 4.1.}
   \figsetgrpend

   \figsetgrpstart
   \figsetgrpnum{4.32}
   \figsetgrptitle{VERTICO CO($2-1$) data products for NGC4536}
   \figsetplot{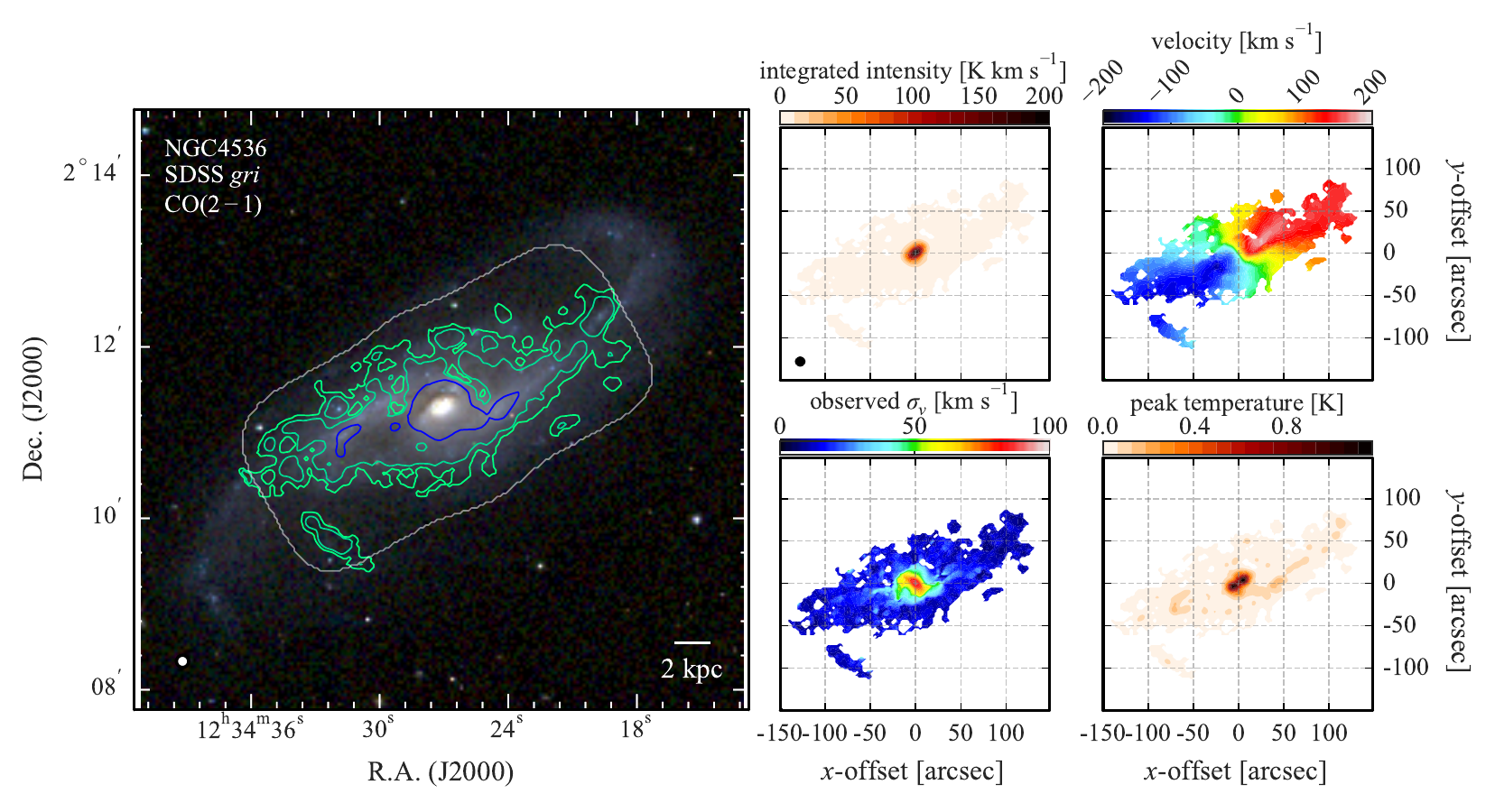}
   \figsetgrpnote{As in Figure 4.1.}
   \figsetgrpend

   \figsetgrpstart
   \figsetgrpnum{4.33}
   \figsetgrptitle{VERTICO CO($2-1$) data products for NGC4548}
   \figsetplot{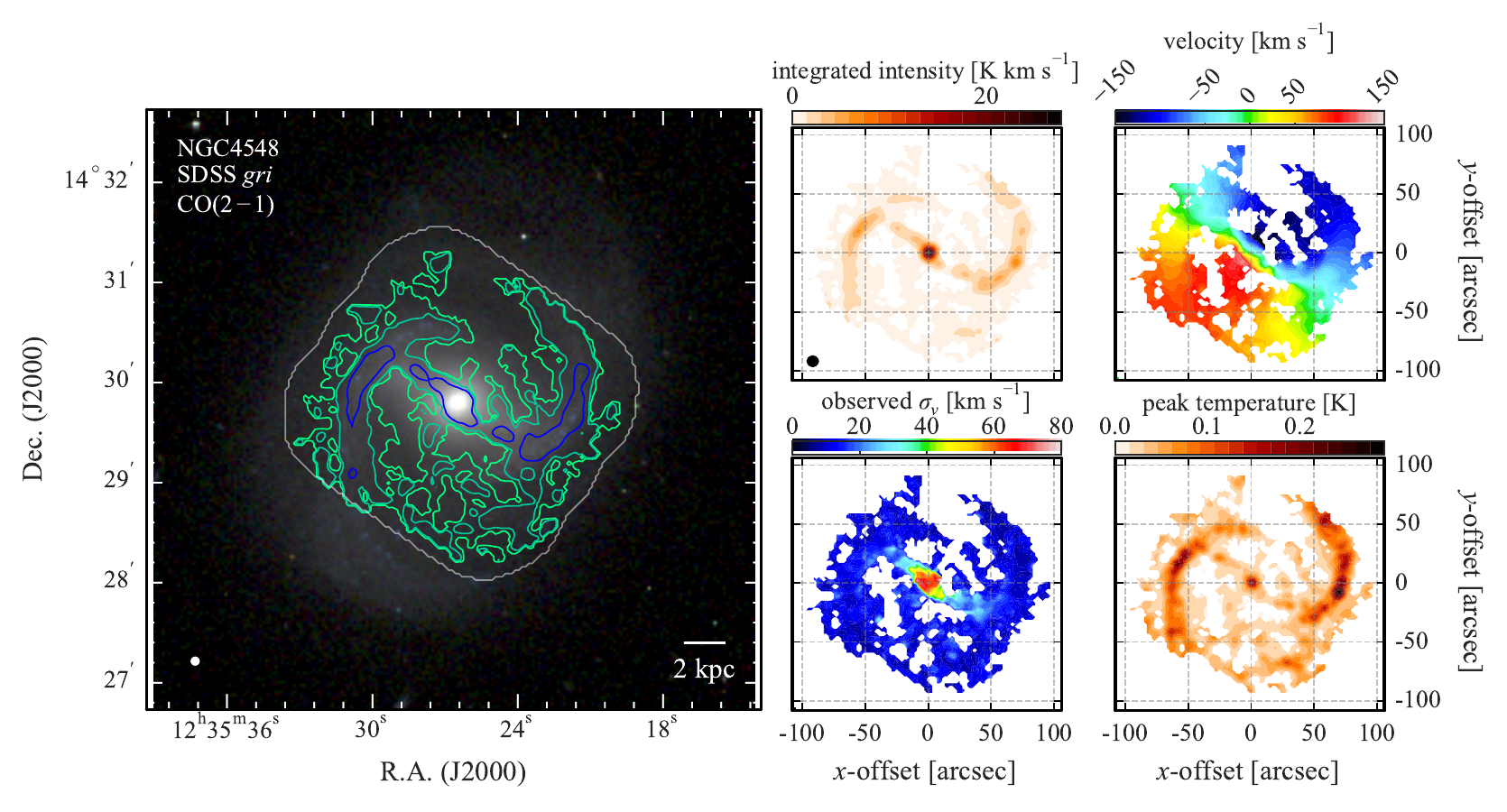}
   \figsetgrpnote{As in Figure 4.1.}
   \figsetgrpend

   \figsetgrpstart
   \figsetgrpnum{4.34}
   \figsetgrptitle{VERTICO CO($2-1$) data products for NGC4561}
   \figsetplot{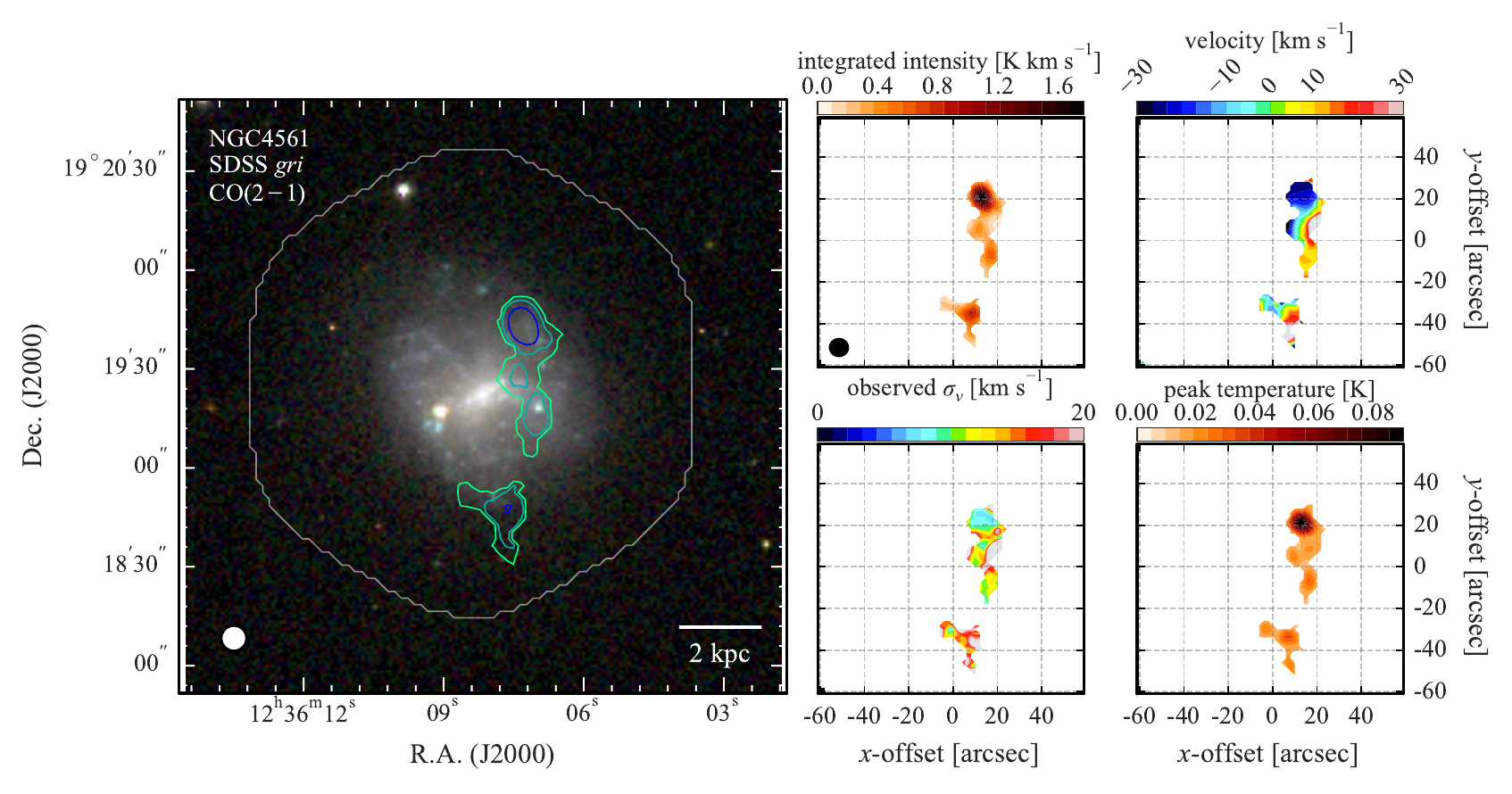}
   \figsetgrpnote{As in Figure 4.1.}
   \figsetgrpend

   \figsetgrpstart
   \figsetgrpnum{4.35}
   \figsetgrptitle{VERTICO CO($2-1$) data products for NGC4567}
   \figsetplot{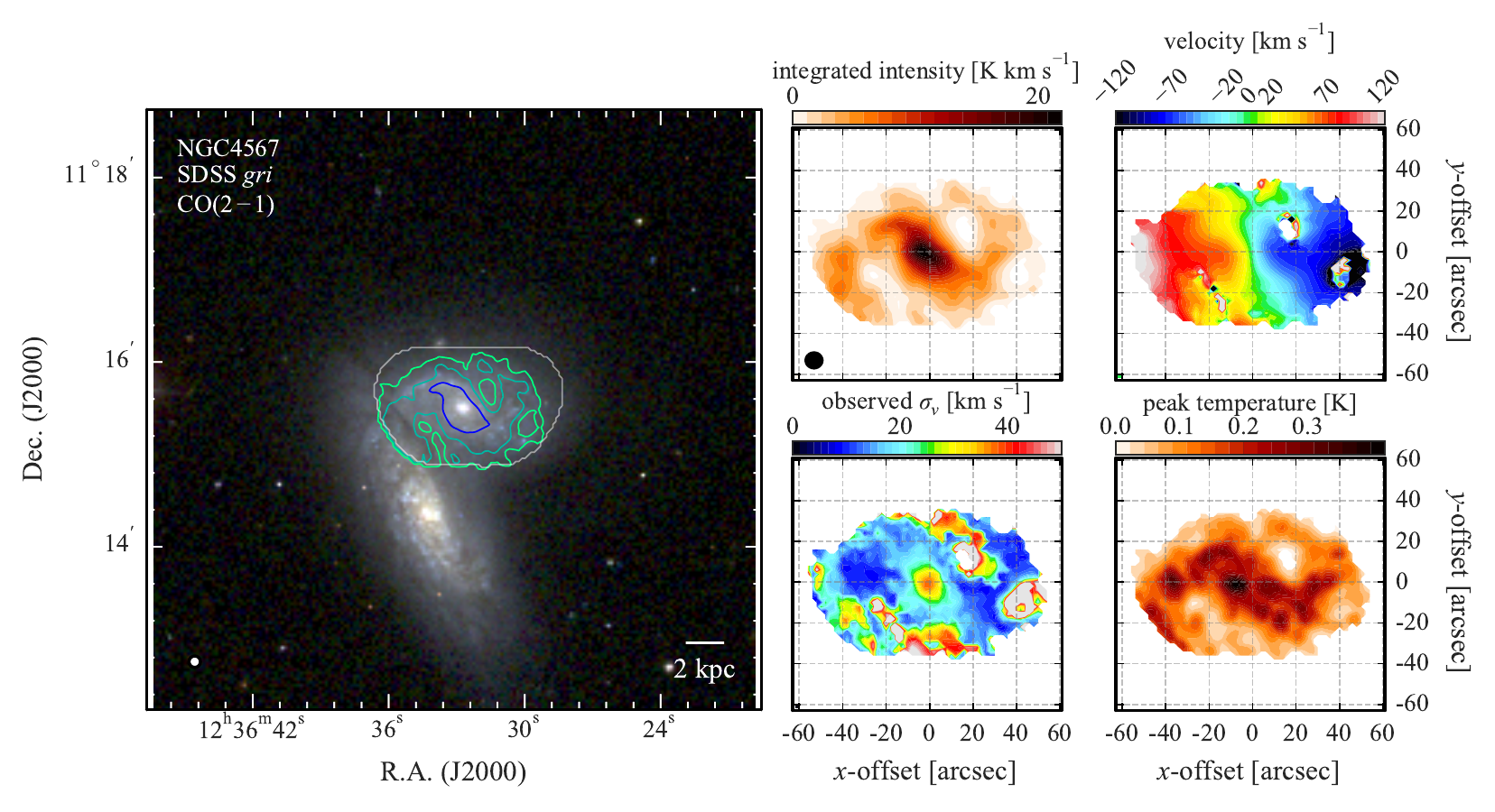}
   \figsetgrpnote{As in Figure 4.1.}
   \figsetgrpend

   \figsetgrpstart
   \figsetgrpnum{4.36}
   \figsetgrptitle{VERTICO CO($2-1$) data products for NGC4568}
   \figsetplot{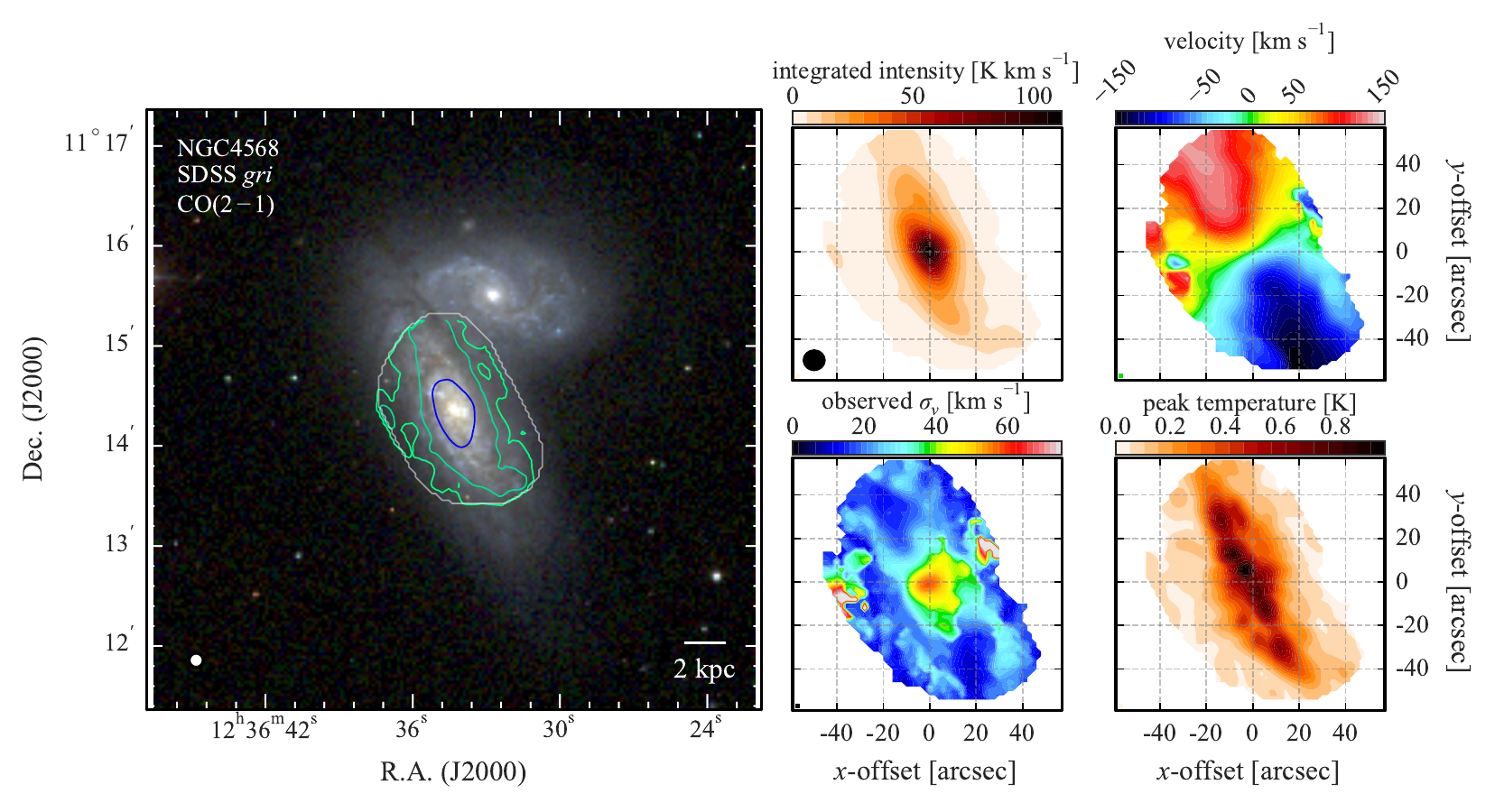}
   \figsetgrpnote{As in Figure 4.1.}
   \figsetgrpend

   \figsetgrpstart
   \figsetgrpnum{4.37}
   \figsetgrptitle{VERTICO CO($2-1$) data products for NGC4569}
   \figsetplot{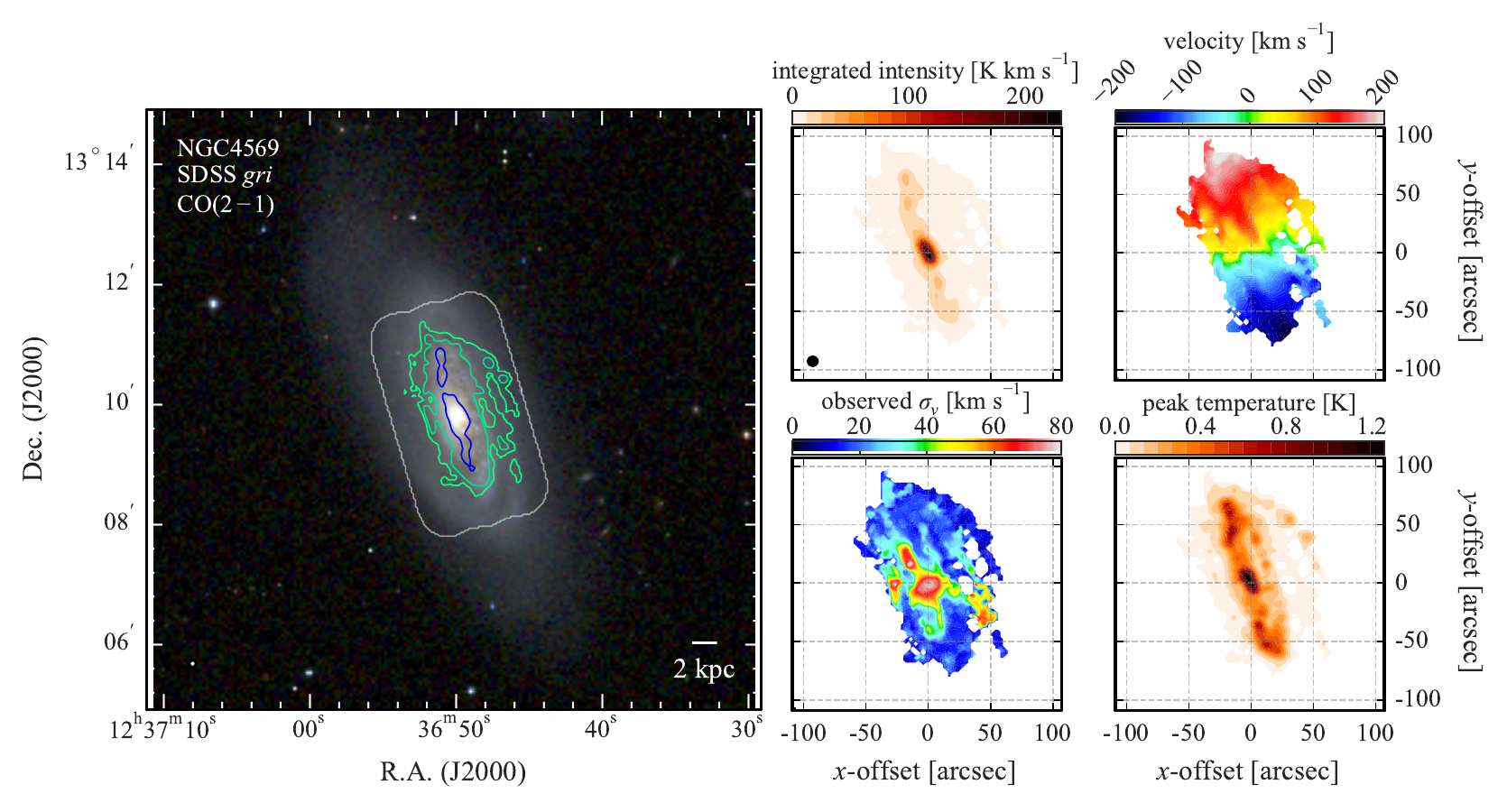}
   \figsetgrpnote{As in Figure 4.1.}
   \figsetgrpend

   \figsetgrpstart
   \figsetgrpnum{4.38}
   \figsetgrptitle{VERTICO CO($2-1$) data products for NGC4579}
   \figsetplot{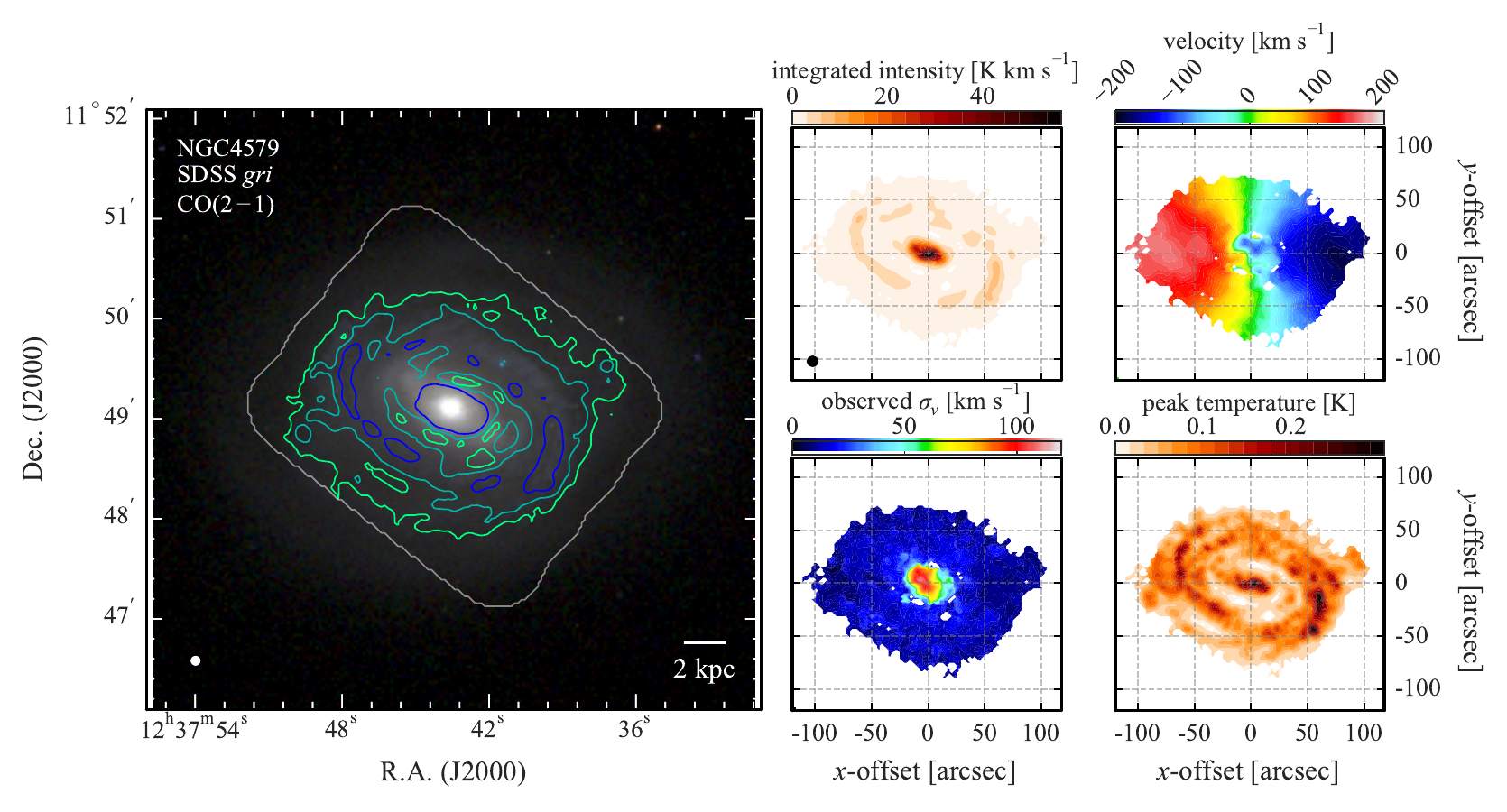}
   \figsetgrpnote{As in Figure 4.1.}
   \figsetgrpend

   \figsetgrpstart
   \figsetgrpnum{4.39}
   \figsetgrptitle{VERTICO CO($2-1$) data products for NGC4580}
   \figsetplot{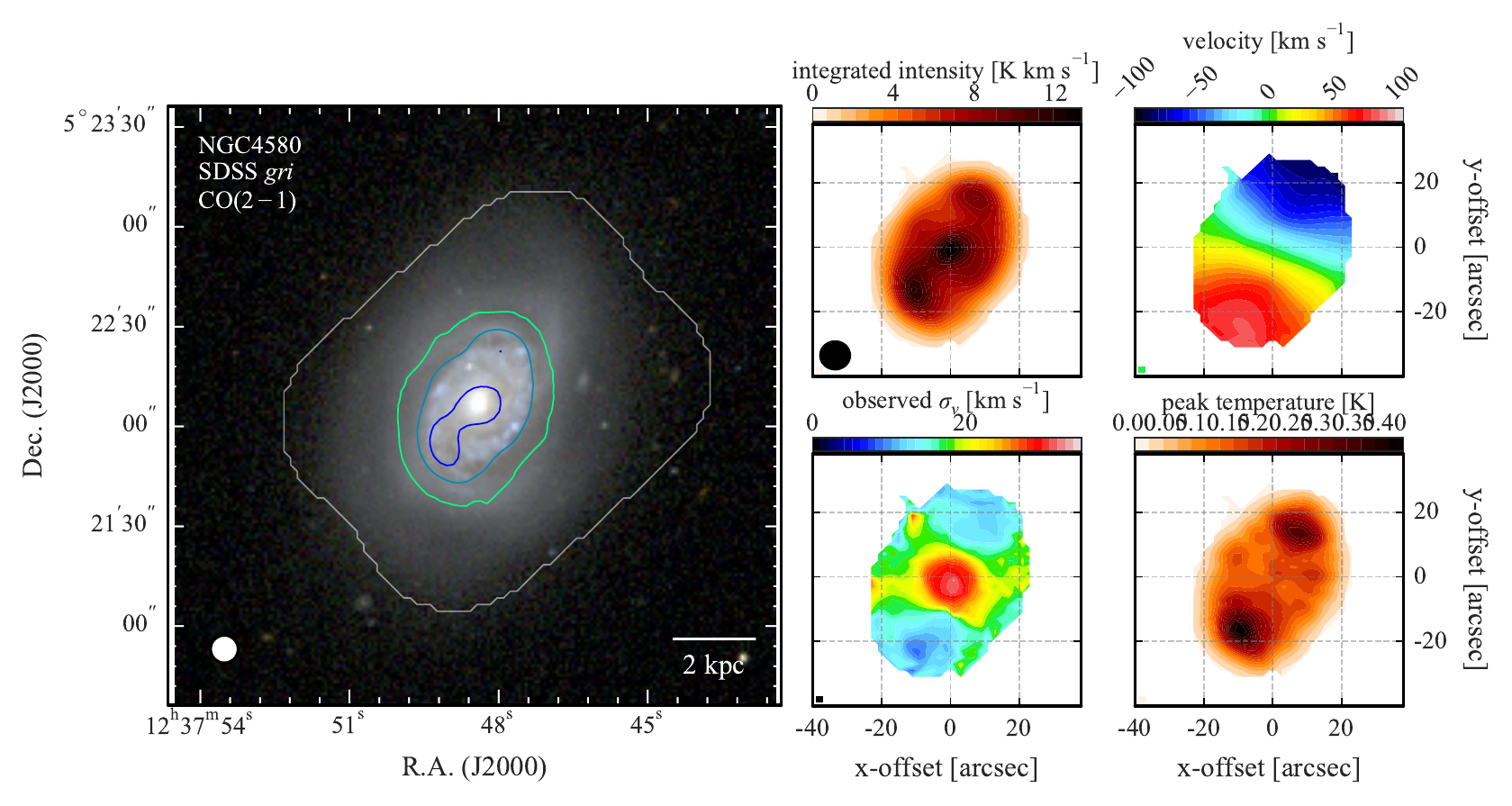}
   \figsetgrpnote{As in Figure 4.1.}
   \figsetgrpend

   \figsetgrpstart
   \figsetgrpnum{4.40}
   \figsetgrptitle{VERTICO CO($2-1$) data products for NGC4606}
   \figsetplot{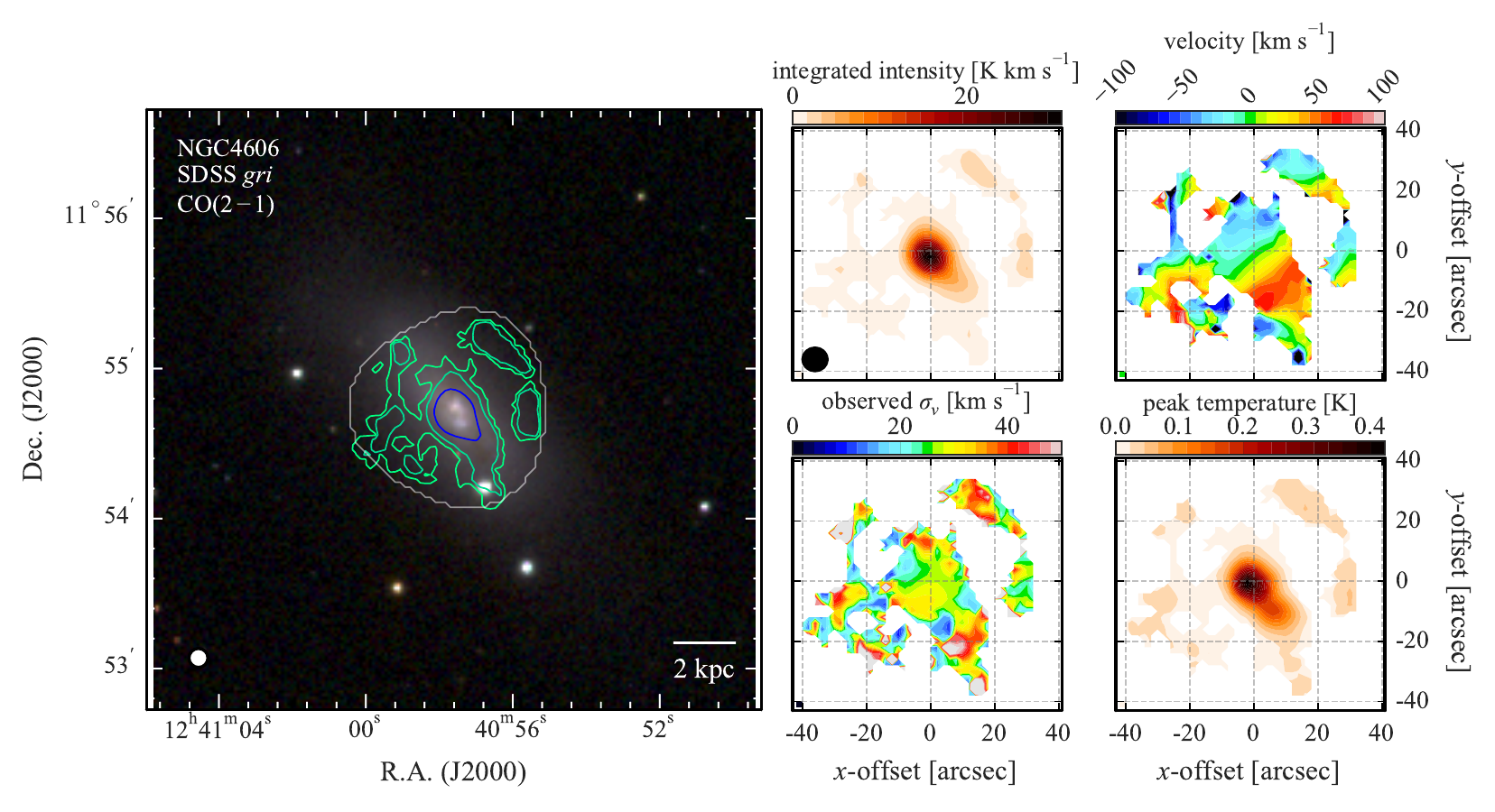}
   \figsetgrpnote{As in Figure 4.1.}
   \figsetgrpend

   \figsetgrpstart
   \figsetgrpnum{4.41}
   \figsetgrptitle{VERTICO CO($2-1$) data products for NGC4607}
   \figsetplot{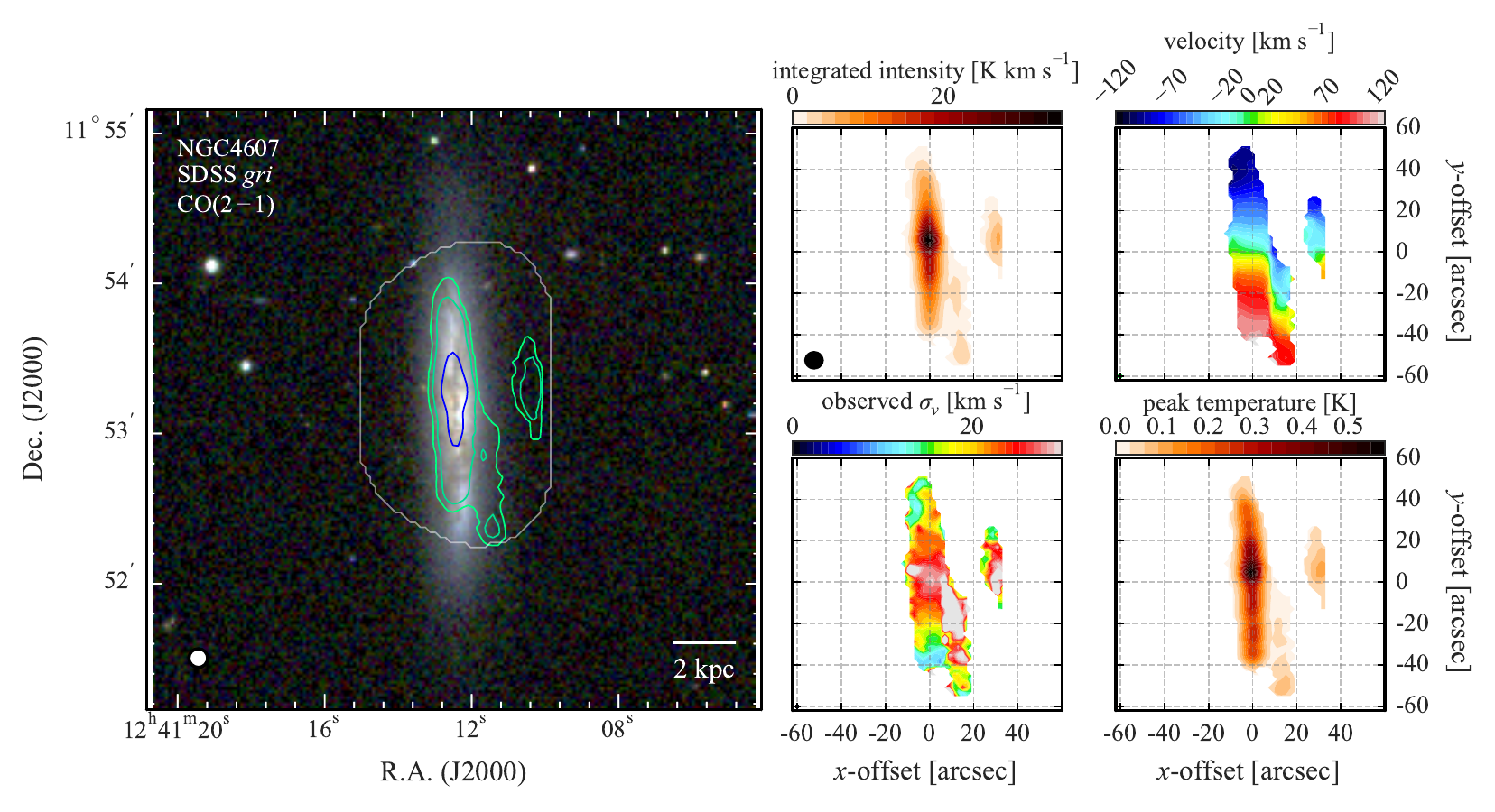}
   \figsetgrpnote{As in Figure 4.1. In the case of NGC4607, we caution against over-interpreting the extended features to the West of the main disk as this is likely related to the PSF pattern of the observations.}
   \figsetgrpend

   \figsetgrpstart
   \figsetgrpnum{4.42}
   \figsetgrptitle{VERTICO CO($2-1$) data products for NGC4651}
   \figsetplot{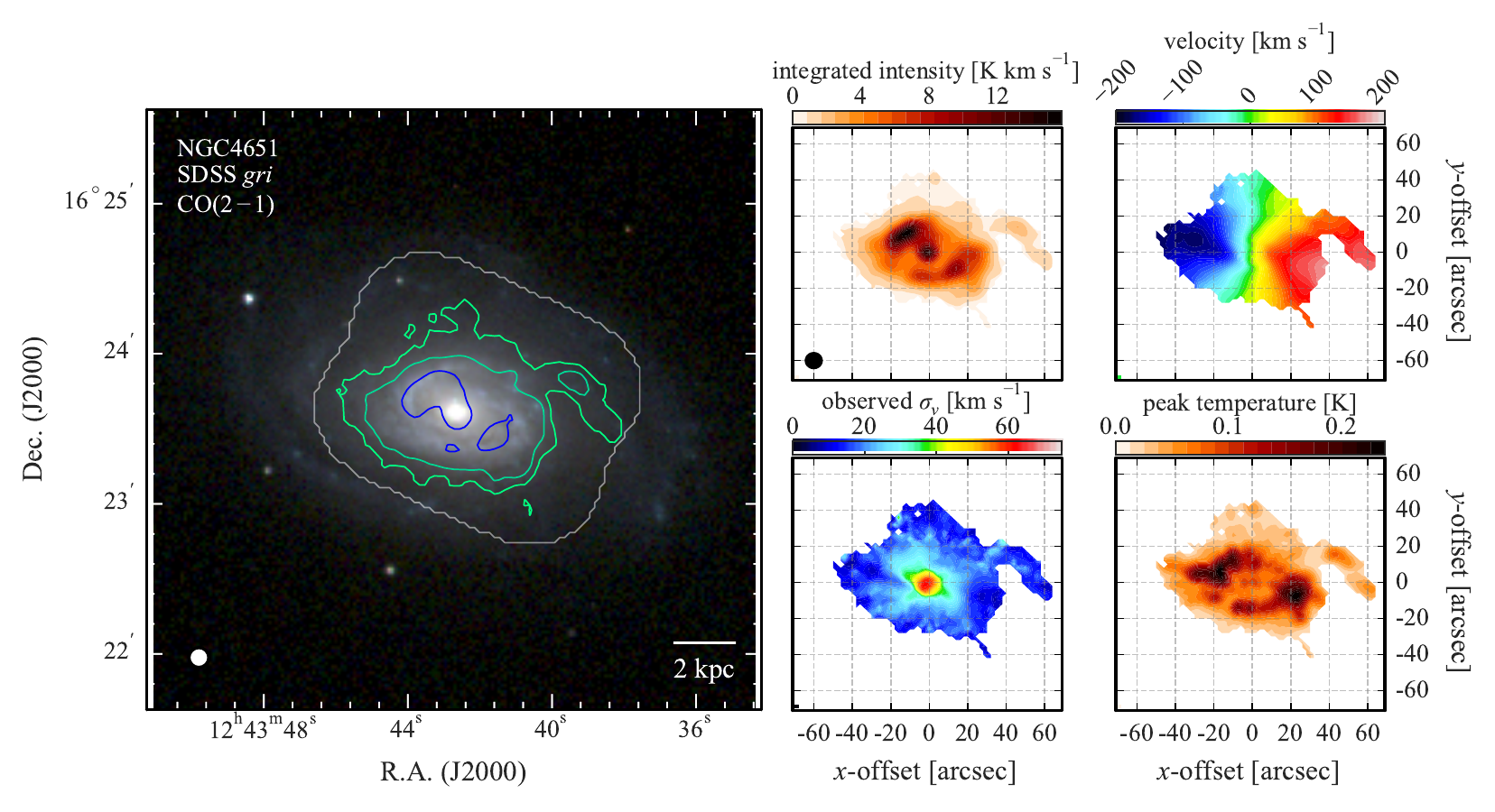}
   \figsetgrpnote{As in Figure 4.1.}
   \figsetgrpend

   \figsetgrpstart
   \figsetgrpnum{4.43}
   \figsetgrptitle{VERTICO CO($2-1$) data products for NGC4654}
   \figsetplot{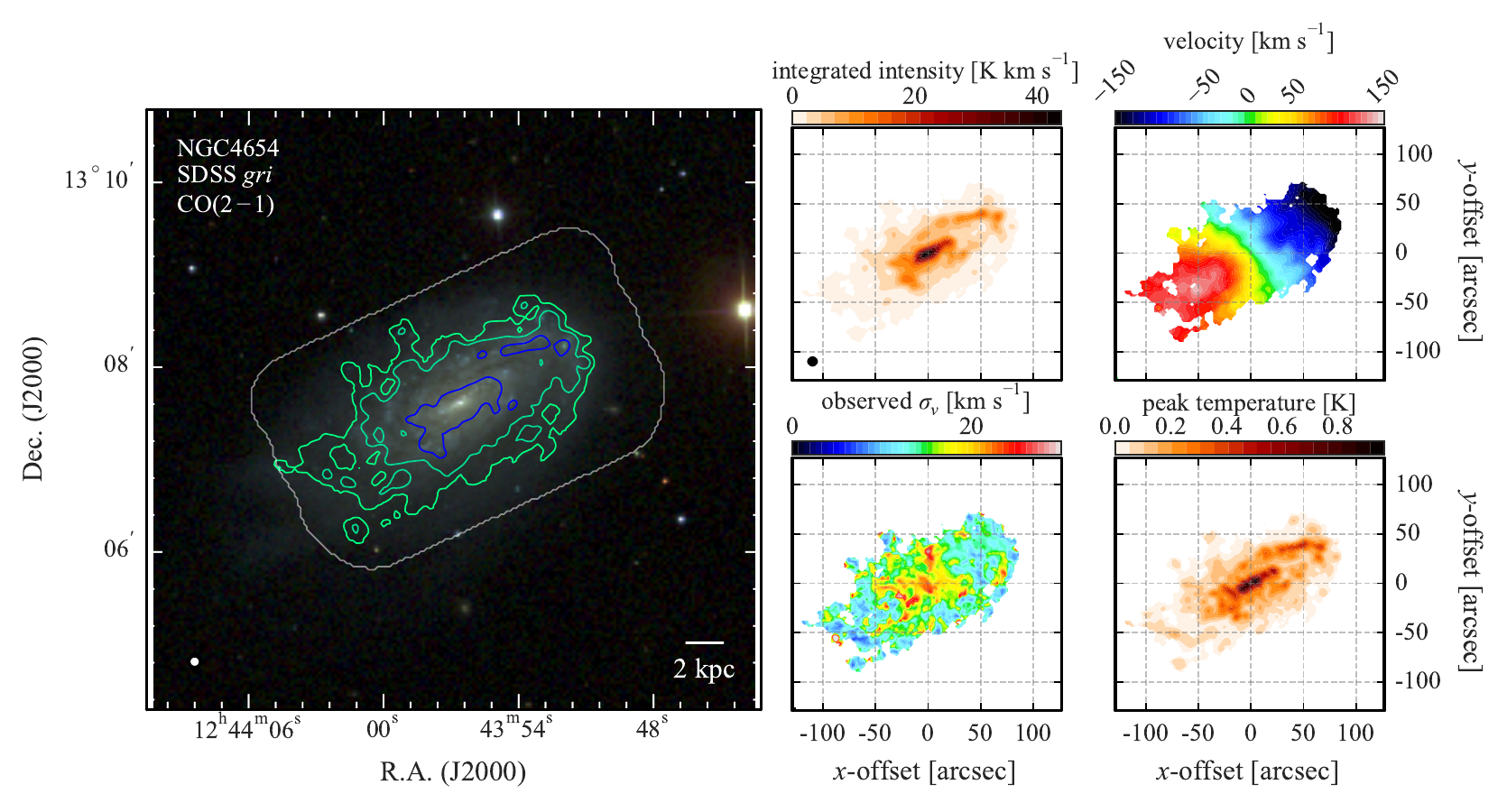}
   \figsetgrpnote{As in Figure 4.1.}
   \figsetgrpend

   \figsetgrpstart
   \figsetgrpnum{4.44}
   \figsetgrptitle{VERTICO CO($2-1$) data products for NGC4689}
   \figsetplot{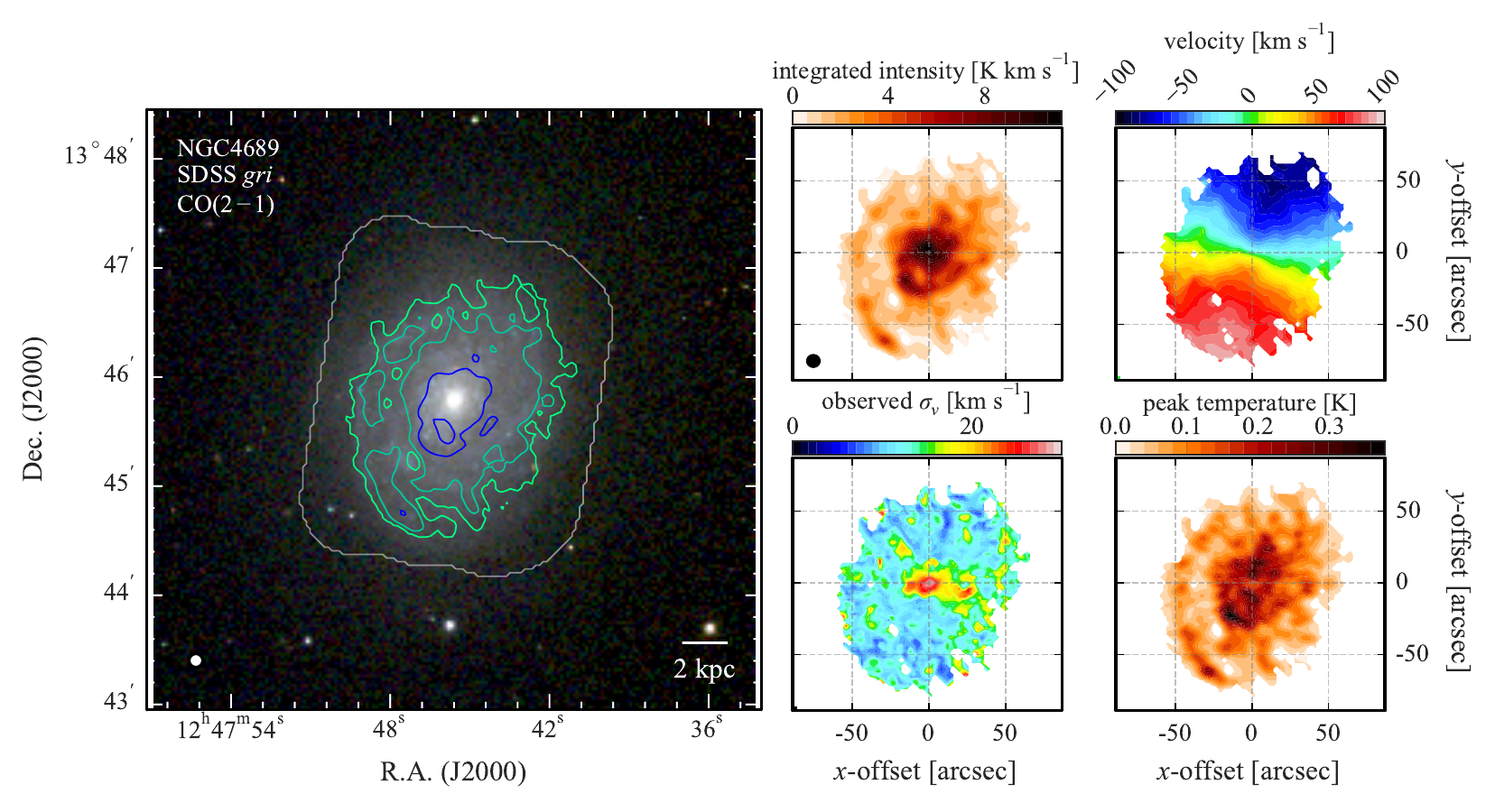}
   \figsetgrpnote{As in Figure 4.1.}
   \figsetgrpend

   \figsetgrpstart
   \figsetgrpnum{4.45}
   \figsetgrptitle{VERTICO CO($2-1$) data products for NGC4694}
   \figsetplot{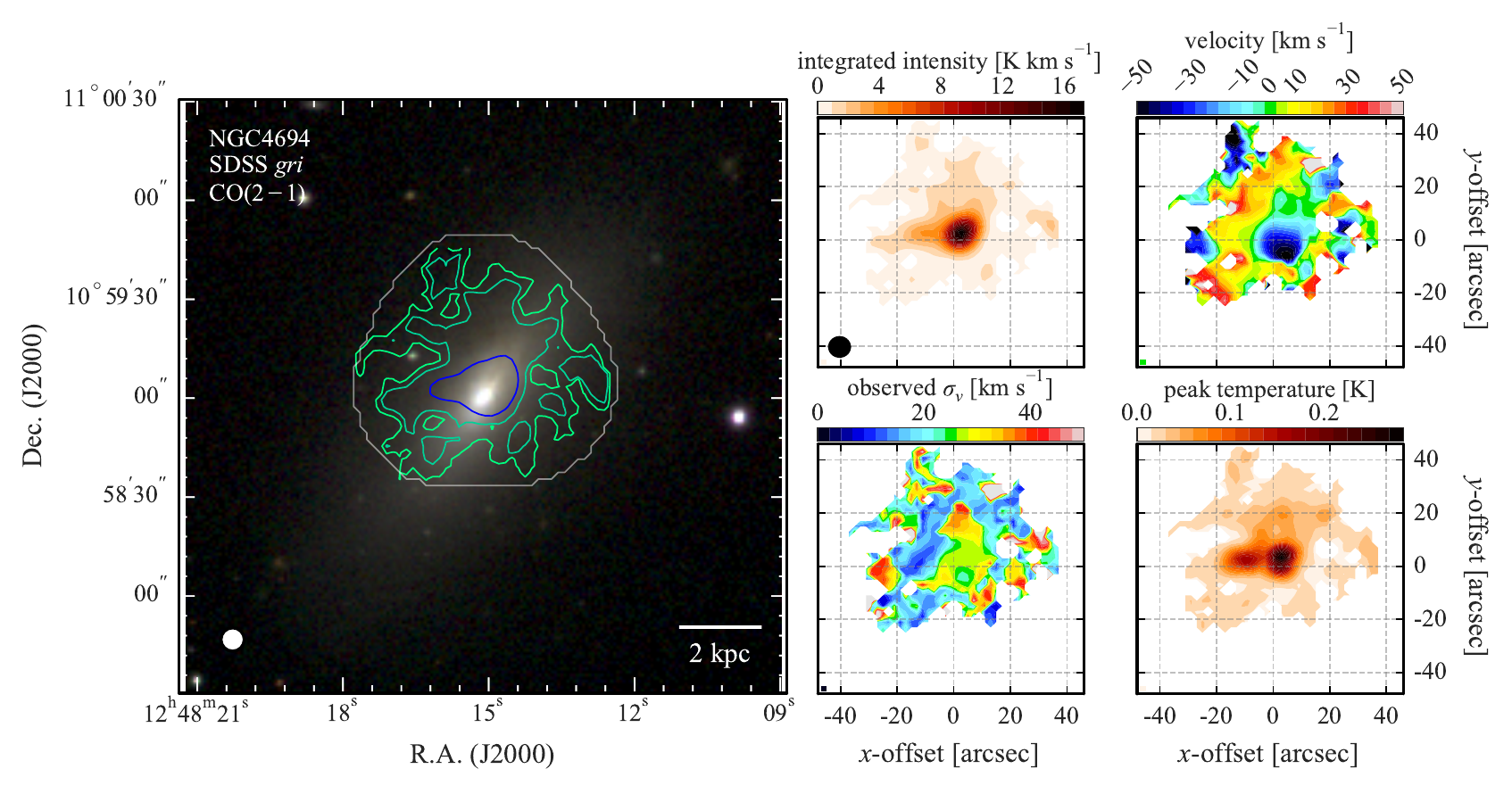}
   \figsetgrpnote{As in Figure 4.1.}
   \figsetgrpend

   \figsetgrpstart
   \figsetgrpnum{4.46}
   \figsetgrptitle{VERTICO CO($2-1$) data products for NGC4698}
   \figsetplot{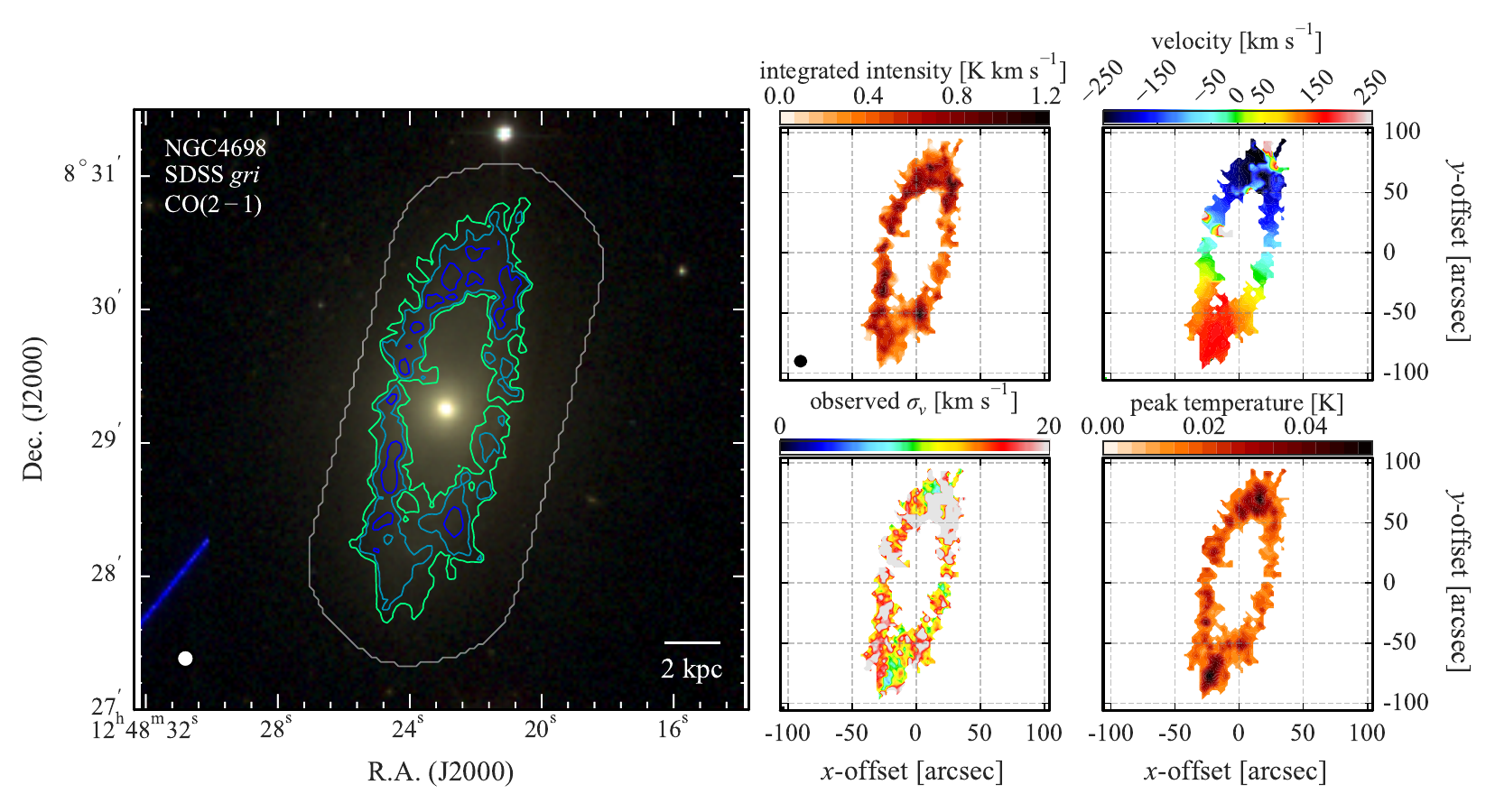}
   \figsetgrpnote{As in Figure 4.1.}
   \figsetgrpend

   \figsetgrpstart
   \figsetgrpnum{4.47}
   \figsetgrptitle{VERTICO CO($2-1$) data products for NGC4713}
   \figsetplot{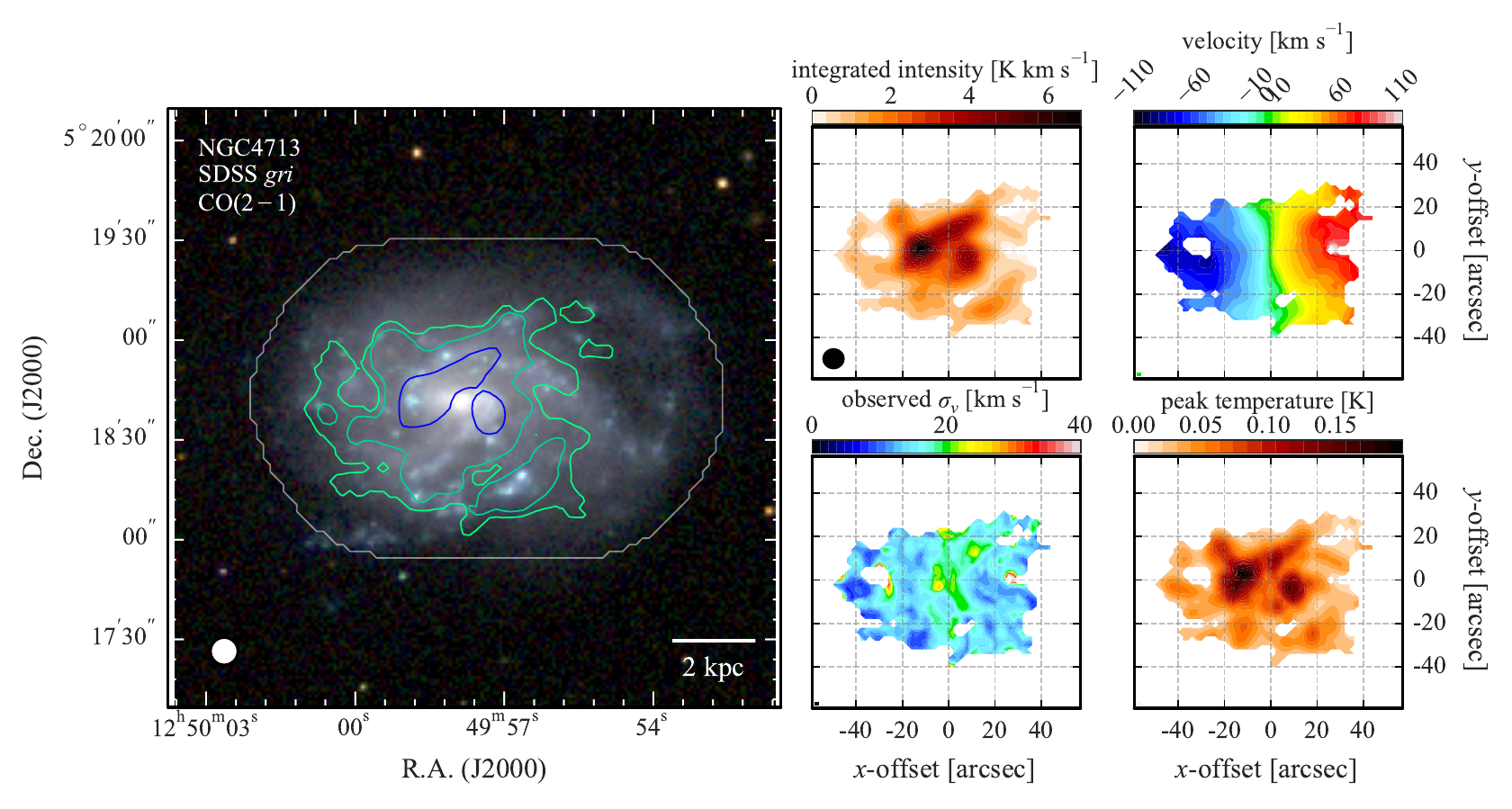}
   \figsetgrpnote{As in Figure 4.1.}
   \figsetgrpend

   \figsetgrpstart
   \figsetgrpnum{4.48}
   \figsetgrptitle{VERTICO CO($2-1$) data products for NGC4772}
   \figsetplot{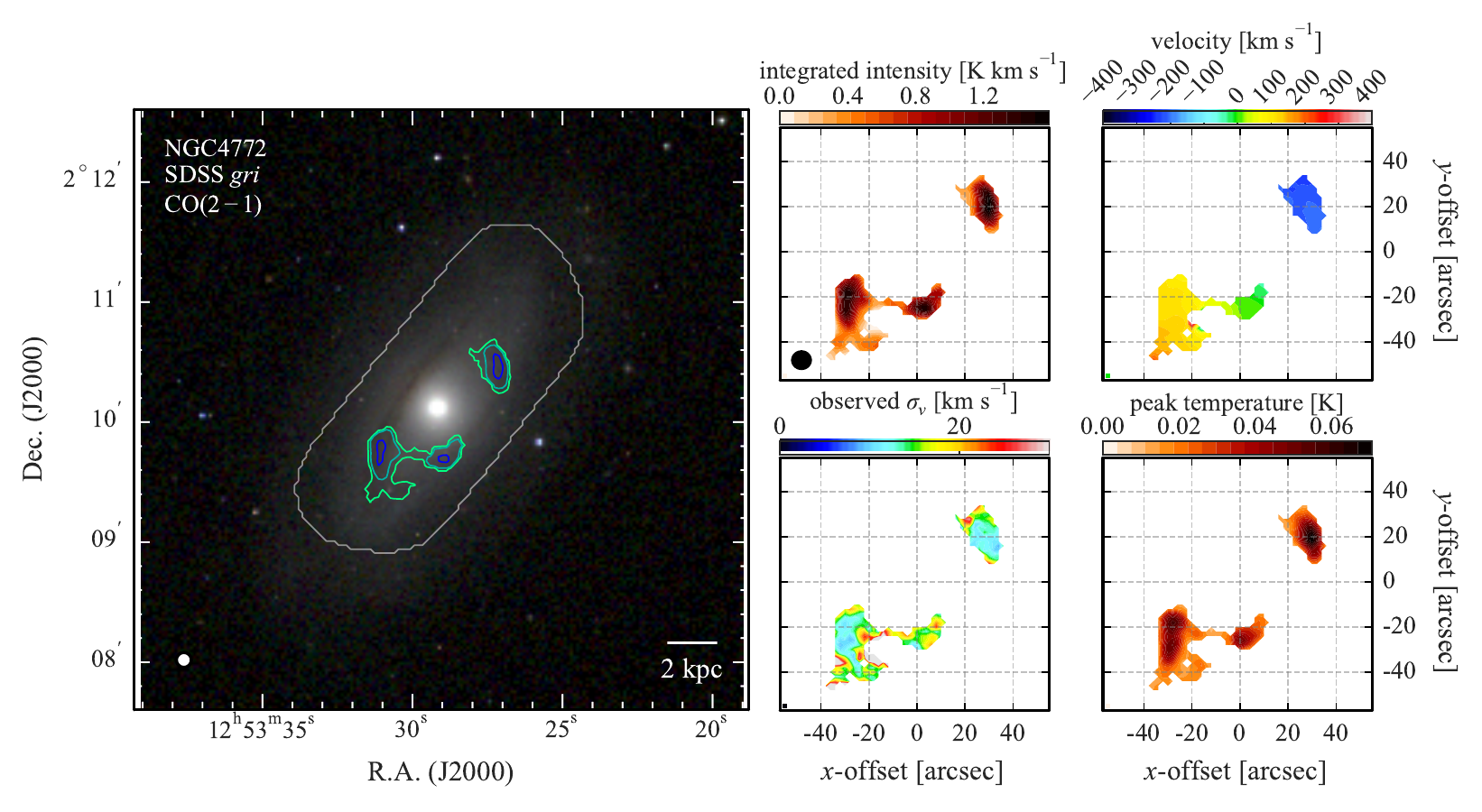}
   \figsetgrpnote{As in Figure 4.1.}
   \figsetgrpend

   \figsetgrpstart
   \figsetgrpnum{4.49}
   \figsetgrptitle{VERTICO CO($2-1$) data products for NGC4808}
   \figsetplot{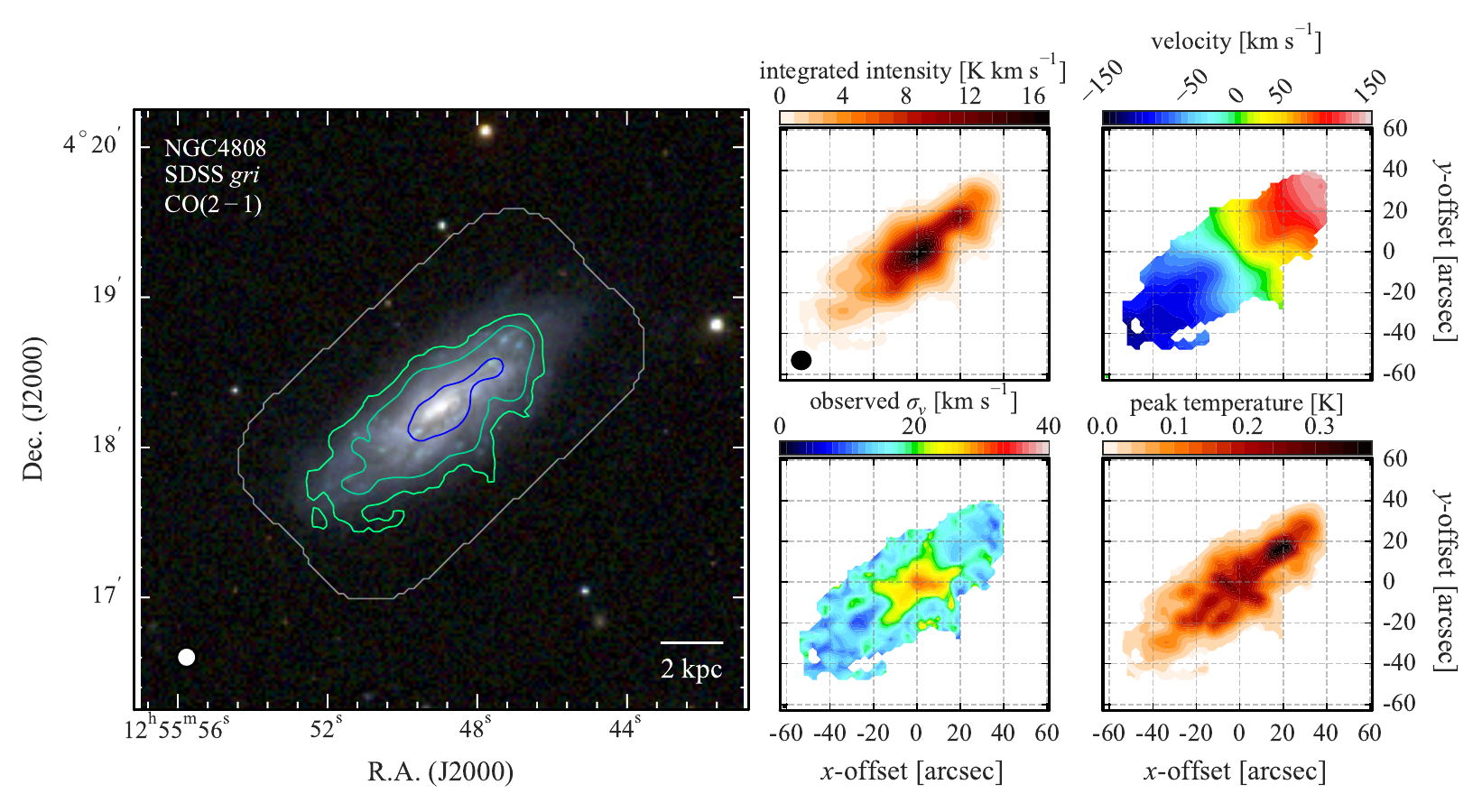}
   \figsetgrpnote{As in Figure 4.1.}
\figsetgrpend

\figsetend

    \figurenum{4.1}
    \begin{figure*}
        \centering
        \includegraphics[width=\linewidth]{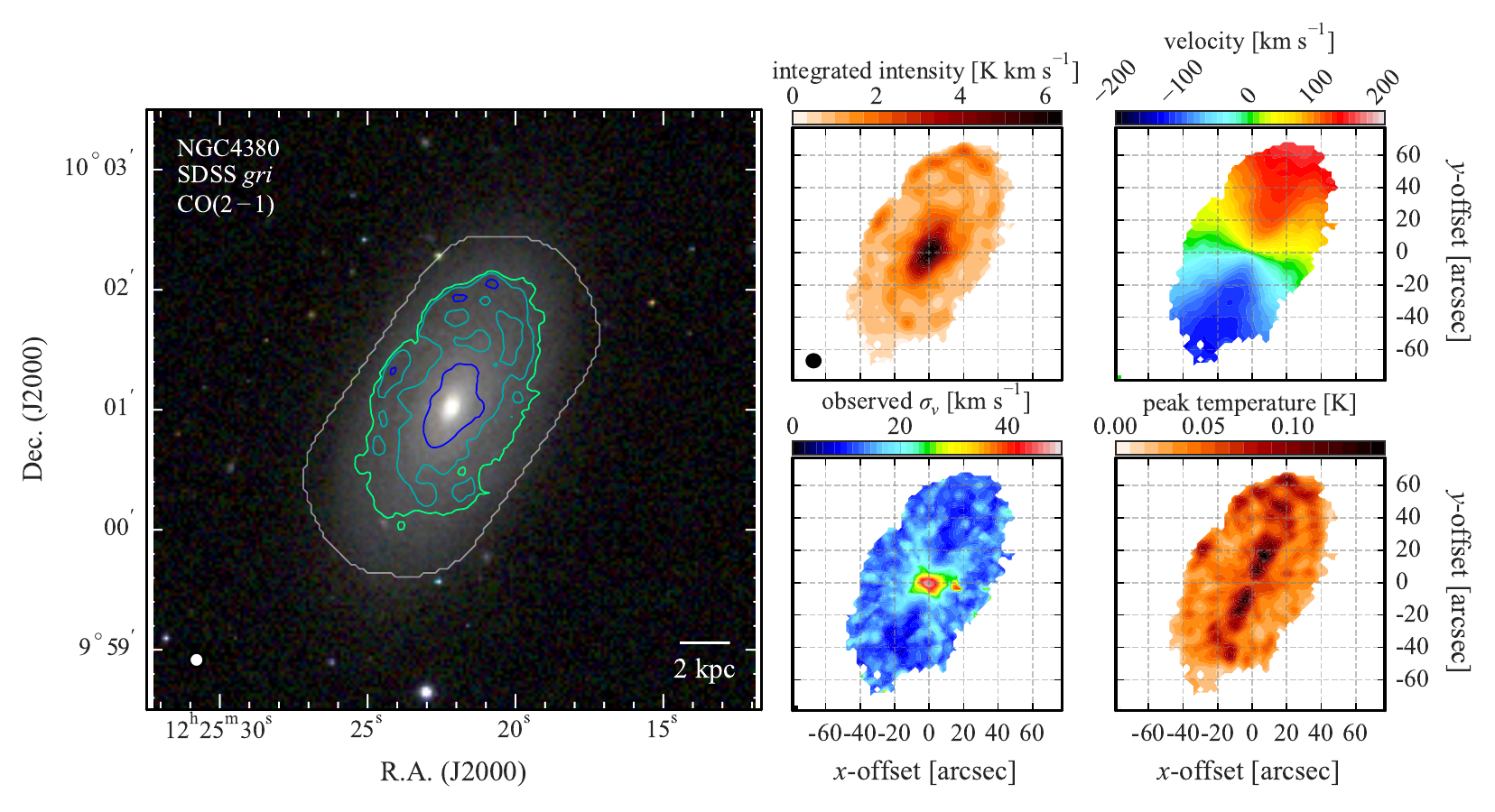}
        \caption{An example of the CO($2-1$) data products available for each
            galaxy in the VERTICO survey. The left panel shows the SDSS $gri$
            composite image for NGC4380 with molecular gas surface brightness
            contours at the 10th, 50th, and 90th percentiles of the
            distribution. The field of view of the ACA observations is defined
            where the primary beam response drops to 50\% and is illustrated by
            the gray line. The rounded synthesized beam is $7.5\arcsec$ in
            diameter and illustrated in the bottom left corner. This beam
            corresponds to $\sim$600 pc at the distance of Virgo (16.5 Mpc). The
            VERTICO CO($2-1$) data products available for each galaxy include
            maps of integrated intensity (upper center panel), the velocity
            field (upper right), observed line width, $\sigma_v$ (lower center),
            and peak temperature (lower left). The $x$- and $y$-axes of each
            moment map shows the angular offset from the optical center listed
            in Table \ref{tab:VERTICO-sample}.}
        \label{figset:NGC4380_panel_plot}
    \end{figure*}

    \begin{figure*}
    \figurenum{4.2}
        \includegraphics[width=\linewidth]{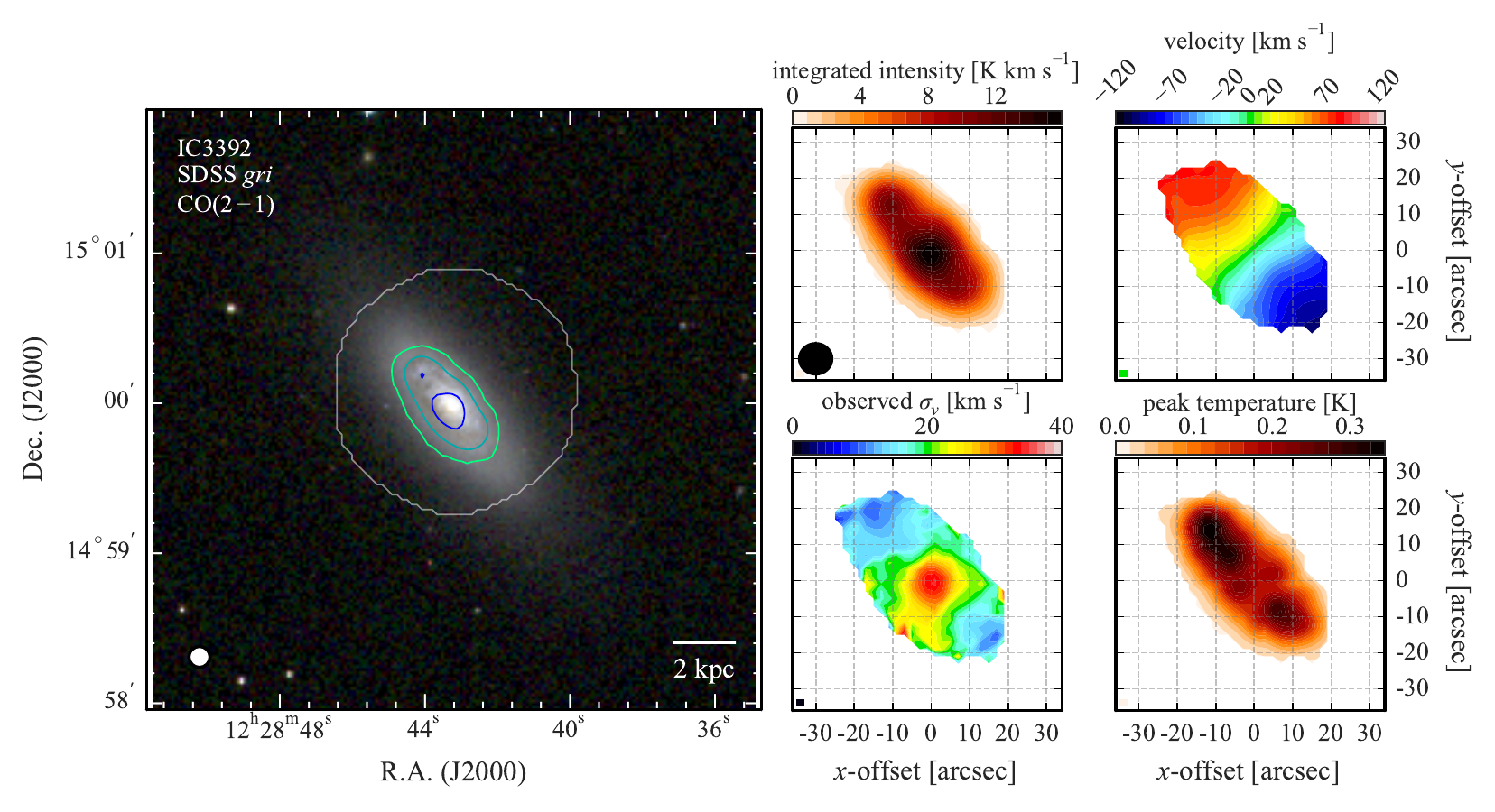}
        \includegraphics[width=\linewidth]{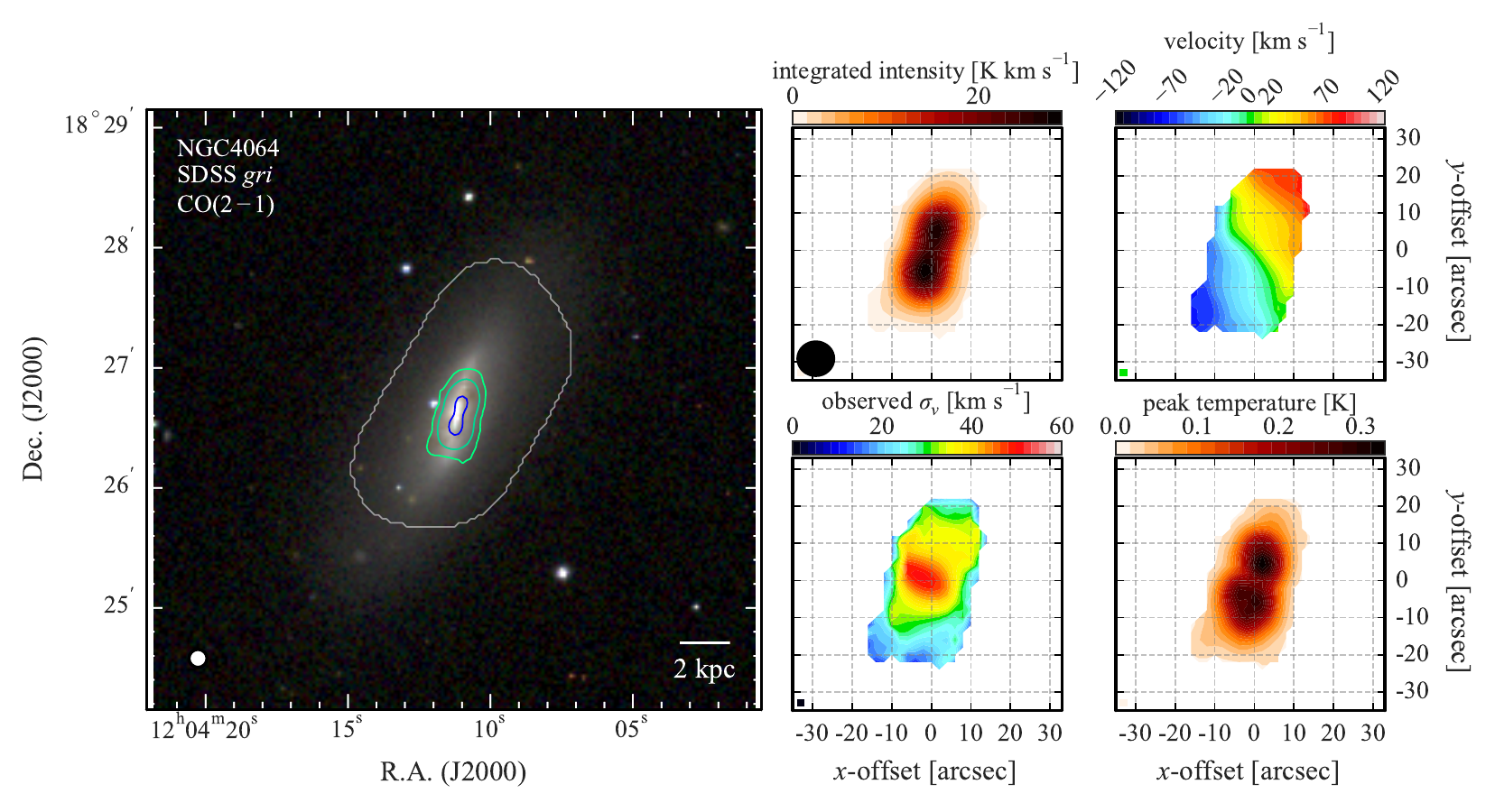}
        \caption{As in Figure 4.1.}
        \label{fig:panelplots_1}
    \end{figure*}

    \begin{figure*}
    \figurenum{4.3}
        \includegraphics[width=\linewidth]{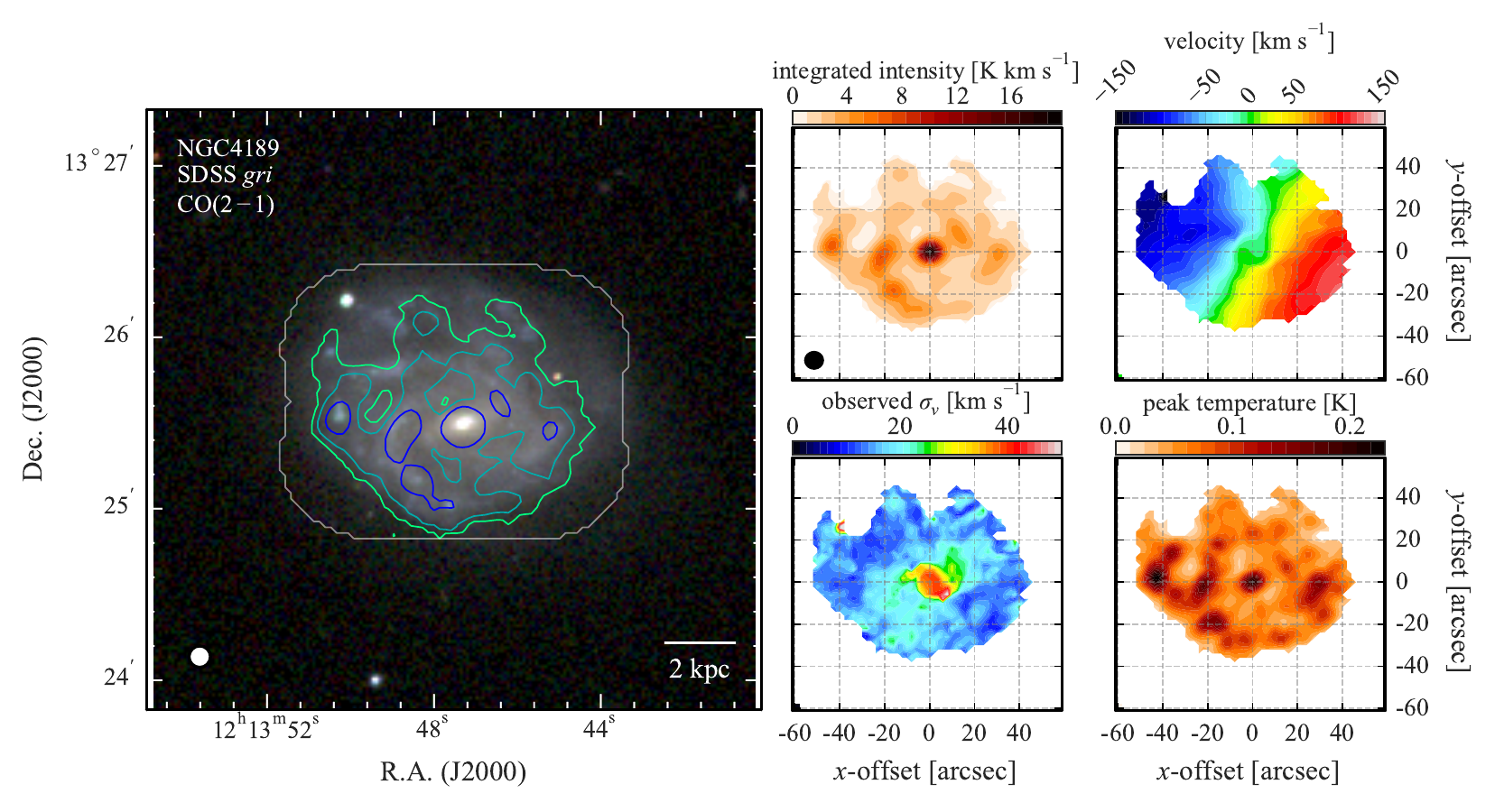}
        \includegraphics[width=\linewidth]{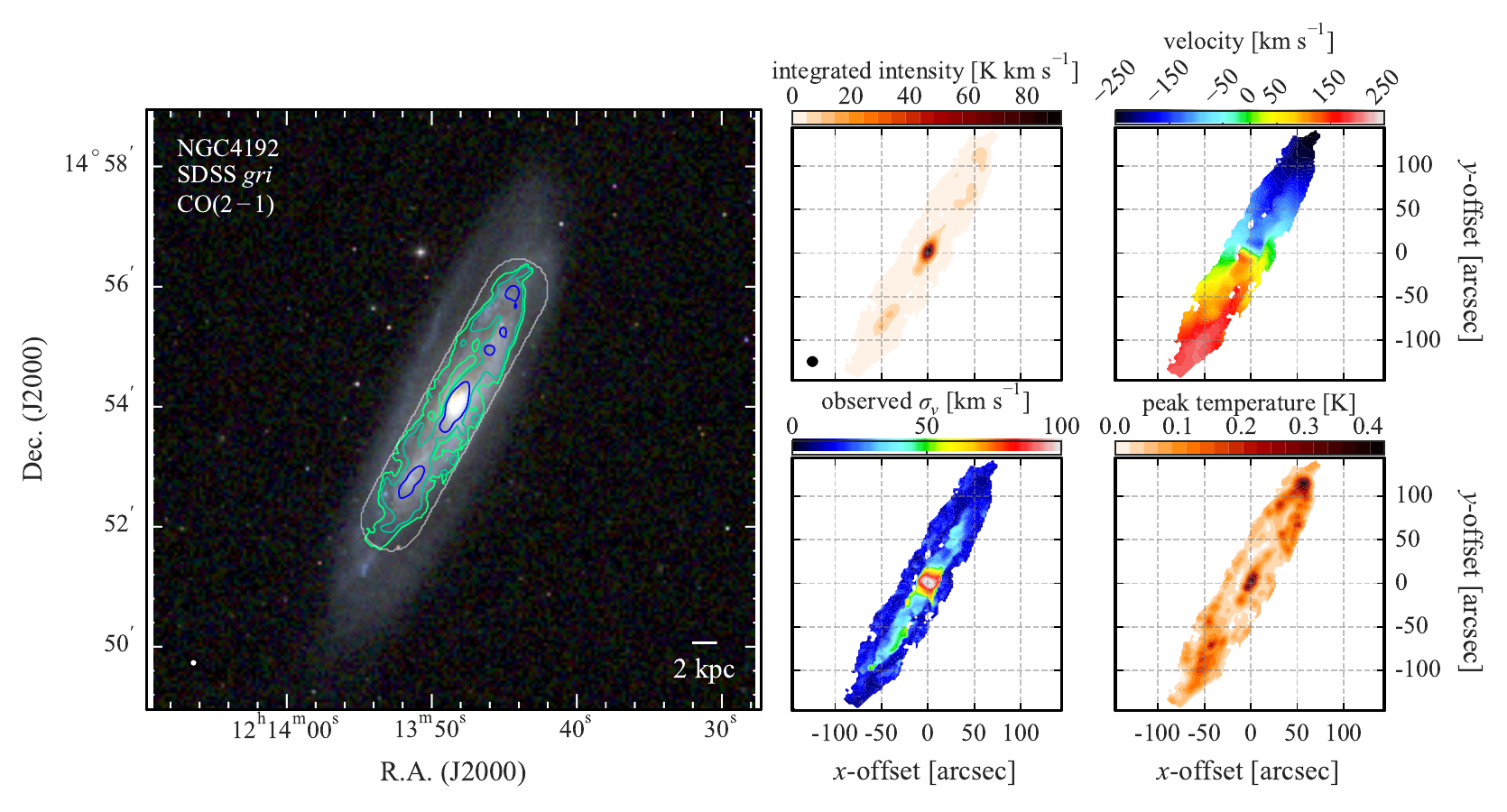}
        \caption{As in Figure 4.1.}
        \label{fig:panelplots_2}
    \end{figure*}

    \begin{figure*}
    \figurenum{4.4}
        \includegraphics[width=\linewidth]{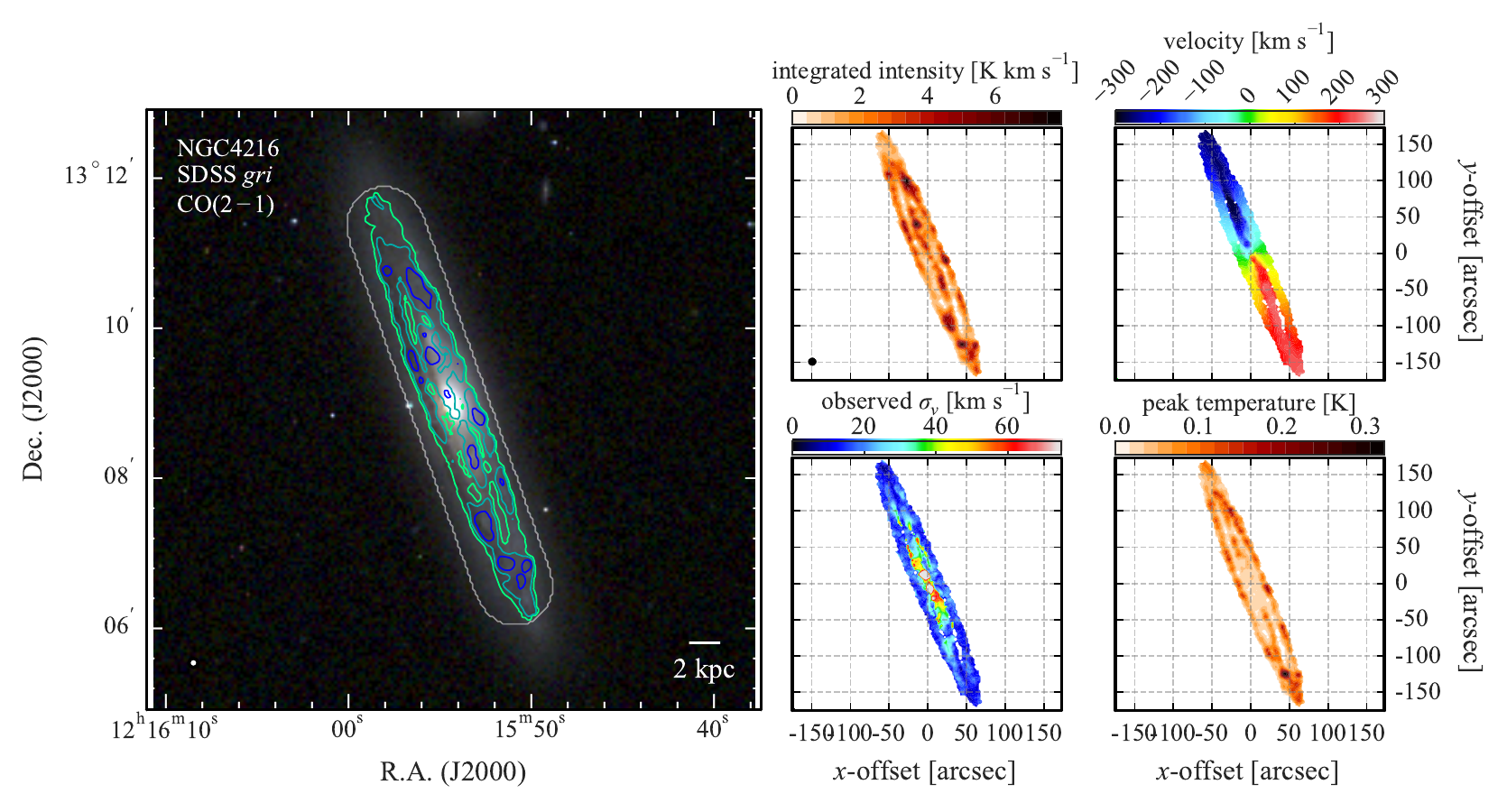}
        \includegraphics[width=\linewidth]{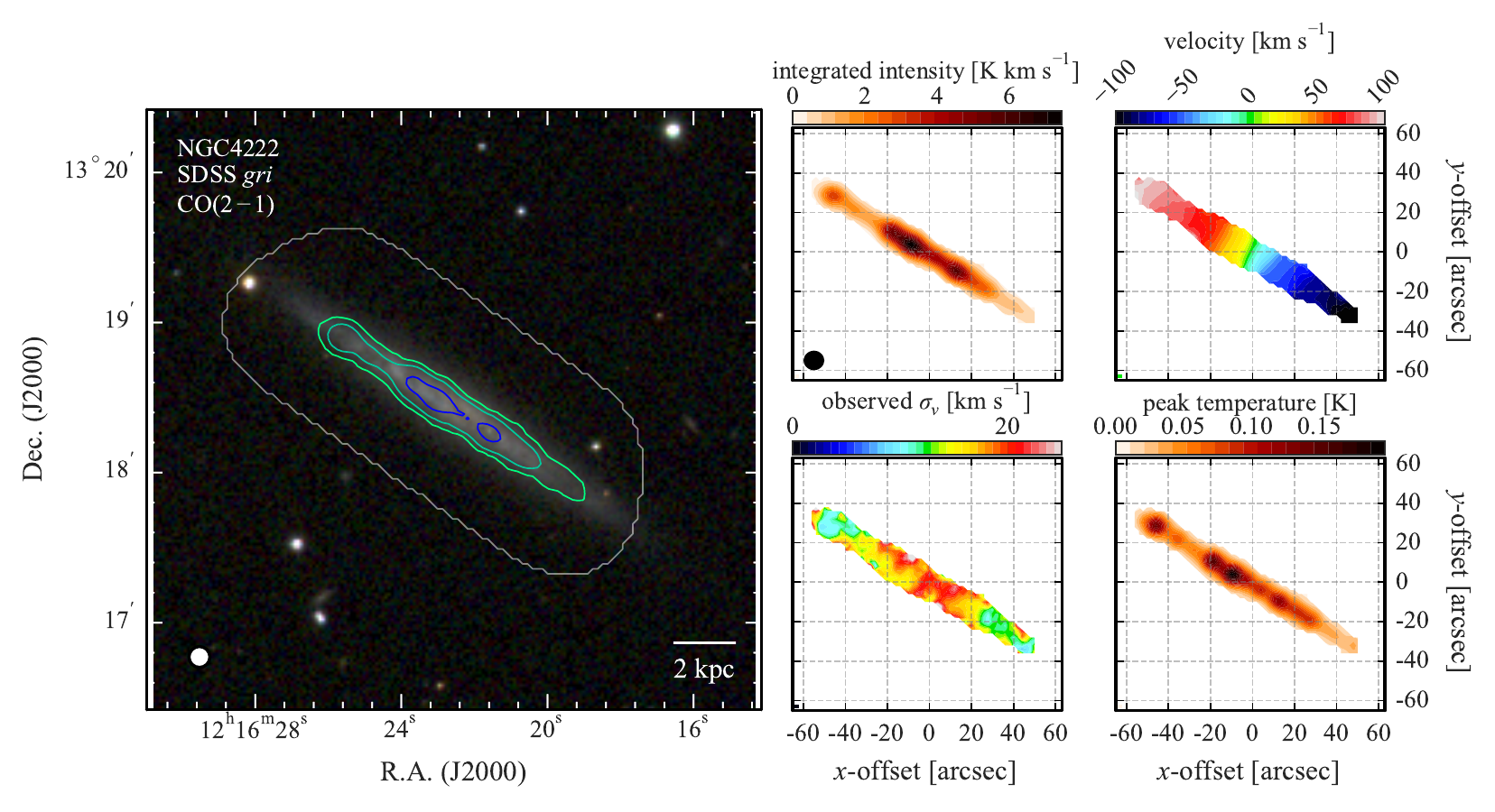}
        \caption{As in Figure 4.1.}
        \label{fig:panelplots_3}
    \end{figure*}

    \begin{figure*}
    \figurenum{4.5}
        \includegraphics[width=\linewidth]{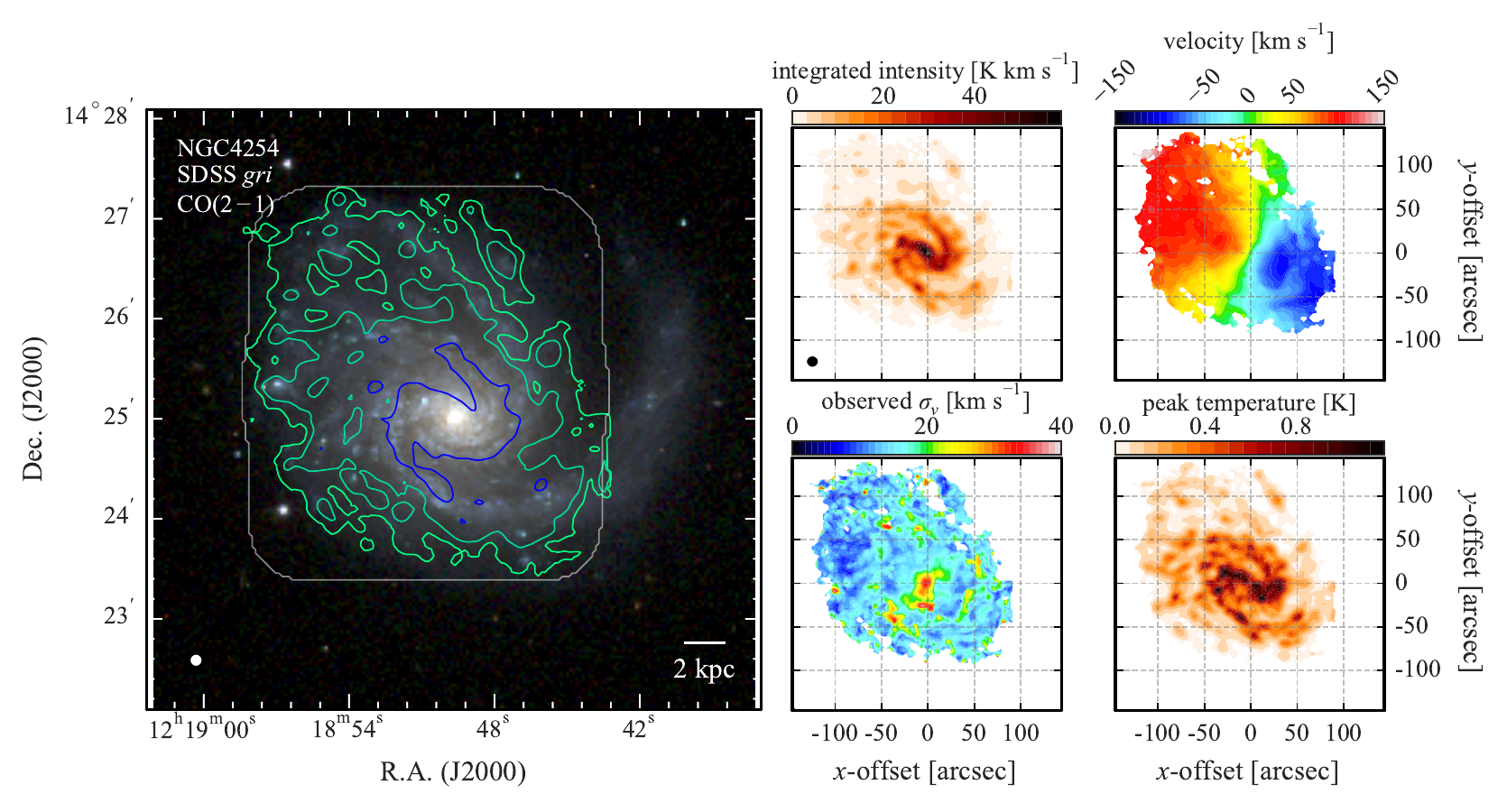}
        \includegraphics[width=\linewidth]{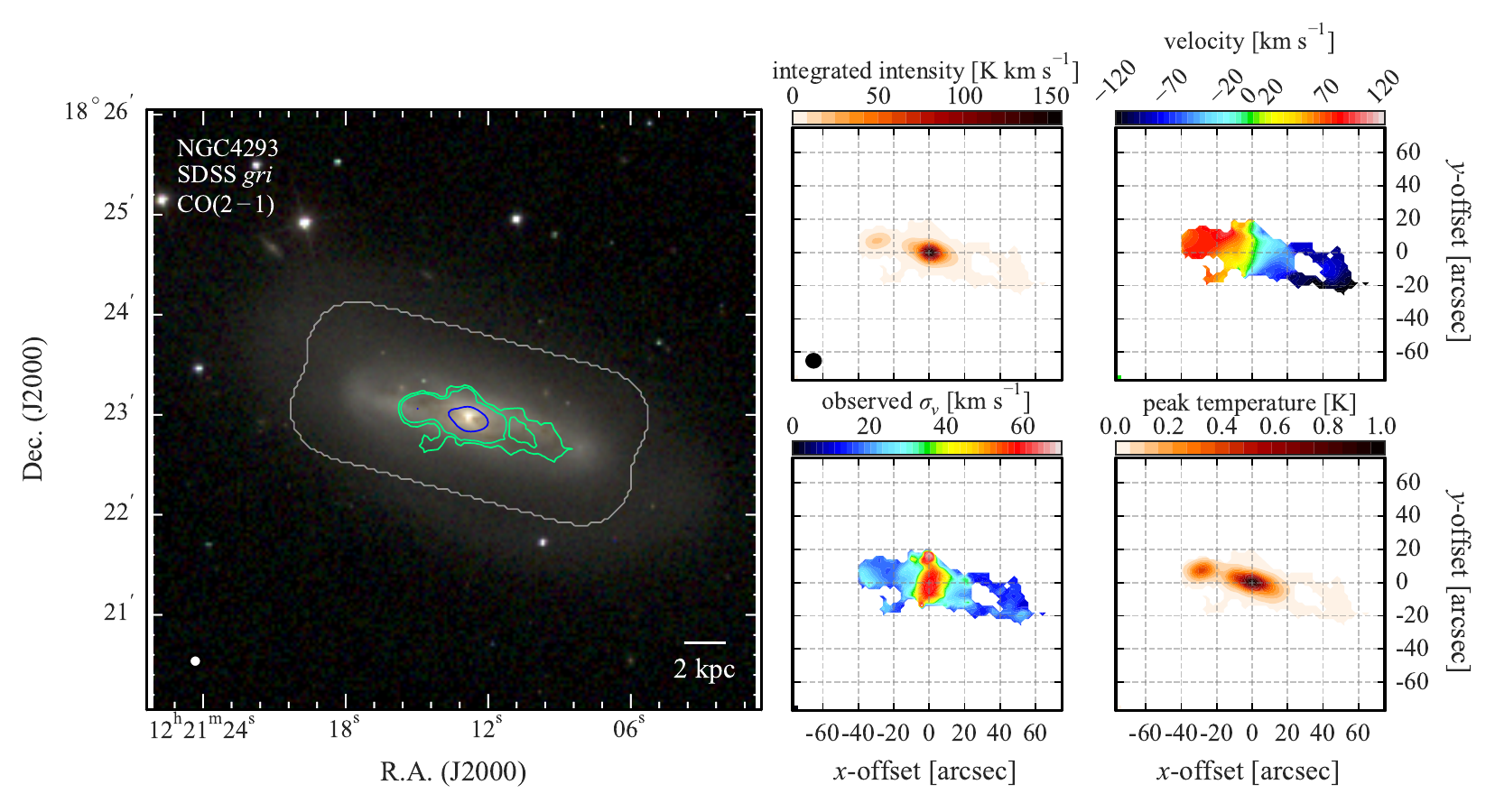}
        \caption{As in Figure 4.1. In the case of NGC4293, we caution against over-interpreting the skinny feature to the South-East of the main disk as this is likely related to the PSF pattern of the observations.}
        \label{fig:panelplots_4}
    \end{figure*}

    \begin{figure*}
    \figurenum{4.6}
        \includegraphics[width=\linewidth]{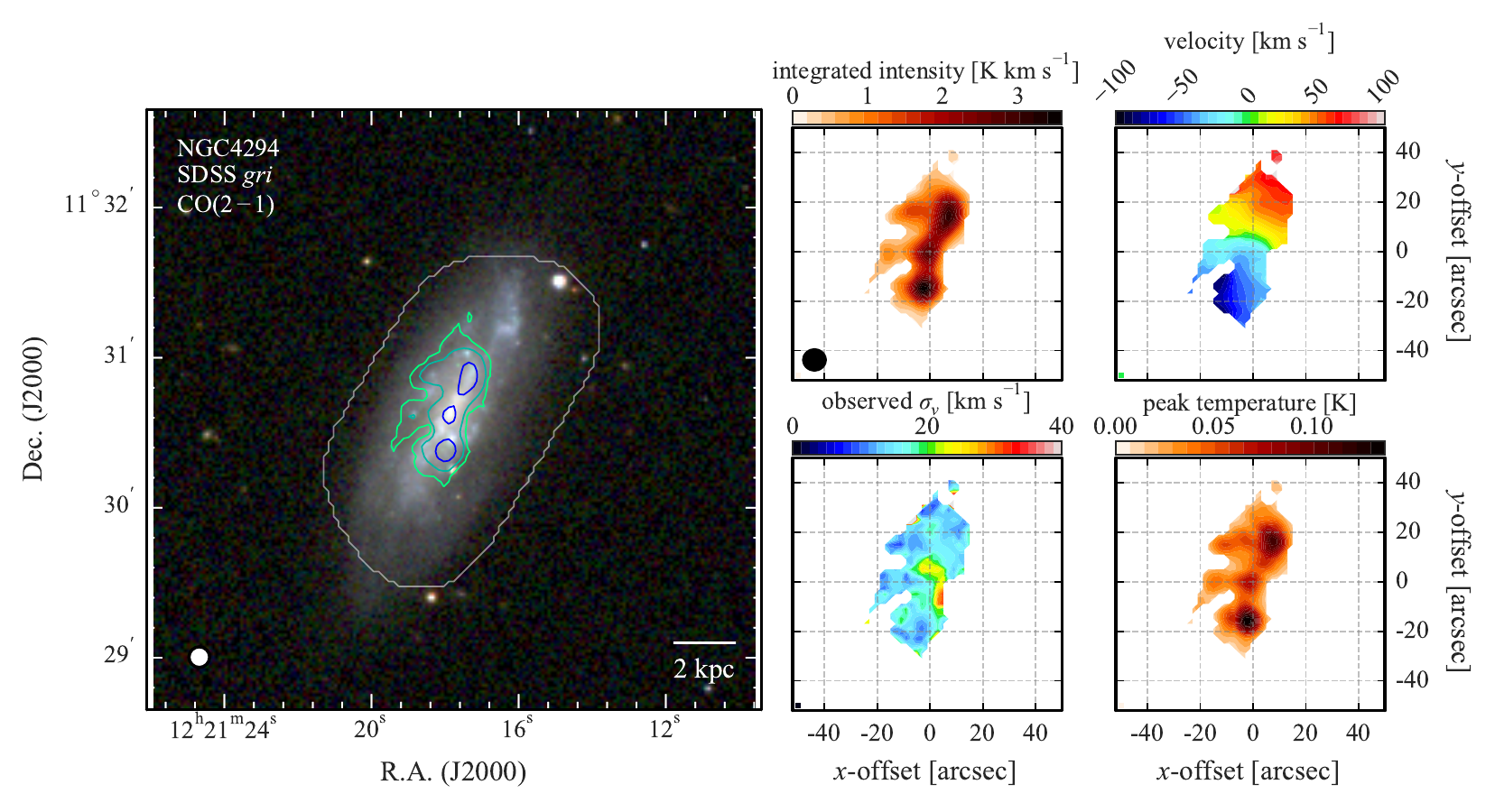}
        \includegraphics[width=\linewidth]{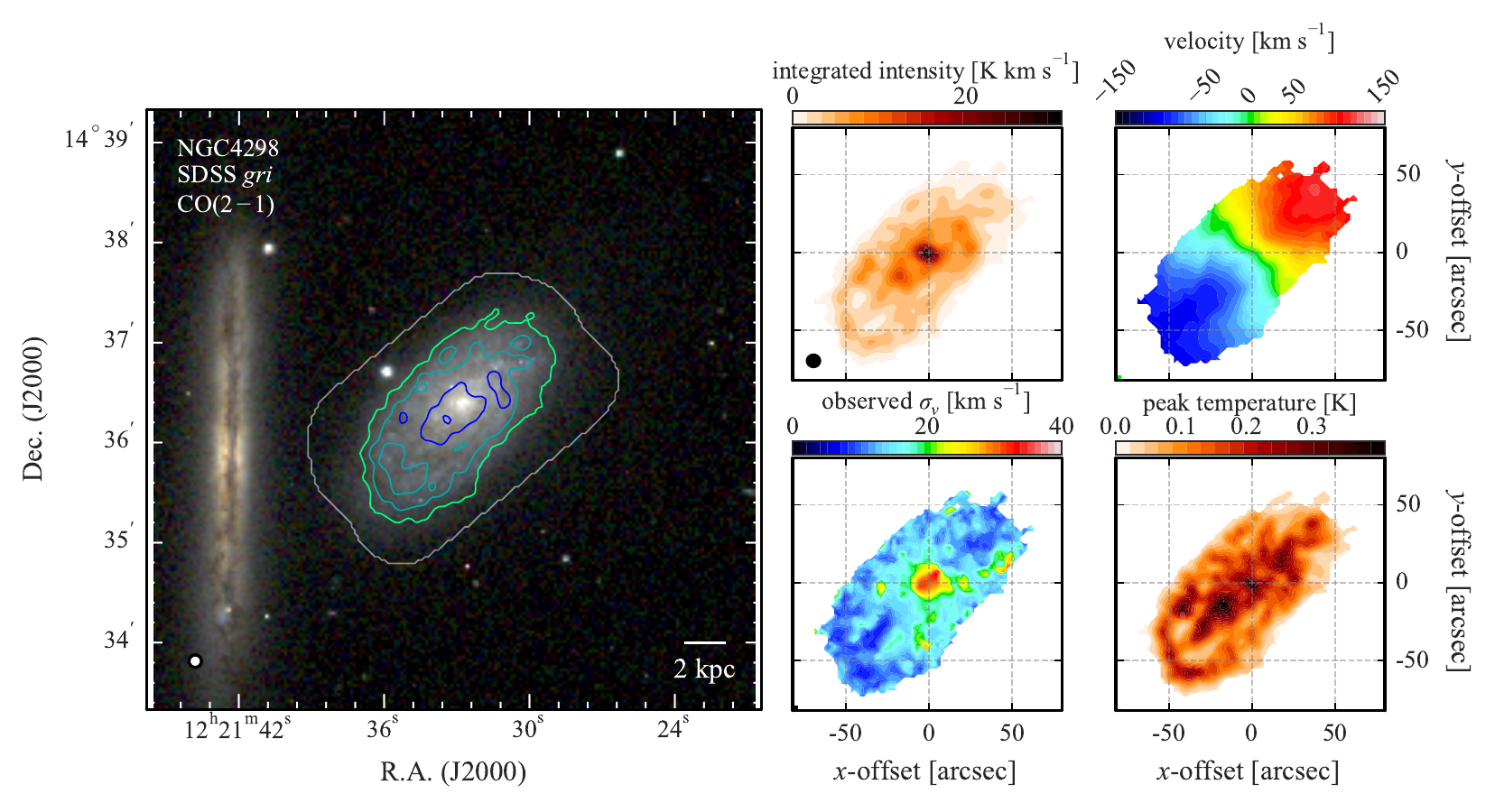}
        \caption{As in Figure 4.1.}
        \label{fig:panelplots_5}
    \end{figure*}

    \begin{figure*}
    \figurenum{4.7}
        \includegraphics[width=\linewidth]{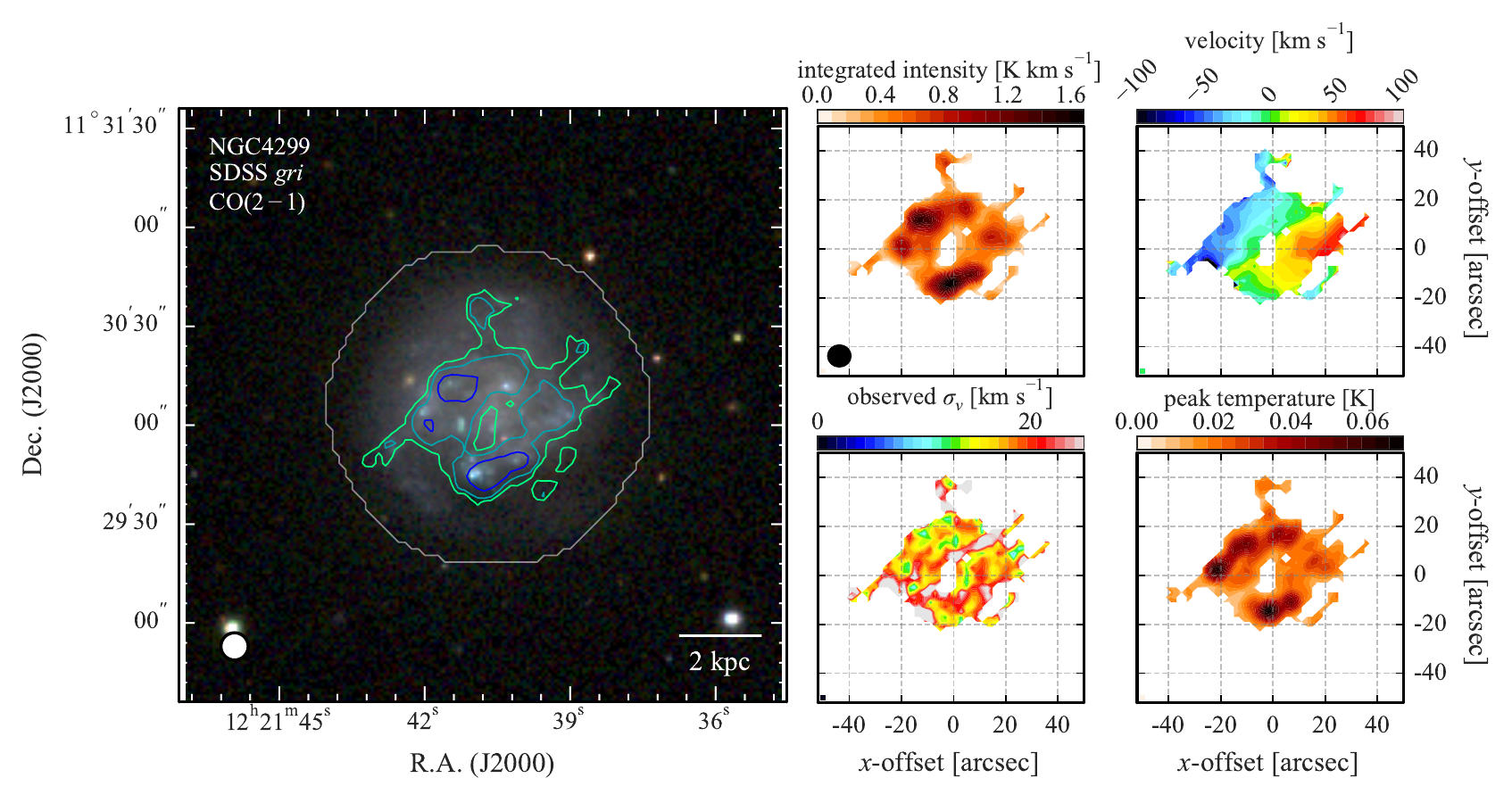}
        \includegraphics[width=\linewidth]{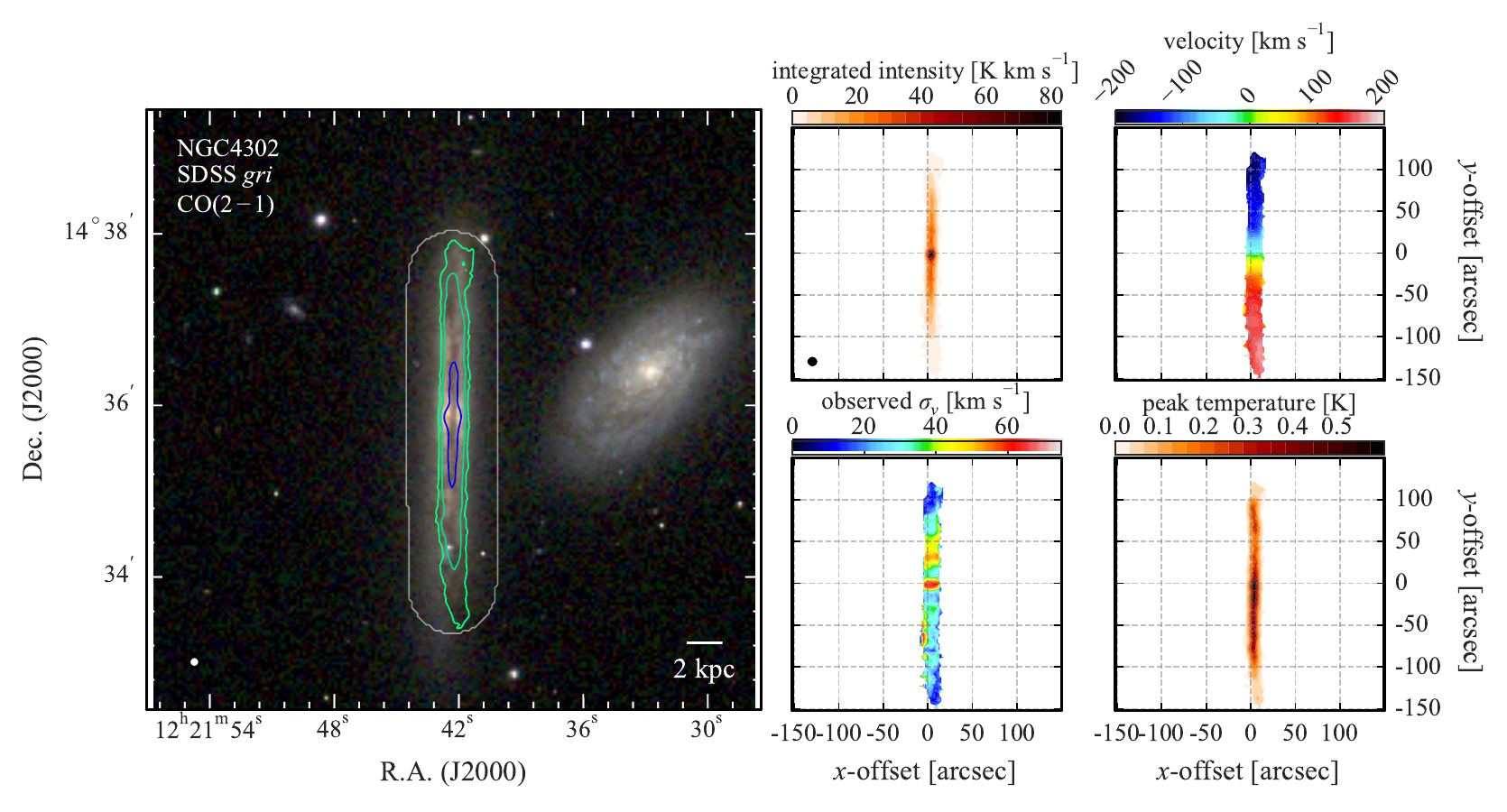}
        \caption{As in Figure 4.1.}
        \label{fig:panelplots_6}
    \end{figure*}

    \begin{figure*}
    \figurenum{4.8}
        \includegraphics[width=\linewidth]{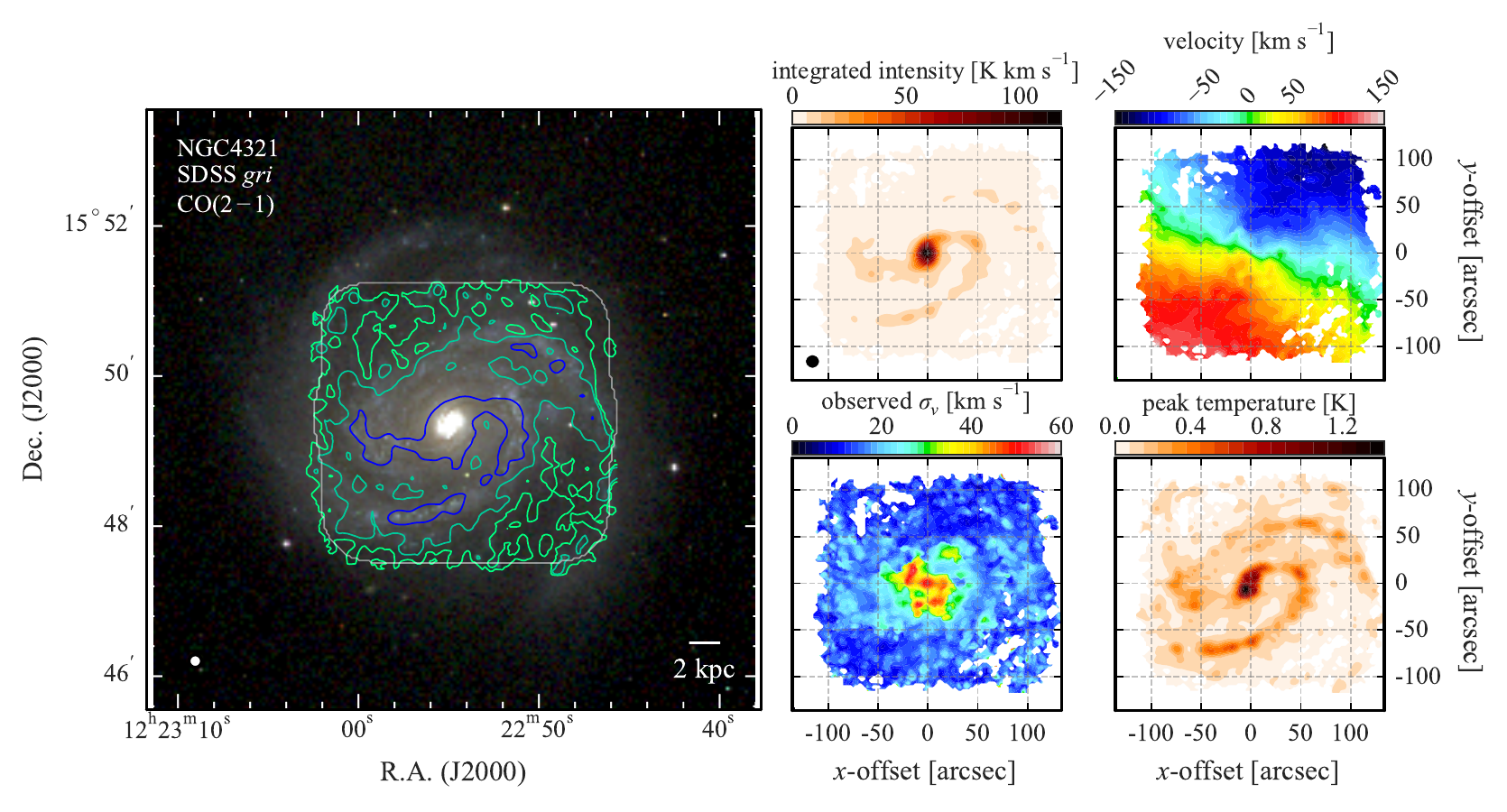}
        \includegraphics[width=\linewidth]{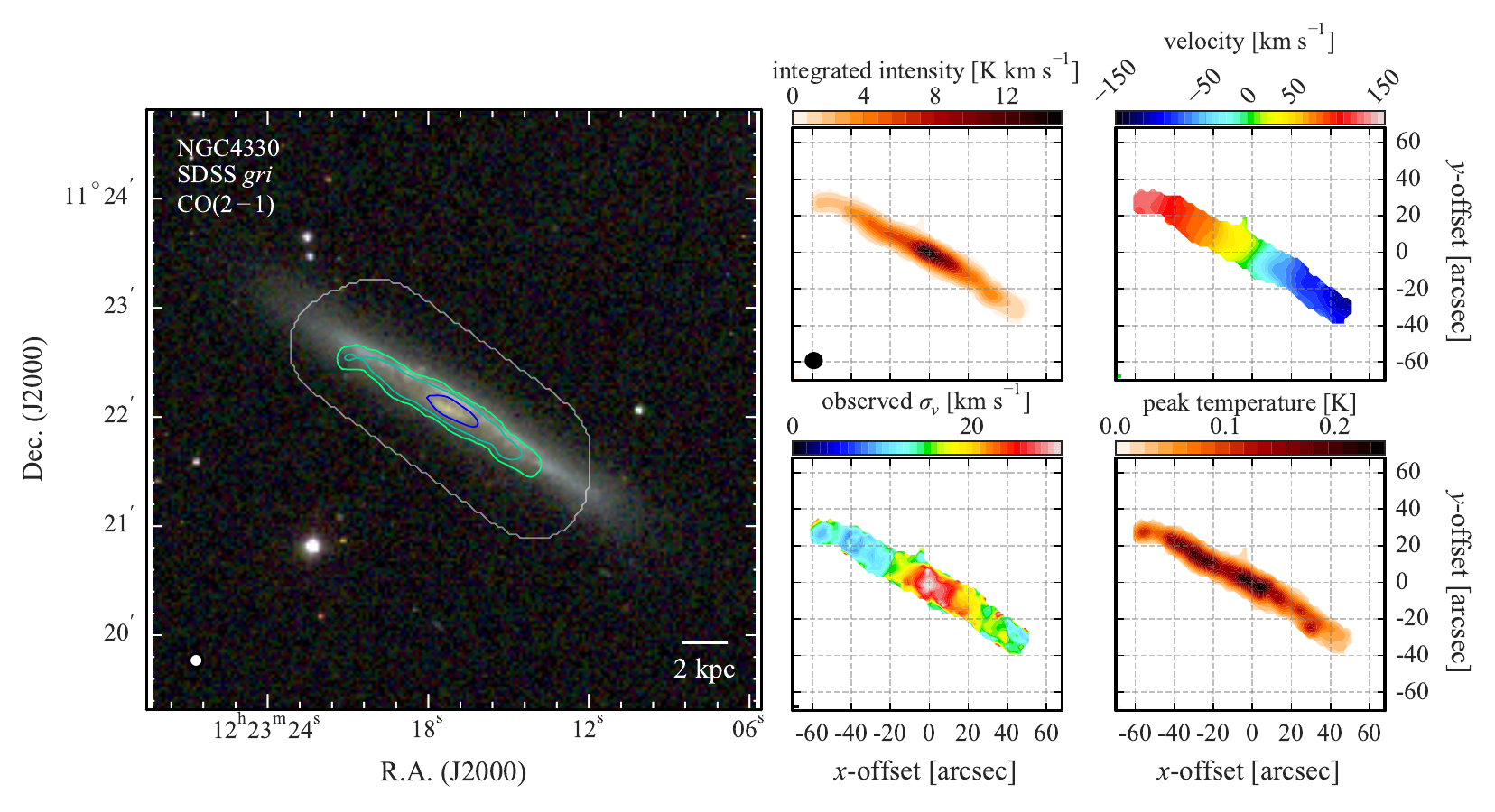}
        \caption{As in Figure 4.1.}
        \label{fig:panelplots_7}
    \end{figure*}

    \begin{figure*}
    \figurenum{4.9}
        \includegraphics[width=\linewidth]{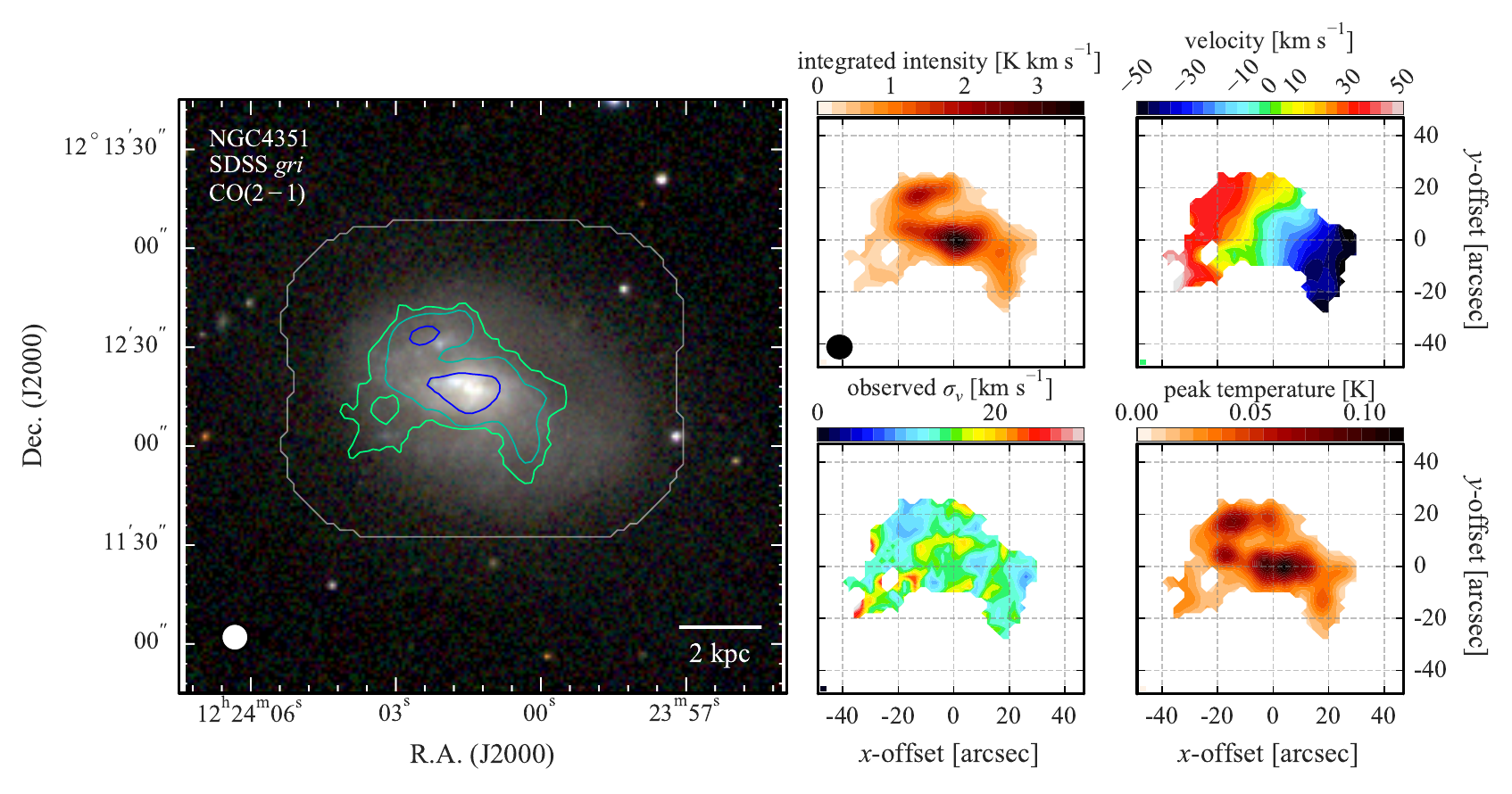}
        \includegraphics[width=\linewidth]{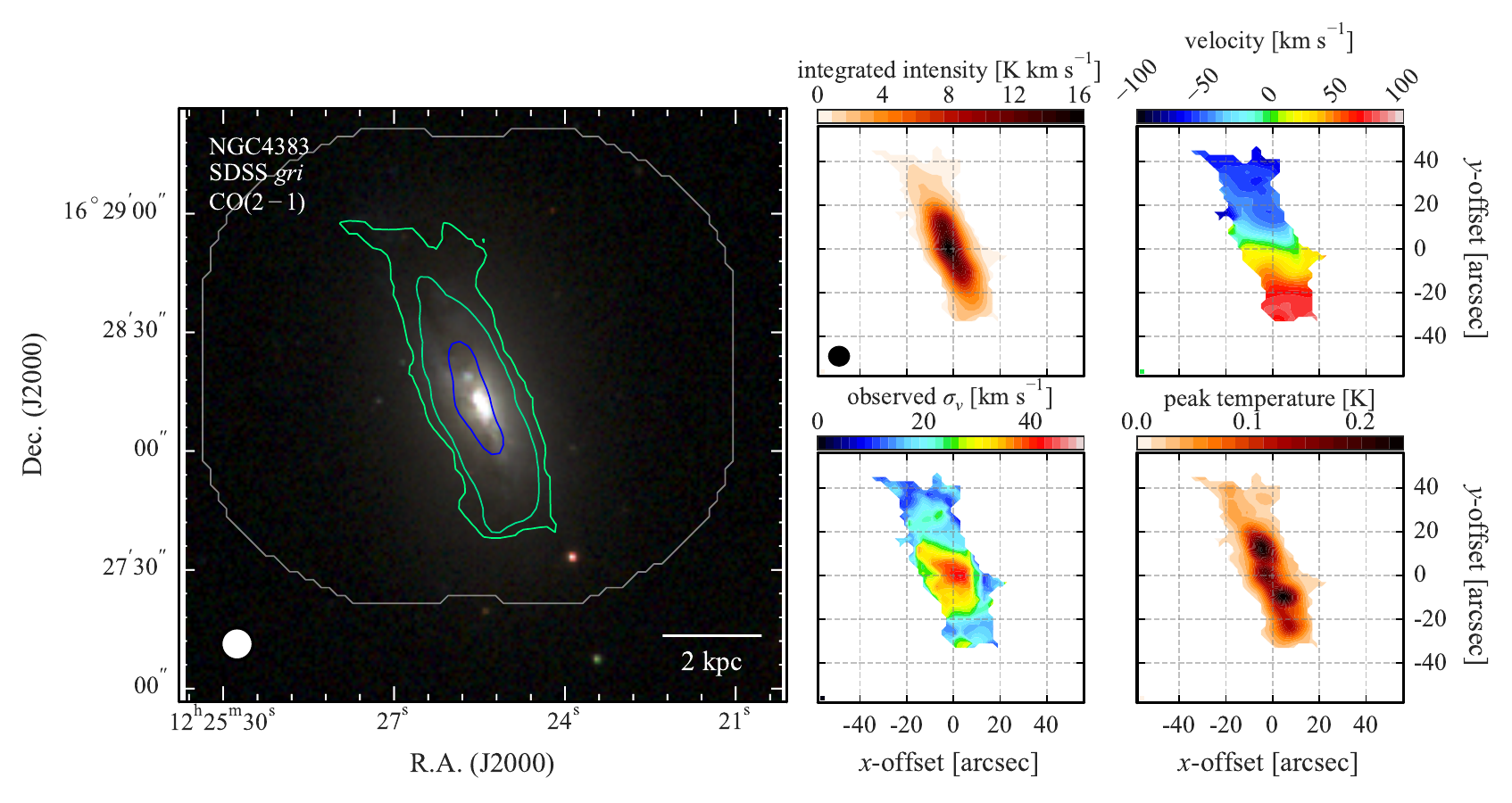}
        \caption{As in Figure 4.1.}
        \label{fig:panelplots_8}
    \end{figure*}

    \begin{figure*}
    \figurenum{4.10}
        \includegraphics[width=\linewidth]{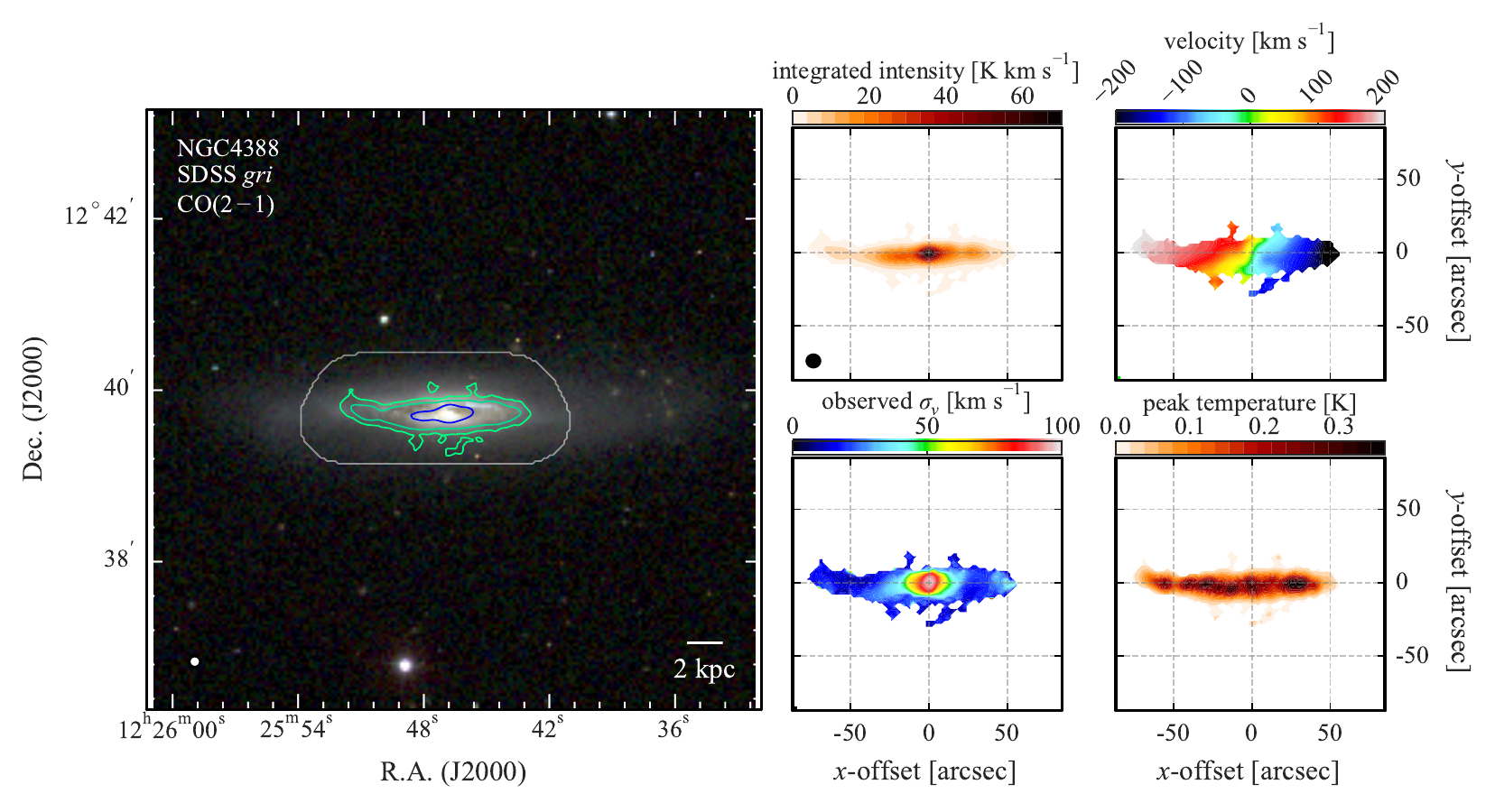}
        \includegraphics[width=\linewidth]{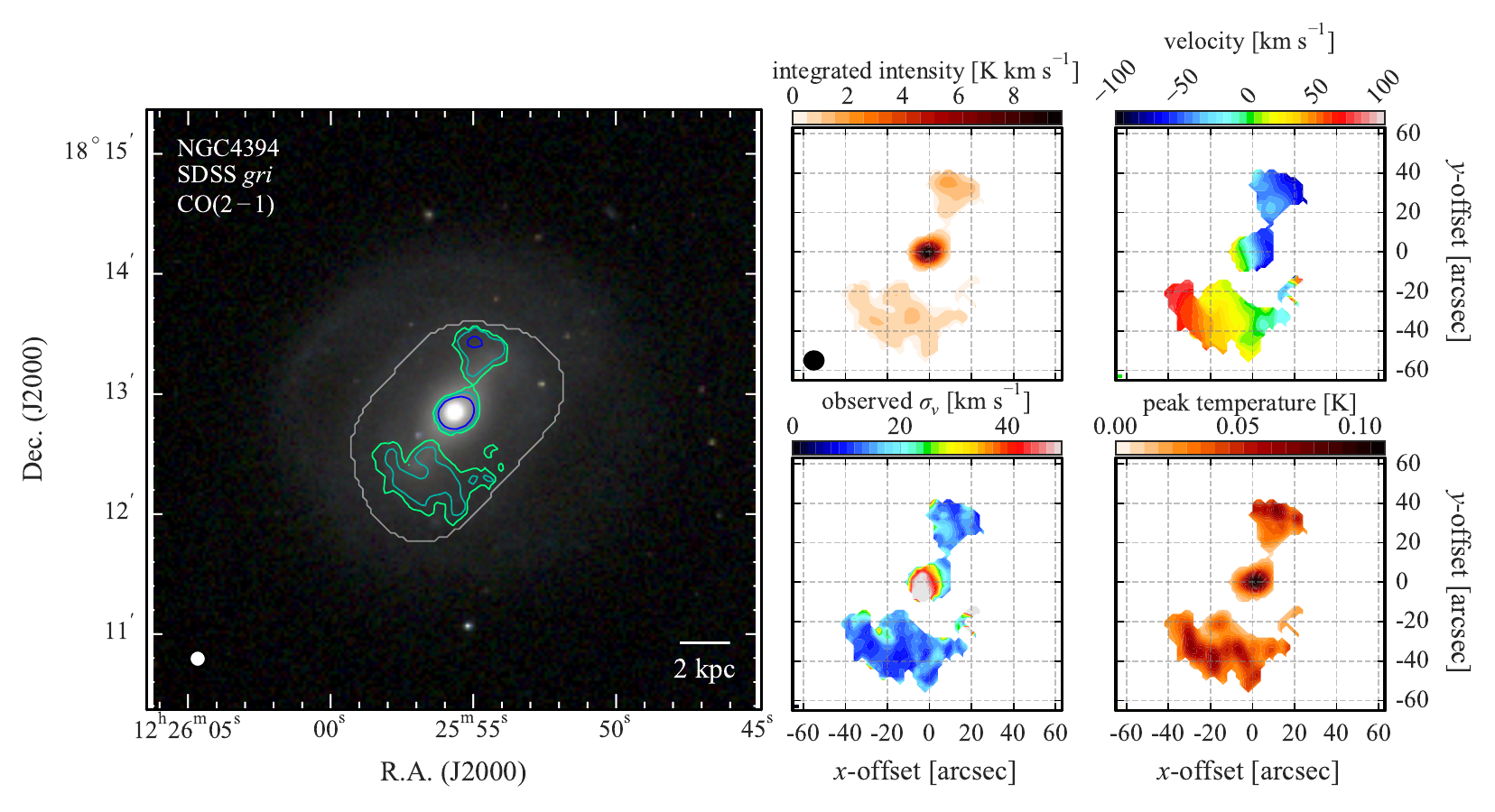}
        \caption{As in Figure 4.1.}
        \label{fig:panelplots_9}
    \end{figure*}

    \begin{figure*}
    \figurenum{4.11}
        \includegraphics[width=\linewidth]{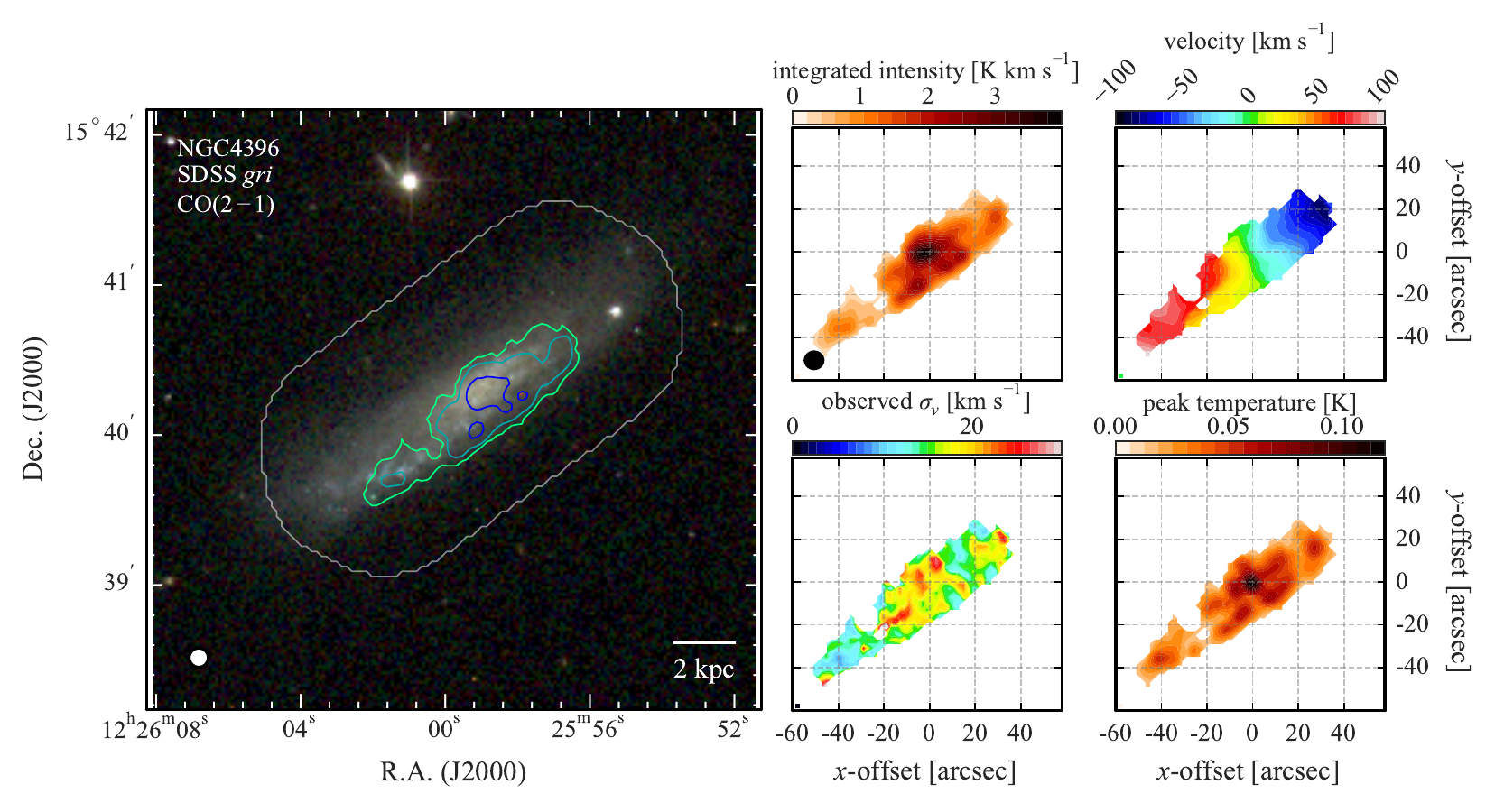}
        \includegraphics[width=\linewidth]{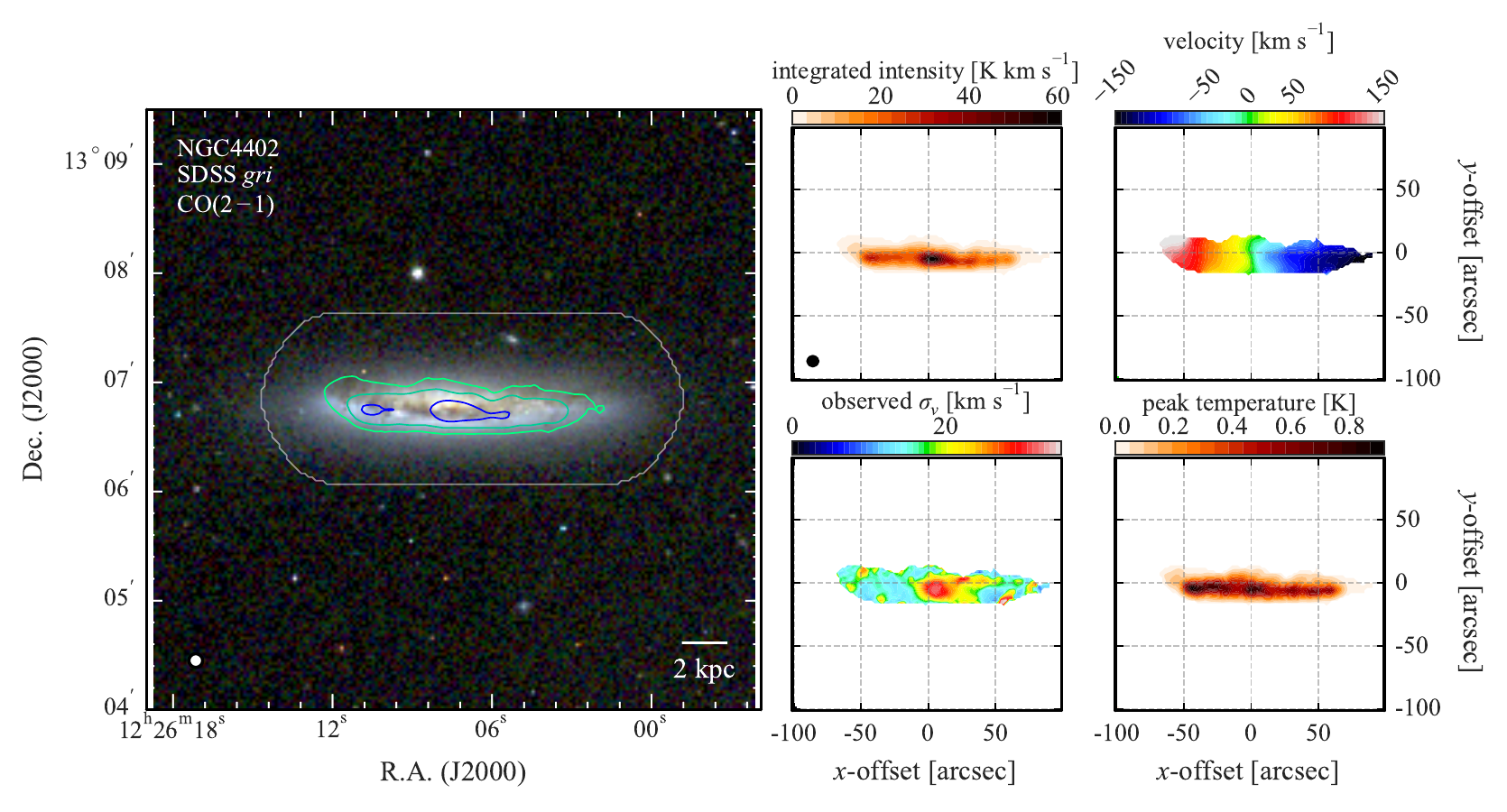}
        \caption{As in Figure 4.1.}
        \label{fig:panelplots_10}
    \end{figure*}

    \begin{figure*}
    \figurenum{4.12}
        \includegraphics[width=\linewidth]{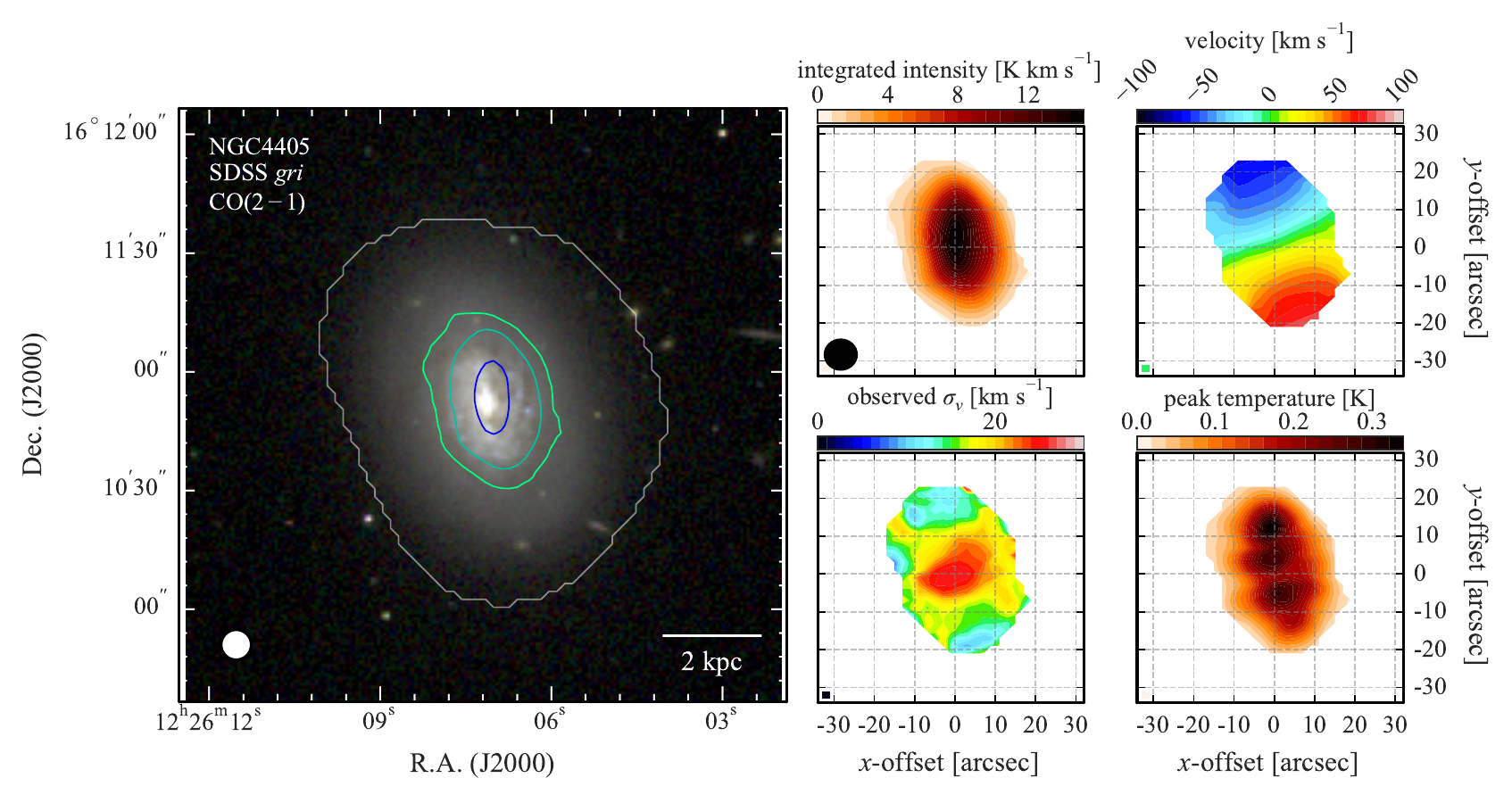}
        \includegraphics[width=\linewidth]{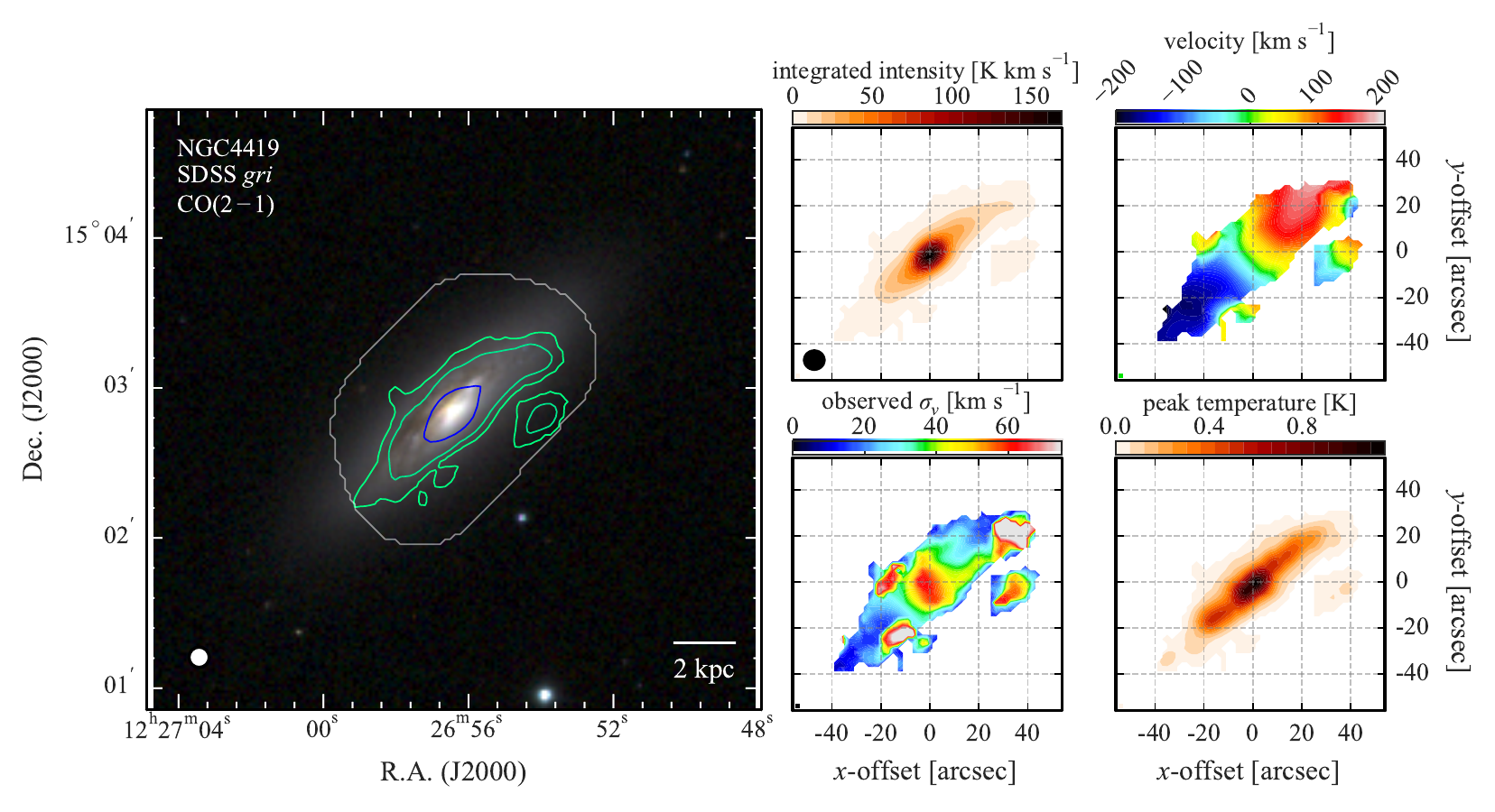}
        \caption{As in Figure 4.1.}
        \label{fig:panelplots_11}
    \end{figure*}

    \begin{figure*}
    \figurenum{4.13}
        \includegraphics[width=\linewidth]{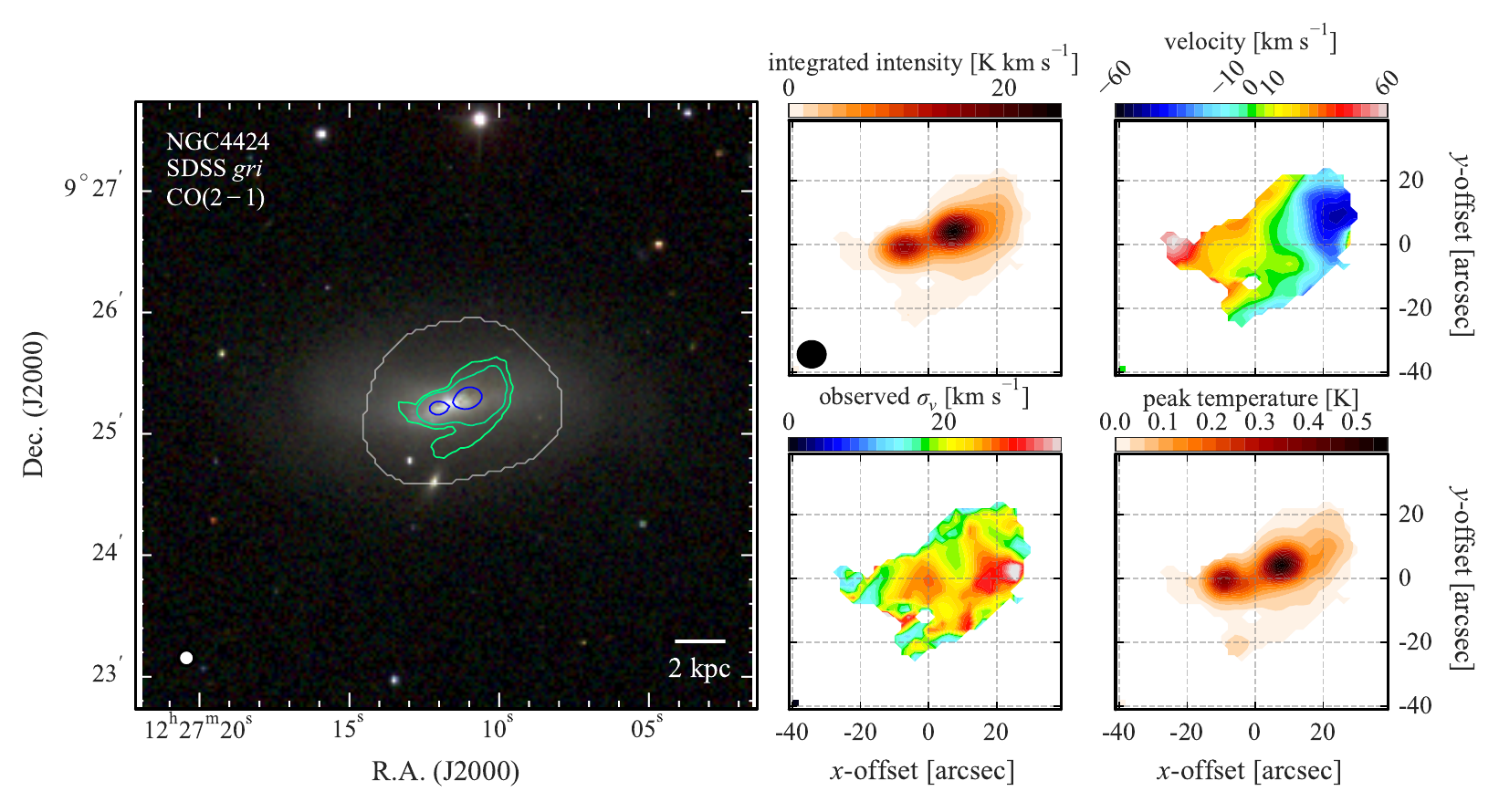}
        \includegraphics[width=\linewidth]{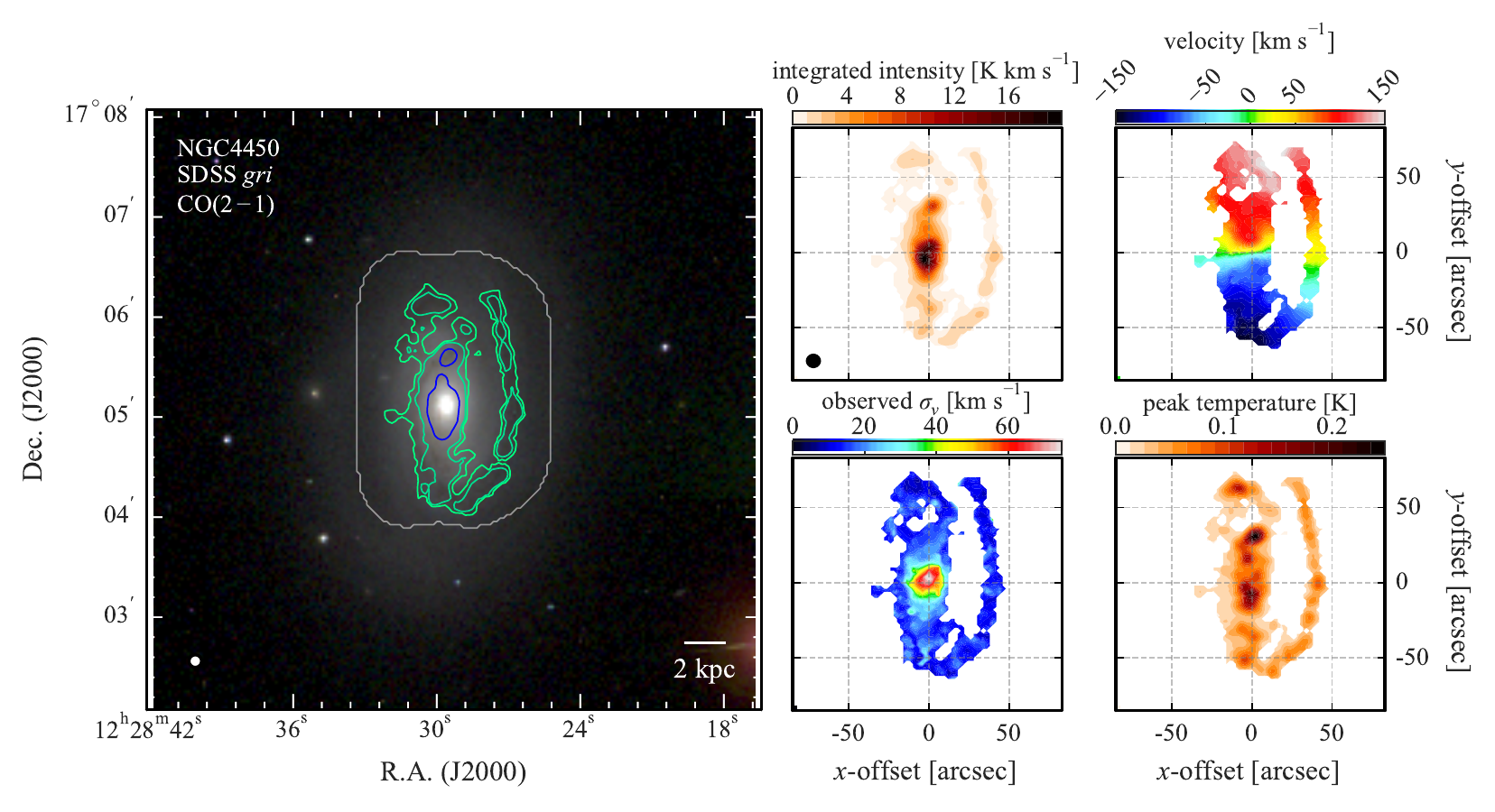}
        \caption{As in Figure 4.1.}
        \label{fig:panelplots_12}
    \end{figure*}

    \begin{figure*}
    \figurenum{4.14}
        \includegraphics[width=\linewidth]{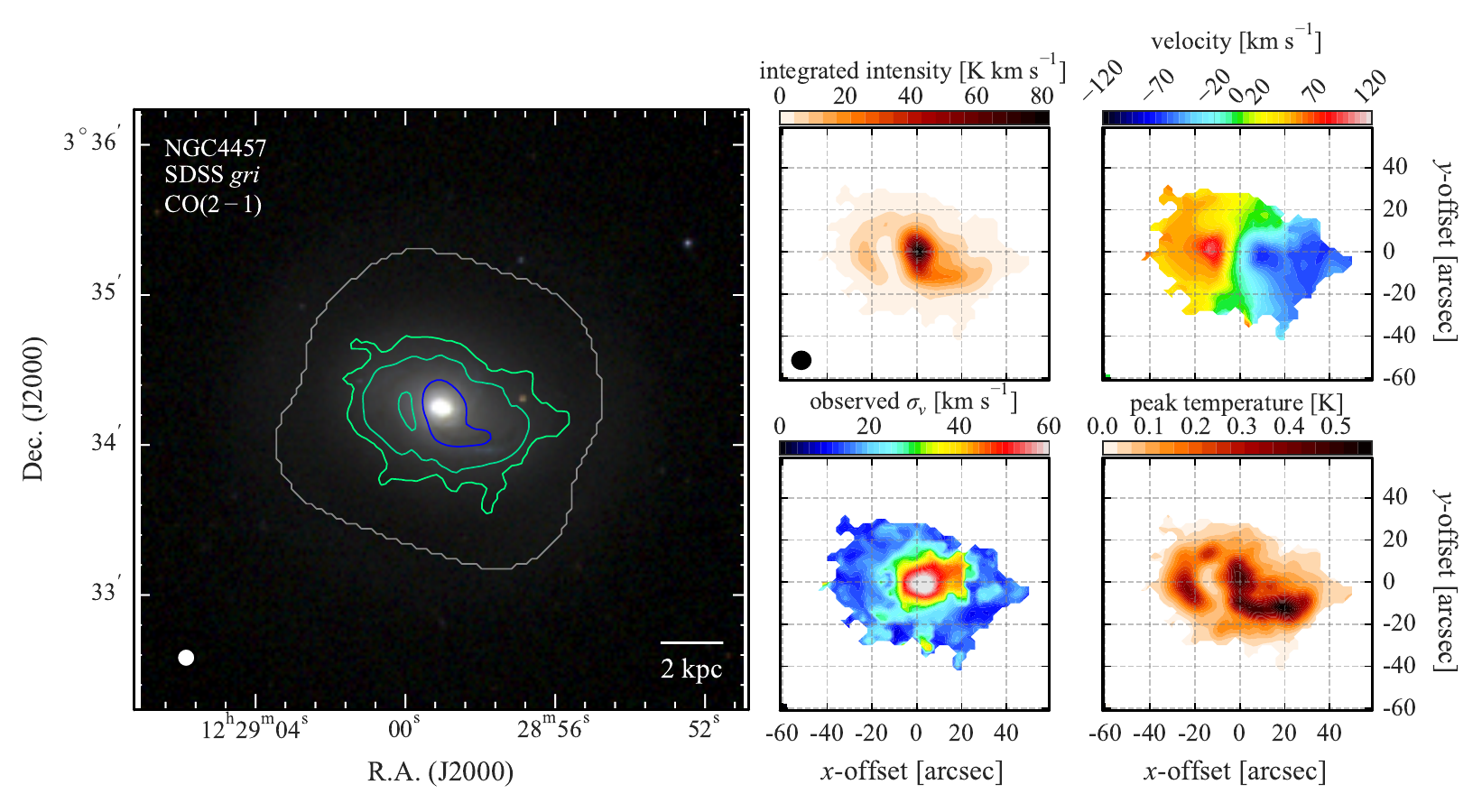}
        \includegraphics[width=\linewidth]{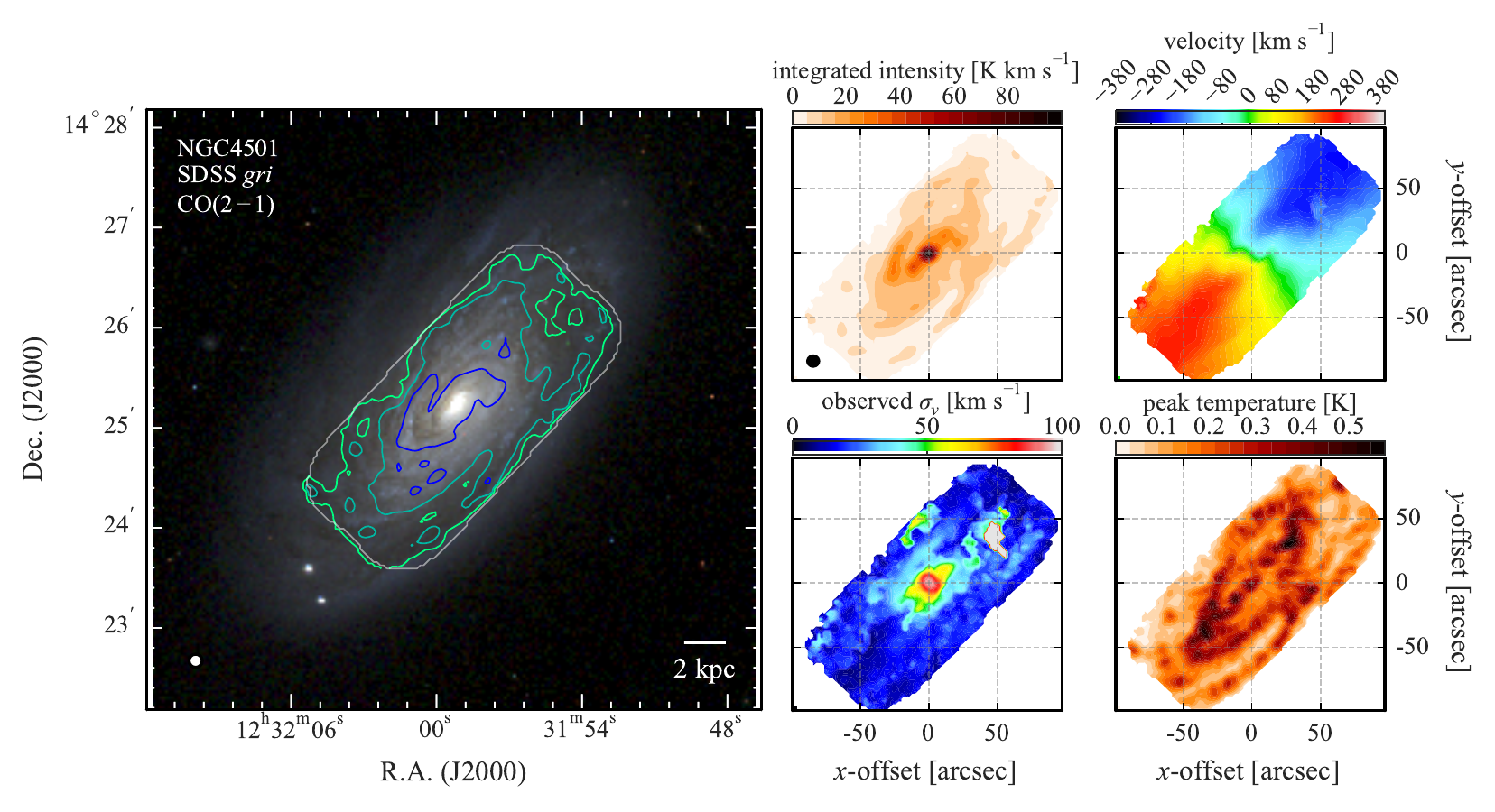}
        \caption{As in Figure 4.1.}
        \label{fig:panelplots_13}
    \end{figure*}

    \begin{figure*}
    \figurenum{4.15}
        \includegraphics[width=\linewidth]{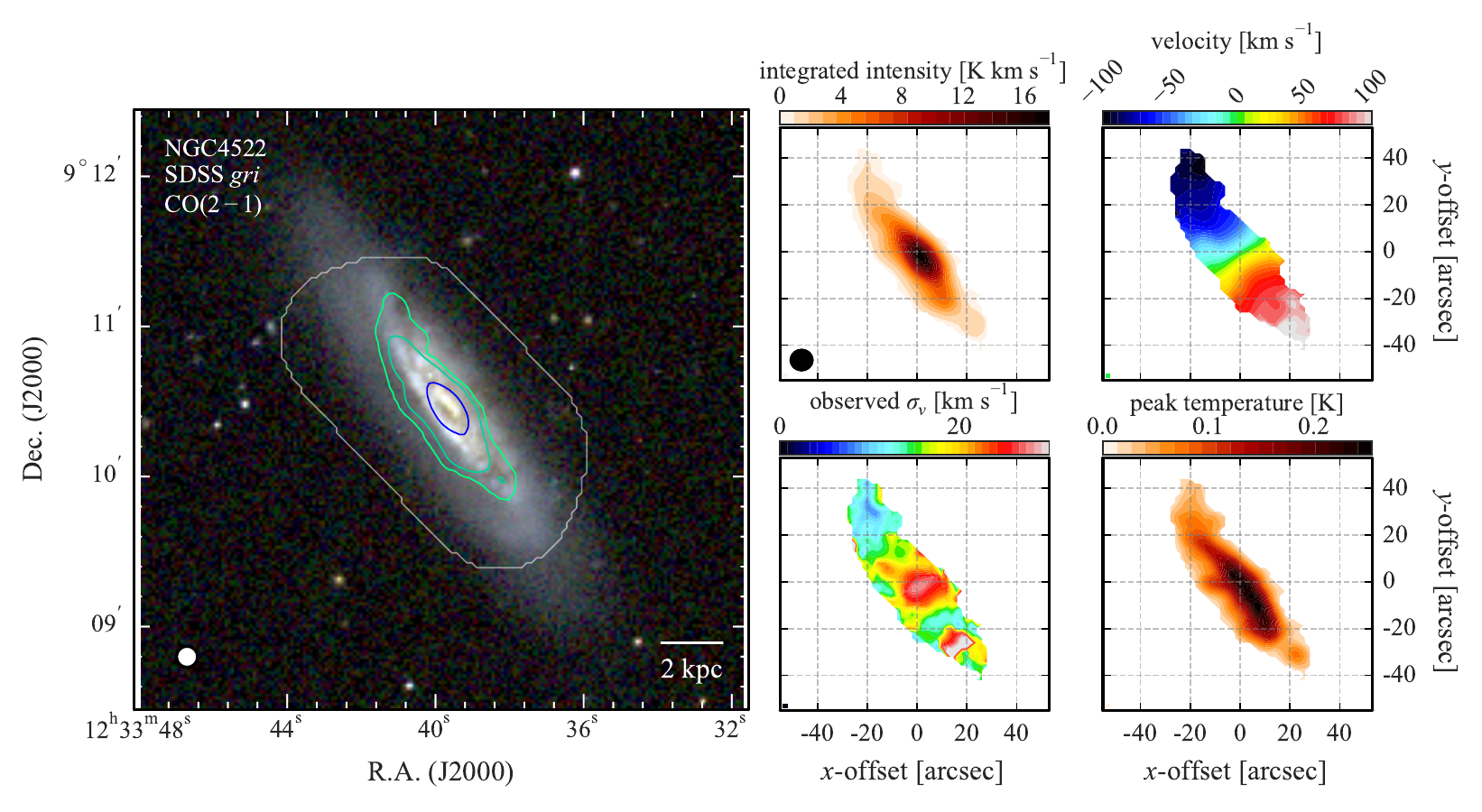}
        \includegraphics[width=\linewidth]{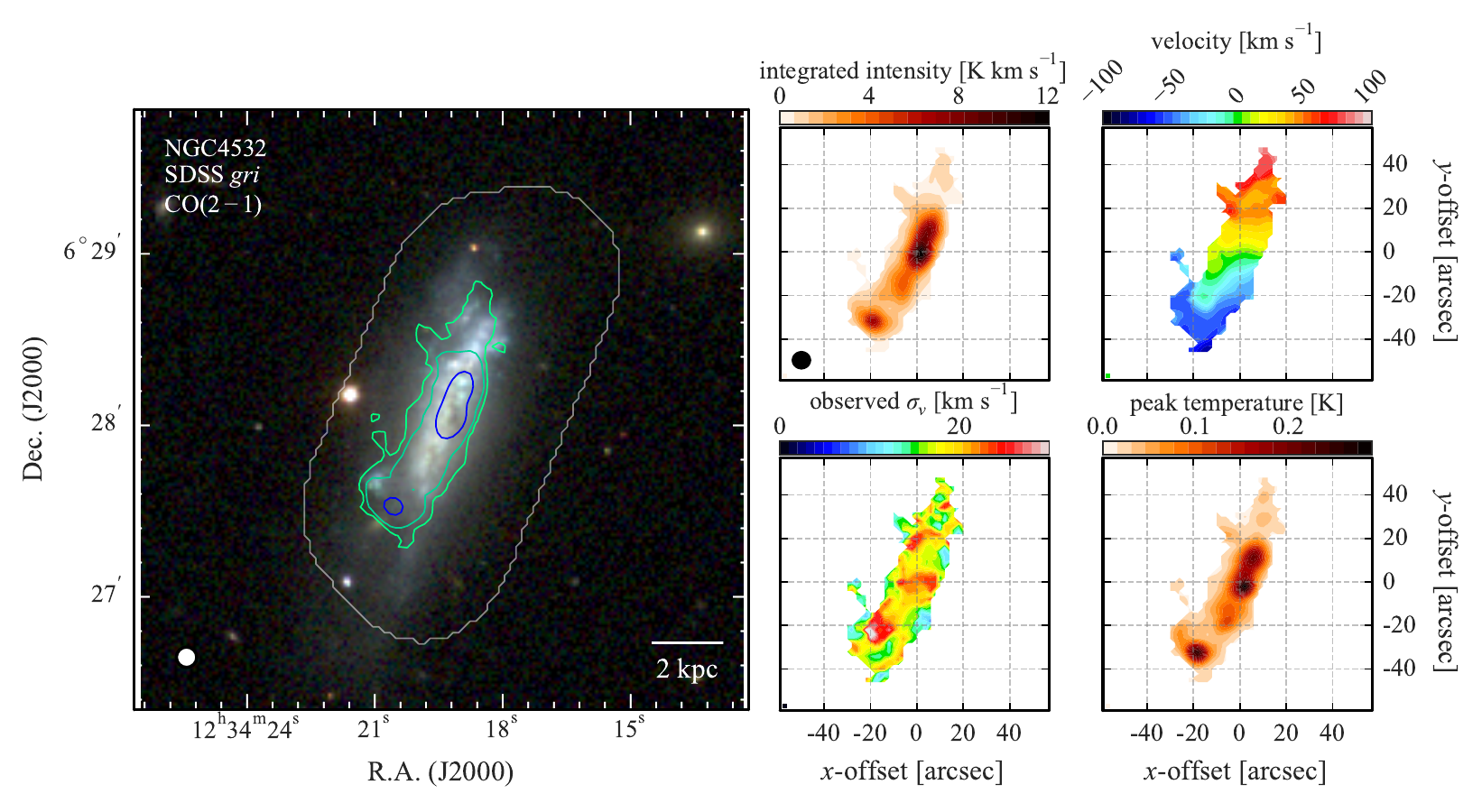}
        \caption{As in Figure 4.1.}
        \label{fig:panelplots_14}
    \end{figure*}

    \begin{figure*}
    \figurenum{4.16}
        \includegraphics[width=\linewidth]{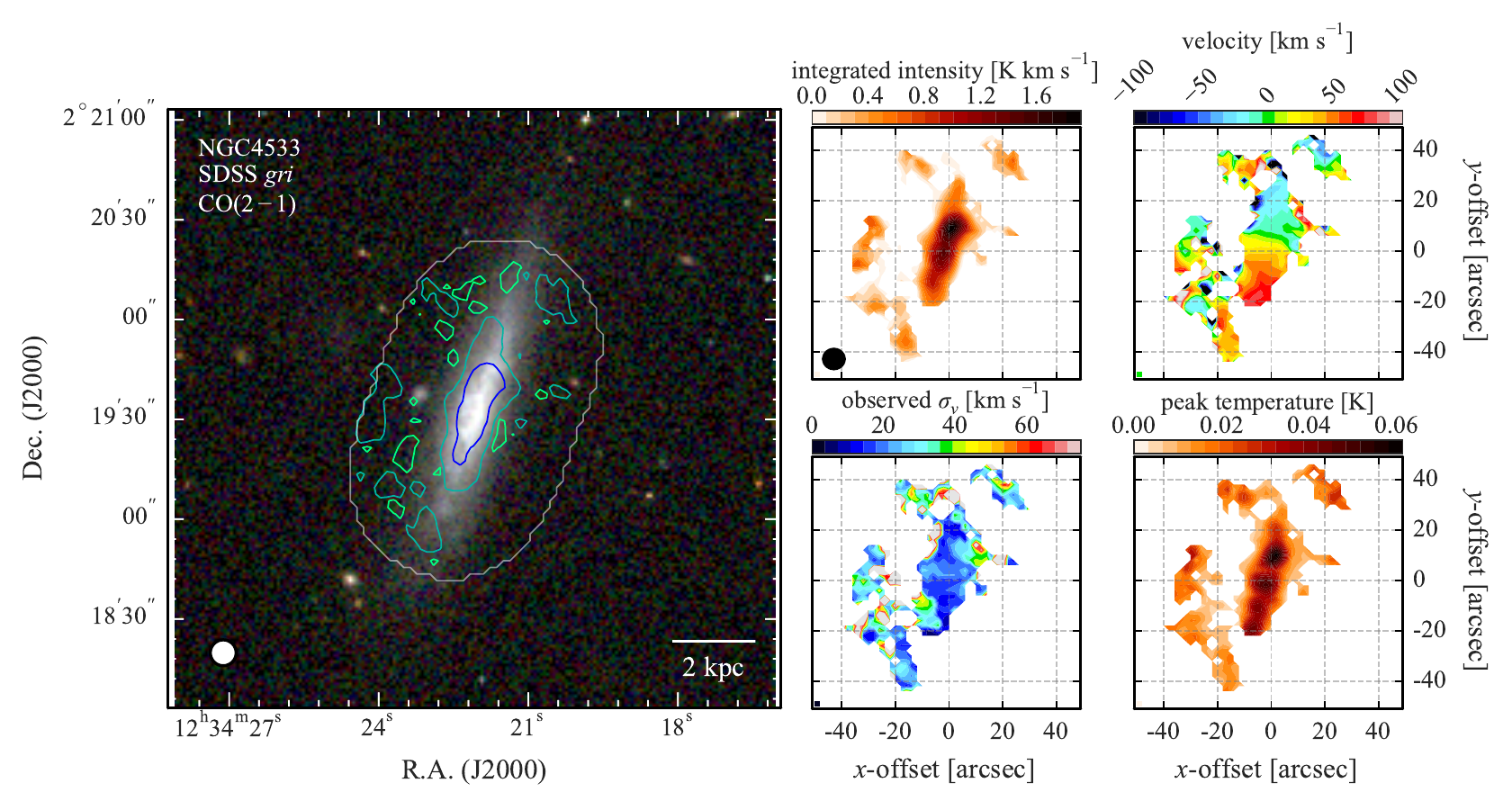}
        \includegraphics[width=\linewidth]{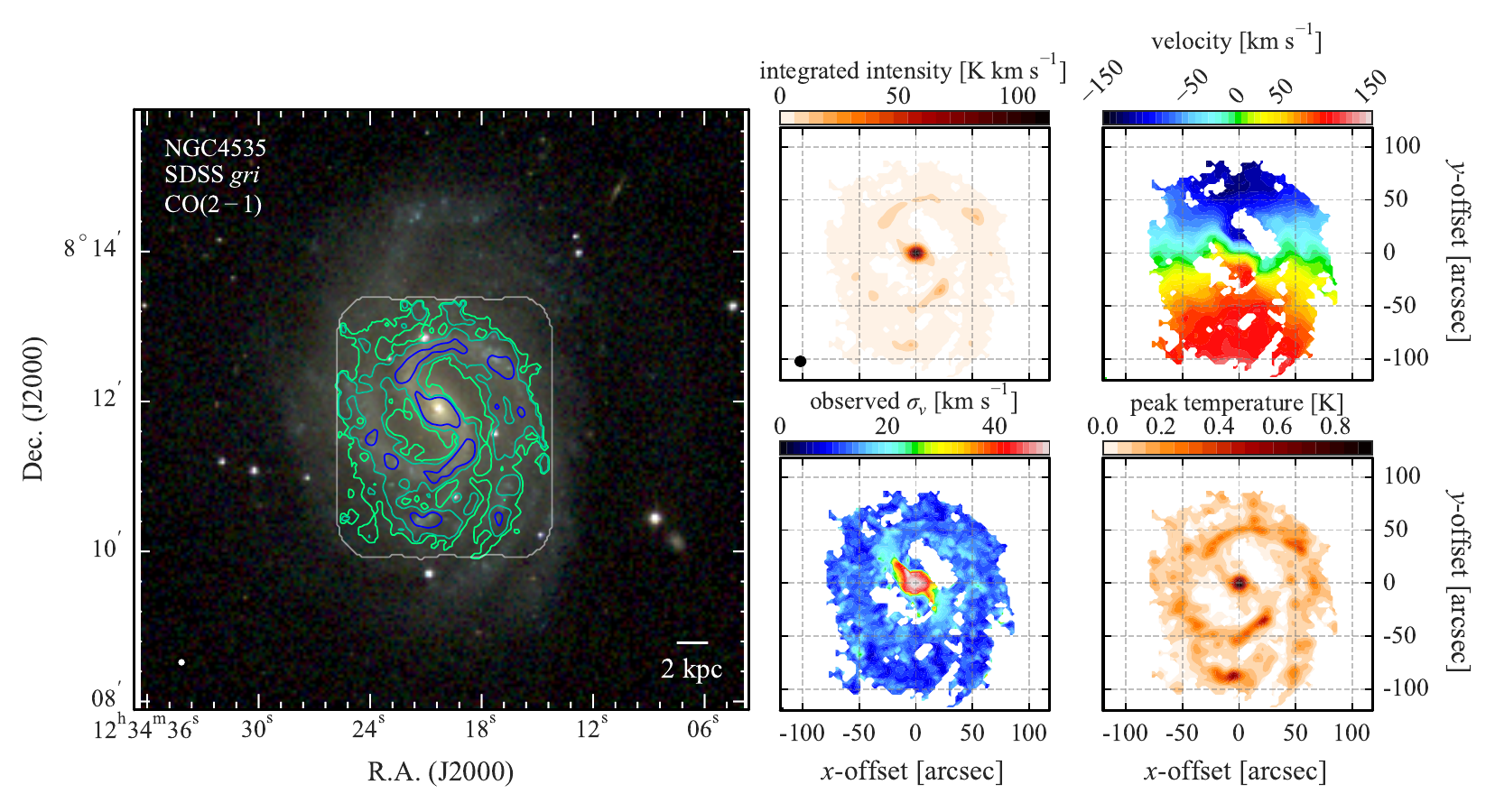}
        \caption{As in Figure 4.1.}
        \label{fig:panelplots_15}
    \end{figure*}

    \begin{figure*}
    \figurenum{4.17}
        \includegraphics[width=\linewidth]{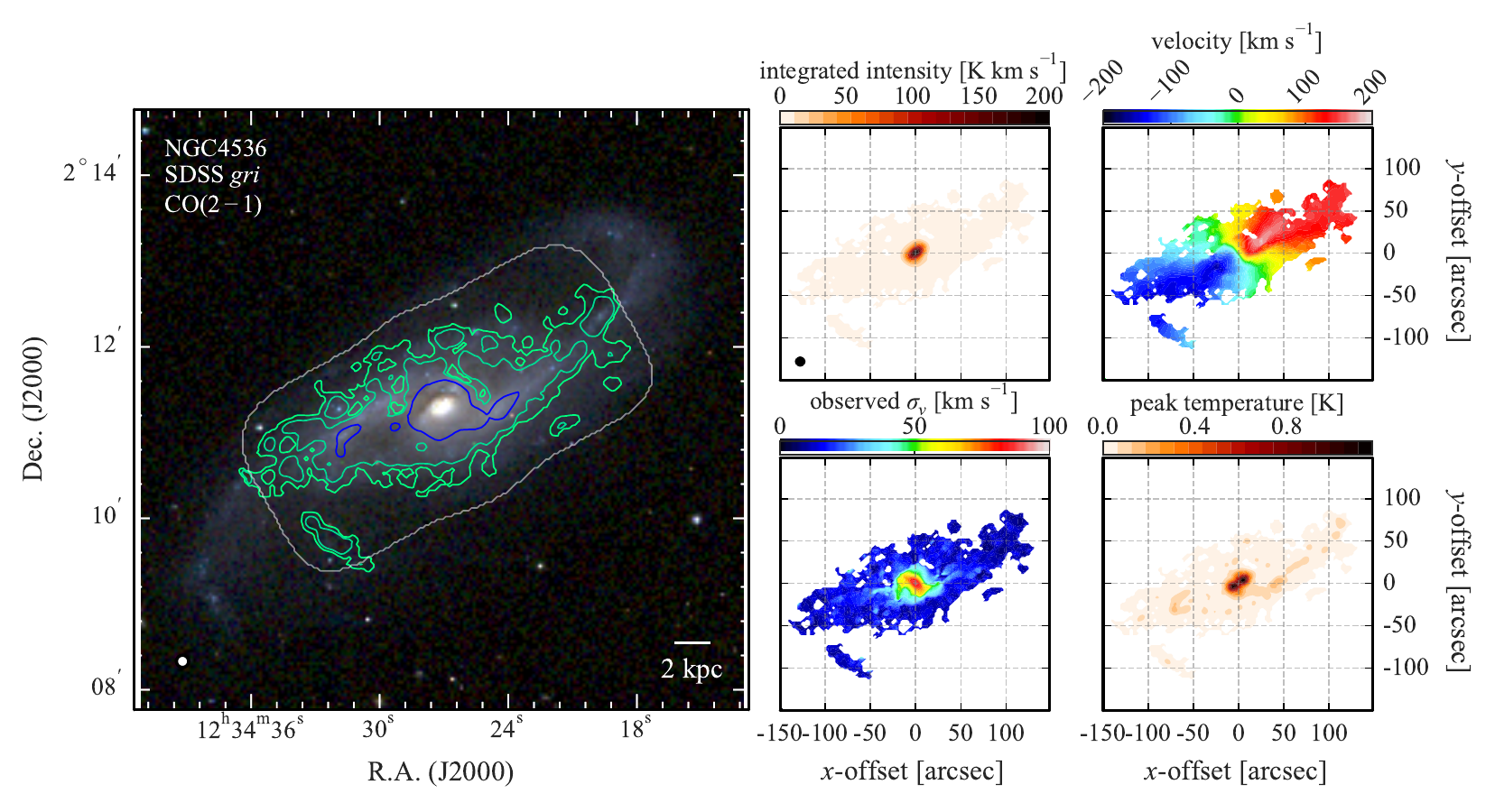}
        \includegraphics[width=\linewidth]{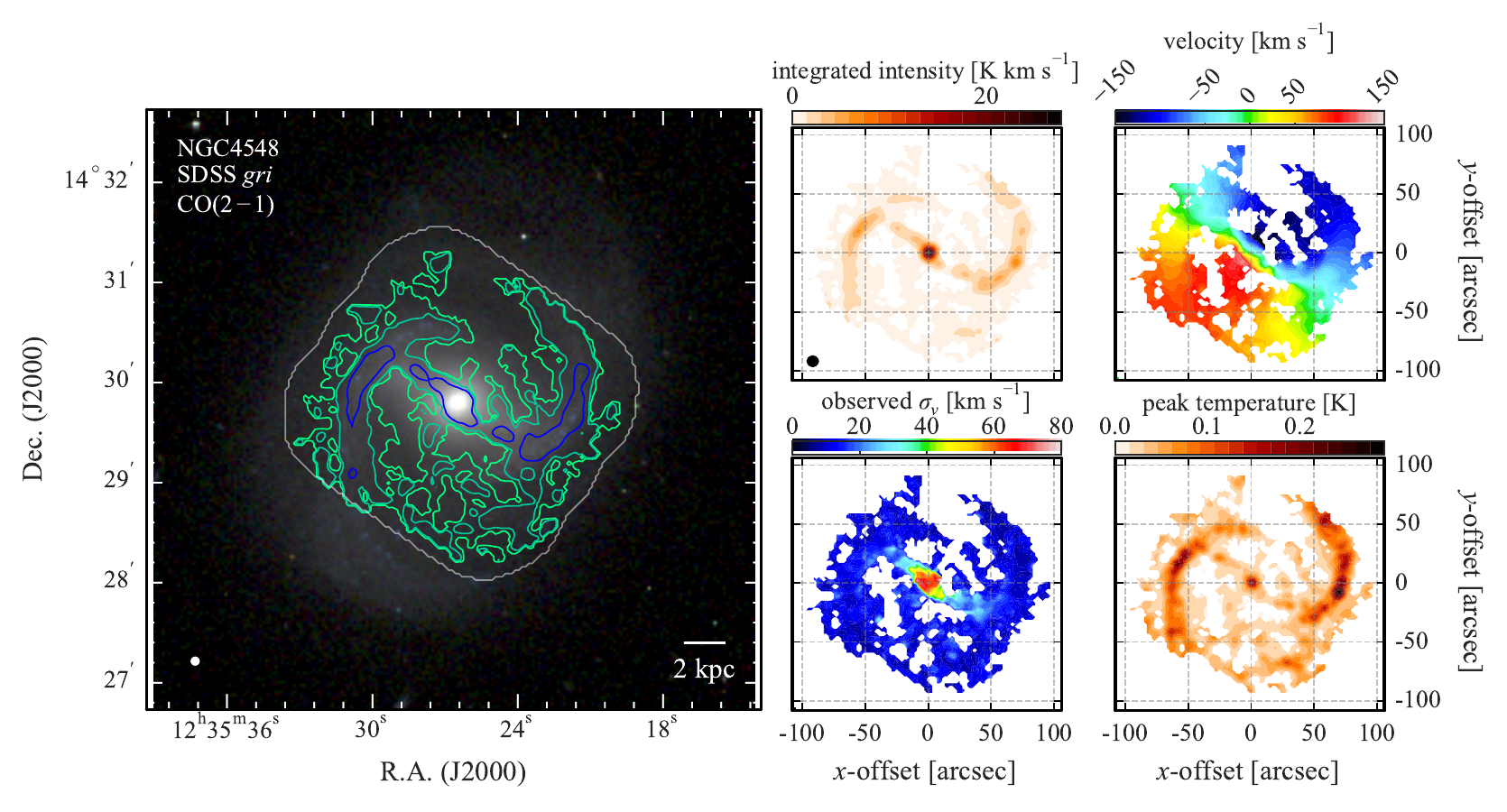}
        \caption{As in Figure 4.1.}
        \label{fig:panelplots_16}
    \end{figure*}

    \begin{figure*}
    \figurenum{4.18}
        \includegraphics[width=\linewidth]{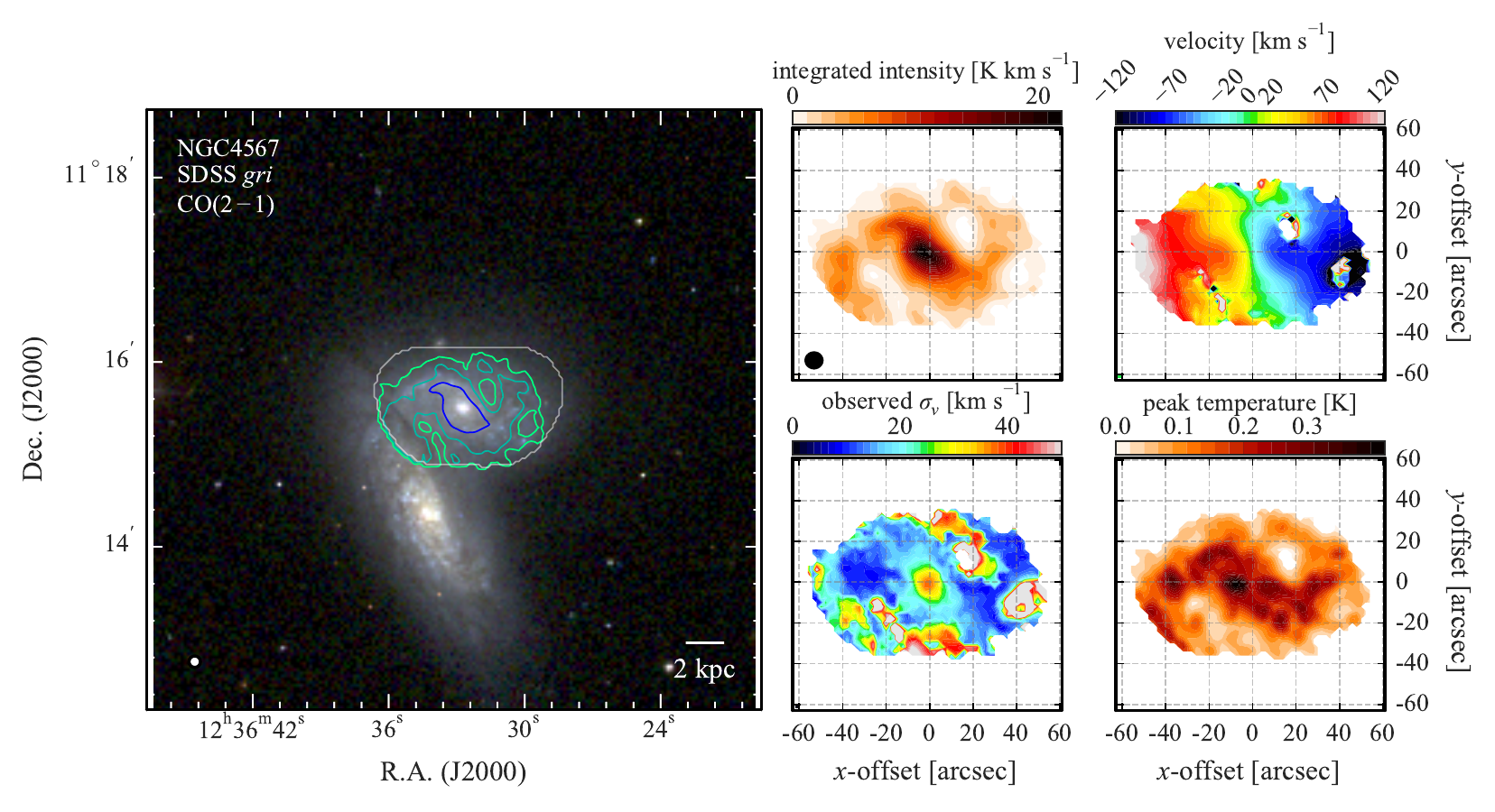}
        \includegraphics[width=\linewidth]{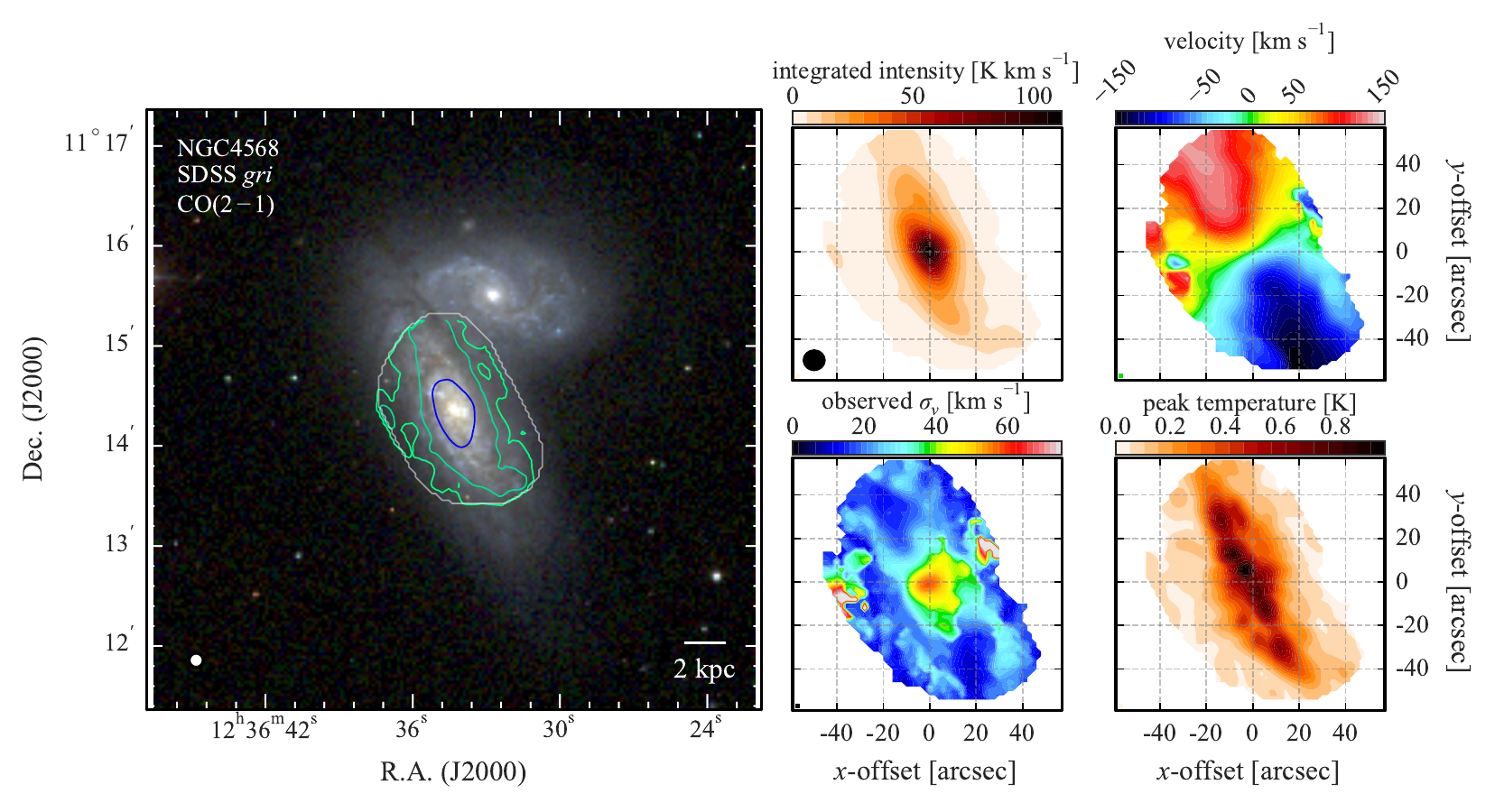}
        \caption{As in Figure 4.1.}
        \label{fig:panelplots_17}
    \end{figure*}

    \begin{figure*}
    \figurenum{4.19}
        \includegraphics[width=\linewidth]{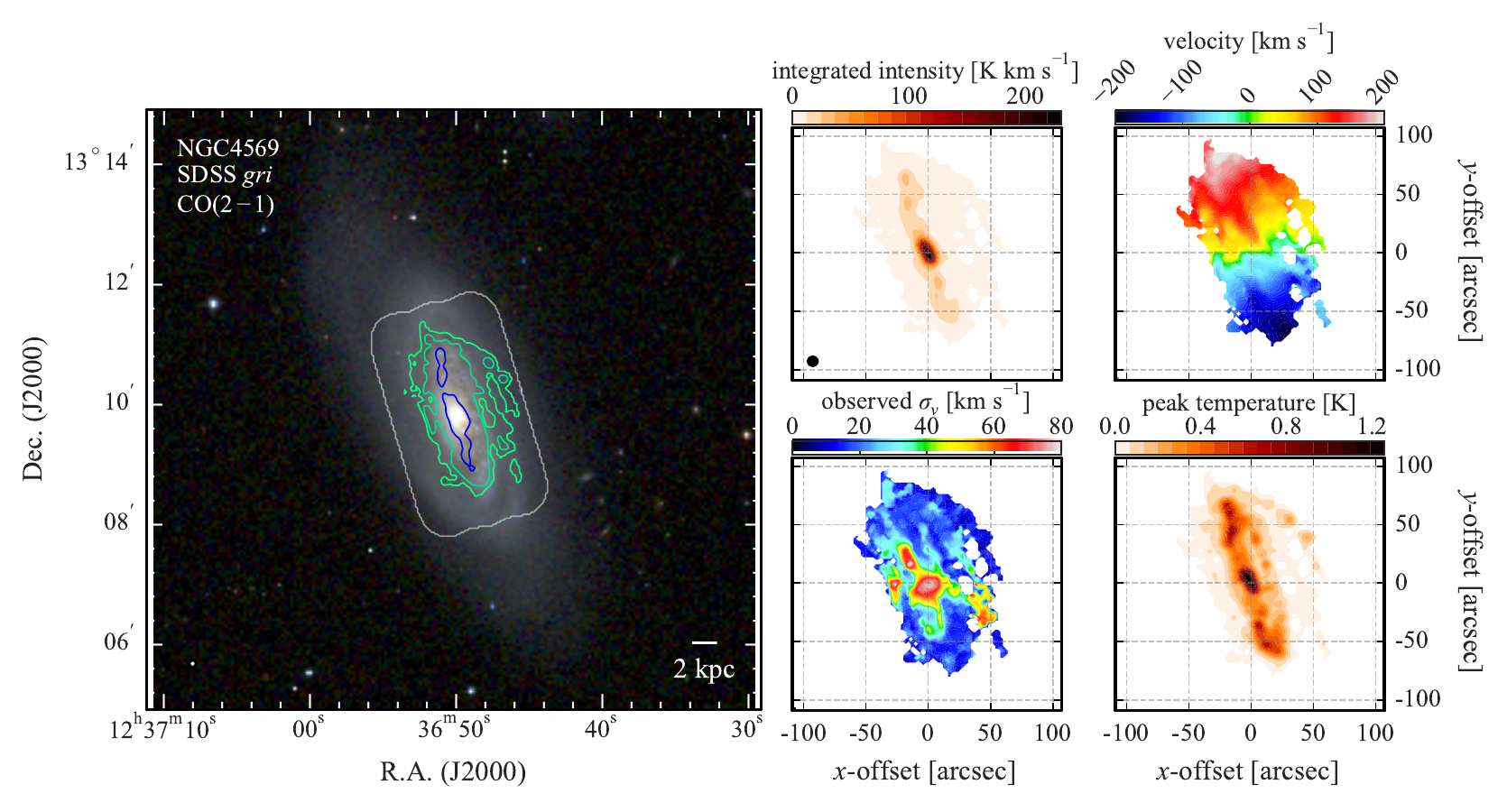}
        \includegraphics[width=\linewidth]{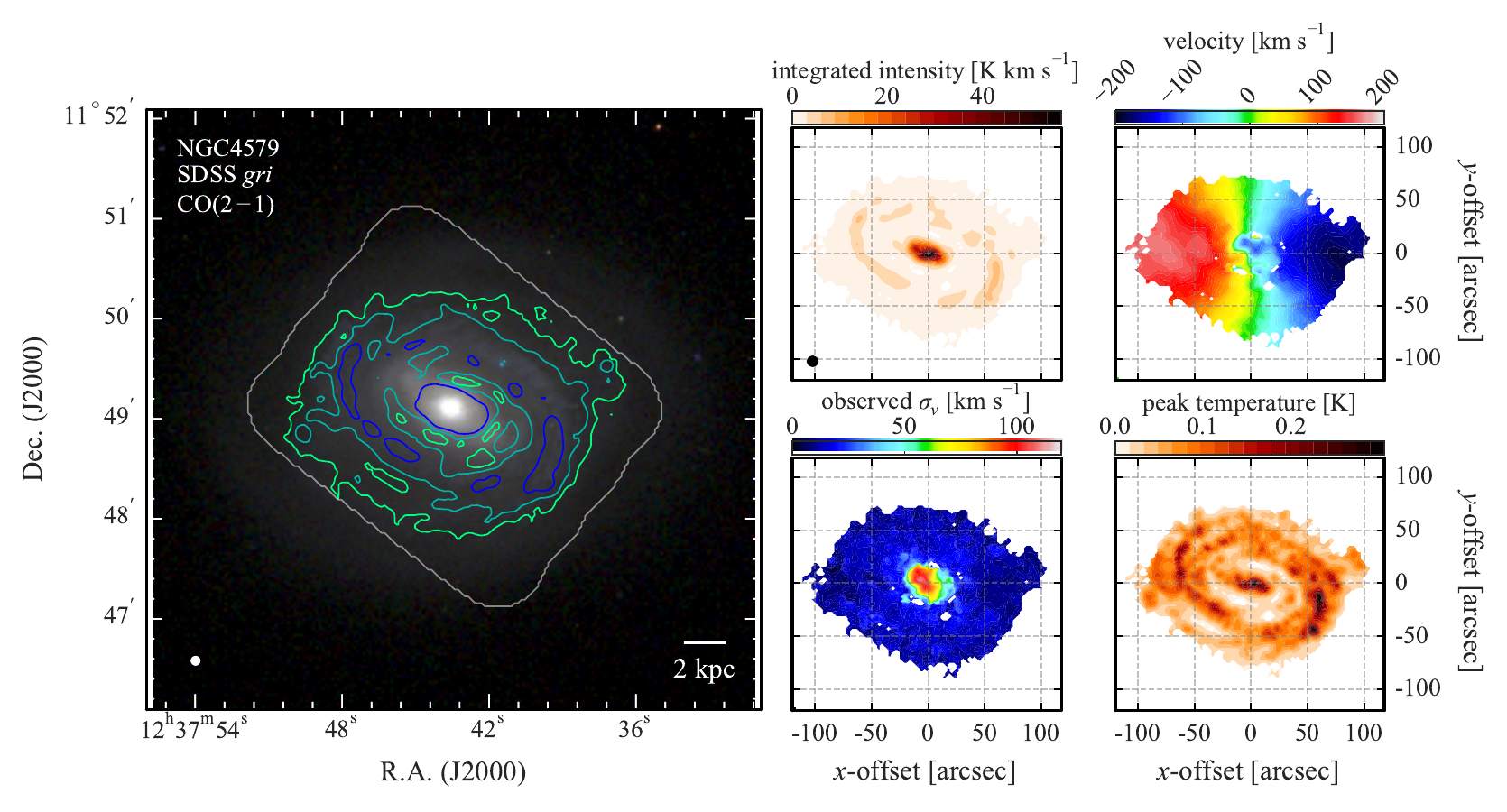}
        \caption{As in Figure 4.1.}
        \label{fig:panelplots_18}
    \end{figure*}

    \begin{figure*}
    \figurenum{4.20}
        \includegraphics[width=\linewidth]{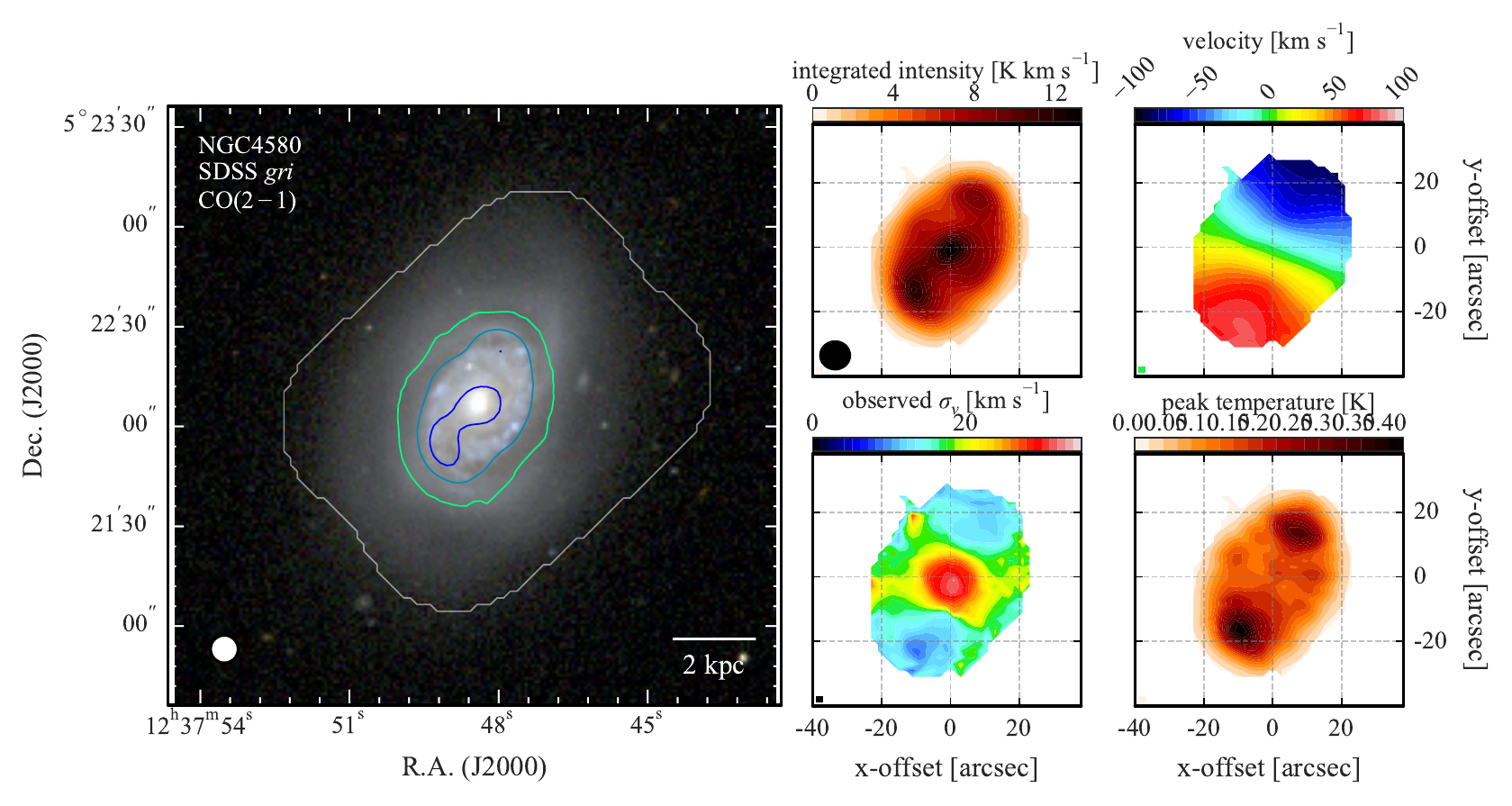}
        \includegraphics[width=\linewidth]{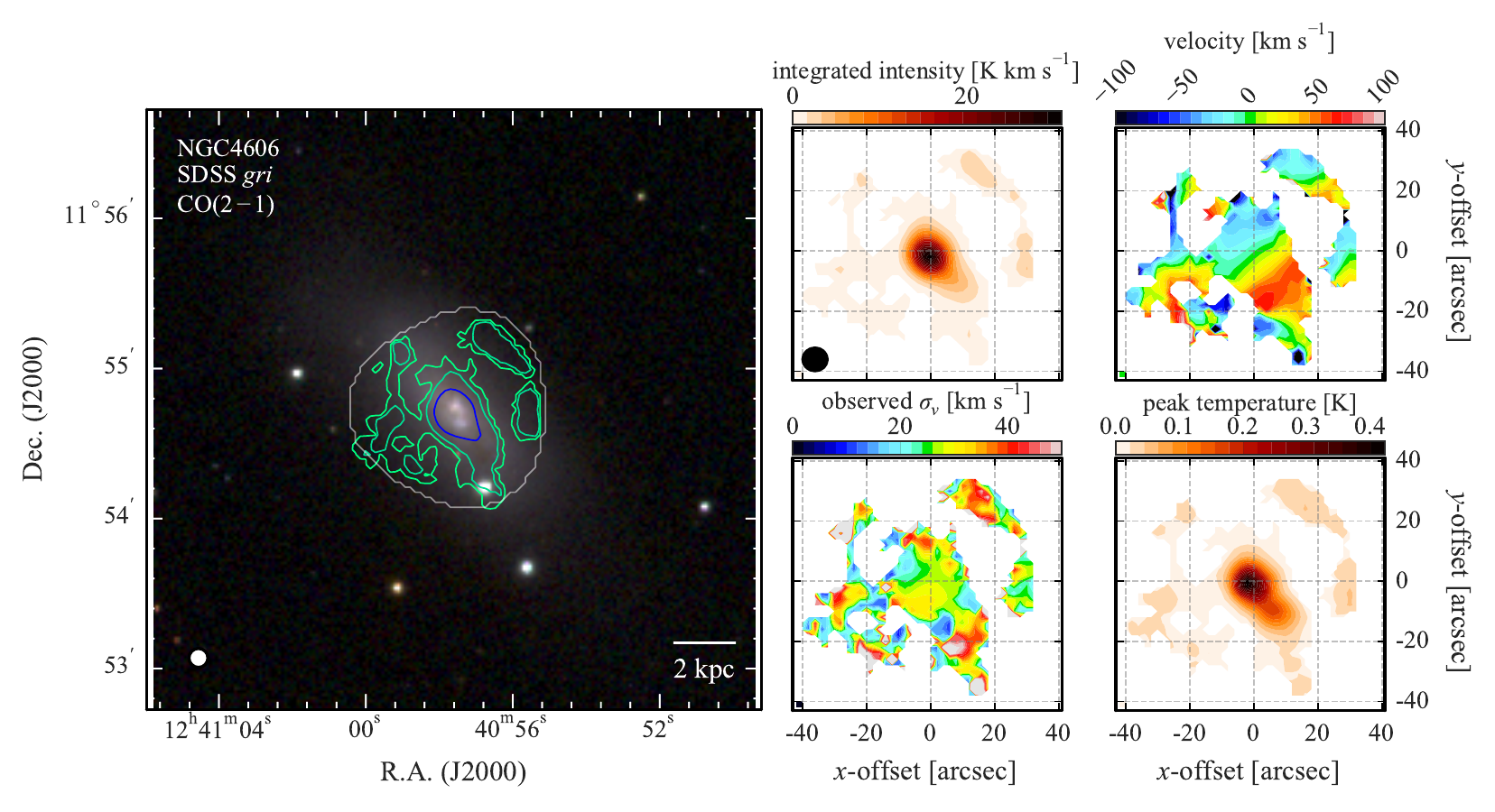}
        \caption{As in Figure 4.1.}
        \label{fig:panelplots_19}
    \end{figure*}

    \begin{figure*}
    \figurenum{4.21}
        \includegraphics[width=\linewidth]{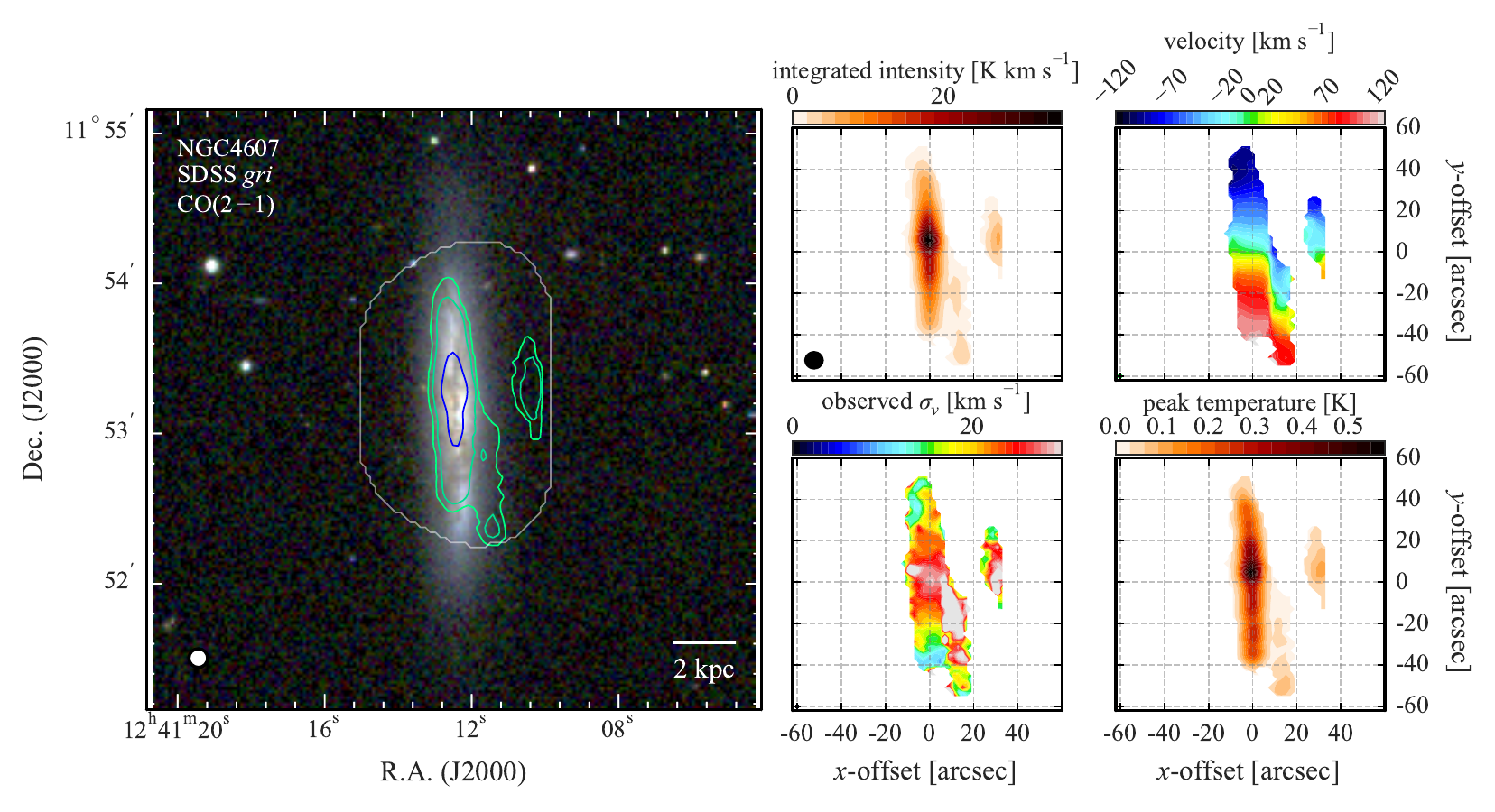}
        \includegraphics[width=\linewidth]{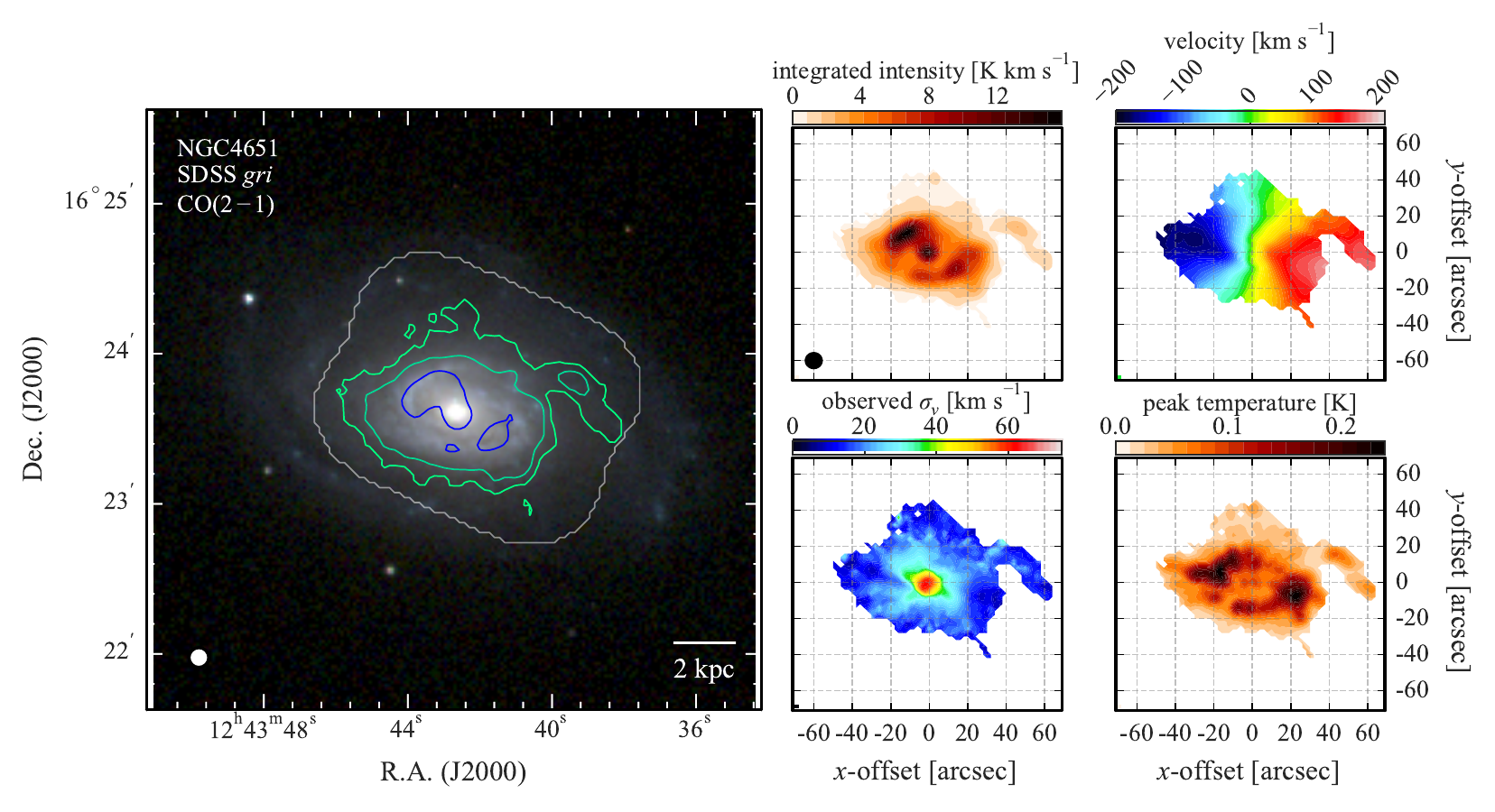}
        \caption{As in Figure 4.1. In the case of NGC4607, we caution against over-interpreting the extended features to the West of the main disk as this is likely related to the PSF pattern of the observations.}
        \label{fig:panelplots_20}
    \end{figure*}

    \begin{figure*}
    \figurenum{4.22}
        \includegraphics[width=\linewidth]{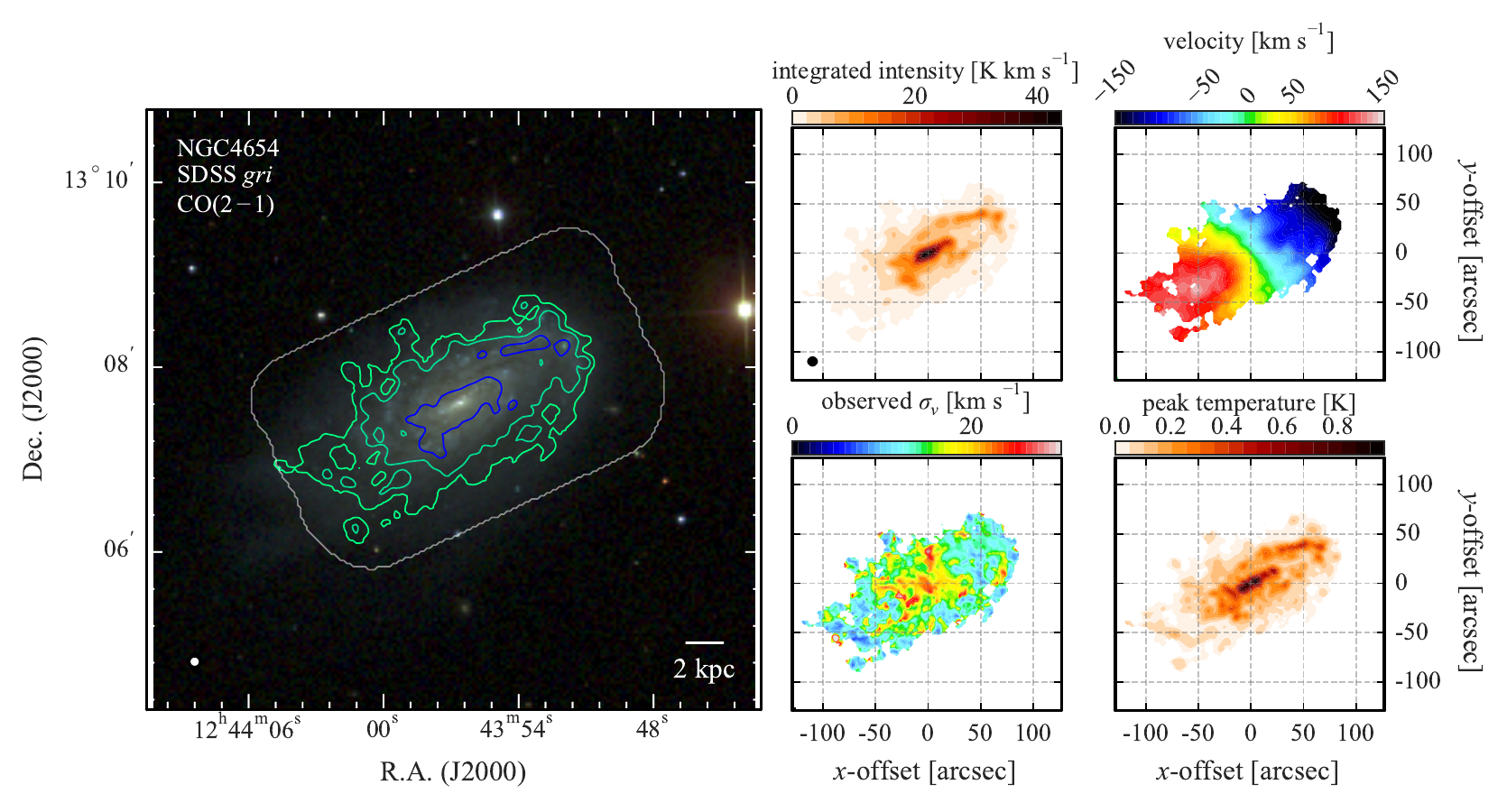}
        \includegraphics[width=\linewidth]{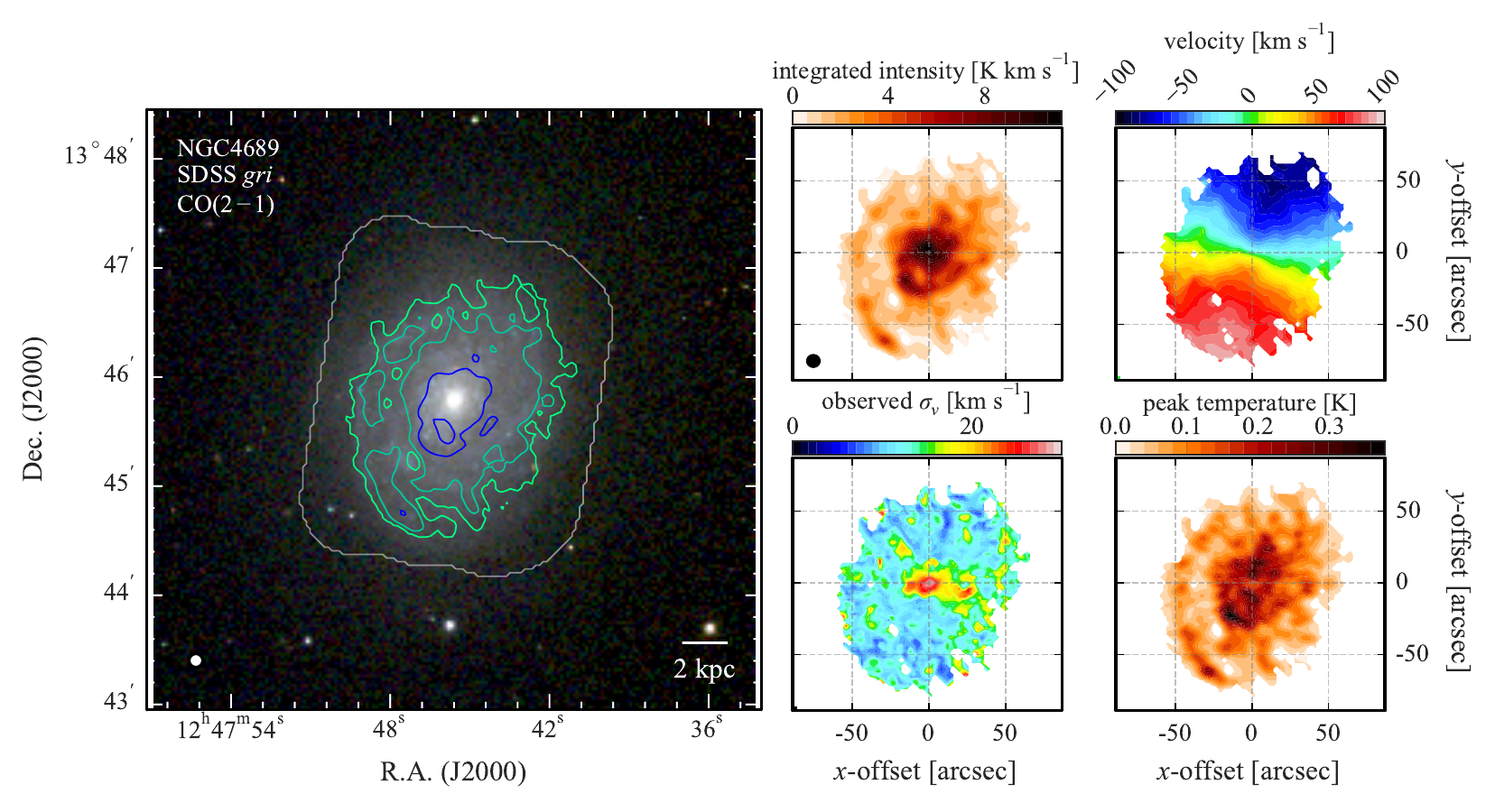}
        \caption{As in Figure 4.1.}
        \label{fig:panelplots_21}
    \end{figure*}

    \begin{figure*}
    \figurenum{4.23}
        \includegraphics[width=\linewidth]{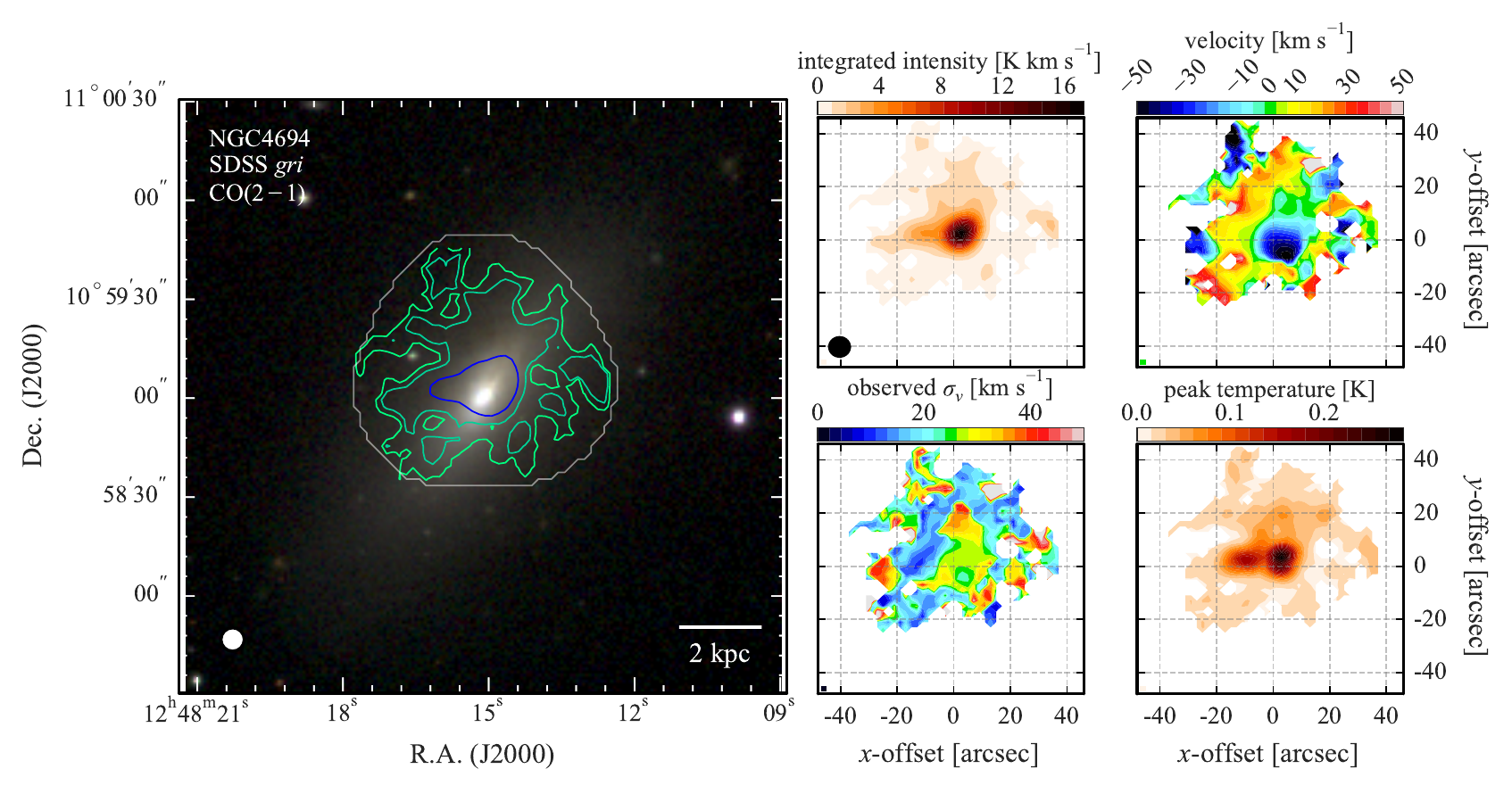}
        \includegraphics[width=\linewidth]{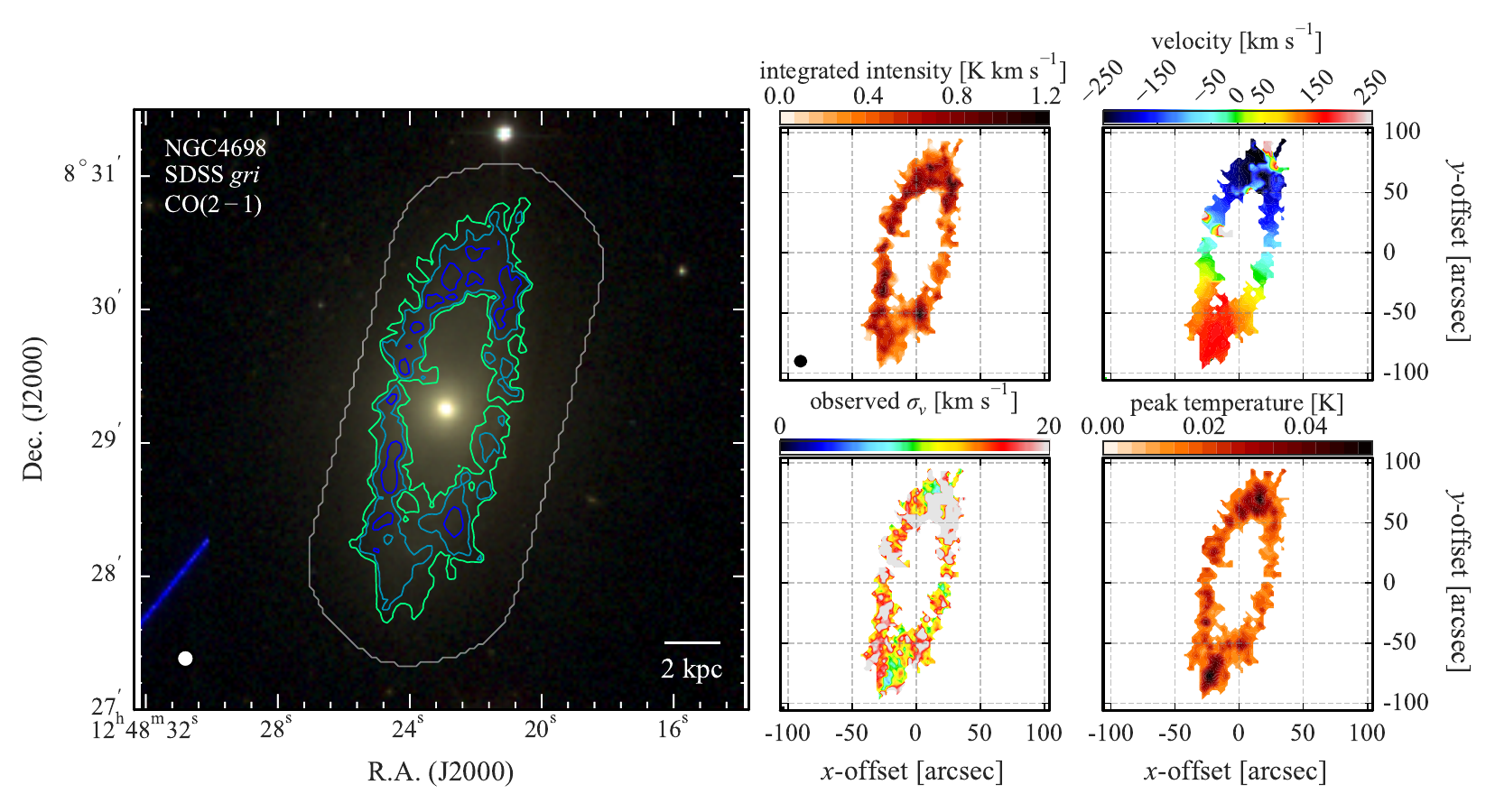}
        \caption{As in Figure 4.1.}
        \label{fig:panelplots_22}
    \end{figure*}

    \begin{figure*}
    \figurenum{4.24}
        \includegraphics[width=\linewidth]{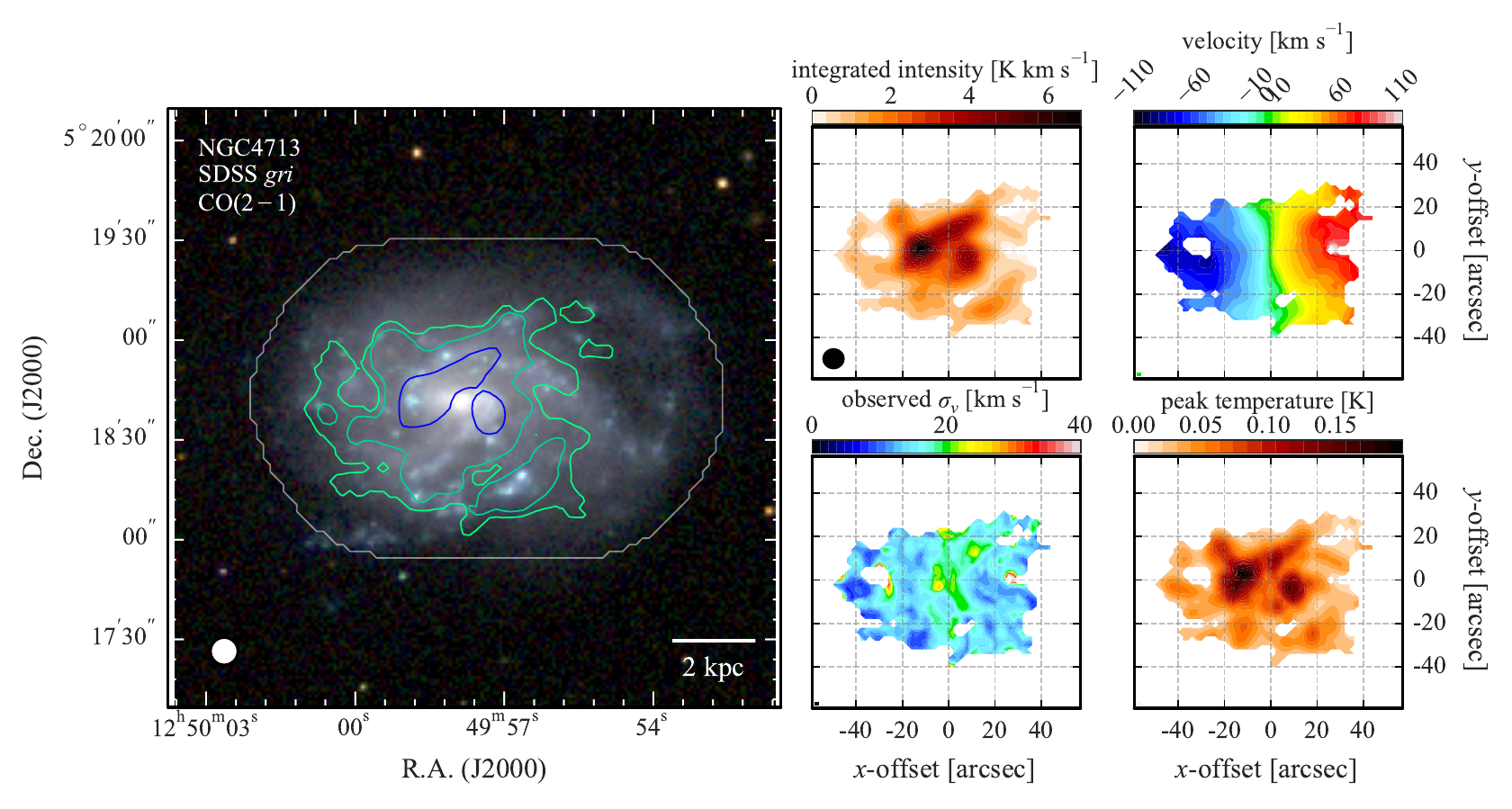}
        \includegraphics[width=\linewidth]{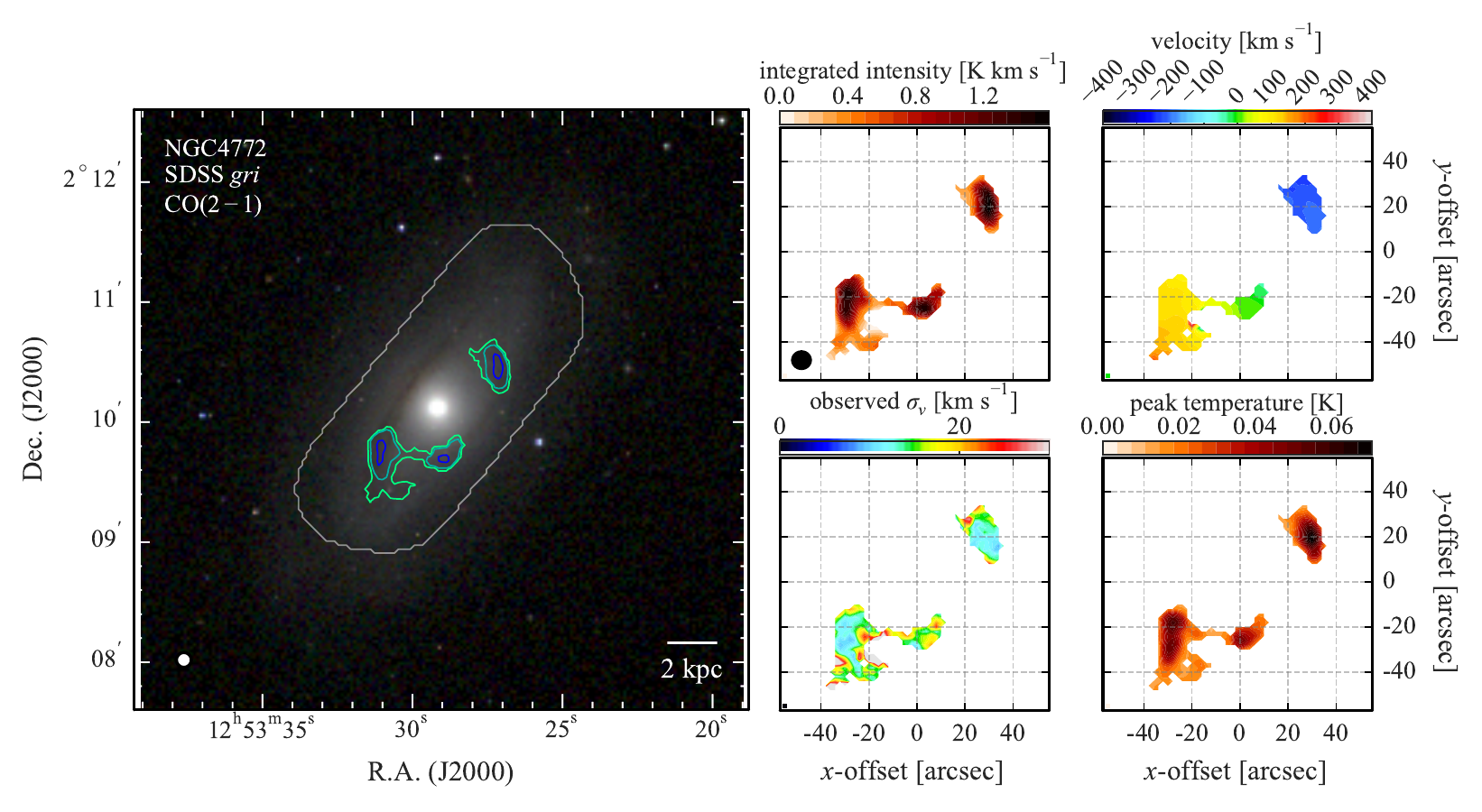}
        \caption{As in Figure 4.1.}
        \label{fig:panelplots_23}
    \end{figure*}

    \begin{figure*}
    \figurenum{4.25}
        \includegraphics[width=\linewidth]{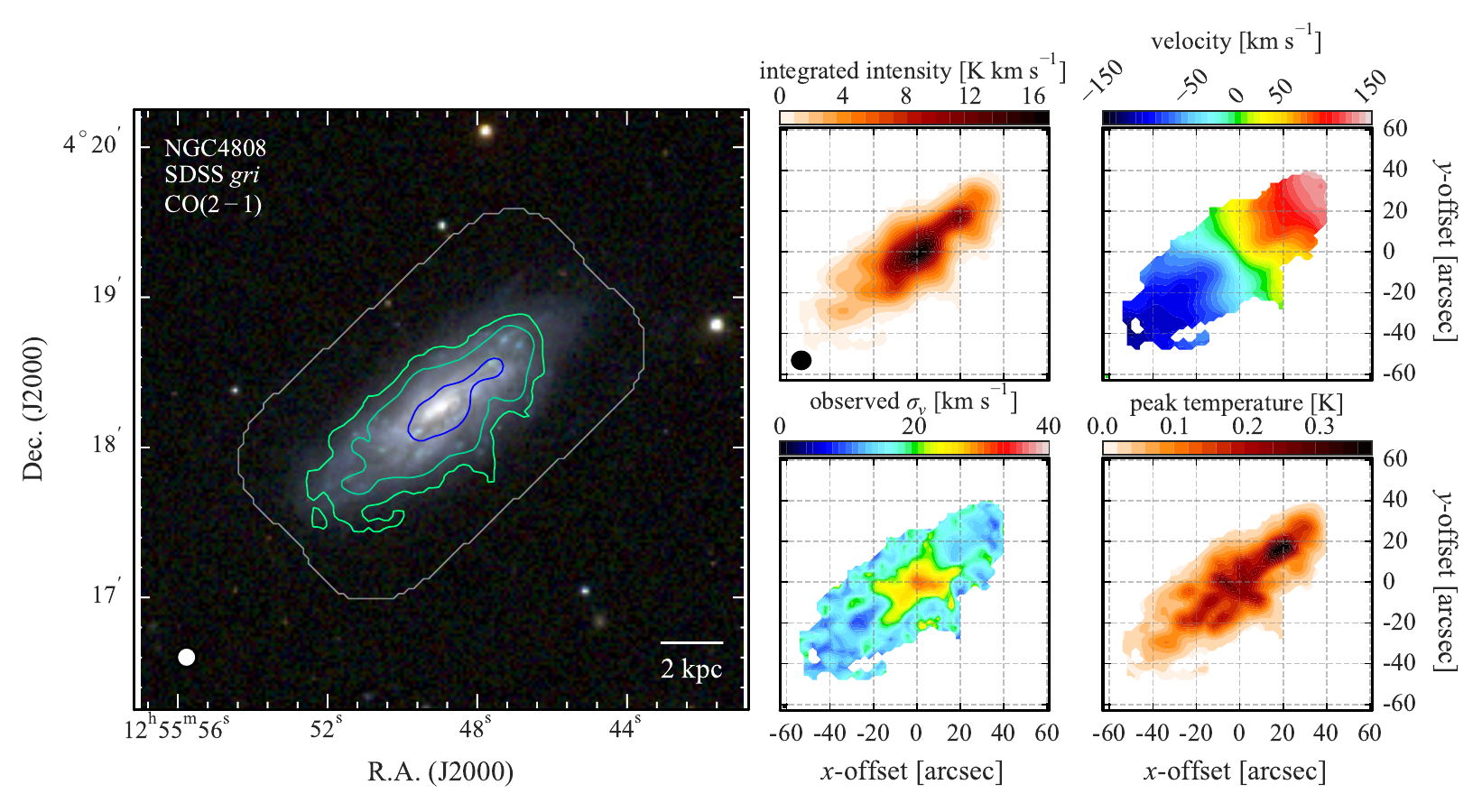}
        \caption{As in Figure 4.1.}
        \label{fig:panelplots_24}
    \end{figure*}

    \newpage

\section{Emission Line Spectra}
\label{app:Spectra}
\begin{figure*}[b]
    \figurenum{13}
	\centering
	\includegraphics[width=0.4\textwidth]{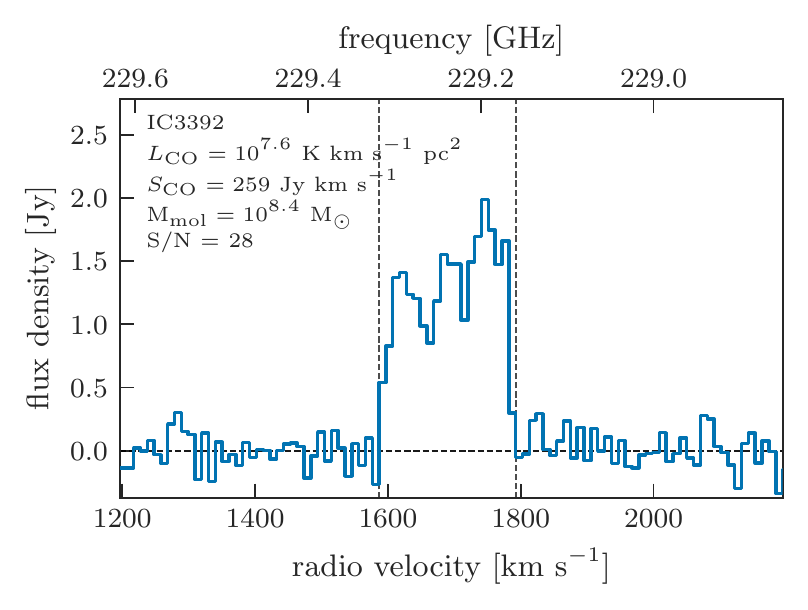}
	\includegraphics[width=0.4\textwidth]{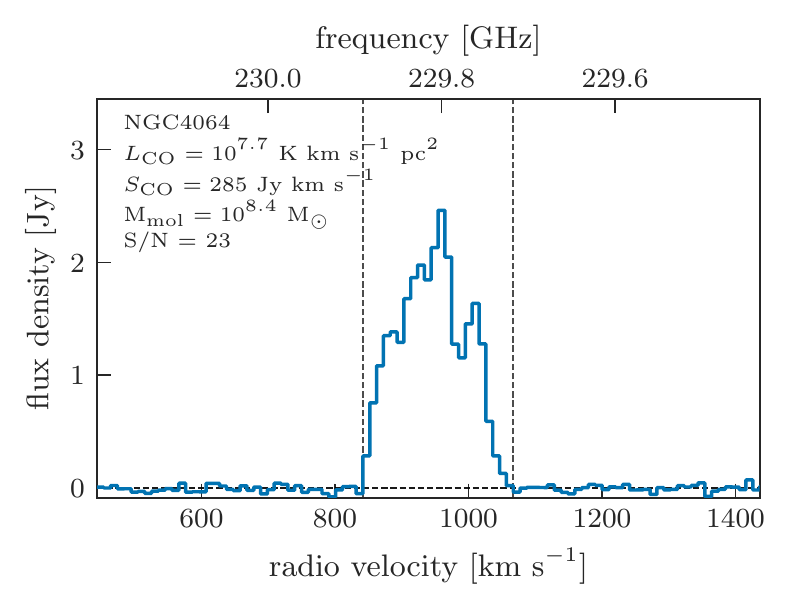}
	\includegraphics[width=0.4\textwidth]{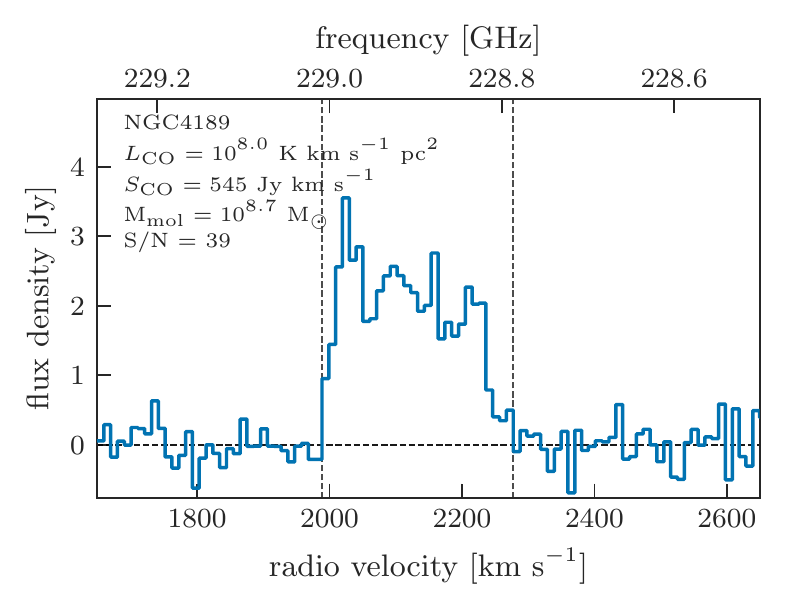}
	\includegraphics[width=0.4\textwidth]{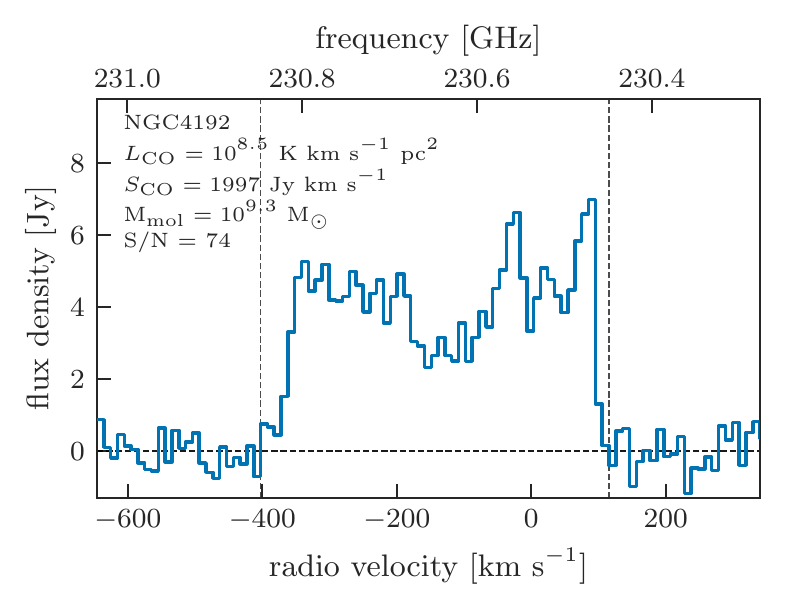}
	\includegraphics[width=0.4\textwidth]{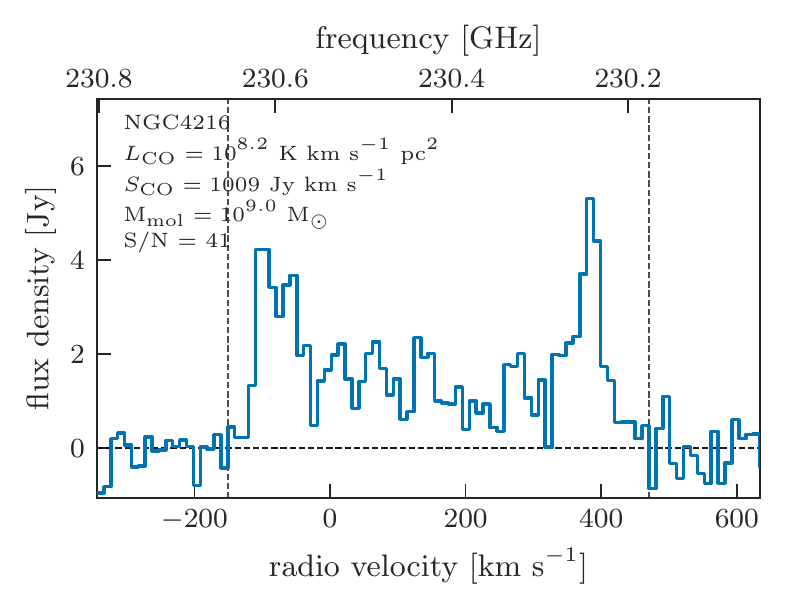}
	\includegraphics[width=0.4\textwidth]{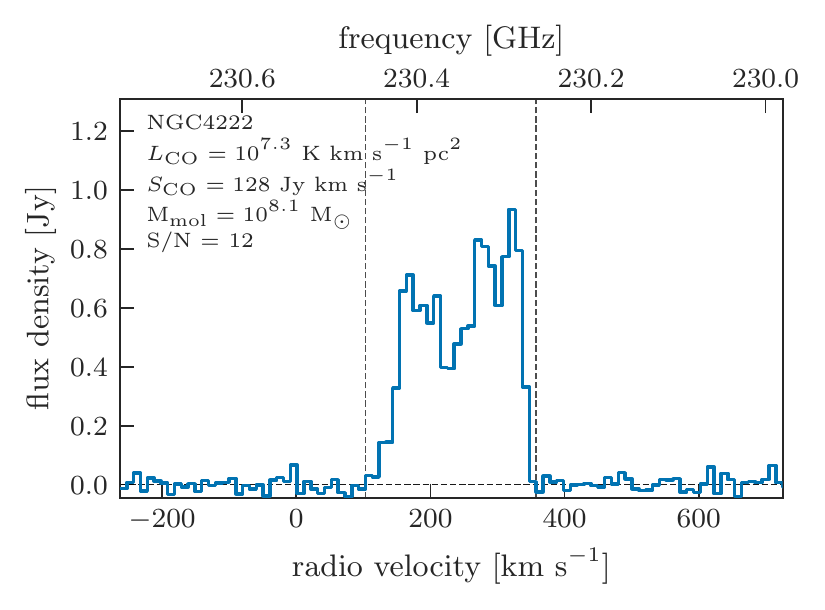}
    \caption{$^{\rm 12}$CO($2-1$) spectra at 10.6~km~s$^{-1}$ resolution.
        The velocity limits used to estimate the velocity width and flux
        density for each galaxy are shown by the dashed vertical lines. The
        galaxy name, integrated CO line luminosity (Eqn. \ref{eq:Lco}),
        integrated flux density, molecular gas mass (Eqn. \ref{eq:Mmol}),
        and signal-to-noise of the spectrum are shown in the upper left
        corner of each panel. \it The complete Figure Set
        containing 51 $^{\rm 12}$CO($2-1$) spectra for all VERTICO
        galaxies is available in below.} 
	\label{fig:spectra}
\end{figure*}

An example of the global CO($2-1$) spectra derived from the masked data cubes described in \ref{sec:COFluxes} are shown in Figure \ref{fig:spectra}. The complete Figure Set of $^{\rm 12}$CO($2-1$) spectra for all 51 VERTICO galaxies is available below.

\figsetstart
\figsetnum{13}
\figsettitle{CO($2-1$) spectra for each galaxy in the VERTICO survey}

\figsetgrpstart
\figsetgrpnum{13.1}
\figsetgrptitle{VERTICO CO($2-1$) spectrum for IC3392}
\figsetplot{IC3392_co21_spec.pdf}
\figsetgrpnote{$^{\rm 12}$CO($2-1$) spectra at 10.6~km~s$^{-1}$ resolution. The velocity
    limits used to estimate the velocity width and flux density for each
    galaxy are shown by the dashed vertical lines. The galaxy name,
    integrated CO line luminosity (Eqn. \ref{eq:Lco}), integrated flux
    density, molecular gas mass (Eqn. \ref{eq:Mmol}), and signal-to-noise of
    the spectrum are shown in the upper left corner of each panel.}
\figsetgrpend

\figsetgrpstart
\figsetgrpnum{13.2}
\figsetgrptitle{VERTICO CO($2-1$) spectrum for NGC4064}
\figsetplot{NGC4064_co21_spec.pdf}
\figsetgrpnote{As in Figure 13.1.}
\figsetgrpend

\figsetgrpstart
\figsetgrpnum{13.3}
\figsetgrptitle{VERTICO CO($2-1$) spectrum for NGC4189}
\figsetplot{NGC4189_co21_spec.pdf}
\figsetgrpnote{As in Figure 13.1.}
\figsetgrpend

\figsetgrpstart
\figsetgrpnum{13.4}
\figsetgrptitle{VERTICO CO($2-1$) spectrum for NGC4192}
\figsetplot{NGC4192_co21_spec.pdf}
\figsetgrpnote{As in Figure 13.1.}
\figsetgrpend

\figsetgrpstart
\figsetgrpnum{13.5}
\figsetgrptitle{VERTICO CO($2-1$) spectrum for NGC4216}
\figsetplot{NGC4216_co21_spec.pdf}
\figsetgrpnote{As in Figure 13.1.}
\figsetgrpend

\figsetgrpstart
\figsetgrpnum{13.6}
\figsetgrptitle{VERTICO CO($2-1$) spectrum for NGC4222}
\figsetplot{NGC4222_co21_spec.pdf}
\figsetgrpnote{As in Figure 13.1.}
\figsetgrpend

\figsetgrpstart
\figsetgrpnum{13.7}
\figsetgrptitle{VERTICO CO($2-1$) spectrum for NGC4254}
\figsetplot{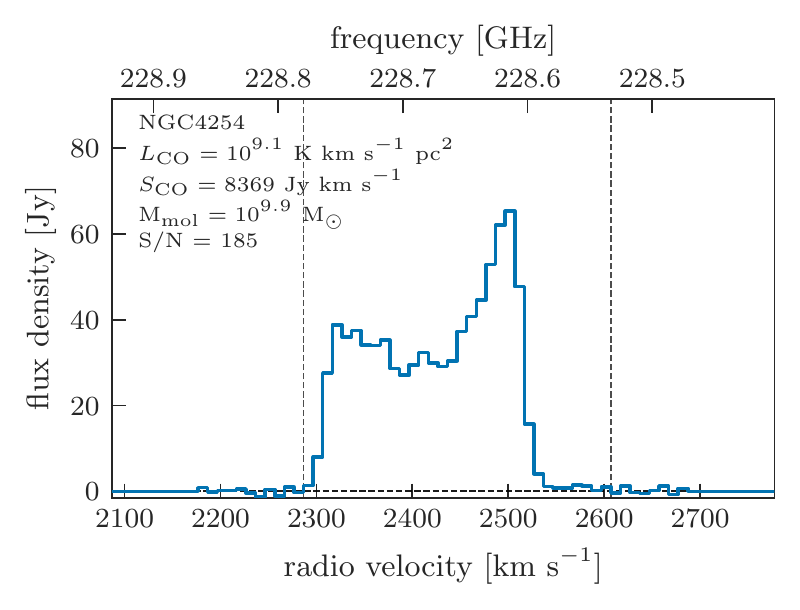}
\figsetgrpnote{As in Figure 13.1.}
\figsetgrpend

\figsetgrpstart
\figsetgrpnum{13.8}
\figsetgrptitle{VERTICO CO($2-1$) spectrum for NGC4293}
\figsetplot{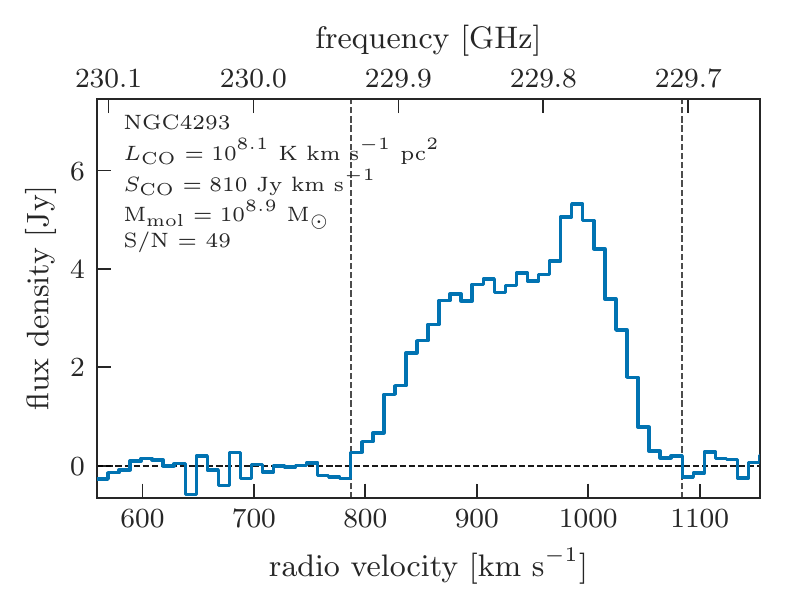}
\figsetgrpnote{As in Figure 13.1.}
\figsetgrpend

\figsetgrpstart
\figsetgrpnum{13.9}
\figsetgrptitle{VERTICO CO($2-1$) spectrum for NGC4294}
\figsetplot{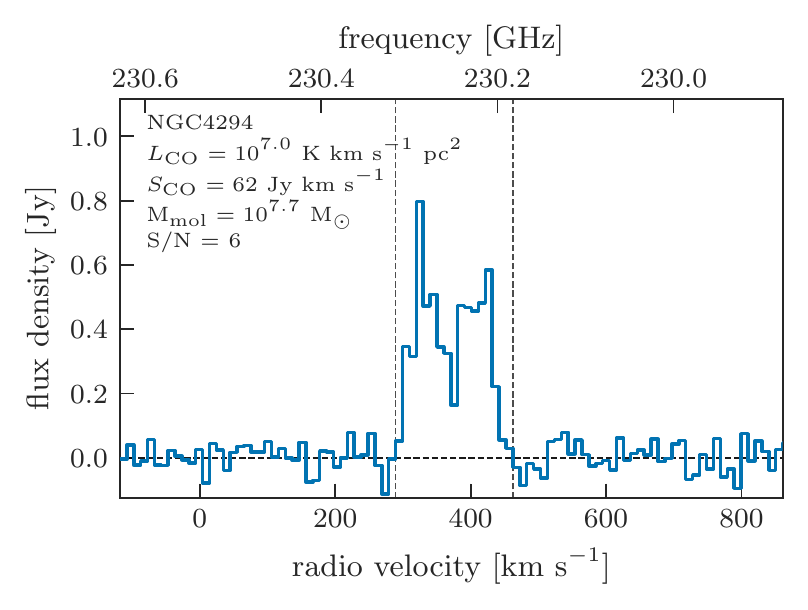}
\figsetgrpnote{As in Figure 13.1.}
\figsetgrpend

\figsetgrpstart
\figsetgrpnum{13.10}
\figsetgrptitle{VERTICO CO($2-1$) spectrum for NGC4298}
\figsetplot{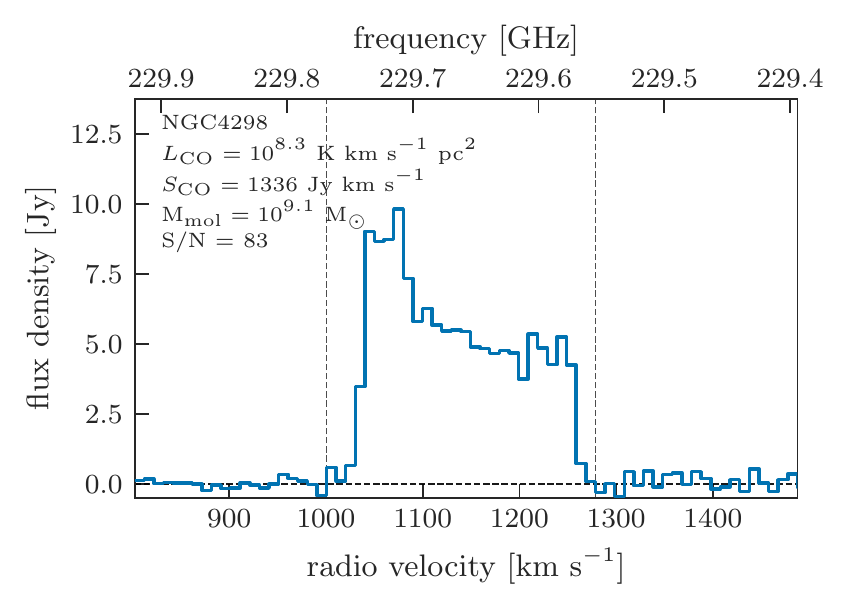}
\figsetgrpnote{As in Figure 13.1.}
\figsetgrpend

\figsetgrpstart
\figsetgrpnum{13.11}
\figsetgrptitle{VERTICO CO($2-1$) spectrum for NGC4299}
\figsetplot{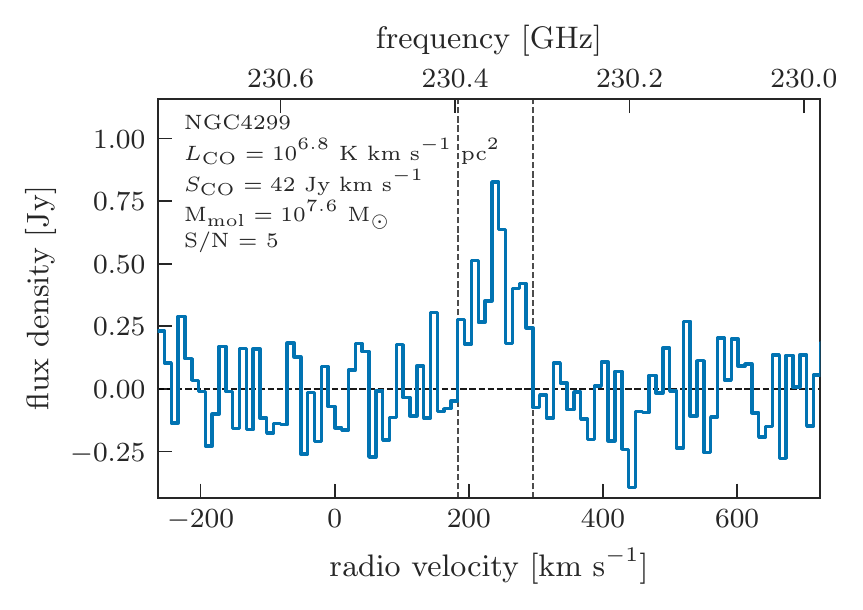}
\figsetgrpnote{As in Figure 13.1.}
\figsetgrpend

\figsetgrpstart
\figsetgrpnum{13.12}
\figsetgrptitle{VERTICO CO($2-1$) spectrum for NGC4302}
\figsetplot{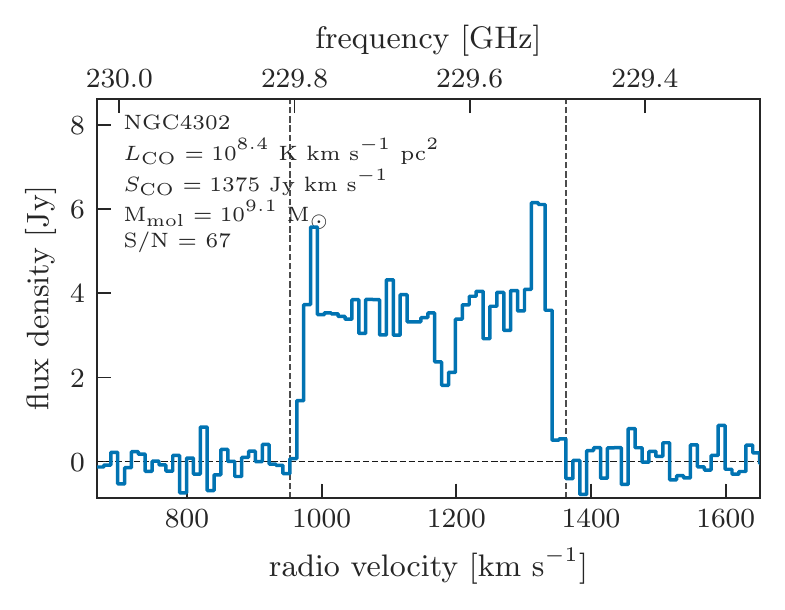}
\figsetgrpnote{As in Figure 13.1.}
\figsetgrpend

\figsetgrpstart
\figsetgrpnum{13.13}
\figsetgrptitle{VERTICO CO($2-1$) spectrum for NGC4321}
\figsetplot{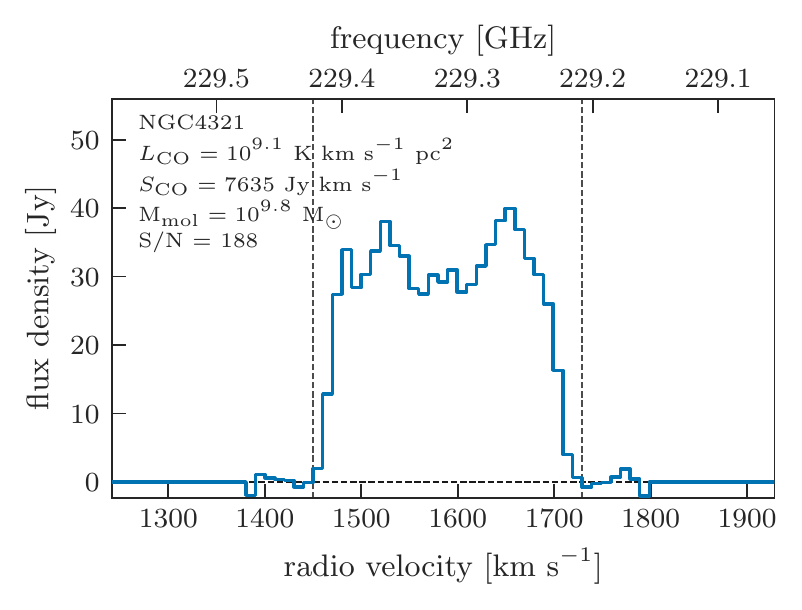}
\figsetgrpnote{As in Figure 13.1.}
\figsetgrpend

\figsetgrpstart
\figsetgrpnum{13.14}
\figsetgrptitle{VERTICO CO($2-1$) spectrum for NGC4330}
\figsetplot{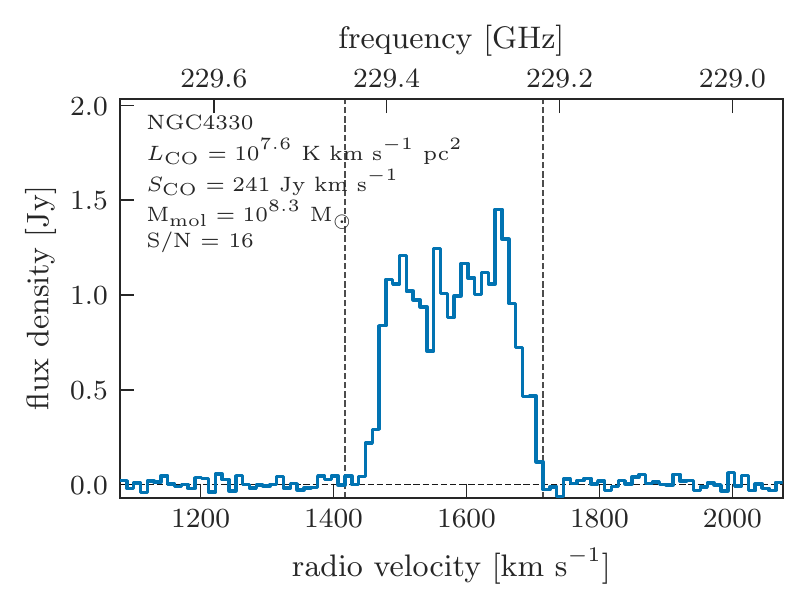}
\figsetgrpnote{As in Figure 13.1.}
\figsetgrpend

\figsetgrpstart
\figsetgrpnum{13.15}
\figsetgrptitle{VERTICO CO($2-1$) spectrum for NGC4351}
\figsetplot{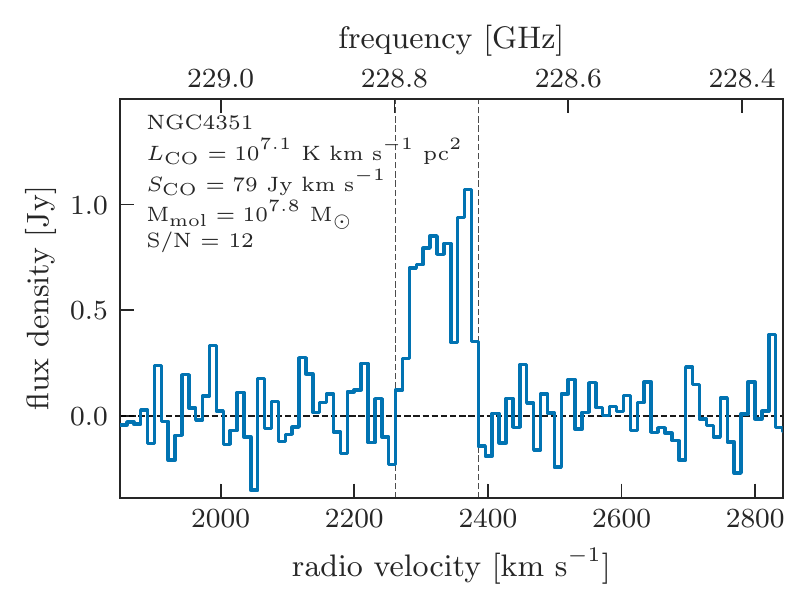}
\figsetgrpnote{As in Figure 13.1.}
\figsetgrpend

\figsetgrpstart
\figsetgrpnum{13.16}
\figsetgrptitle{VERTICO CO($2-1$) spectrum for NGC4380}
\figsetplot{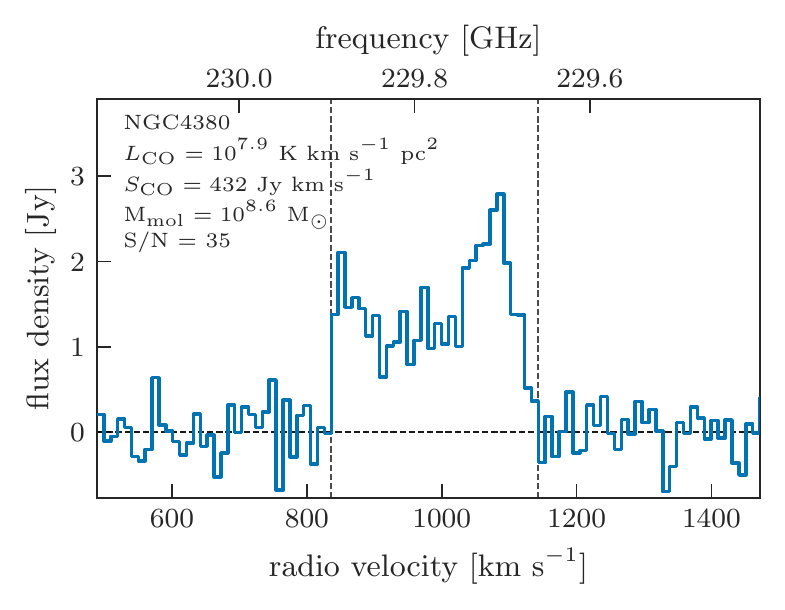}
\figsetgrpnote{As in Figure 13.1.}
\figsetgrpend

\figsetgrpstart
\figsetgrpnum{13.17}
\figsetgrptitle{VERTICO CO($2-1$) spectrum for NGC4383}
\figsetplot{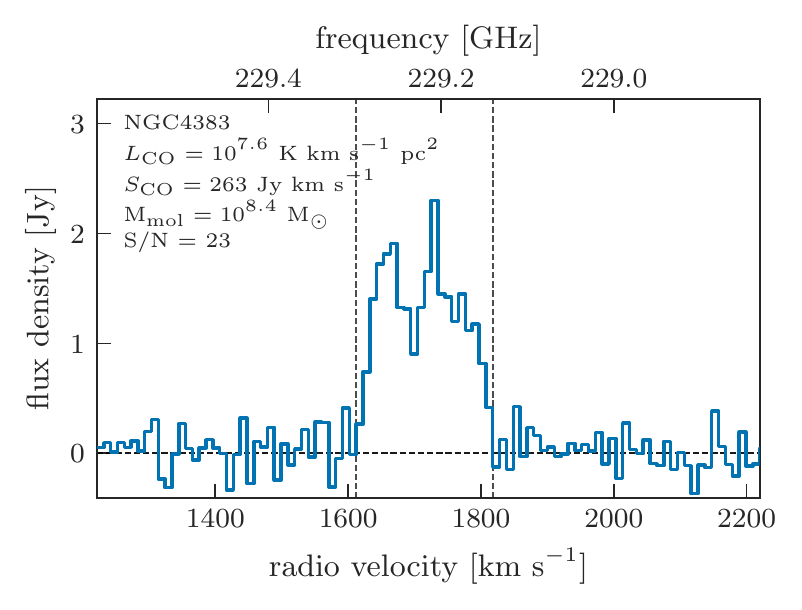}
\figsetgrpnote{As in Figure 13.1.}
\figsetgrpend

\figsetgrpstart
\figsetgrpnum{13.18}
\figsetgrptitle{VERTICO CO($2-1$) spectrum for NGC4388}
\figsetplot{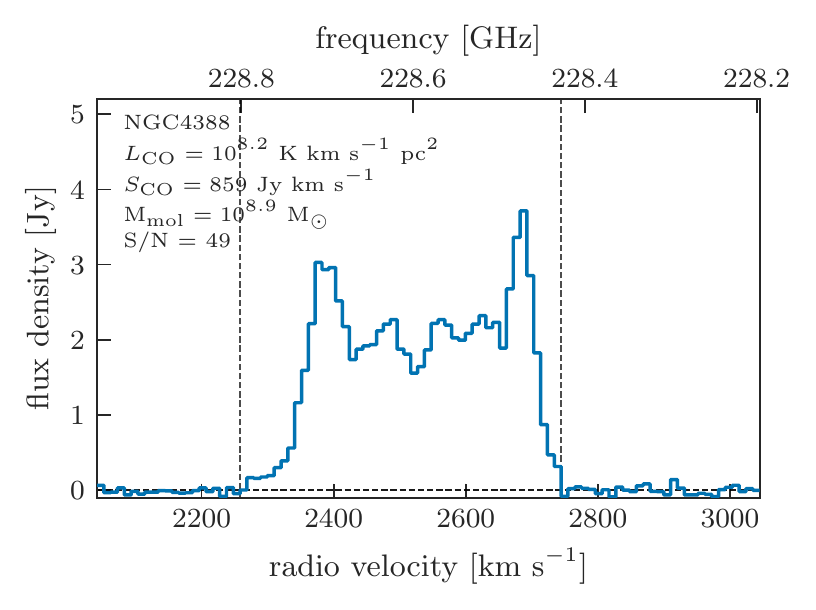}
\figsetgrpnote{As in Figure 13.1.}
\figsetgrpend

\figsetgrpstart
\figsetgrpnum{13.19}
\figsetgrptitle{VERTICO CO($2-1$) spectrum for NGC4394}
\figsetplot{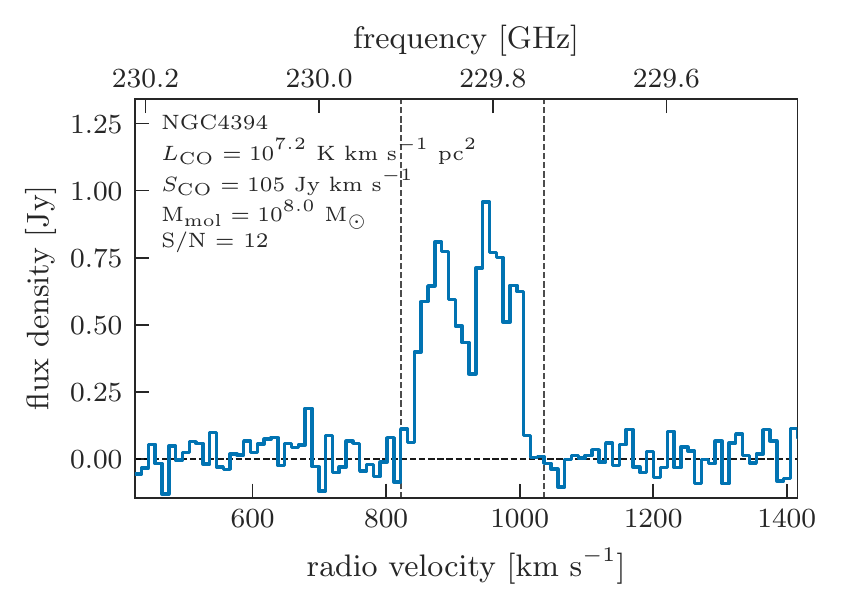}
\figsetgrpnote{As in Figure 13.1.}
\figsetgrpend

\figsetgrpstart
\figsetgrpnum{13.20}
\figsetgrptitle{VERTICO CO($2-1$) spectrum for NGC4396}
\figsetplot{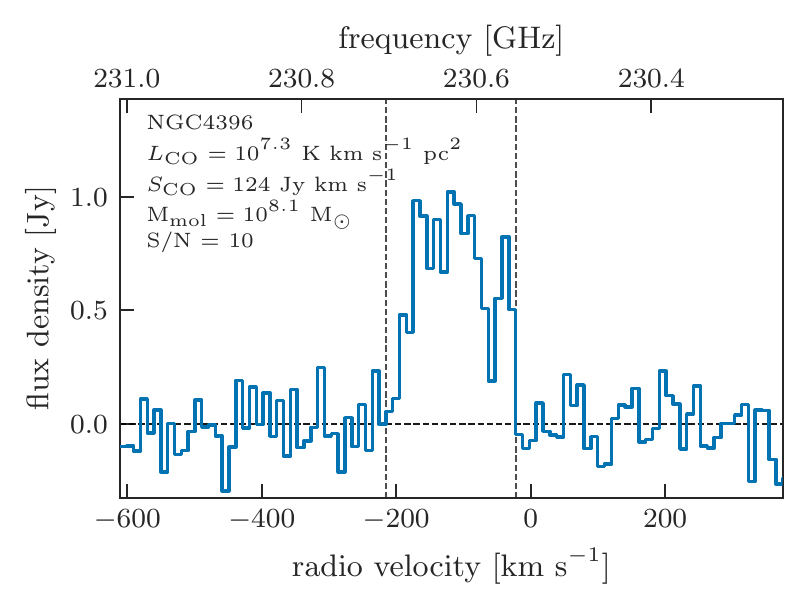}
\figsetgrpnote{As in Figure 13.1.}
\figsetgrpend

\figsetgrpstart
\figsetgrpnum{13.21}
\figsetgrptitle{VERTICO CO($2-1$) spectrum for NGC4402}
\figsetplot{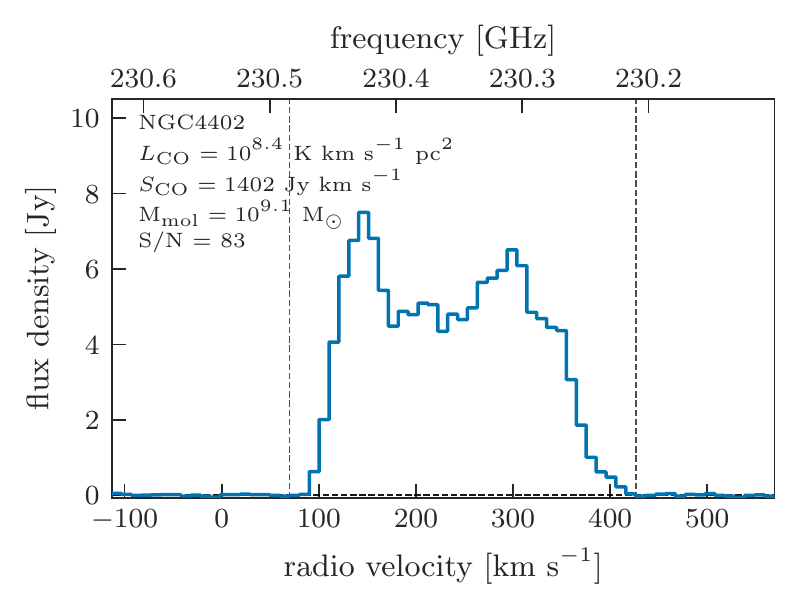}
\figsetgrpnote{As in Figure 13.1.}
\figsetgrpend

\figsetgrpstart
\figsetgrpnum{13.22}
\figsetgrptitle{VERTICO CO($2-1$) spectrum for NGC4405}
\figsetplot{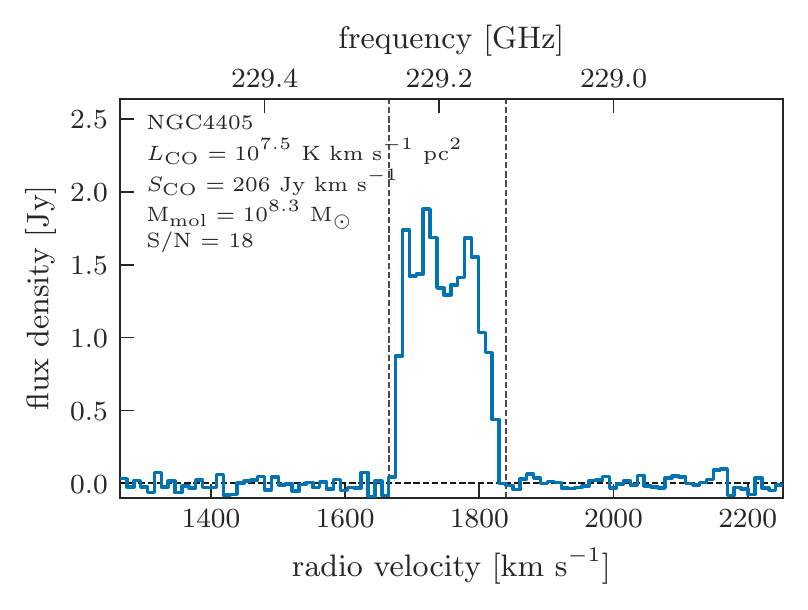}
\figsetgrpnote{As in Figure 13.1.}
\figsetgrpend

\figsetgrpstart
\figsetgrpnum{13.23}
\figsetgrptitle{VERTICO CO($2-1$) spectrum for NGC4419}
\figsetplot{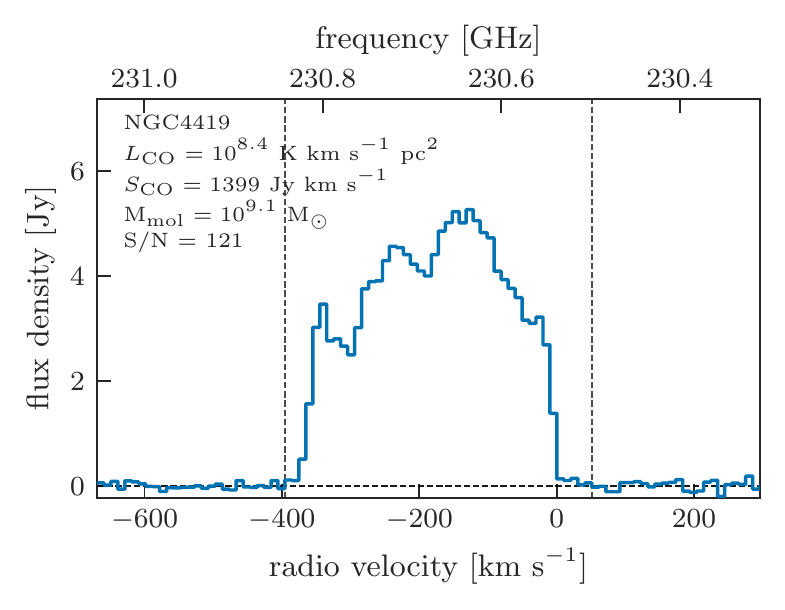}
\figsetgrpnote{As in Figure 13.1.}
\figsetgrpend

\figsetgrpstart
\figsetgrpnum{13.24}
\figsetgrptitle{VERTICO CO($2-1$) spectrum for NGC4424}
\figsetplot{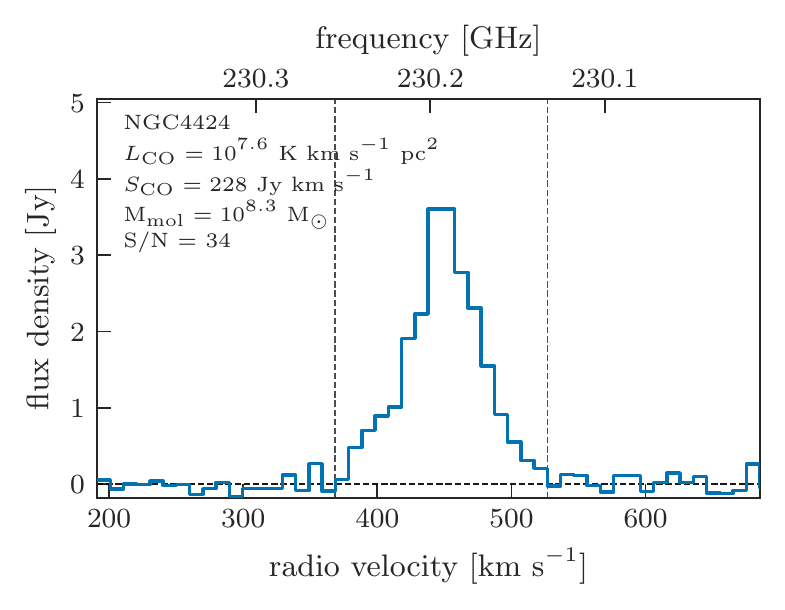}
\figsetgrpnote{As in Figure 13.1.}
\figsetgrpend

\figsetgrpstart
\figsetgrpnum{13.25}
\figsetgrptitle{VERTICO CO($2-1$) spectrum for NGC4450}
\figsetplot{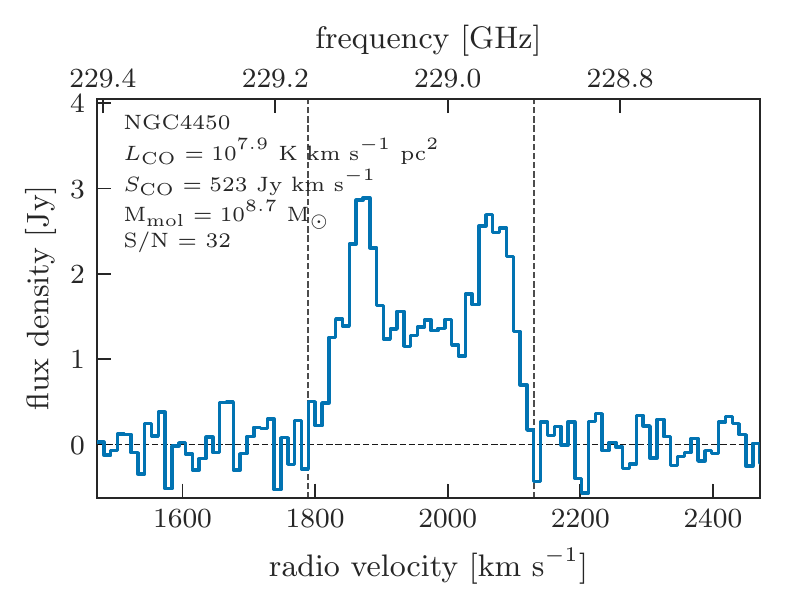}
\figsetgrpnote{As in Figure 13.1.}
\figsetgrpend

\figsetgrpstart
\figsetgrpnum{13.26}
\figsetgrptitle{VERTICO CO($2-1$) spectrum for NGC4457}
\figsetplot{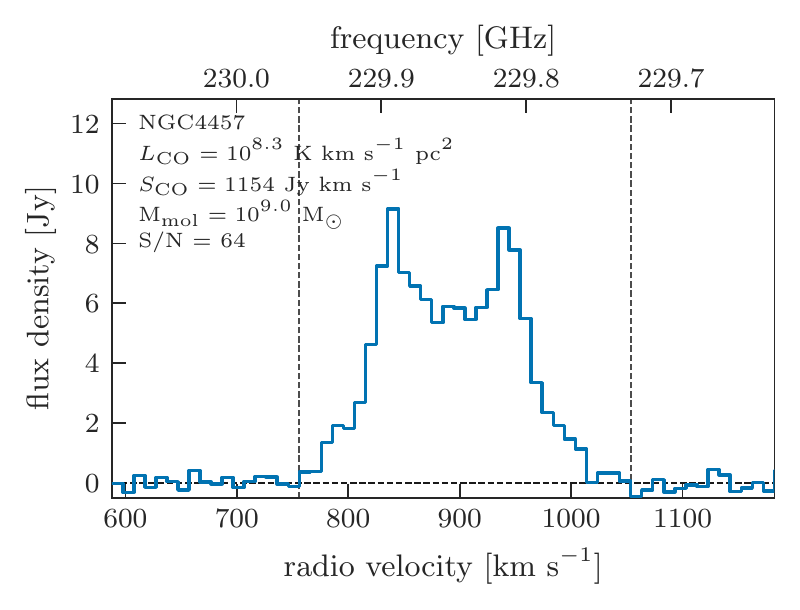}
\figsetgrpnote{As in Figure 13.1.}
\figsetgrpend

\figsetgrpstart
\figsetgrpnum{13.27}
\figsetgrptitle{VERTICO CO($2-1$) spectrum for NGC4501}
\figsetplot{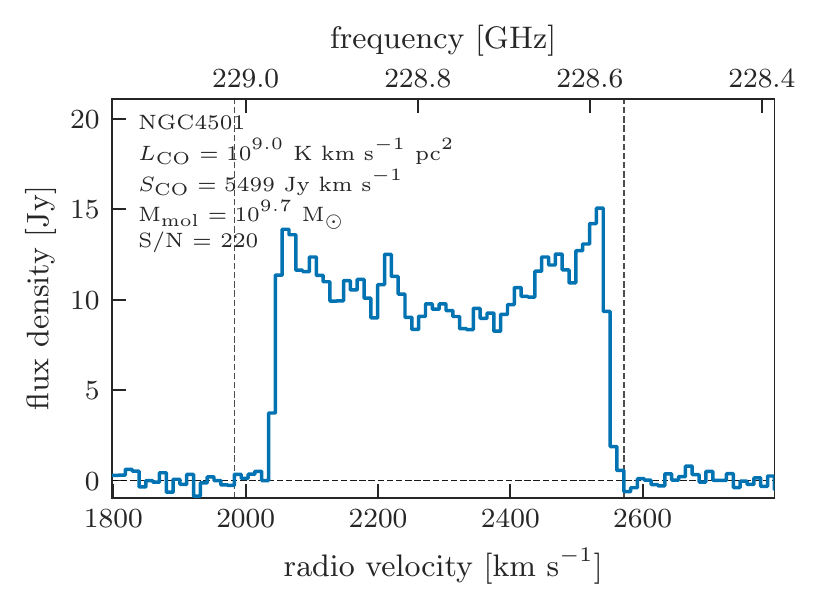}
\figsetgrpnote{As in Figure 13.1.}
\figsetgrpend

\figsetgrpstart
\figsetgrpnum{13.28}
\figsetgrptitle{VERTICO CO($2-1$) spectrum for NGC4522}
\figsetplot{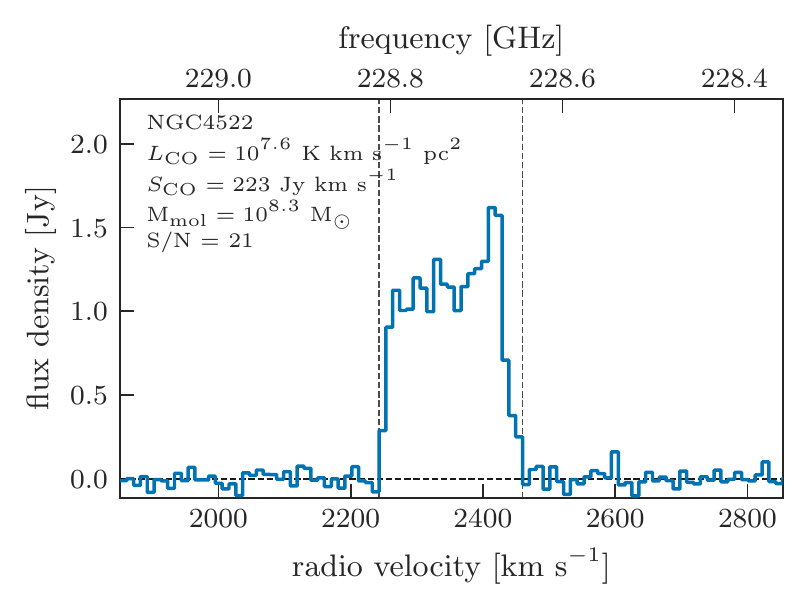}
\figsetgrpnote{As in Figure 13.1.}
\figsetgrpend

\figsetgrpstart
\figsetgrpnum{13.29}
\figsetgrptitle{VERTICO CO($2-1$) spectrum for NGC4532}
\figsetplot{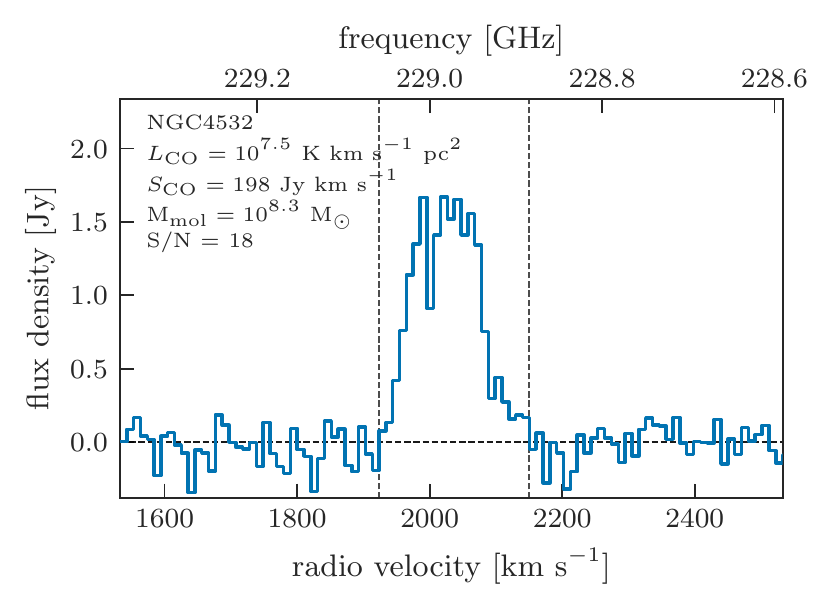}
\figsetgrpnote{As in Figure 13.1.}
\figsetgrpend

\figsetgrpstart
\figsetgrpnum{13.30}
\figsetgrptitle{VERTICO CO($2-1$) spectrum for NGC4533}
\figsetplot{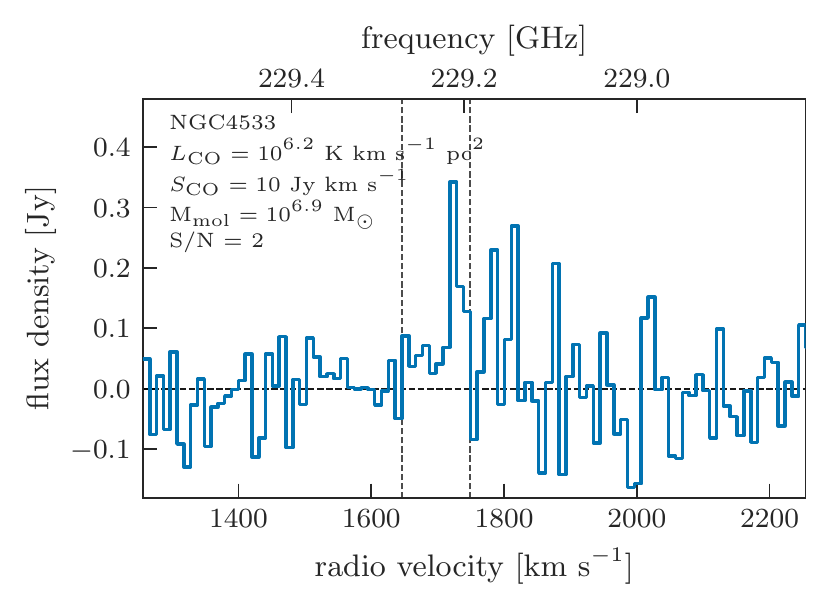}
\figsetgrpnote{As in Figure 13.1.}
\figsetgrpend

\figsetgrpstart
\figsetgrpnum{13.31}
\figsetgrptitle{VERTICO CO($2-1$) spectrum for NGC4535}
\figsetplot{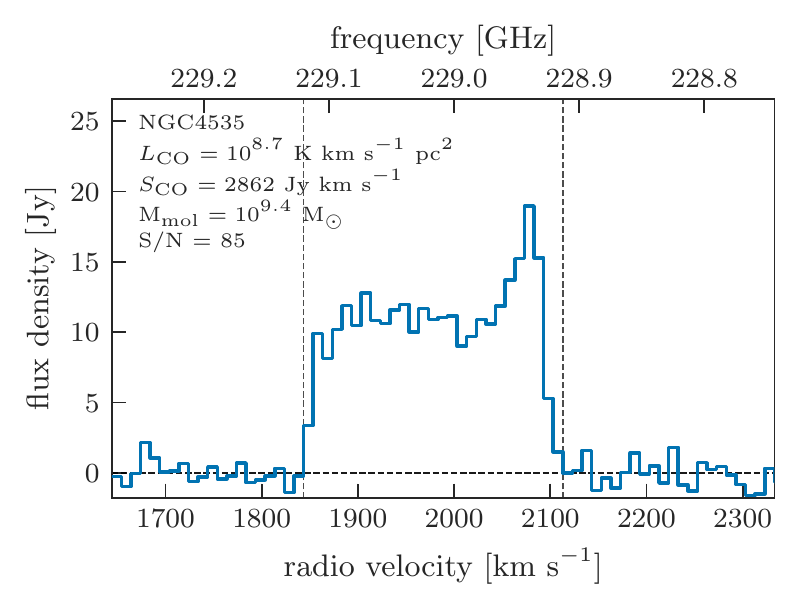}
\figsetgrpnote{As in Figure 13.1.}
\figsetgrpend

\figsetgrpstart
\figsetgrpnum{13.32}
\figsetgrptitle{VERTICO CO($2-1$) spectrum for NGC4536}
\figsetplot{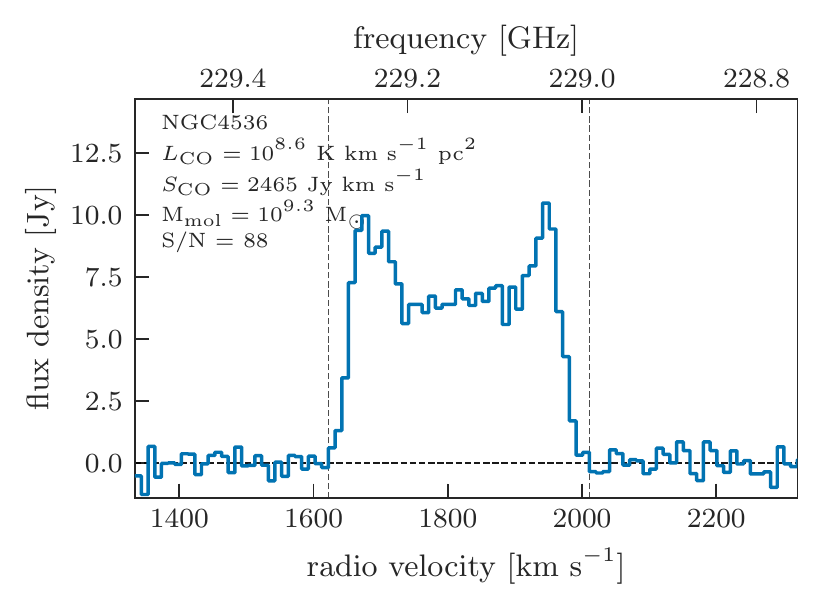}
\figsetgrpnote{As in Figure 13.1.}
\figsetgrpend

\figsetgrpstart
\figsetgrpnum{13.33}
\figsetgrptitle{VERTICO CO($2-1$) spectrum for NGC4548}
\figsetplot{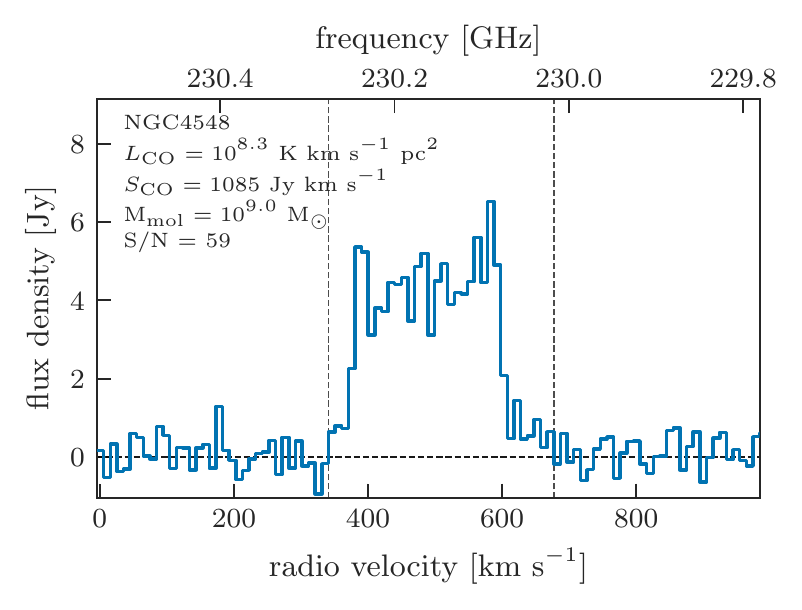}
\figsetgrpnote{As in Figure 13.1.}
\figsetgrpend

\figsetgrpstart
\figsetgrpnum{13.34}
\figsetgrptitle{VERTICO CO($2-1$) spectrum for NGC4561}
\figsetplot{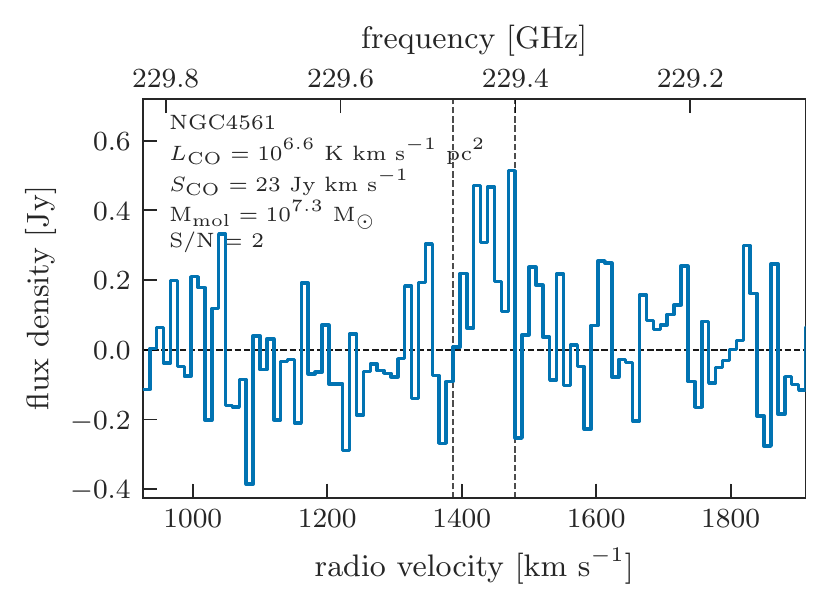}
\figsetgrpnote{As in Figure 13.1.}
\figsetgrpend

\figsetgrpstart
\figsetgrpnum{13.35}
\figsetgrptitle{VERTICO CO($2-1$) spectrum for NGC4567}
\figsetplot{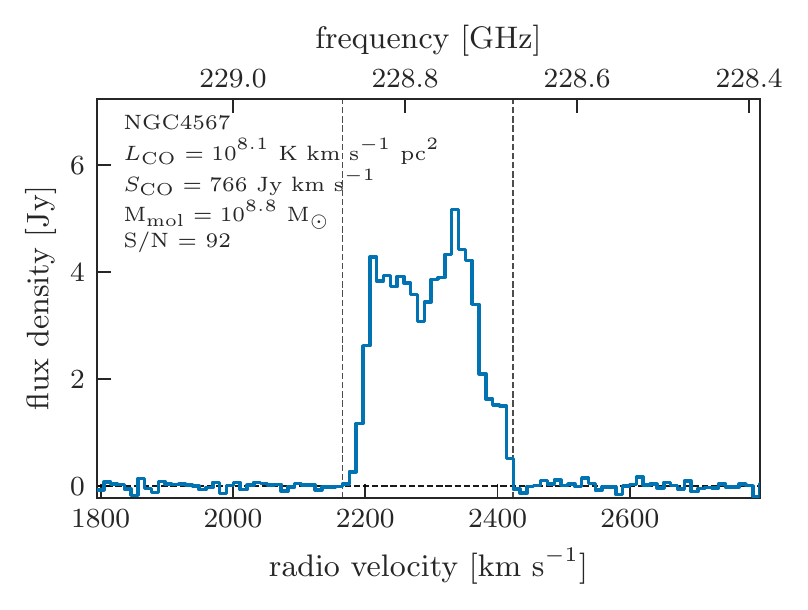}
\figsetgrpnote{As in Figure 13.1.}
\figsetgrpend

\figsetgrpstart
\figsetgrpnum{13.36}
\figsetgrptitle{VERTICO CO($2-1$) spectrum for NGC4568}
\figsetplot{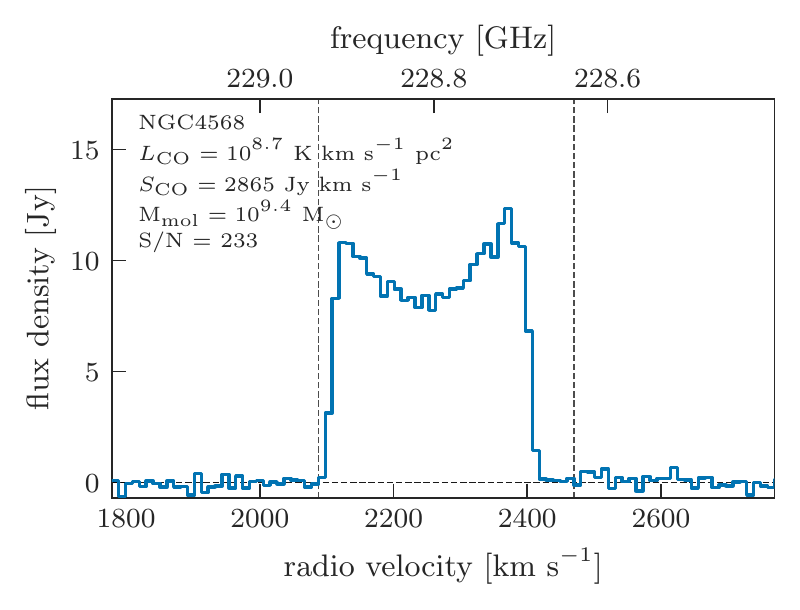}
\figsetgrpnote{As in Figure 13.1.}
\figsetgrpend

\figsetgrpstart
\figsetgrpnum{13.37}
\figsetgrptitle{VERTICO CO($2-1$) spectrum for NGC4569}
\figsetplot{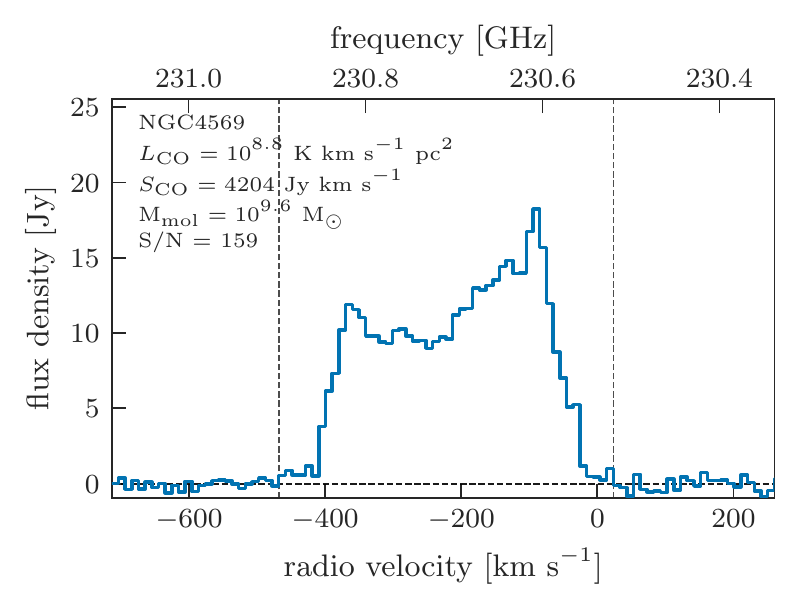}
\figsetgrpnote{As in Figure 13.1.}
\figsetgrpend

\figsetgrpstart
\figsetgrpnum{13.38}
\figsetgrptitle{VERTICO CO($2-1$) spectrum for NGC4579}
\figsetplot{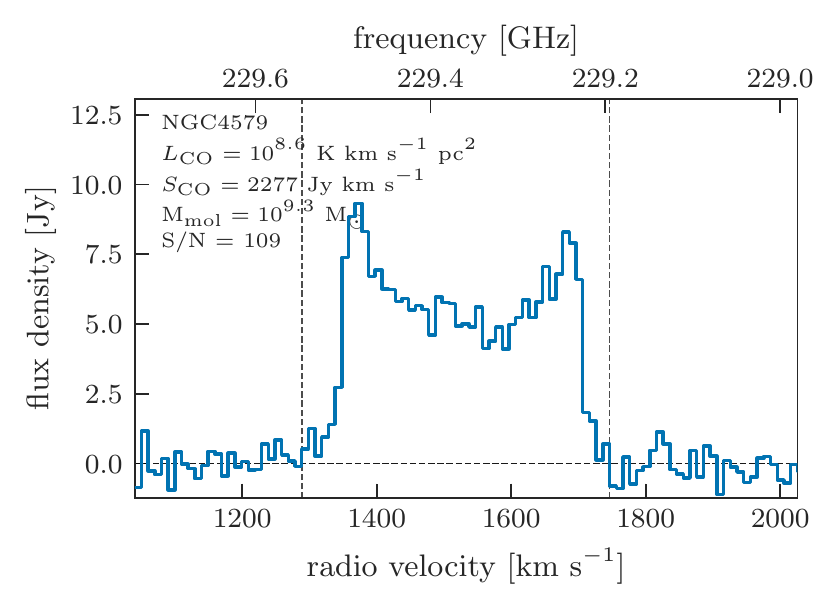}
\figsetgrpnote{As in Figure 13.1.}
\figsetgrpend

\figsetgrpstart
\figsetgrpnum{13.39}
\figsetgrptitle{VERTICO CO($2-1$) spectrum for NGC4580}
\figsetplot{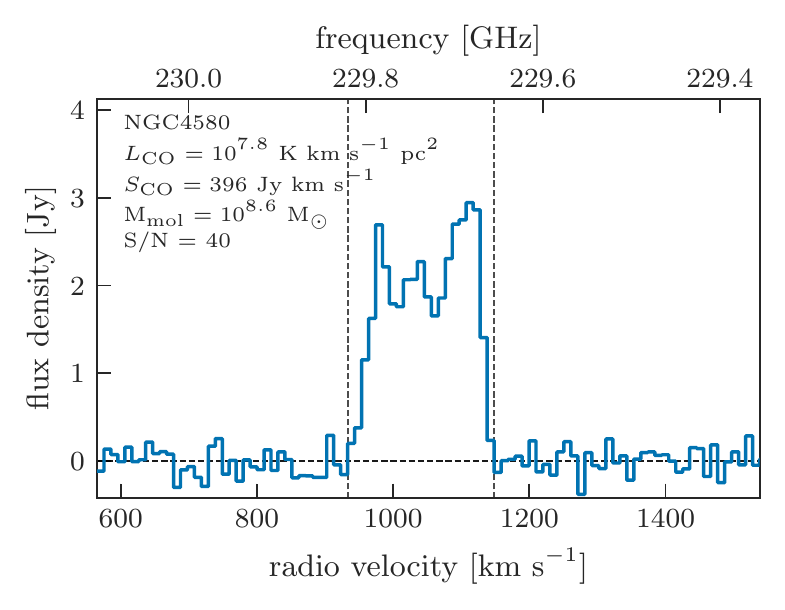}
\figsetgrpnote{As in Figure 13.1.}
\figsetgrpend

\figsetgrpstart
\figsetgrpnum{13.40}
\figsetgrptitle{VERTICO CO($2-1$) spectrum for NGC4606}
\figsetplot{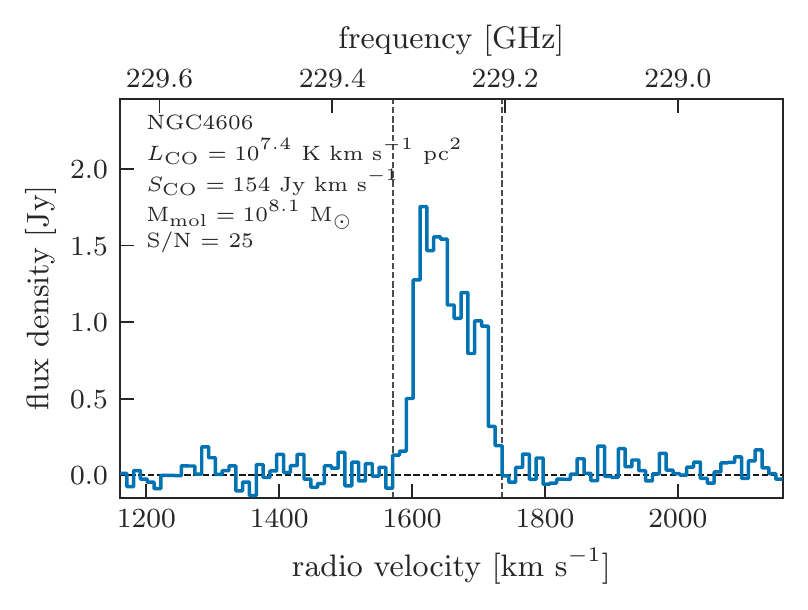}
\figsetgrpnote{As in Figure 13.1.}
\figsetgrpend

\figsetgrpstart
\figsetgrpnum{13.41}
\figsetgrptitle{VERTICO CO($2-1$) spectrum for NGC4607}
\figsetplot{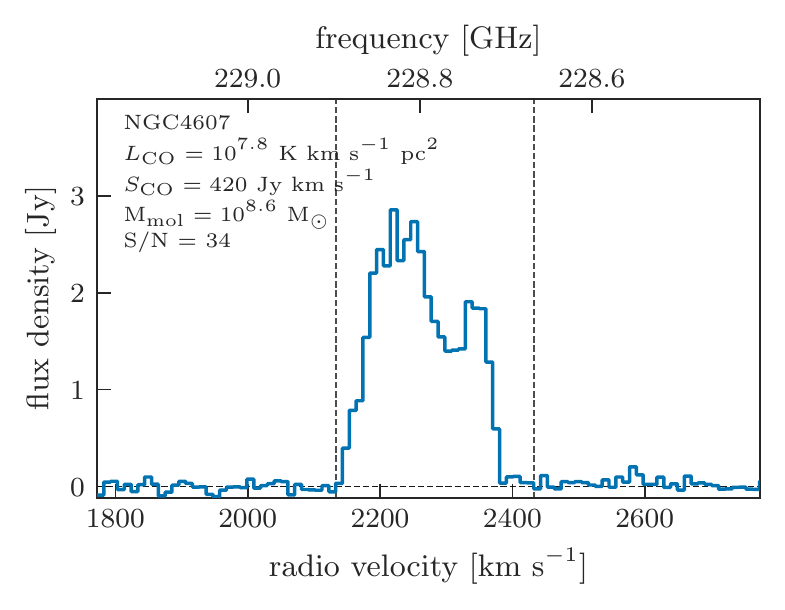}
\figsetgrpnote{As in Figure 13.1.}
\figsetgrpend

\figsetgrpstart
\figsetgrpnum{13.42}
\figsetgrptitle{VERTICO CO($2-1$) spectrum for NGC4651}
\figsetplot{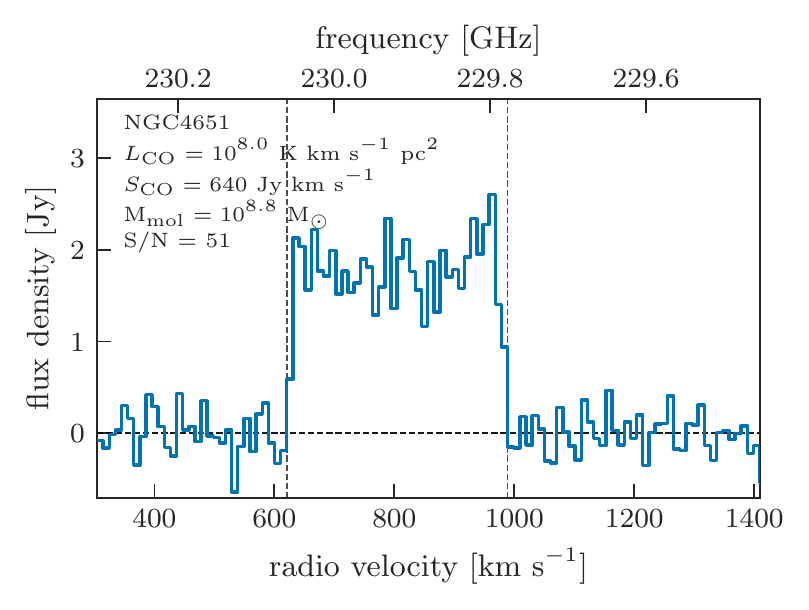}
\figsetgrpnote{As in Figure 13.1.}
\figsetgrpend

\figsetgrpstart
\figsetgrpnum{13.43}
\figsetgrptitle{VERTICO CO($2-1$) spectrum for NGC4654}
\figsetplot{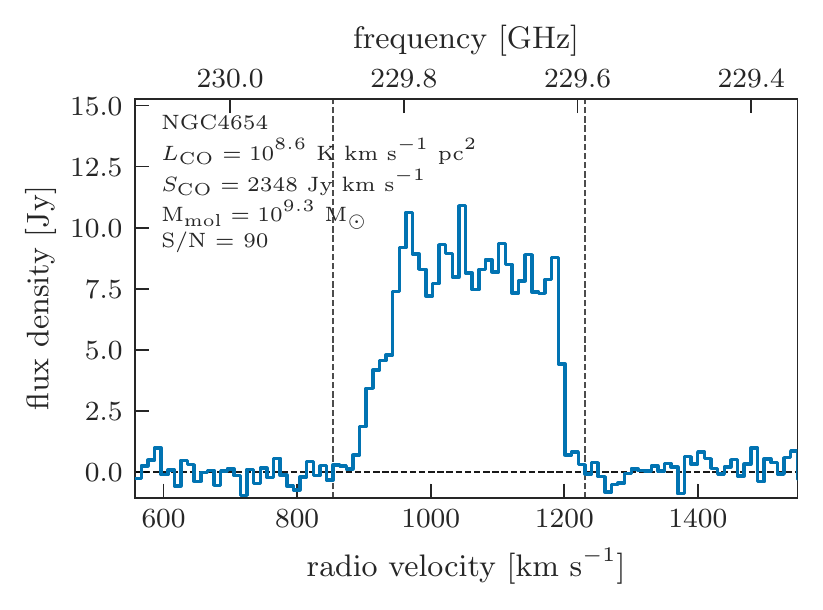}
\figsetgrpnote{As in Figure 13.1.}
\figsetgrpend

\figsetgrpstart
\figsetgrpnum{13.44}
\figsetgrptitle{VERTICO CO($2-1$) spectrum for NGC4689}
\figsetplot{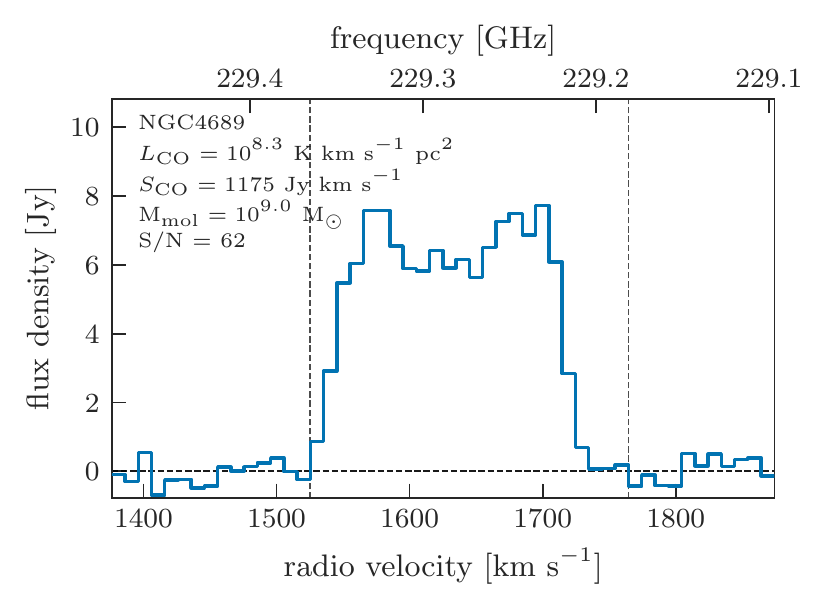}
\figsetgrpnote{As in Figure 13.1.}
\figsetgrpend

\figsetgrpstart
\figsetgrpnum{13.45}
\figsetgrptitle{VERTICO CO($2-1$) spectrum for NGC4694}
\figsetplot{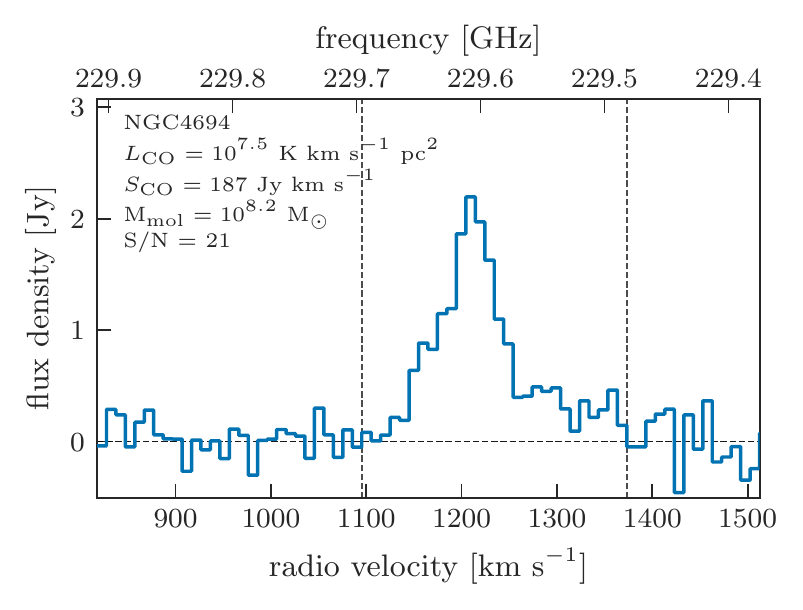}
\figsetgrpnote{As in Figure 13.1.}
\figsetgrpend

\figsetgrpstart
\figsetgrpnum{13.46}
\figsetgrptitle{VERTICO CO($2-1$) spectrum for NGC4698}
\figsetplot{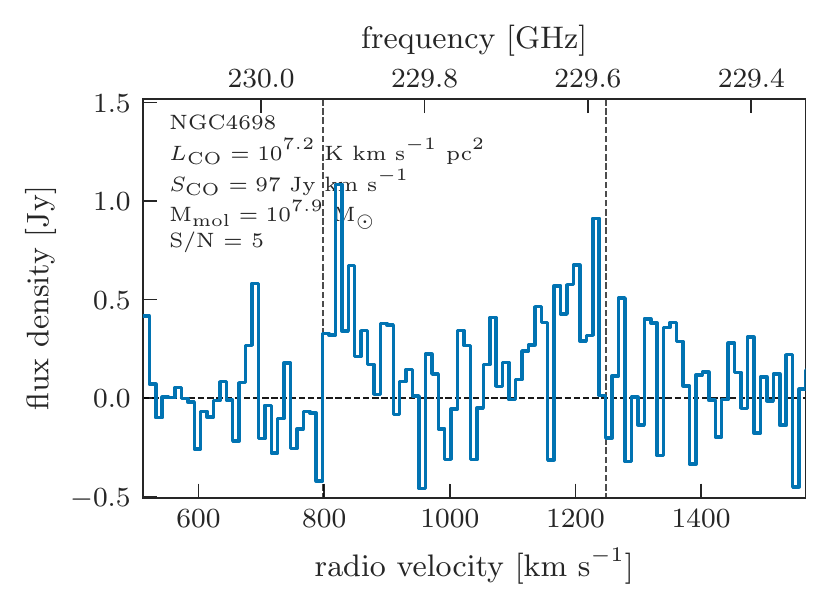}
\figsetgrpnote{As in Figure 13.1.}
\figsetgrpend

\figsetgrpstart
\figsetgrpnum{13.47}
\figsetgrptitle{VERTICO CO($2-1$) spectrum for NGC4713}
\figsetplot{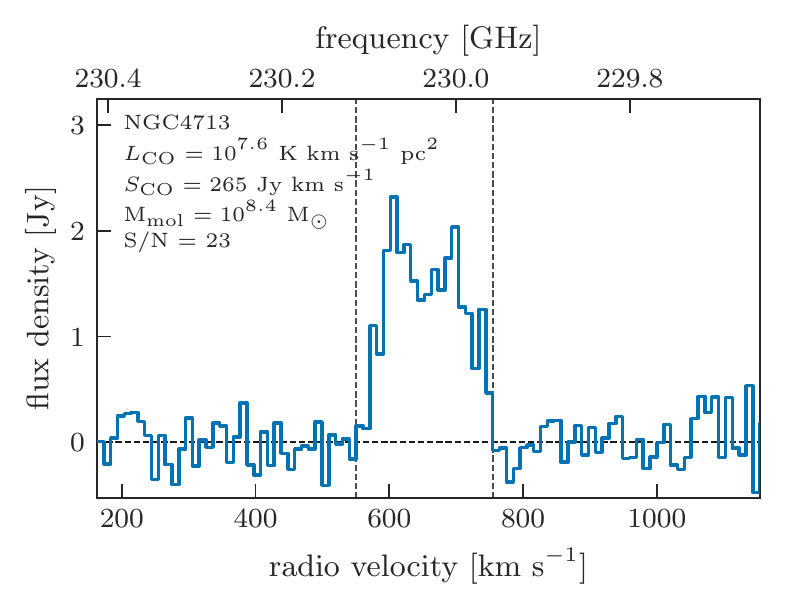}
\figsetgrpnote{As in Figure 13.1.}
\figsetgrpend

\figsetgrpstart
\figsetgrpnum{13.48}
\figsetgrptitle{VERTICO CO($2-1$) spectrum for NGC4772}
\figsetplot{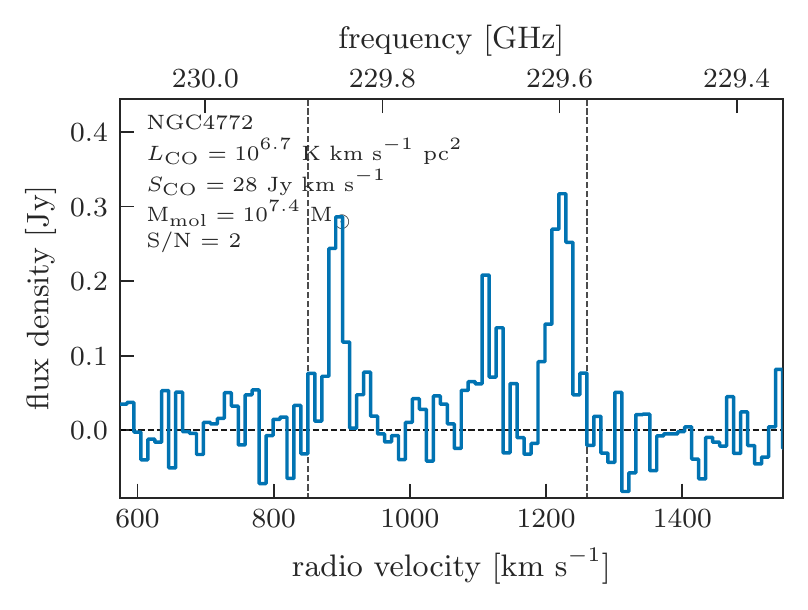}
\figsetgrpnote{As in Figure 13.1.}
\figsetgrpend

\figsetgrpstart
\figsetgrpnum{13.49}
\figsetgrptitle{VERTICO CO($2-1$) spectrum for NGC4808}
\figsetplot{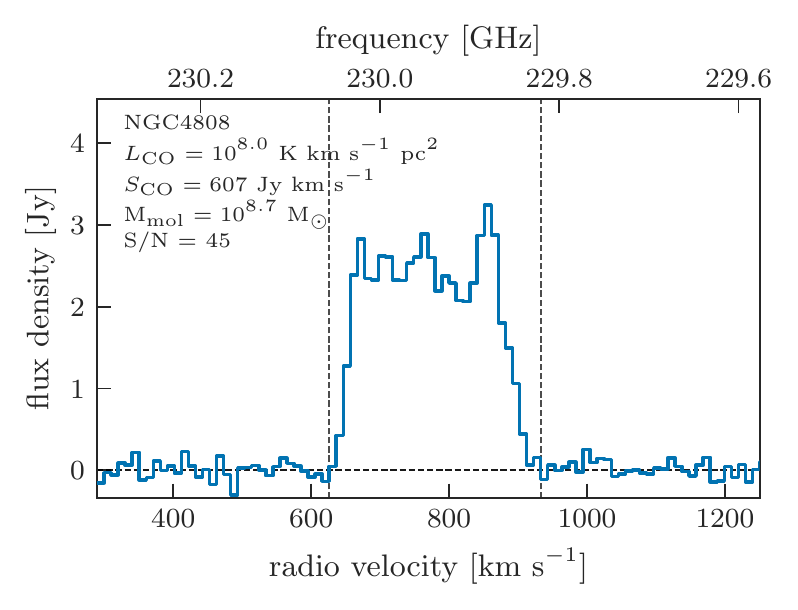}
\figsetgrpnote{As in Figure 13.1.}
\figsetgrpend

\figsetgrpstart
\figsetgrpnum{13.50}
\figsetgrptitle{VERTICO CO($2-1$) spectrum for IC3418}
\figsetplot{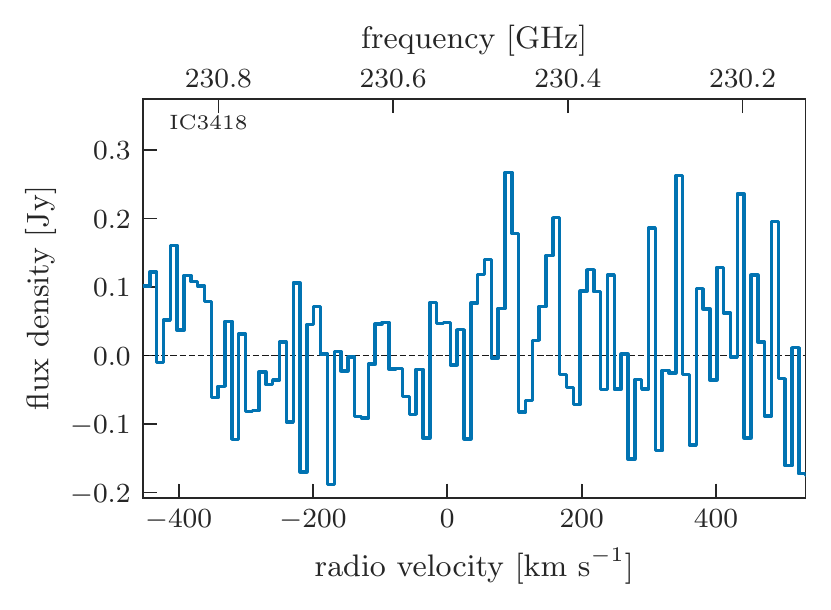}
\figsetgrpnote{As in Figure 13.1.}
\figsetgrpend

\figsetgrpstart
\figsetgrpnum{13.51}
\figsetgrptitle{VERTICO CO($2-1$) spectrum for VCC1581}
\figsetplot{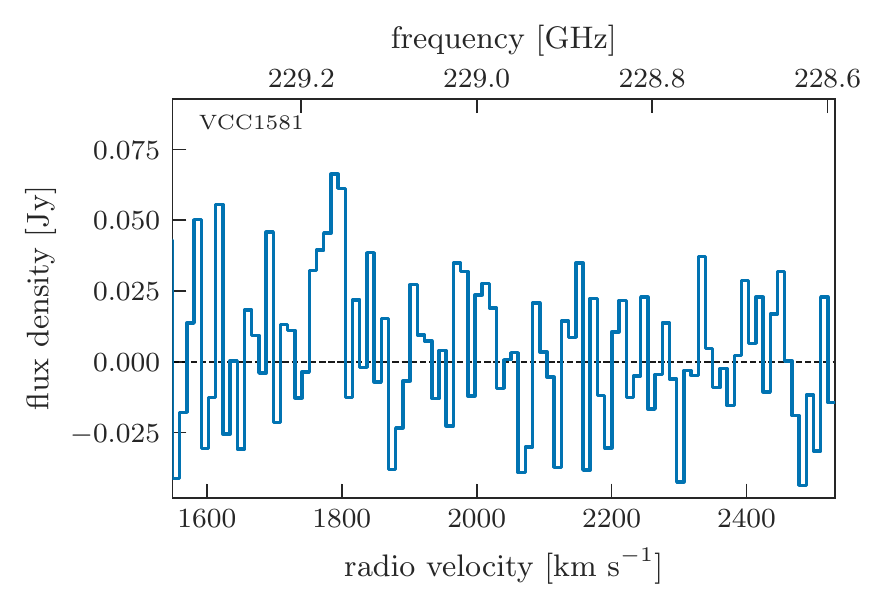}
\figsetgrpnote{As in Figure 13.1.}
\figsetgrpend

\figsetend

%
%
 \begin{figure*}
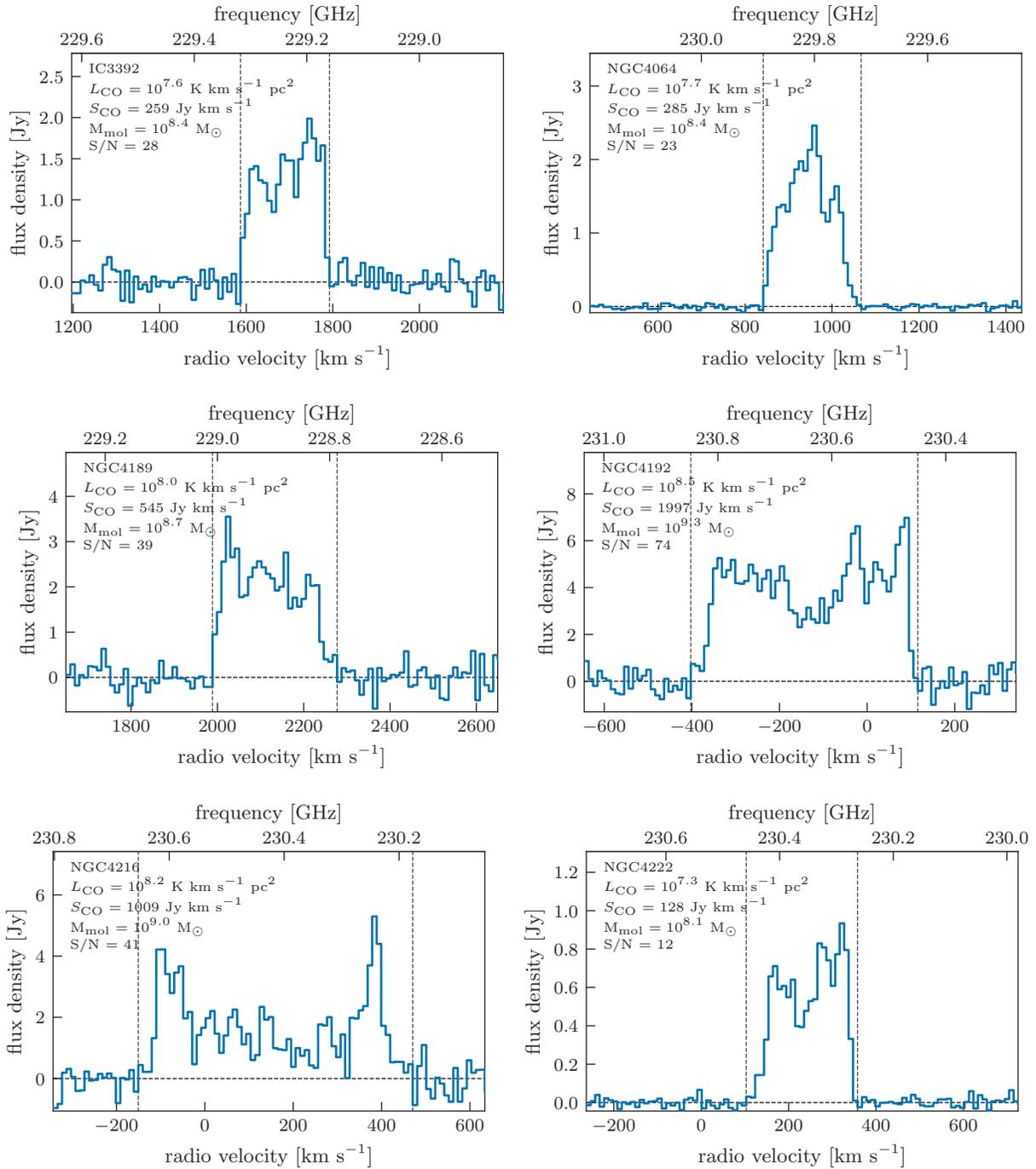

     \figurenum{13.1}
     \centering
     \includegraphics{IC3392_co21_spec.pdf}
     \includegraphics{NGC4064_co21_spec.pdf}
     \includegraphics{NGC4189_co21_spec.pdf}
     \includegraphics{NGC4192_co21_spec.pdf}
     \includegraphics{NGC4216_co21_spec.pdf}
     \includegraphics{NGC4222_co21_spec.pdf}
     \caption{As in Figure 13.1.}
     \label{fig:spectra_1}
 \end{figure*}

 \begin{figure*}
     \figurenum{13.2}
     \centering
     \includegraphics{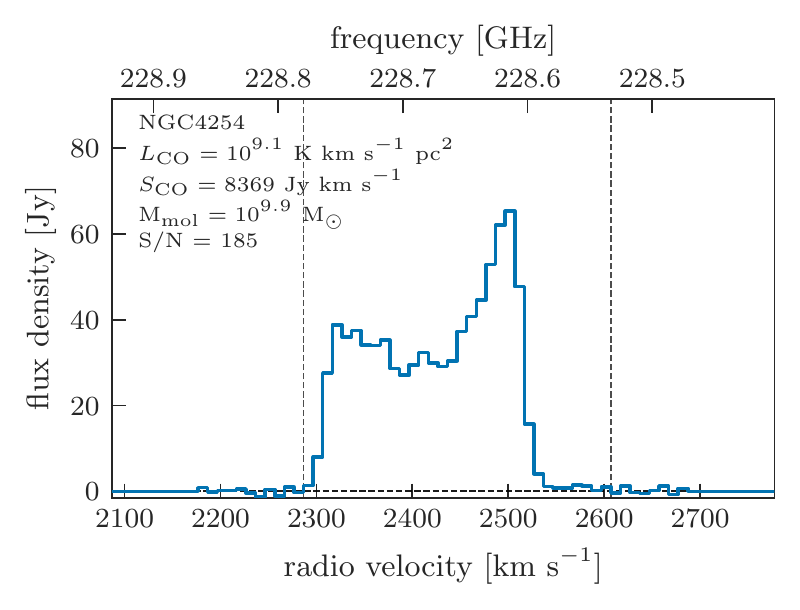}
     \includegraphics{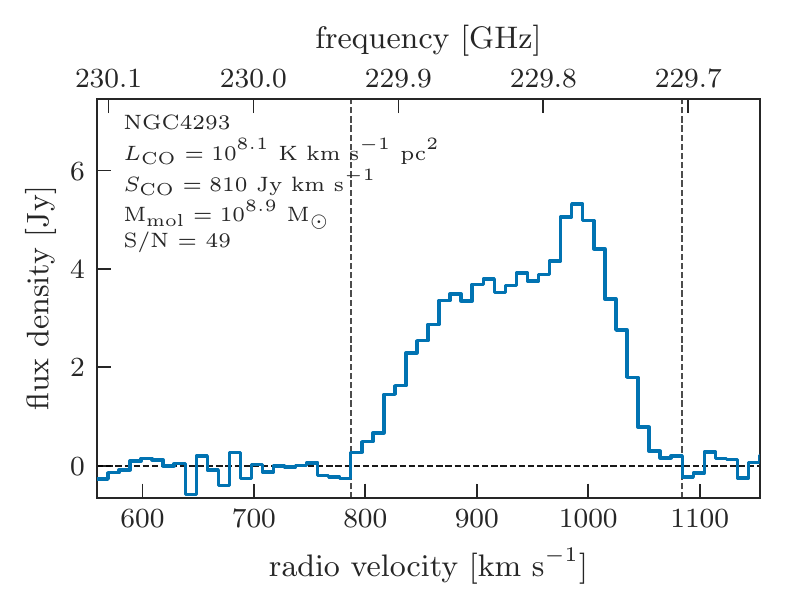}
     \includegraphics{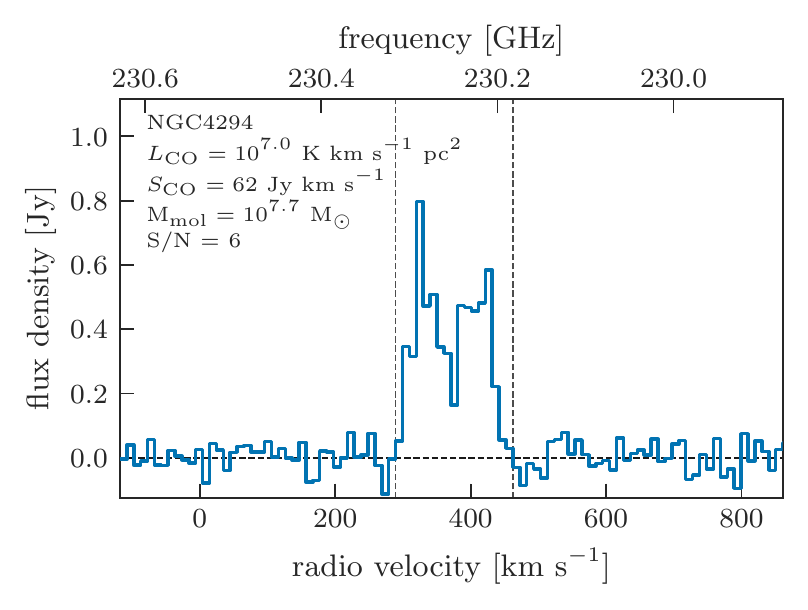}
     \includegraphics{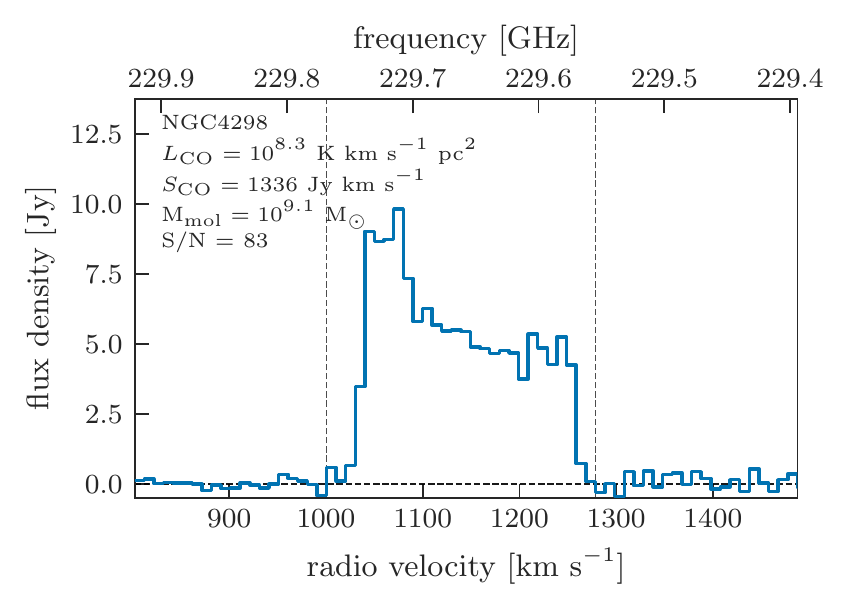}
     \includegraphics{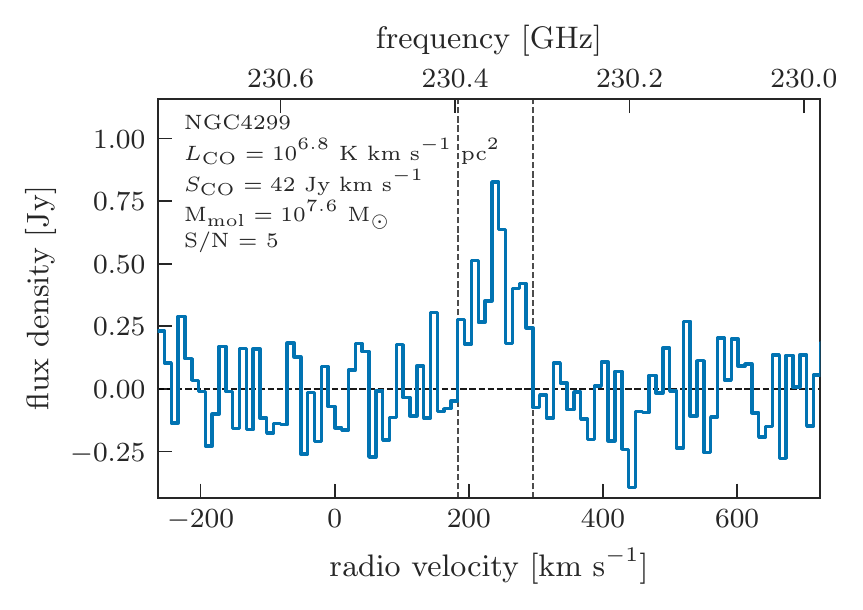}
     \includegraphics{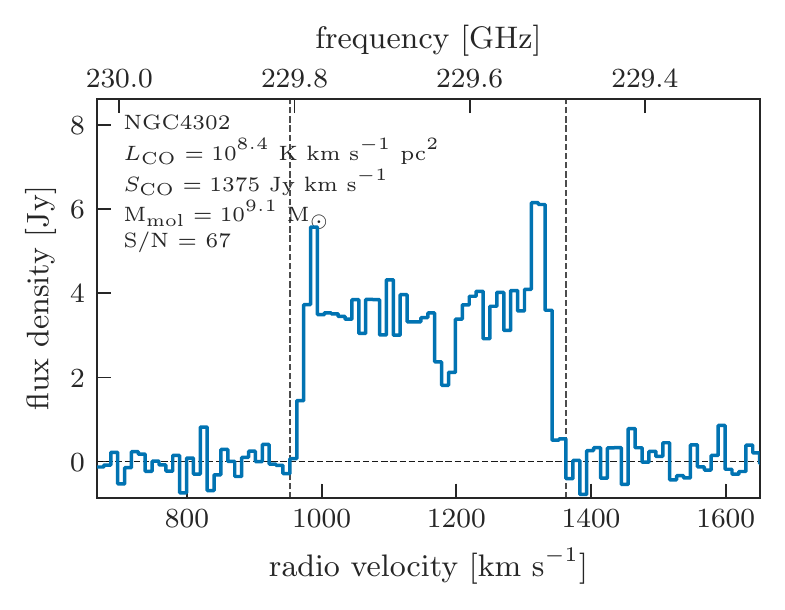}
     \caption{As Figure \ref{fig:spectra_1}.}
     \label{fig:spectra_2}
 \end{figure*}

 \begin{figure*}
     \figurenum{13.3}
     \centering
     \includegraphics{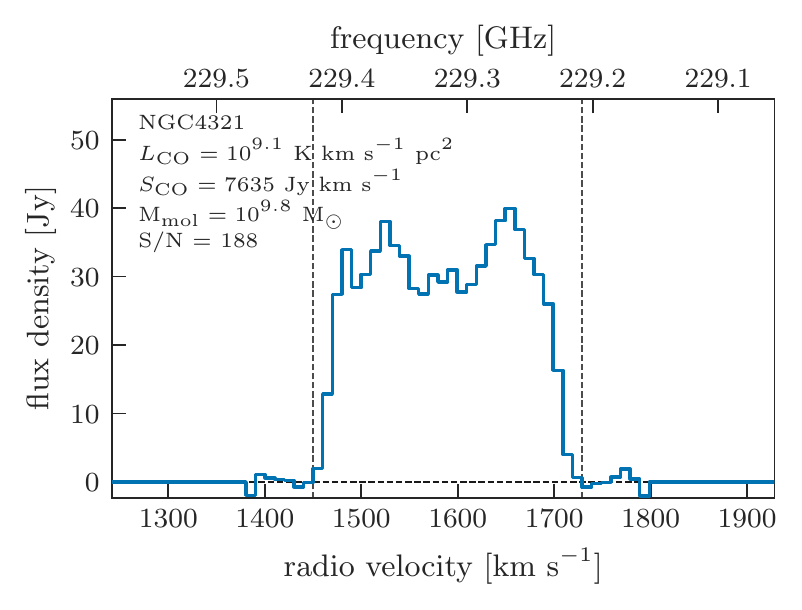}
     \includegraphics{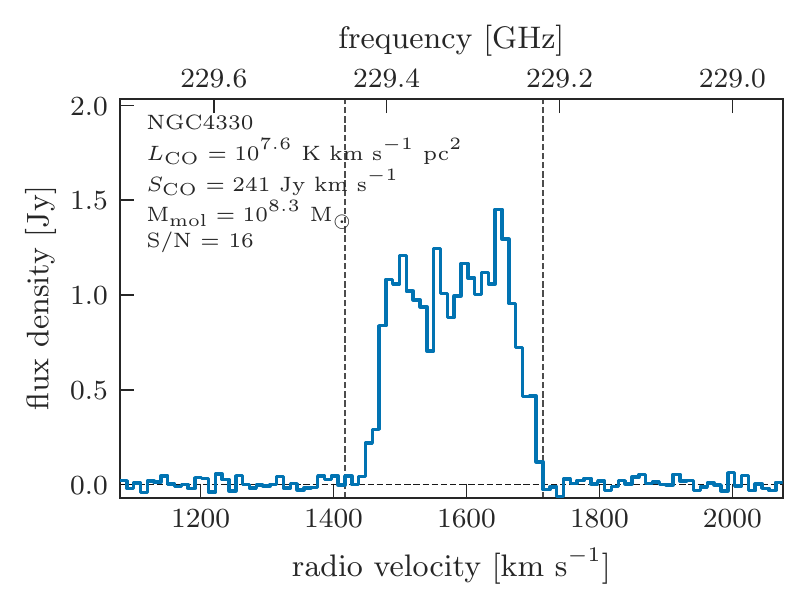}
     \includegraphics{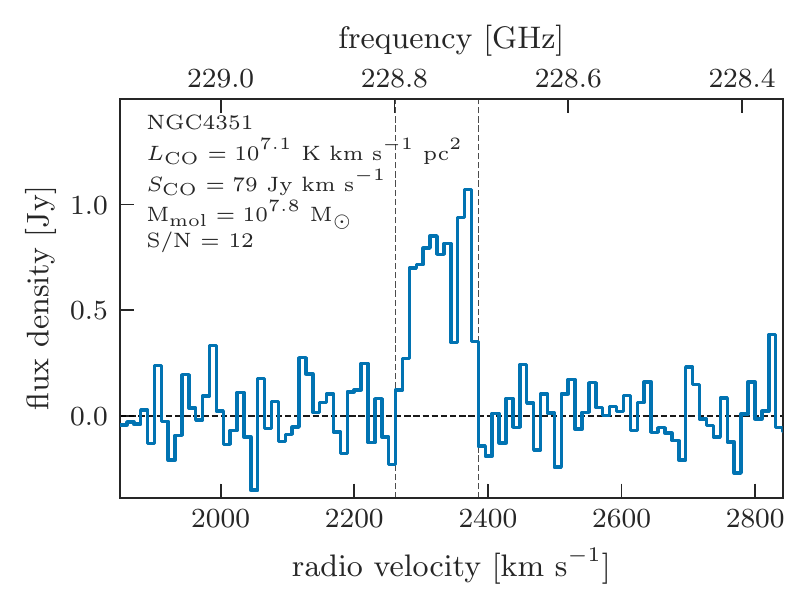}
     \includegraphics{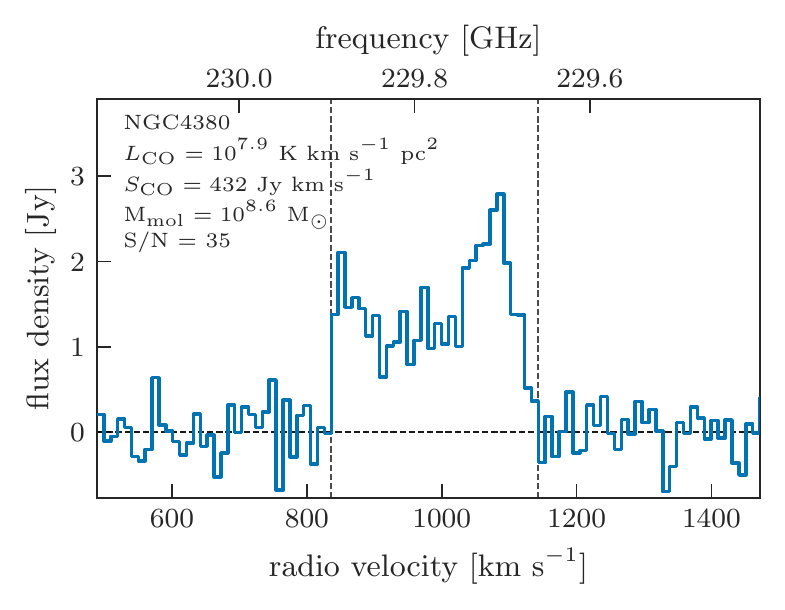}
     \includegraphics{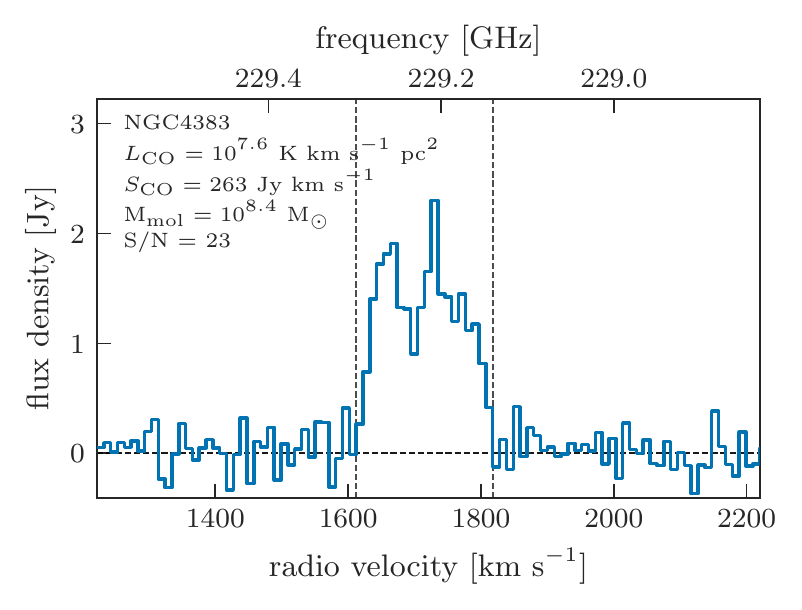}
     \includegraphics{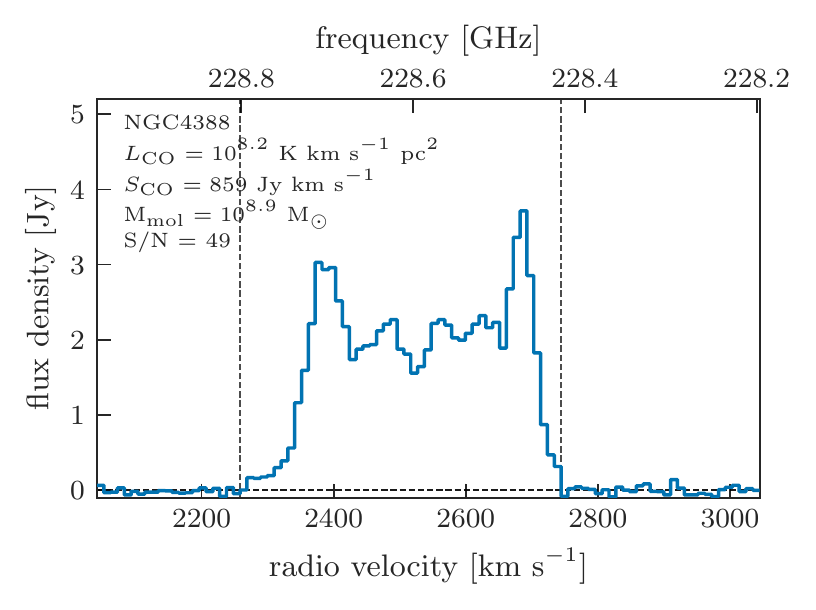}
     \caption{As Figure \ref{fig:spectra_1}.}
     \label{fig:spectra_3}
 \end{figure*}

 \begin{figure*}
     \figurenum{13.4}
     \centering
     \includegraphics{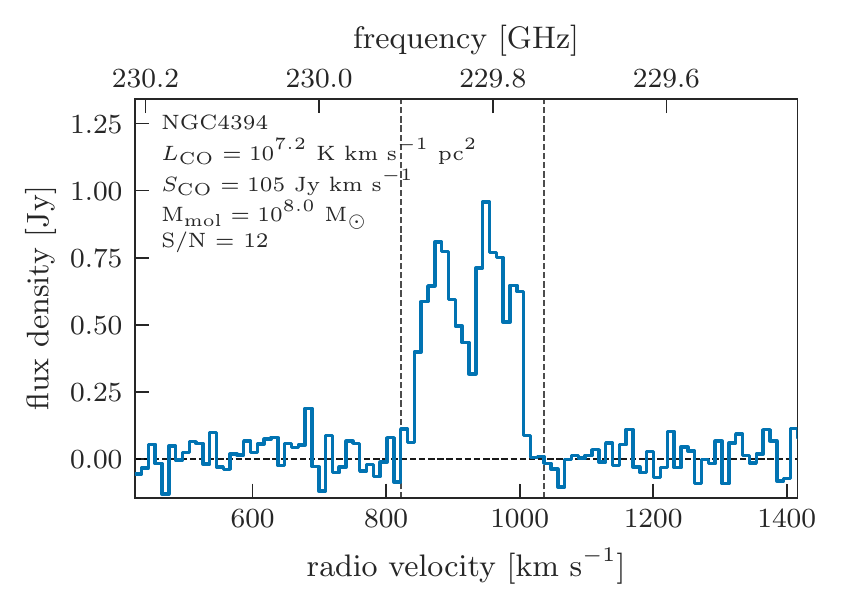}
     \includegraphics{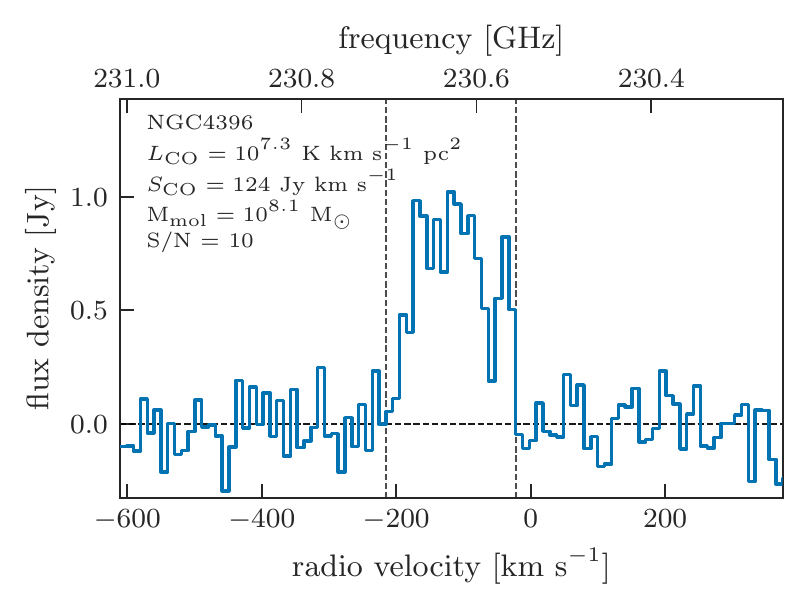}
     \includegraphics{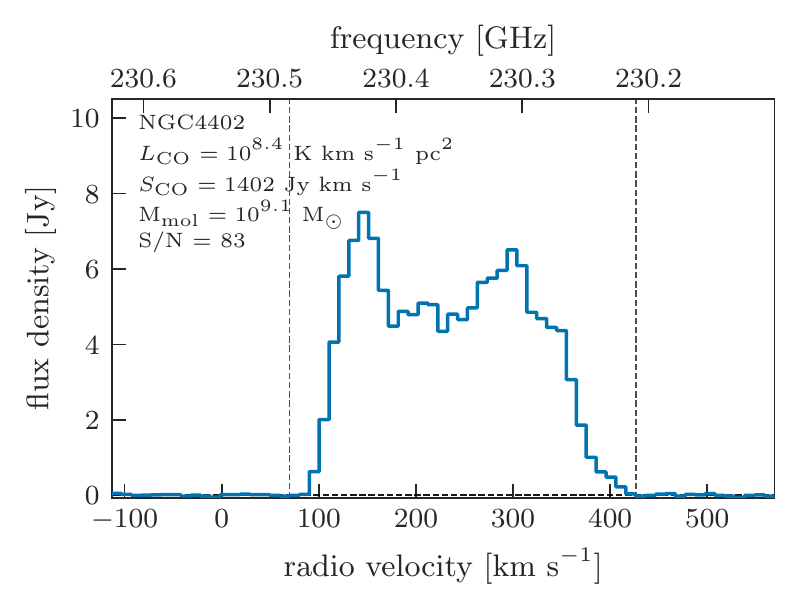}
     \includegraphics{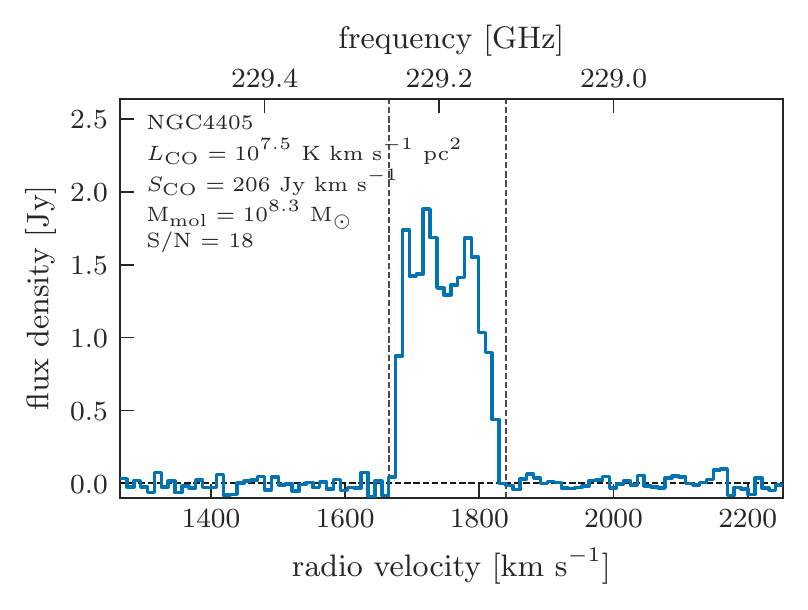}
     \includegraphics{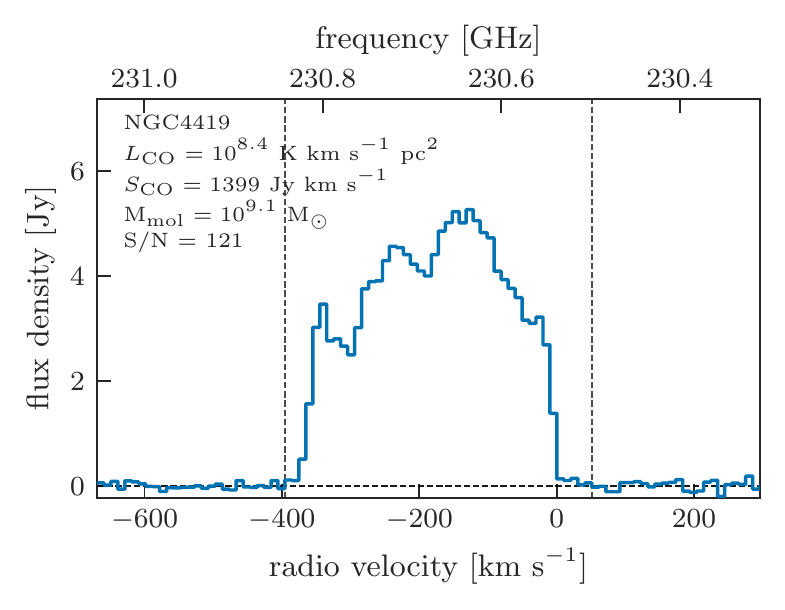}
     \includegraphics{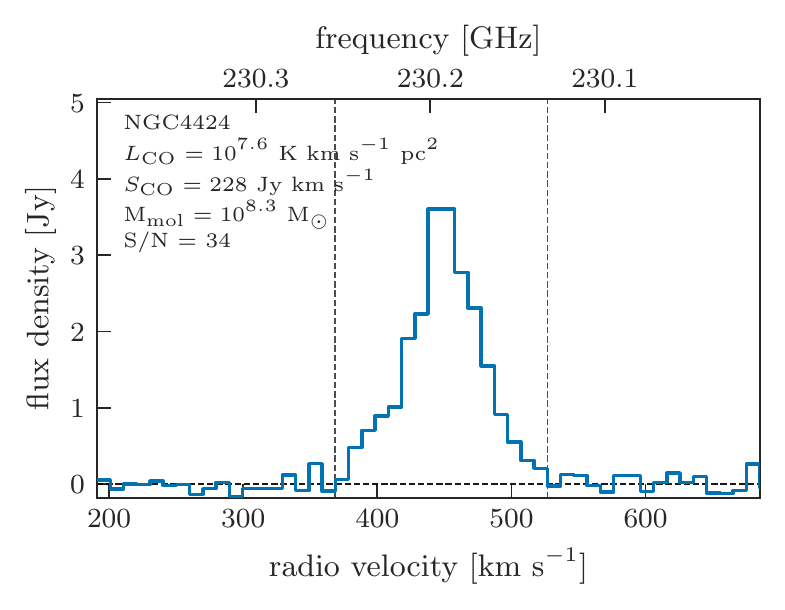}
     \caption{As Figure \ref{fig:spectra_1}.}
     \label{fig:spectra_4}
 \end{figure*}

 \begin{figure*}
     \figurenum{13.5}
     \centering
     \includegraphics{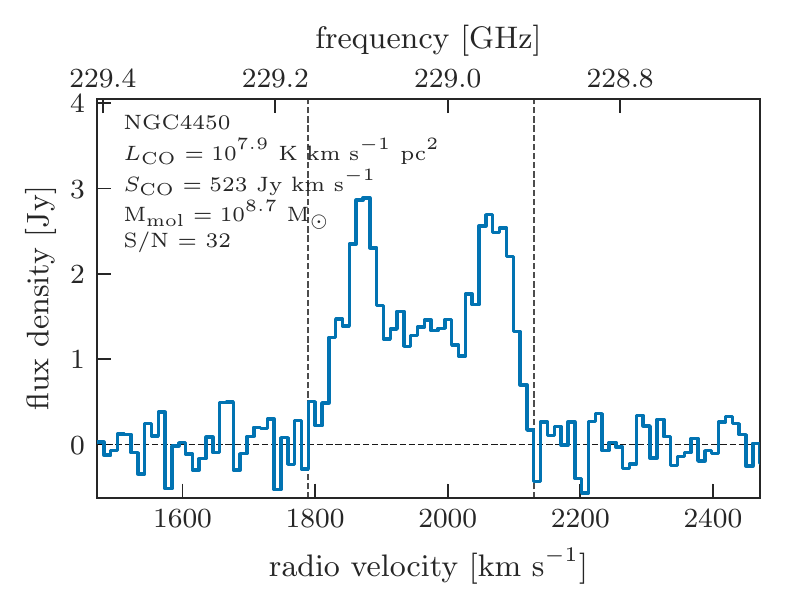}
     \includegraphics{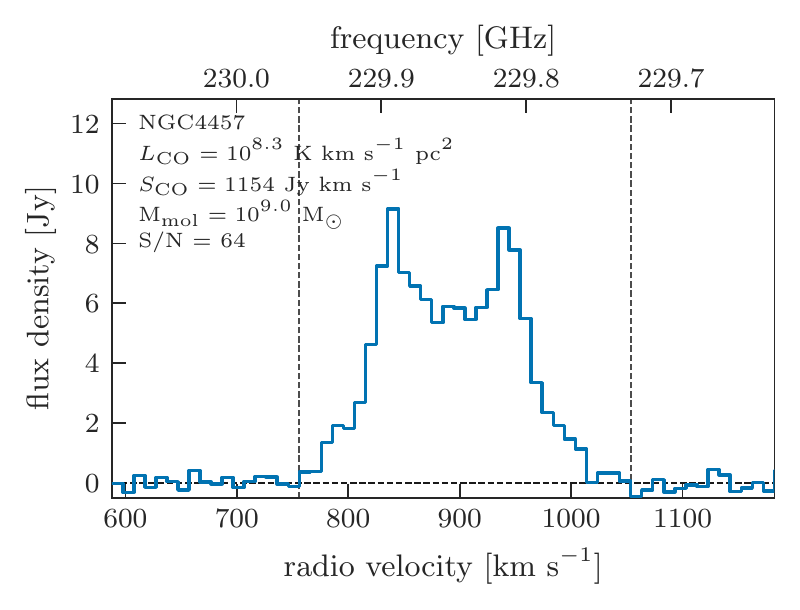}
     \includegraphics{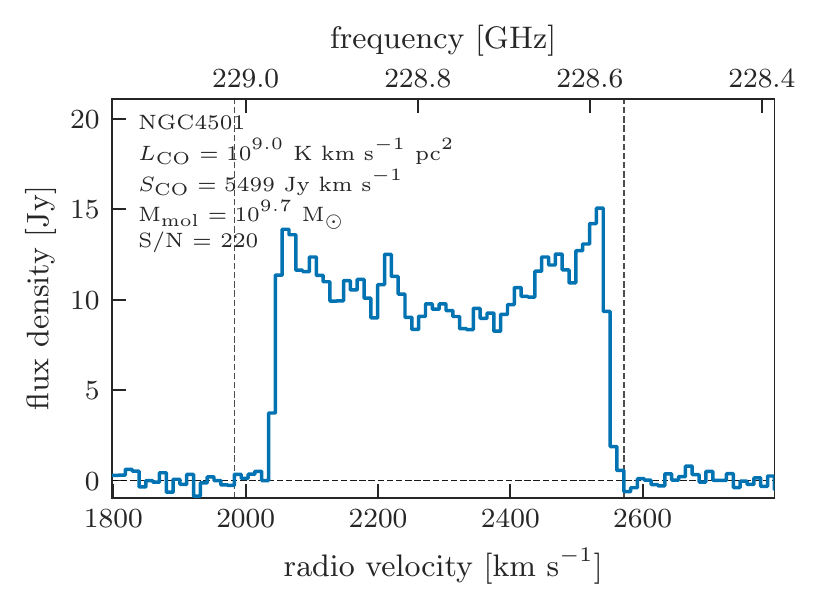}
     \includegraphics{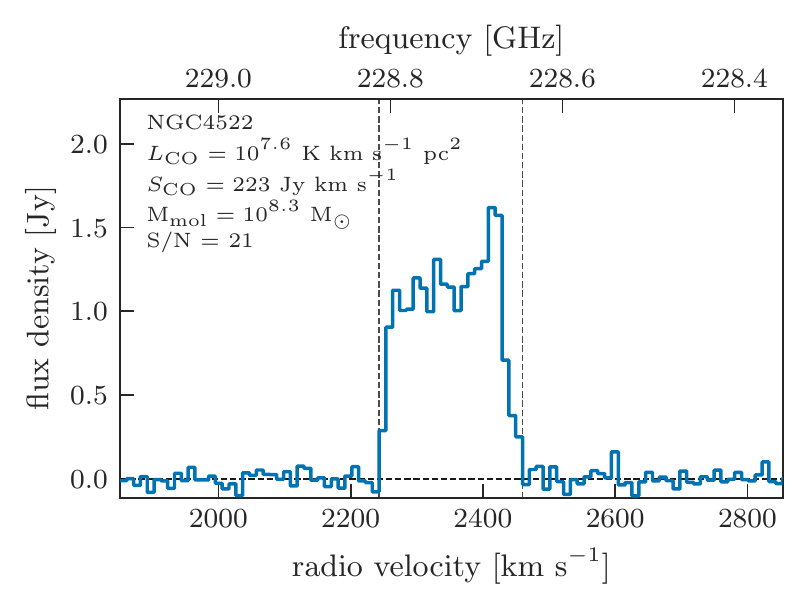}
     \includegraphics{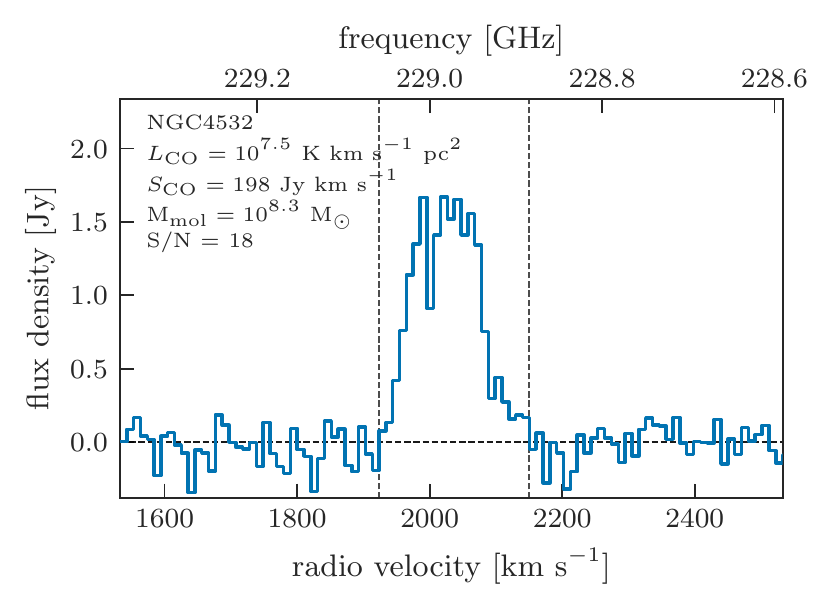}
     \includegraphics{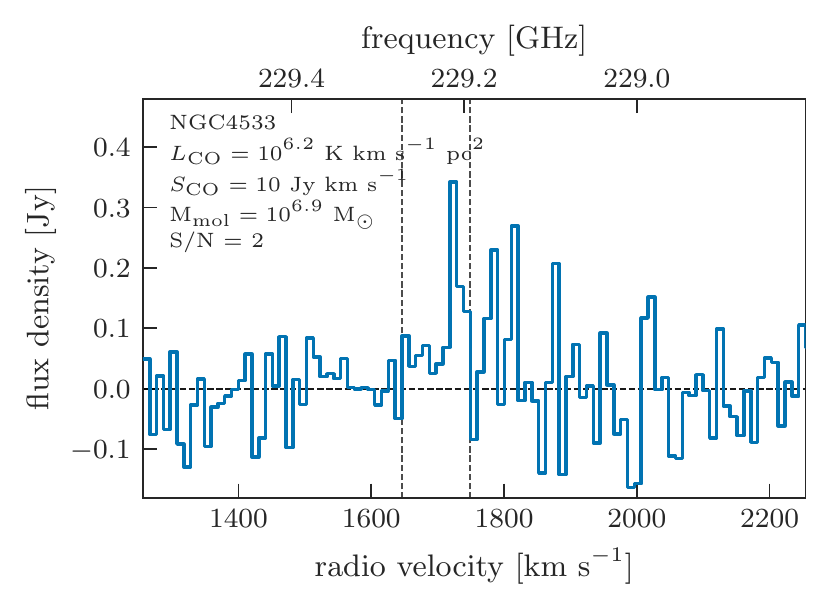}
     \caption{As Figure \ref{fig:spectra_1}.}
     \label{fig:spectra_5}
 \end{figure*}

 \begin{figure*}
     \figurenum{13.6}
     \centering
     \includegraphics{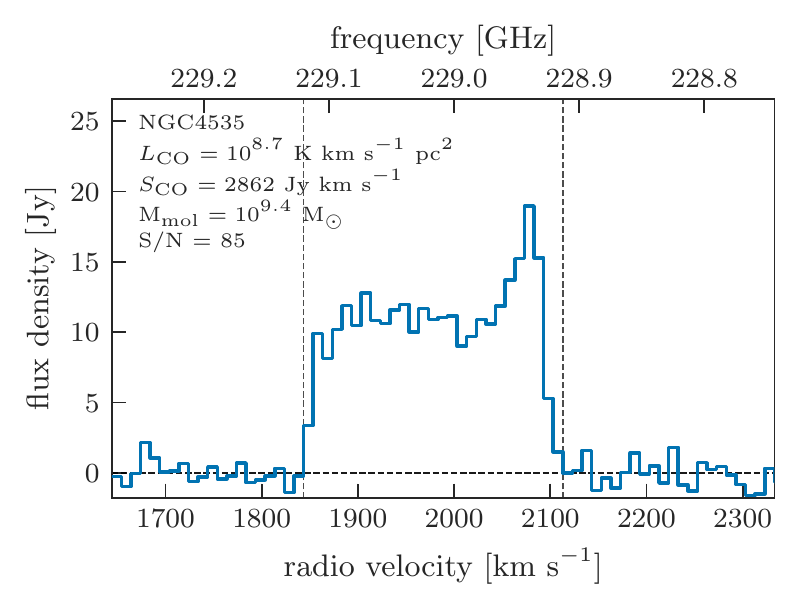}
     \includegraphics{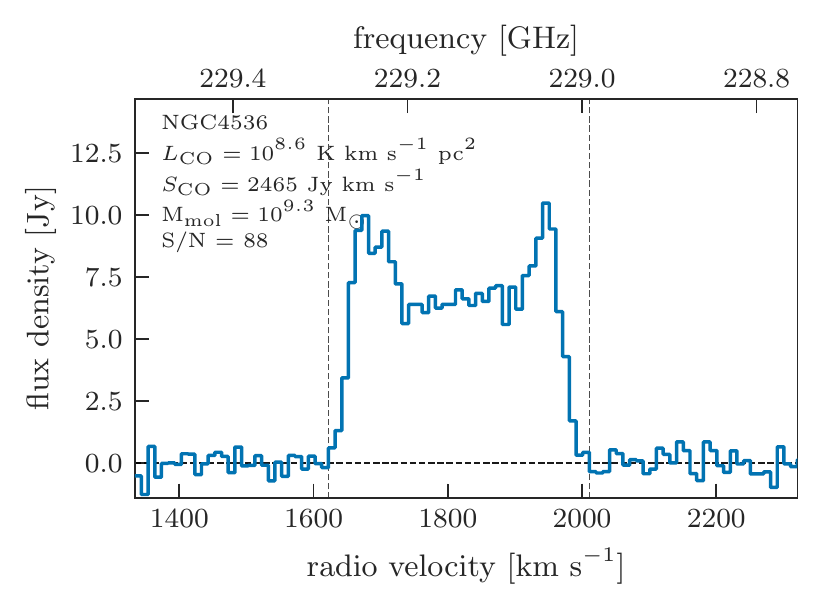}
     \includegraphics{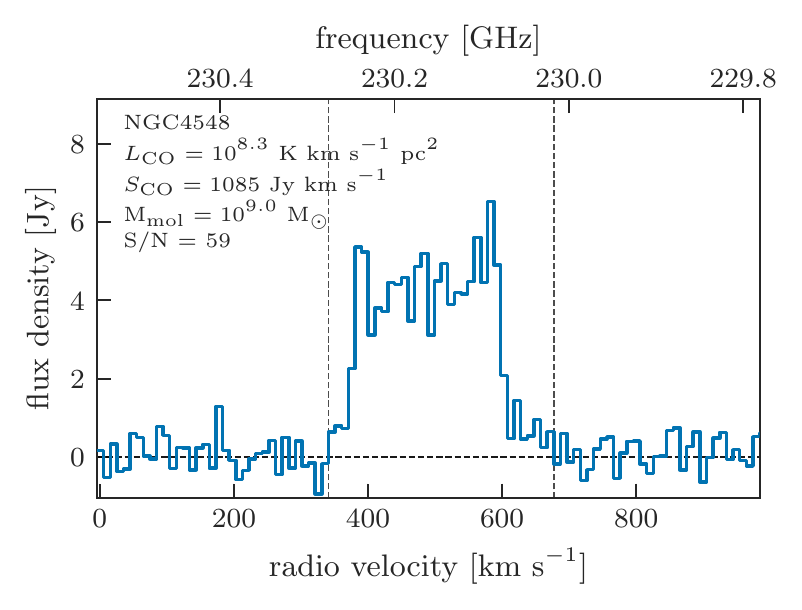}
     \includegraphics{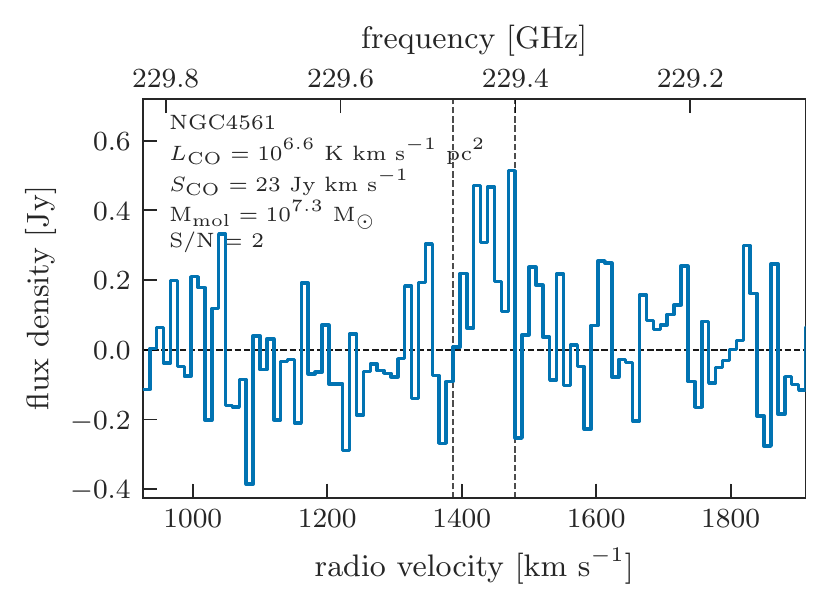}
     \includegraphics{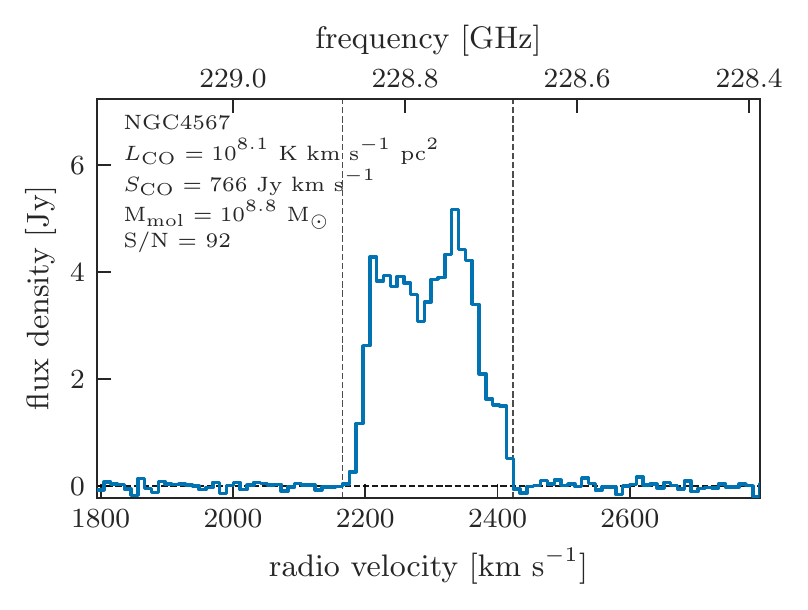}
     \includegraphics{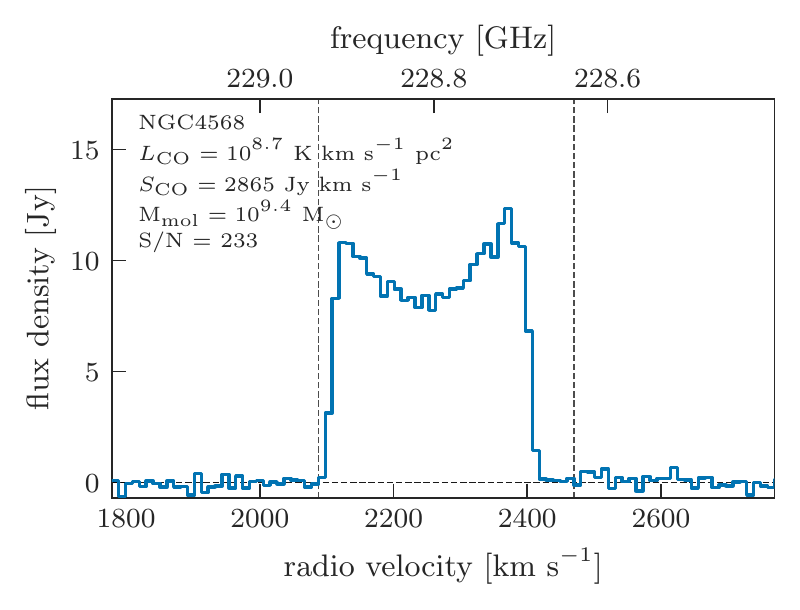}
     \caption{As Figure \ref{fig:spectra_1}.}
     \label{fig:spectra_6}
 \end{figure*}

 \begin{figure*}
     \figurenum{13.7}
     \centering
     \includegraphics{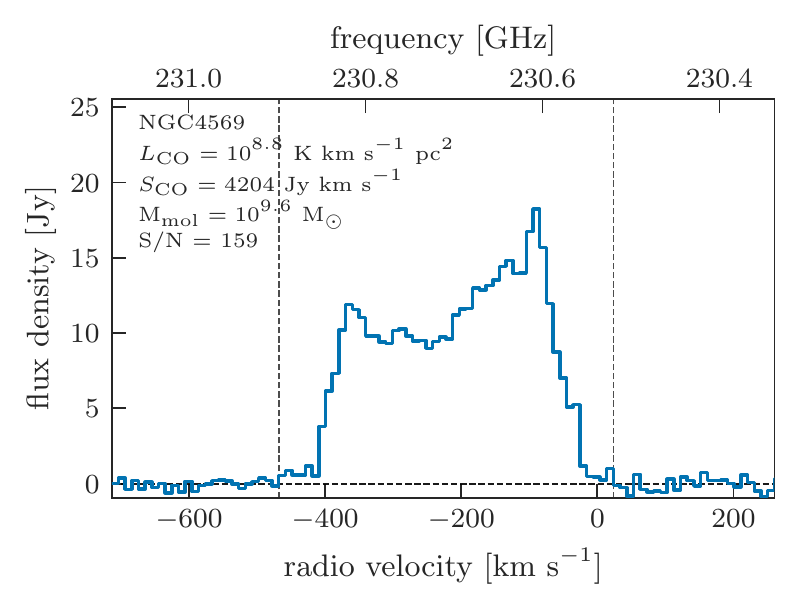}
     \includegraphics{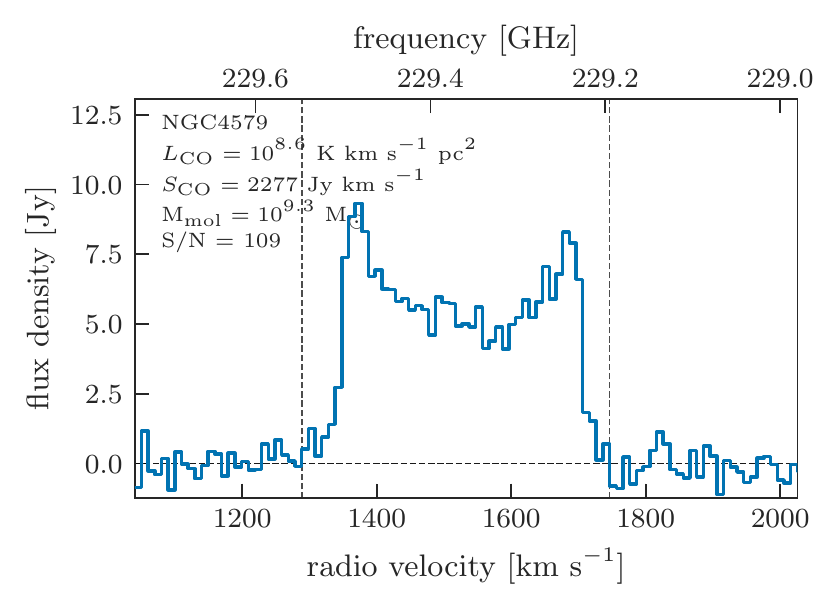}
     \includegraphics{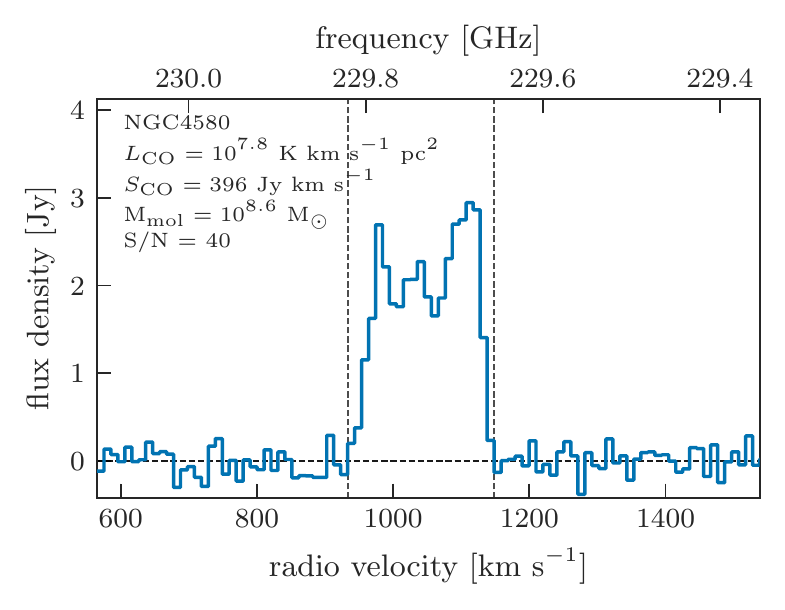}
     \includegraphics{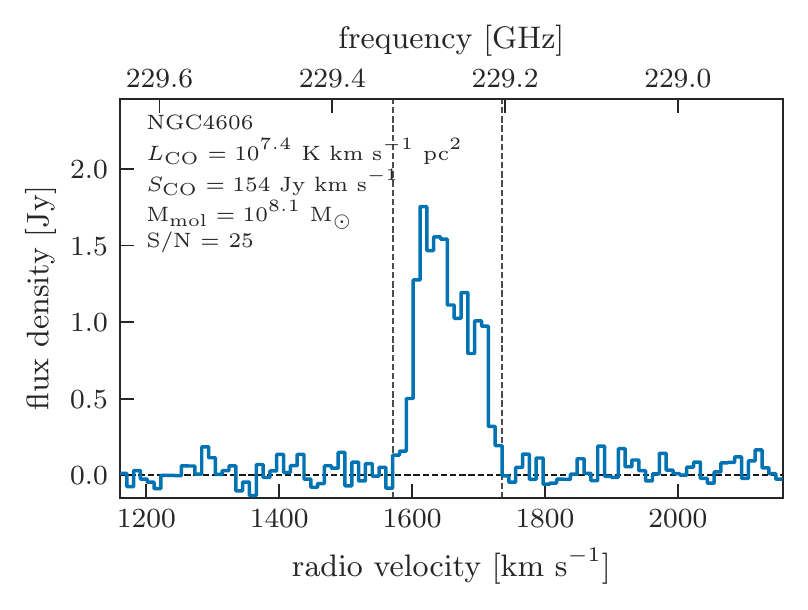}
     \includegraphics{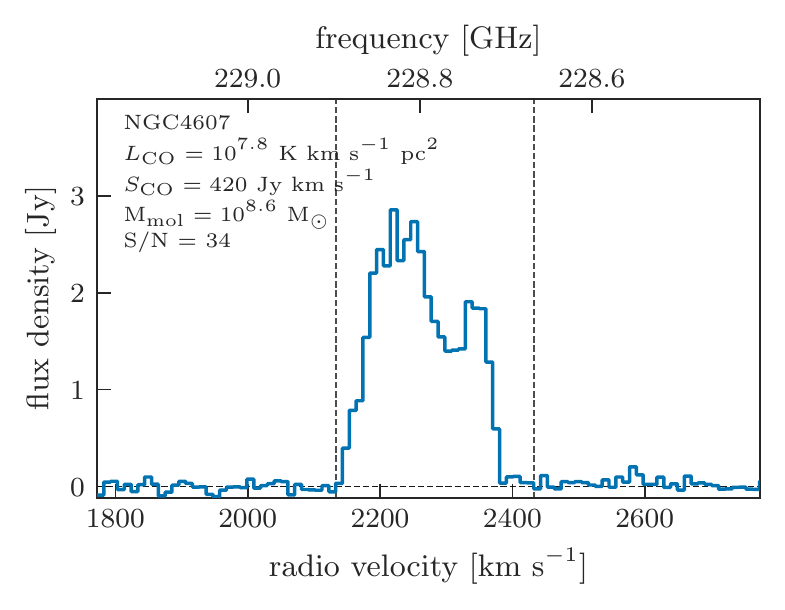}
     \includegraphics{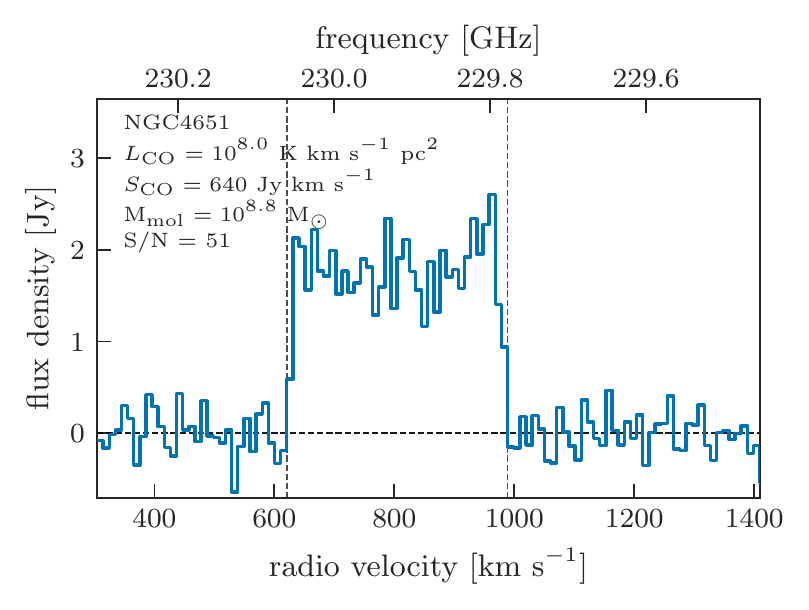}
     \caption{As Figure \ref{fig:spectra_1}.}
     \label{fig:spectra_7}
 \end{figure*}

 \begin{figure*}
     \figurenum{13.8}
     \centering
     \includegraphics{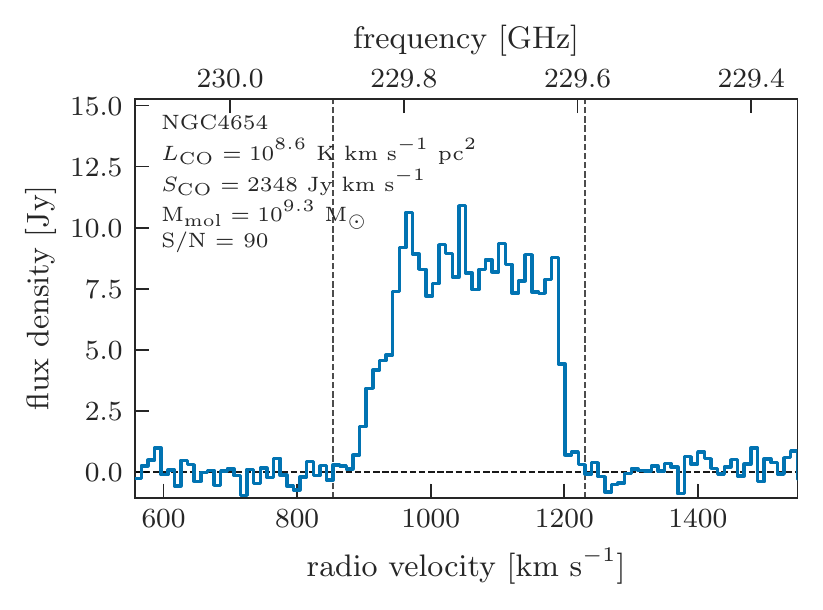}
     \includegraphics{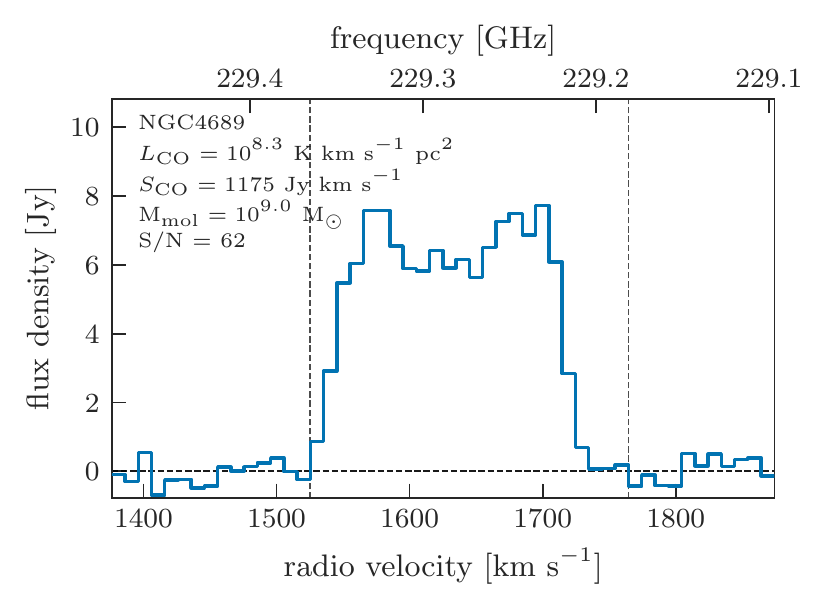}
     \includegraphics{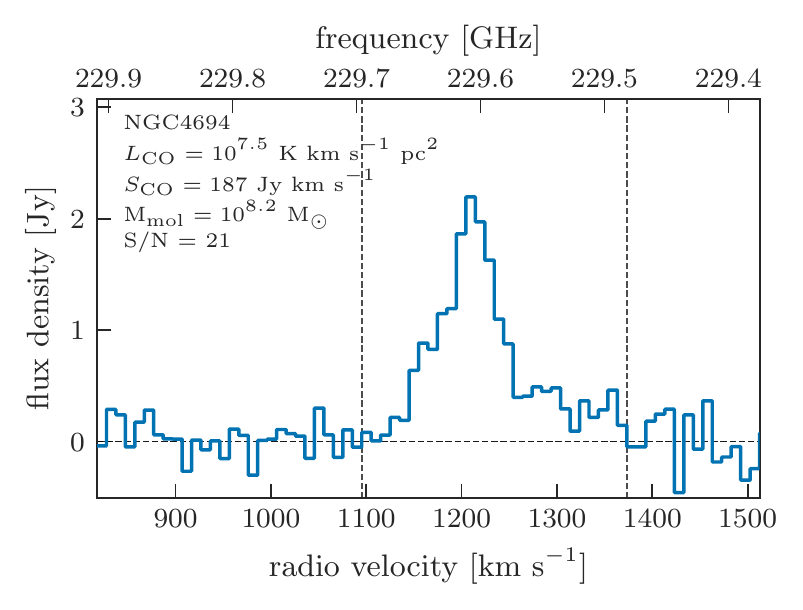}
     \includegraphics{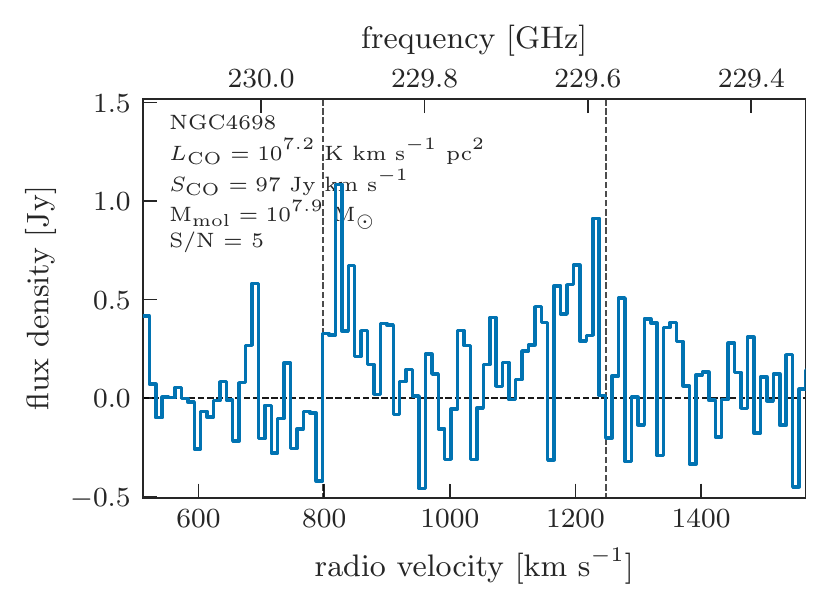}
     \caption{As Figure \ref{fig:spectra_1}.}
     \label{fig:spectra_8}
 \end{figure*}

 \begin{figure*}
     \figurenum{13.9}
     \centering
     \includegraphics{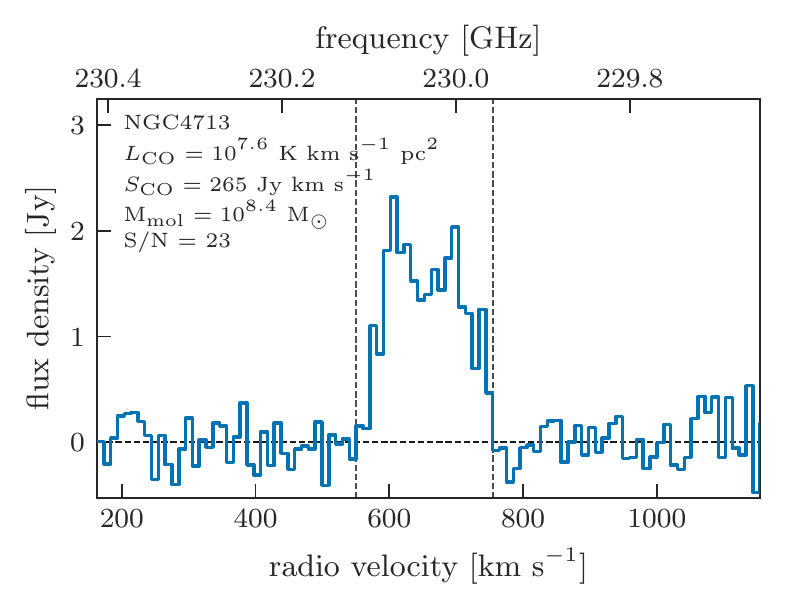}
     \includegraphics{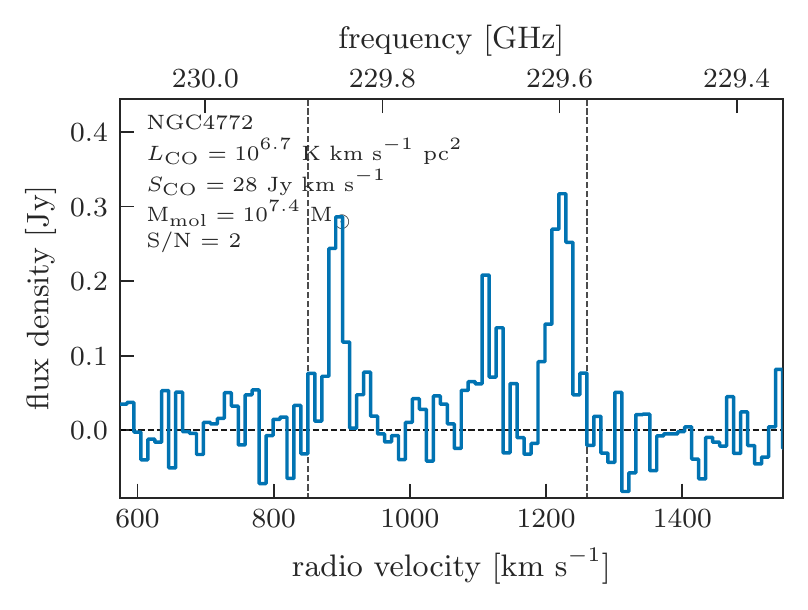}
     \includegraphics{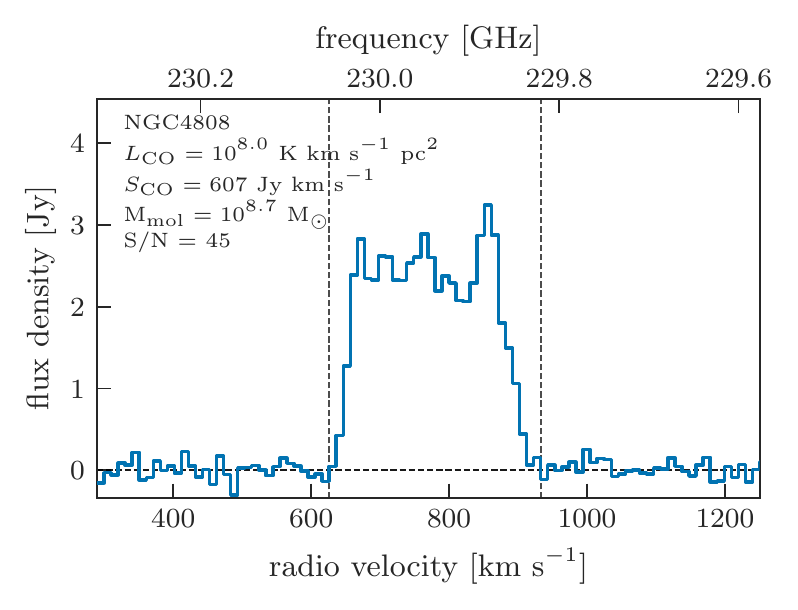}
     \caption{As Figure \ref{fig:spectra_1}.}
     \label{fig:spectra_9}
 \end{figure*}

 \begin{figure*}
     \figurenum{13.10}
     \centering
     \includegraphics{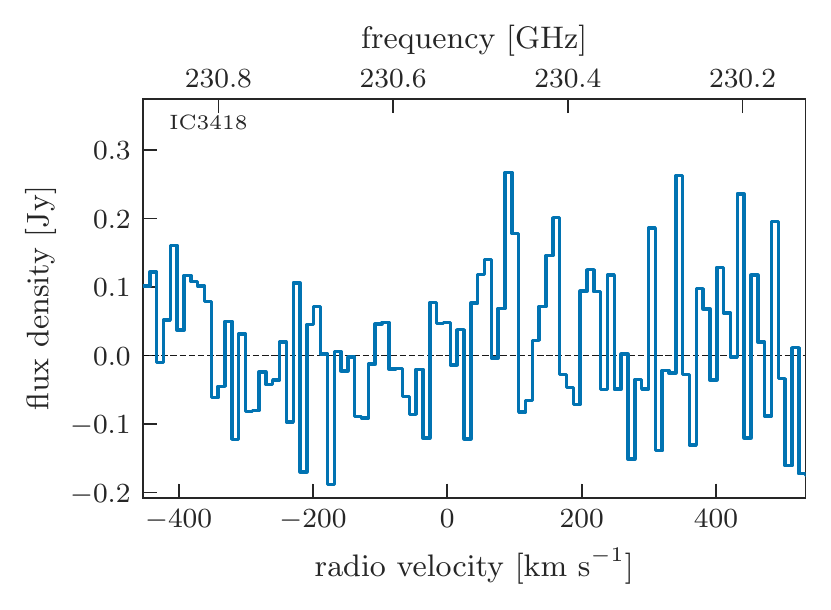}
     \includegraphics{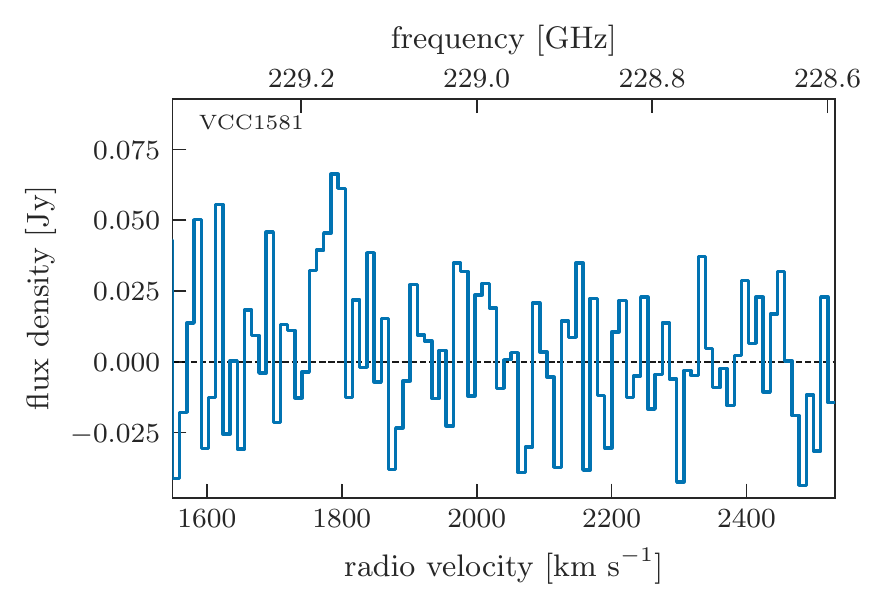}
     \caption{Spectra covering the expected frequency of $^{\rm 12}$CO($2-1$) calculated from the velocity listed in Table \ref{tab:VERTICO-sample} for the two VERTICO non-detections, IC3418 and VCC1581. The velocity resolution is 10.6~km~s$^{-1}$.}
     \label{fig:spectra_10}
 \end{figure*}

\bibliography{refs}{}

\begin{thebibliography}{}
\expandafter\ifx\csname natexlab\endcsname\relax\def\natexlab#1{#1}\fi
\providecommand{\url}[1]{\href{#1}{#1}}
\providecommand{\dodoi}[1]{doi:~\href{http://doi.org/#1}{\nolinkurl{#1}}}
\providecommand{\doeprint}[1]{\href{http://ascl.net/#1}{\nolinkurl{http://ascl.net/#1}}}
\providecommand{\doarXiv}[1]{\href{https://arxiv.org/abs/#1}{\nolinkurl{https://arxiv.org/abs/#1}}}

\bibitem[{{Accurso} {et~al.}(2017){Accurso}, {Saintonge}, {Catinella},
  {Cortese}, {Dav{\'e}}, {Dunsheath}, {Genzel}, {Gracia-Carpio}, {Heckman},
  {Jimmy}, {Kramer}, {Li}, {Lutz}, {Schiminovich}, {Schuster}, {Sternberg},
  {Sturm}, {Tacconi}, {Tran}, \& {Wang}}]{Accurso2017}
{Accurso}, G., {Saintonge}, A., {Catinella}, B., {et~al.} 2017, \mnras, 470,
  4750, \dodoi{10.1093/mnras/stx1556}

\bibitem[{{Alam} {et~al.}(2015){Alam}, {Albareti}, {Allende Prieto}, {Anders},
  {Anderson}, {Anderton}, {Andrews}, {Armengaud}, {Aubourg}, {Bailey}, {Basu},
  {Bautista}, {Beaton}, {Beers}, {Bender}, {Berlind}, {Beutler}, {Bhardwaj},
  {Bird}, {Bizyaev}, {Blake}, {Blanton}, {Blomqvist}, {Bochanski}, {Bolton},
  {Bovy}, {Shelden Bradley}, {Brandt}, {Brauer}, {Brinkmann}, {Brown},
  {Brownstein}, {Burden}, {Burtin}, {Busca}, {Cai}, {Capozzi}, {Carnero
  Rosell}, {Carr}, {Carrera}, {Chambers}, {Chaplin}, {Chen}, {Chiappini},
  {Chojnowski}, {Chuang}, {Clerc}, {Comparat}, {Covey}, {Croft}, {Cuesta},
  {Cunha}, {da Costa}, {Da Rio}, {Davenport}, {Dawson}, {De Lee}, {Delubac},
  {Deshpande}, {Dhital}, {Dutra-Ferreira}, {Dwelly}, {Ealet}, {Ebelke},
  {Edmondson}, {Eisenstein}, {Ellsworth}, {Elsworth}, {Epstein}, {Eracleous},
  {Escoffier}, {Esposito}, {Evans}, {Fan}, {Fern{\'a}ndez-Alvar}, {Feuillet},
  {Filiz Ak}, {Finley}, {Finoguenov}, {Flaherty}, {Fleming}, {Font-Ribera},
  {Foster}, {Frinchaboy}, {Galbraith-Frew}, {Garc{\'\i}a},
  {Garc{\'\i}a-Hern{\'a}ndez}, {Garc{\'\i}a P{\'e}rez}, {Gaulme}, {Ge},
  {G{\'e}nova-Santos}, {Georgakakis}, {Ghezzi}, {Gillespie}, {Girardi},
  {Goddard}, {Gontcho}, {Gonz{\'a}lez Hern{\'a}ndez}, {Grebel}, {Green},
  {Grieb}, {Grieves}, {Gunn}, {Guo}, {Harding}, {Hasselquist}, {Hawley},
  {Hayden}, {Hearty}, {Hekker}, {Ho}, {Hogg}, {Holley-Bockelmann}, {Holtzman},
  {Honscheid}, {Huber}, {Huehnerhoff}, {Ivans}, {Jiang}, {Johnson},
  {Kinemuchi}, {Kirkby}, {Kitaura}, {Klaene}, {Knapp}, {Kneib}, {Koenig},
  {Lam}, {Lan}, {Lang}, {Laurent}, {Le Goff}, {Leauthaud}, {Lee}, {Lee},
  {Licquia}, {Liu}, {Long}, {L{\'o}pez-Corredoira}, {Lorenzo-Oliveira},
  {Lucatello}, {Lundgren}, {Lupton}, {Mack}, {Mahadevan}, {Maia}, {Majewski},
  {Malanushenko}, {Malanushenko}, {Manchado}, {Manera}, {Mao}, {Maraston},
  {Marchwinski}, {Margala}, {Martell}, {Martig}, {Masters}, {Mathur},
  {McBride}, {McGehee}, {McGreer}, {McMahon}, {M{\'e}nard}, {Menzel},
  {Merloni}, {M{\'e}sz{\'a}ros}, {Miller}, {Miralda-Escud{\'e}}, {Miyatake},
  {Montero-Dorta}, {More}, {Morganson}, {Morice-Atkinson}, {Morrison},
  {Mosser}, {Muna}, {Myers}, {Nandra}, {Newman}, {Neyrinck}, {Nguyen},
  {Nichol}, {Nidever}, {Noterdaeme}, {Nuza}, {O'Connell}, {O'Connell},
  {O'Connell}, {Ogando}, {Olmstead}, {Oravetz}, {Oravetz}, {Osumi}, {Owen},
  {Padgett}, {Padmanabhan}, {Paegert}, {Palanque-Delabrouille}, {Pan},
  {Parejko}, {P{\^a}ris}, {Park}, {Pattarakijwanich}, {Pellejero-Ibanez},
  {Pepper}, {Percival}, {P{\'e}rez-Fournon}, {P{\textasciiacute}rez-Ra`fols},
  {Petitjean}, {Pieri}, {Pinsonneault}, {Porto de Mello}, {Prada}, {Prakash},
  {Price-Whelan}, {Protopapas}, {Raddick}, {Rahman}, {Reid}, {Rich}, {Rix},
  {Robin}, {Rockosi}, {Rodrigues}, {Rodr{\'\i}guez-Torres}, {Roe}, {Ross},
  {Ross}, {Rossi}, {Ruan}, {Rubi{\~n}o-Mart{\'\i}n}, {Rykoff},
  {Salazar-Albornoz}, {Salvato}, {Samushia}, {S{\'a}nchez}, {Santiago},
  {Sayres}, {Schiavon}, {Schlegel}, {Schmidt}, {Schneider}, {Schultheis},
  {Schwope}, {Sc{\'o}ccola}, {Scott}, {Sellgren}, {Seo}, {Serenelli}, {Shane},
  {Shen}, {Shetrone}, {Shu}, {Silva Aguirre}, {Sivarani}, {Skrutskie},
  {Slosar}, {Smith}, {Sobreira}, {Souto}, {Stassun}, {Steinmetz}, {Stello},
  {Strauss}, {Streblyanska}, {Suzuki}, {Swanson}, {Tan}, {Tayar}, {Terrien},
  {Thakar}, {Thomas}, {Thomas}, {Thompson}, {Tinker}, {Tojeiro}, {Troup},
  {Vargas-Maga{\~n}a}, {Vazquez}, {Verde}, {Viel}, {Vogt}, {Wake}, {Wang},
  {Weaver}, {Weinberg}, {Weiner}, {White}, {Wilson}, {Wisniewski},
  {Wood-Vasey}, {Ye`che}, {York}, {Zakamska}, {Zamora}, {Zasowski}, {Zehavi},
  {Zhao}, {Zheng}, {Zhou}, {Zhou}, {Zou}, \& {Zhu}}]{Alam2015}
{Alam}, S., {Albareti}, F.~D., {Allende Prieto}, C., {et~al.} 2015, \apjs, 219,
  12, \dodoi{10.1088/0067-0049/219/1/12}

\bibitem[{{Astropy Collaboration} {et~al.}(2013){Astropy Collaboration},
  {Robitaille}, {Tollerud}, {Greenfield}, {Droettboom}, {Bray}, {Aldcroft},
  {Davis}, {Ginsburg}, {Price-Whelan}, {Kerzendorf}, {Conley}, {Crighton},
  {Barbary}, {Muna}, {Ferguson}, {Grollier}, {Parikh}, {Nair}, {Unther},
  {Deil}, {Woillez}, {Conseil}, {Kramer}, {Turner}, {Singer}, {Fox}, {Weaver},
  {Zabalza}, {Edwards}, {Azalee Bostroem}, {Burke}, {Casey}, {Crawford},
  {Dencheva}, {Ely}, {Jenness}, {Labrie}, {Lim}, {Pierfederici}, {Pontzen},
  {Ptak}, {Refsdal}, {Servillat}, \& {Streicher}}]{astropy:2013}
{Astropy Collaboration}, {Robitaille}, T.~P., {Tollerud}, E.~J., {et~al.} 2013,
  \aap, 558, A33, \dodoi{10.1051/0004-6361/201322068}

\bibitem[{{Balogh} {et~al.}(1999){Balogh}, {Morris}, {Yee}, {Carlberg}, \&
  {Ellingson}}]{Balogh1999}
{Balogh}, M.~L., {Morris}, S.~L., {Yee}, H.~K.~C., {Carlberg}, R.~G., \&
  {Ellingson}, E. 1999, \apj, 527, 54, \dodoi{10.1086/308056}

\bibitem[{{Balogh} {et~al.}(2000){Balogh}, {Navarro}, \& {Morris}}]{Balogh2000}
{Balogh}, M.~L., {Navarro}, J.~F., \& {Morris}, S.~L. 2000, \apj, 540, 113,
  \dodoi{10.1086/309323}

\bibitem[{{Balogh} {et~al.}(1998){Balogh}, {Schade}, {Morris}, {Yee},
  {Carlberg}, \& {Ellingson}}]{Balogh1998}
{Balogh}, M.~L., {Schade}, D., {Morris}, S.~L., {et~al.} 1998, \apjl, 504, L75,
  \dodoi{10.1086/311576}

\bibitem[{{Bekki} {et~al.}(2002){Bekki}, {Couch}, \& {Shioya}}]{Bekki2002}
{Bekki}, K., {Couch}, W.~J., \& {Shioya}, Y. 2002, \apj, 577, 651,
  \dodoi{10.1086/342221}

\bibitem[{{Bigiel} \& {Blitz}(2012)}]{Bigiel2012}
{Bigiel}, F., \& {Blitz}, L. 2012, \apj, 756, 183,
  \dodoi{10.1088/0004-637X/756/2/183}

\bibitem[{{B{\"o}hringer} {et~al.}(1994){B{\"o}hringer}, {Briel}, {Schwarz},
  {Voges}, {Hartner}, \& {Tr{\"u}mper}}]{Bohringer1994}
{B{\"o}hringer}, H., {Briel}, U.~G., {Schwarz}, R.~A., {et~al.} 1994, \nat,
  368, 828, \dodoi{10.1038/368828a0}

\bibitem[{{Bolatto} {et~al.}(2013){Bolatto}, {Wolfire}, \&
  {Leroy}}]{Bolatto2013}
{Bolatto}, A.~D., {Wolfire}, M., \& {Leroy}, A.~K. 2013, \araa, 51, 207,
  \dodoi{10.1146/annurev-astro-082812-140944}

\bibitem[{{Bolatto} {et~al.}(2017){Bolatto}, {Wong}, {Utomo}, {Blitz}, {Vogel},
  {S{\'a}nchez}, {Barrera-Ballesteros}, {Cao}, {Colombo}, {Dannerbauer},
  {Garc{\'\i}a-Benito}, {Herrera-Camus}, {Husemann}, {Kalinova}, {Leroy},
  {Leung}, {Levy}, {Mast}, {Ostriker}, {Rosolowsky}, {Sandstrom}, {Teuben},
  {van de Ven}, \& {Walter}}]{Bolatto2017}
{Bolatto}, A.~D., {Wong}, T., {Utomo}, D., {et~al.} 2017, \apj, 846, 159,
  \dodoi{10.3847/1538-4357/aa86aa}

\bibitem[{{Boselli} {et~al.}(2014){Boselli}, {Cortese}, \&
  {Boquien}}]{Boselli2014a}
{Boselli}, A., {Cortese}, L., \& {Boquien}, M. 2014, \aap, 564, A65,
  \dodoi{10.1051/0004-6361/201322311}

\bibitem[{{Boselli} \& {Gavazzi}(2006)}]{Boselli2006}
{Boselli}, A., \& {Gavazzi}, G. 2006, \pasp, 118, 517, \dodoi{10.1086/500691}

\bibitem[{{Boselli} {et~al.}(1997){Boselli}, {Gavazzi}, {Lequeux}, {Buat},
  {Casoli}, {Dickey}, \& {Donas}}]{Boselli1997}
{Boselli}, A., {Gavazzi}, G., {Lequeux}, J., {et~al.} 1997, \aap, 327, 522

\bibitem[{{Boselli} {et~al.}(2002){Boselli}, {Lequeux}, \&
  {Gavazzi}}]{Boselli2002}
{Boselli}, A., {Lequeux}, J., \& {Gavazzi}, G. 2002, \aap, 384, 33,
  \dodoi{10.1051/0004-6361:20011747}

\bibitem[{{Boselli} {et~al.}(2010){Boselli}, {Eales}, {Cortese}, {Bendo},
  {Chanial}, {Buat}, {Davies}, {Auld}, {Rigby}, {Baes}, {Barlow}, {Bock},
  {Bradford}, {Castro-Rodriguez}, {Charlot}, {Clements}, {Cormier}, {Dwek},
  {Elbaz}, {Galametz}, {Galliano}, {Gear}, {Glenn}, {Gomez}, {Griffin}, {Hony},
  {Isaak}, {Levenson}, {Lu}, {Madden}, {O'Halloran}, {Okamura}, {Oliver},
  {Page}, {Panuzzo}, {Papageorgiou}, {Parkin}, {Perez-Fournon}, {Pohlen},
  {Rangwala}, {Roussel}, {Rykala}, {Sacchi}, {Sauvage}, {Schulz}, {Schirm},
  {Smith}, {Spinoglio}, {Stevens}, {Symeonidis}, {Vaccari}, {Vigroux},
  {Wilson}, {Wozniak}, {Wright}, \& {Zeilinger}}]{Boselli2010}
{Boselli}, A., {Eales}, S., {Cortese}, L., {et~al.} 2010, \pasp, 122, 261,
  \dodoi{10.1086/651535}

\bibitem[{{Boselli} {et~al.}(2011){Boselli}, {Boissier}, {Heinis}, {Cortese},
  {Ilbert}, {Hughes}, {Cucciati}, {Davies}, {Ferrarese}, {Giovanelli},
  {Haynes}, {Baes}, {Balkowski}, {Brosch}, {Chapman}, {Charmandaris},
  {Clemens}, {Dariush}, {De Looze}, {di Serego Alighieri}, {Duc}, {Durrell},
  {Emsellem}, {Erben}, {Fritz}, {Garcia-Appadoo}, {Gavazzi}, {Grossi},
  {Jord{\'a}n}, {Hess}, {Huertas-Company}, {Hunt}, {Kent}, {Lambas},
  {Lan{\c{c}}on}, {MacArthur}, {Madden}, {Magrini}, {Mei}, {Momjian}, {Olowin},
  {Papastergis}, {Smith}, {Solanes}, {Spector}, {Spekkens}, {Taylor},
  {Valotto}, {van Driel}, {Verstappen}, {Vlahakis}, {Vollmer}, \&
  {Xilouris}}]{Boselli2011}
{Boselli}, A., {Boissier}, S., {Heinis}, S., {et~al.} 2011, \aap, 528, A107,
  \dodoi{10.1051/0004-6361/201016389}

\bibitem[{{Boselli} {et~al.}(2018){Boselli}, {Fossati}, {Ferrarese},
  {Boissier}, {Consolandi}, {Longobardi}, {Amram}, {Balogh}, {Barmby},
  {Boquien}, {Boulanger}, {Braine}, {Buat}, {Burgarella}, {Combes}, {Contini},
  {Cortese}, {C{\^o}t{\'e}}, {C{\^o}t{\'e}}, {Cuillandre}, {Drissen}, {Epinat},
  {Fumagalli}, {Gallagher}, {Gavazzi}, {Gomez-Lopez}, {Gwyn}, {Harris},
  {Hensler}, {Koribalski}, {Marcelin}, {McConnachie}, {Miville-Deschenes},
  {Navarro}, {Patton}, {Peng}, {Plana}, {Prantzos}, {Robert}, {Roediger},
  {Roehlly}, {Russeil}, {Salome}, {Sanchez-Janssen}, {Serra}, {Spekkens},
  {Sun}, {Taylor}, {Tonnesen}, {Vollmer}, {Willis}, {Wozniak}, {Burdullis},
  {Devost}, {Mahoney}, {Manset}, {Petric}, {Prunet}, \&
  {Withington}}]{Boselli2018}
{Boselli}, A., {Fossati}, M., {Ferrarese}, L., {et~al.} 2018, \aap, 614, A56,
  \dodoi{10.1051/0004-6361/201732407}

\bibitem[{Bradley {et~al.}(2020)Bradley, Sip{\H o}cz, Robitaille, Tollerud,
  Vin{\'{\i}}cius, Deil, Barbary, Wilson, Busko, G{\"u}nther, Cara, Conseil,
  Bostroem, Droettboom, Bray, Bratholm, Lim, Barentsen, Craig, Pascual, Perren,
  Greco, Donath, de~Val-Borro, Kerzendorf, Bach, Weaver, D'Eugenio, Souchereau,
  \& Ferreira}]{larry_bradley_2020_4044744}
Bradley, L., Sip{\H o}cz, B., Robitaille, T., {et~al.} 2020, astropy/photutils:
  1.0.0, 1.0.0,  Zenodo, \dodoi{10.5281/zenodo.4044744}

\bibitem[{{Briggs}(1995)}]{Briggs1995}
{Briggs}, D.~S. 1995, in Bulletin of the American Astronomical Society,
  Vol.~27, American Astronomical Society Meeting Abstracts, 1444

\bibitem[{{Brown} {et~al.}(2017){Brown}, {Catinella}, {Cortese}, {Lagos},
  {Dav{\'e}}, {Kilborn}, {Haynes}, {Giovanelli}, \&
  {Rafieferantsoa}}]{Brown2017}
{Brown}, T., {Catinella}, B., {Cortese}, L., {et~al.} 2017, \mnras, 466, 1275,
  \dodoi{10.1093/mnras/stw2991}

\bibitem[{{Cappellari} {et~al.}(2013){Cappellari}, {Scott}, {Alatalo}, {Blitz},
  {Bois}, {Bournaud}, {Bureau}, {Crocker}, {Davies}, {Davis}, {de Zeeuw},
  {Duc}, {Emsellem}, {Khochfar}, {Krajnovi{\'c}}, {Kuntschner}, {McDermid},
  {Morganti}, {Naab}, {Oosterloo}, {Sarzi}, {Serra}, {Weijmans}, \&
  {Young}}]{Cappellari2013}
{Cappellari}, M., {Scott}, N., {Alatalo}, K., {et~al.} 2013, \mnras, 432, 1709,
  \dodoi{10.1093/mnras/stt562}

\bibitem[{{Catinella} {et~al.}(2018){Catinella}, {Saintonge}, {Janowiecki},
  {Cortese}, {Dav{\'e}}, {Lemonias}, {Cooper}, {Schiminovich}, {Hummels},
  {Fabello}, {Ger{\'e}b}, {Kilborn}, \& {Wang}}]{Catinella2018}
{Catinella}, B., {Saintonge}, A., {Janowiecki}, S., {et~al.} 2018, \mnras, 476,
  875, \dodoi{10.1093/mnras/sty089}

\bibitem[{{Chabrier}(2003)}]{Chabrier2003}
{Chabrier}, G. 2003, \pasp, 115, 763, \dodoi{10.1086/376392}

\bibitem[{{Chamaraux} {et~al.}(1980){Chamaraux}, {Balkowski}, \&
  {Gerard}}]{Chamaraux1980}
{Chamaraux}, P., {Balkowski}, C., \& {Gerard}, E. 1980, \aap, 83, 38

\bibitem[{{Chown} {et~al.}(2019){Chown}, {Li}, {Athanassoula}, {Li}, {Wilson},
  {Lin}, {Mo}, {Parker}, \& {Xiao}}]{Chown2019}
{Chown}, R., {Li}, C., {Athanassoula}, E., {et~al.} 2019, \mnras, 484, 5192,
  \dodoi{10.1093/mnras/stz349}

\bibitem[{{Chung} {et~al.}(2009{\natexlab{a}}){Chung}, {van Gorkom}, {Kenney},
  {Crowl}, \& {Vollmer}}]{Chung2009b}
{Chung}, A., {van Gorkom}, J.~H., {Kenney}, J. D.~P., {Crowl}, H., \&
  {Vollmer}, B. 2009{\natexlab{a}}, \aj, 138, 1741,
  \dodoi{10.1088/0004-6256/138/6/1741}

\bibitem[{{Chung} {et~al.}(2009{\natexlab{b}}){Chung}, {Rhee}, {Kim}, {Yun},
  {Heyer}, \& {Young}}]{Chung2009a}
{Chung}, E.~J., {Rhee}, M.-H., {Kim}, H., {et~al.} 2009{\natexlab{b}}, \apjs,
  184, 199, \dodoi{10.1088/0067-0049/184/2/199}

\bibitem[{{Chung} {et~al.}(2017){Chung}, {Yun}, {Verheijen}, \&
  {Chung}}]{Chung2017}
{Chung}, E.~J., {Yun}, M.~S., {Verheijen}, M. A.~W., \& {Chung}, A. 2017, \apj,
  843, 50, \dodoi{10.3847/1538-4357/aa756b}

\bibitem[{{Ciesla} {et~al.}(2012){Ciesla}, {Boselli}, {Smith}, {Bendo},
  {Cortese}, {Eales}, {Bianchi}, {Boquien}, {Buat}, {Davies}, {Pohlen},
  {Zibetti}, {Baes}, {Cooray}, {De Looze}, {di Serego Alighieri}, {Galametz},
  {Gomez}, {Lebouteiller}, {Madden}, {Pappalardo}, {Remy}, {Spinoglio},
  {Vaccari}, {Auld}, \& {Clements}}]{Ciesla2012}
{Ciesla}, L., {Boselli}, A., {Smith}, M.~W.~L., {et~al.} 2012, \aap, 543, A161,
  \dodoi{10.1051/0004-6361/201219216}

\bibitem[{{Clark} {et~al.}(2018){Clark}, {Verstocken}, {Bianchi}, {Fritz},
  {Viaene}, {Smith}, {Baes}, {Casasola}, {Cassara}, {Davies}, {De Looze}, {De
  Vis}, {Evans}, {Galametz}, {Jones}, {Lianou}, {Madden}, {Mosenkov}, \&
  {Xilouris}}]{Clark2018}
{Clark}, C.~J.~R., {Verstocken}, S., {Bianchi}, S., {et~al.} 2018, \aap, 609,
  A37, \dodoi{10.1051/0004-6361/201731419}

\bibitem[{Comrie {et~al.}(2020)Comrie, Wang, Ford, Moraghan, Hsu, Pińska,
  Chiang, Jan, Simmonds, Chang, \& Lin}]{angus_comrie_2020_3746095}
Comrie, A., Wang, K.-S., Ford, P., {et~al.} 2020, {CARTA: The Cube Analysis and
  Rendering Tool for Astronomy}, 1.3.0,  Zenodo, \dodoi{10.5281/zenodo.3746095}

\bibitem[{{Corbelli} {et~al.}(2012){Corbelli}, {Bianchi}, {Cortese},
  {Giovanardi}, {Magrini}, {Pappalardo}, {Boselli}, {Bendo}, {Davies},
  {Grossi}, {Madden}, {Smith}, {Vlahakis}, {Auld}, {Baes}, {De Looze}, {Fritz},
  {Pohlen}, \& {Verstappen}}]{Corbelli2012}
{Corbelli}, E., {Bianchi}, S., {Cortese}, L., {et~al.} 2012, \aap, 542, A32,
  \dodoi{10.1051/0004-6361/201117329}

\bibitem[{{Cortese} {et~al.}(2011){Cortese}, {Catinella}, {Boissier},
  {Boselli}, \& {Heinis}}]{Cortese2011}
{Cortese}, L., {Catinella}, B., {Boissier}, S., {Boselli}, A., \& {Heinis}, S.
  2011, \mnras, 415, 1797, \dodoi{10.1111/j.1365-2966.2011.18822.x}

\bibitem[{{Cortese} {et~al.}(2021){Cortese}, {Catinella}, \&
  {Smith}}]{Cortese2021}
{Cortese}, L., {Catinella}, B., \& {Smith}, R. 2021, arXiv e-prints,
  arXiv:2104.02193.
\newblock \doarXiv{2104.02193}

\bibitem[{{Cortese} {et~al.}(2012){Cortese}, {Boissier}, {Boselli}, {Bendo},
  {Buat}, {Davies}, {Eales}, {Heinis}, {Isaak}, \& {Madden}}]{Cortese2012}
{Cortese}, L., {Boissier}, S., {Boselli}, A., {et~al.} 2012, \aap, 544, A101,
  \dodoi{10.1051/0004-6361/201219312}

\bibitem[{{Cowie} \& {Songaila}(1977)}]{Cowie1977}
{Cowie}, L.~L., \& {Songaila}, A. 1977, \nat, 266, 501,
  \dodoi{10.1038/266501a0}

\bibitem[{{Cramer} {et~al.}(2020){Cramer}, {Kenney}, {Cortes}, {Cortes P.~C.},
  {Vlahakis}, {J{\'a}chym}, {Pompei}, \& {Rubio}}]{Cramer2020}
{Cramer}, W.~J., {Kenney}, J.~D.~P., {Cortes}, J.~R., {et~al.} 2020, \apj, 901,
  95, \dodoi{10.3847/1538-4357/abaf54}

\bibitem[{{Cramer} {et~al.}(2019){Cramer}, {Kenney}, {Sun}, {Crowl}, {Yagi},
  {J{\'a}chym}, {Roediger}, \& {Waldron}}]{Cramer2019}
{Cramer}, W.~J., {Kenney}, J.~D.~P., {Sun}, M., {et~al.} 2019, \apj, 870, 63,
  \dodoi{10.3847/1538-4357/aaefff}

\bibitem[{{Davies} {et~al.}(2010){Davies}, {Baes}, {Bendo}, {Bianchi},
  {Bomans}, {Boselli}, {Clemens}, {Corbelli}, {Cortese}, {Dariush}, {De Looze},
  {di Serego Alighieri}, {Fadda}, {Fritz}, {Garcia-Appadoo}, {Gavazzi},
  {Giovanardi}, {Grossi}, {Hughes}, {Hunt}, {Jones}, {Madden}, {Pierini},
  {Pohlen}, {Sabatini}, {Smith}, {Verstappen}, {Vlahakis}, {Xilouris}, \&
  {Zibetti}}]{Davies2010}
{Davies}, J.~I., {Baes}, M., {Bendo}, G.~J., {et~al.} 2010, \aap, 518, L48,
  \dodoi{10.1051/0004-6361/201014571}

\bibitem[{{Davies} \& {Lewis}(1973)}]{Davies1973}
{Davies}, R.~D., \& {Lewis}, B.~M. 1973, \mnras, 165, 231,
  \dodoi{10.1093/mnras/165.2.231}

\bibitem[{{Davis} {et~al.}(2013){Davis}, {Alatalo}, {Bureau}, {Cappellari},
  {Scott}, {Young}, {Blitz}, {Crocker}, {Bayet}, {Bois}, {Bournaud}, {Davies},
  {de Zeeuw}, {Duc}, {Emsellem}, {Khochfar}, {Krajnovi{\'c}}, {Kuntschner},
  {Lablanche}, {McDermid}, {Morganti}, {Naab}, {Oosterloo}, {Sarzi}, {Serra},
  \& {Weijmans}}]{Davis2013}
{Davis}, T.~A., {Alatalo}, K., {Bureau}, M., {et~al.} 2013, \mnras, 429, 534,
  \dodoi{10.1093/mnras/sts353}

\bibitem[{{den Brok} {et~al.}(2021){den Brok}, {Chatzigiannakis}, {Bigiel},
  {Puschnig}, {Barnes}, {Leroy}, {Jim{\'e}nez-Donaire}, {Usero}, {Schinnerer},
  {Rosolowsky}, {Faesi}, {Grasha}, {Hughes}, {Kruijssen}, {Liu}, {Neumann},
  {Pety}, {Querejeta}, {Saito}, {Schruba}, \& {Stuber}}]{denBrok2021}
{den Brok}, J.~S., {Chatzigiannakis}, D., {Bigiel}, F., {et~al.} 2021, \mnras,
  504, 3221, \dodoi{10.1093/mnras/stab859}

\bibitem[{{Ebeling} {et~al.}(2014){Ebeling}, {Stephenson}, \&
  {Edge}}]{Ebeling2014}
{Ebeling}, H., {Stephenson}, L.~N., \& {Edge}, A.~C. 2014, \apjl, 781, L40,
  \dodoi{10.1088/2041-8205/781/2/L40}

\bibitem[{{Ebeling} {et~al.}(2006){Ebeling}, {White}, \&
  {Rangarajan}}]{Ebeling2006}
{Ebeling}, H., {White}, D.~A., \& {Rangarajan}, F.~V.~N. 2006, \mnras, 368, 65,
  \dodoi{10.1111/j.1365-2966.2006.10135.x}

\bibitem[{{Elbaz} {et~al.}(2007){Elbaz}, {Daddi}, {Le Borgne}, {Dickinson},
  {Alexander}, {Chary}, {Starck}, {Brandt}, {Kitzbichler}, {MacDonald},
  {Nonino}, {Popesso}, {Stern}, \& {Vanzella}}]{Elbaz2007}
{Elbaz}, D., {Daddi}, E., {Le Borgne}, D., {et~al.} 2007, \aap, 468, 33,
  \dodoi{10.1051/0004-6361:20077525}

\bibitem[{{Ferrarese} {et~al.}(2012){Ferrarese}, {C{\^o}t{\'e}}, {Cuillandre},
  {Gwyn}, {Peng}, {MacArthur}, {Duc}, {Boselli}, {Mei}, {Erben}, {McConnachie},
  {Durrell}, {Mihos}, {Jord{\'a}n}, {Lan{\c{c}}on}, {Puzia}, {Emsellem},
  {Balogh}, {Blakeslee}, {van Waerbeke}, {Gavazzi}, {Vollmer}, {Kavelaars},
  {Woods}, {Ball}, {Boissier}, {Courteau}, {Ferriere}, {Gavazzi},
  {Hildebrandt}, {Hudelot}, {Huertas-Company}, {Liu}, {McLaughlin}, {Mellier},
  {Milkeraitis}, {Schade}, {Balkowski}, {Bournaud}, {Carlberg}, {Chapman},
  {Hoekstra}, {Peng}, {Sawicki}, {Simard}, {Taylor}, {Tully}, {van Driel},
  {Wilson}, {Burdullis}, {Mahoney}, \& {Manset}}]{Ferrarese2012}
{Ferrarese}, L., {C{\^o}t{\'e}}, P., {Cuillandre}, J.-C., {et~al.} 2012, \apjs,
  200, 4, \dodoi{10.1088/0067-0049/200/1/4}

\bibitem[{{Fruscione} {et~al.}(2006){Fruscione}, {McDowell}, {Allen},
  {Brickhouse}, {Burke}, {Davis}, {Durham}, {Elvis}, {Galle}, {Harris},
  {Huenemoerder}, {Houck}, {Ishibashi}, {Karovska}, {Nicastro}, {Noble},
  {Nowak}, {Primini}, {Siemiginowska}, {Smith}, \& {Wise}}]{Fruscione2006}
{Fruscione}, A., {McDowell}, J.~C., {Allen}, G.~E., {et~al.} 2006, in Society
  of Photo-Optical Instrumentation Engineers (SPIE) Conference Series, Vol.
  6270, Society of Photo-Optical Instrumentation Engineers (SPIE) Conference
  Series, ed. D.~R. {Silva} \& R.~E. {Doxsey}, 62701V,
  \dodoi{10.1117/12.671760}

\bibitem[{{Fujita}(1998)}]{Fujita1998}
{Fujita}, Y. 1998, \apj, 509, 587, \dodoi{10.1086/306518}

\bibitem[{{Fumagalli} {et~al.}(2014){Fumagalli}, {Fossati}, {Hau}, {Gavazzi},
  {Bower}, {Sun}, \& {Boselli}}]{Fumagalli2014}
{Fumagalli}, M., {Fossati}, M., {Hau}, G. K.~T., {et~al.} 2014, \mnras, 445,
  4335, \dodoi{10.1093/mnras/stu2092}

\bibitem[{{Fumagalli} {et~al.}(2011){Fumagalli}, {Gavazzi}, {Scaramella}, \&
  {Franzetti}}]{Fumagalli2011}
{Fumagalli}, M., {Gavazzi}, G., {Scaramella}, R., \& {Franzetti}, P. 2011,
  \aap, 528, A46, \dodoi{10.1051/0004-6361/201015463}

\bibitem[{{Gaudet} {et~al.}(2010){Gaudet}, {Hill}, {Armstrong}, {Ball},
  {Burke}, {Chapel}, {Chapin}, {Damian}, {Dowler}, {Gable}, {Goliath},
  {Ghiurea}, {Fabbro}, {Gwyn}, {Jenkins}, {Kavelaars}, {Major}, {Ouellette},
  {Paterson}, {Peddle}, {Penfold-Brown}, {Pritchet}, {Schade}, {Sobie},
  {Woods}, {Yeung}, \& {Zhang}}]{Gaudet2010}
{Gaudet}, S., {Hill}, N., {Armstrong}, P., {et~al.} 2010, in Society of
  Photo-Optical Instrumentation Engineers (SPIE) Conference Series, Vol. 7740,
  Software and Cyberinfrastructure for Astronomy, ed. N.~M. {Radziwill} \&
  A.~{Bridger}, 77401I, \dodoi{10.1117/12.858026}

\bibitem[{{Gavazzi} {et~al.}(1999){Gavazzi}, {Boselli}, {Scodeggio}, {Pierini},
  \& {Belsole}}]{Gavazzi1999b}
{Gavazzi}, G., {Boselli}, A., {Scodeggio}, M., {Pierini}, D., \& {Belsole}, E.
  1999, \mnras, 304, 595, \dodoi{10.1046/j.1365-8711.1999.02350.x}

\bibitem[{{Gavazzi} {et~al.}(2008){Gavazzi}, {Giovanelli}, {Haynes}, {Fabello},
  {Fumagalli}, {Kent}, {Koopmann}, {Brosch}, {Hoffman}, {Salzer}, \&
  {Boselli}}]{Gavazzi2008}
{Gavazzi}, G., {Giovanelli}, R., {Haynes}, M.~P., {et~al.} 2008, \aap, 482, 43,
  \dodoi{10.1051/0004-6361:200809382}

\bibitem[{{Giovanelli} \& {Haynes}(1985)}]{Giovanelli1985}
{Giovanelli}, R., \& {Haynes}, M.~P. 1985, \apj, 292, 404,
  \dodoi{10.1086/163170}

\bibitem[{{Girardi} {et~al.}(1998){Girardi}, {Giuricin}, {Mardirossian},
  {Mezzetti}, \& {Boschin}}]{Girardi1998}
{Girardi}, M., {Giuricin}, G., {Mardirossian}, F., {Mezzetti}, M., \&
  {Boschin}, W. 1998, \apj, 505, 74, \dodoi{10.1086/306157}

\bibitem[{{G{\'o}mez} {et~al.}(2003){G{\'o}mez}, {Nichol}, {Miller}, {Balogh},
  {Goto}, {Zabludoff}, {Romer}, {Bernardi}, {Sheth}, {Hopkins}, {Castander},
  {Connolly}, {Schneider}, {Brinkmann}, {Lamb}, {SubbaRao}, \&
  {York}}]{Gomez2003}
{G{\'o}mez}, P.~L., {Nichol}, R.~C., {Miller}, C.~J., {et~al.} 2003, \apj, 584,
  210, \dodoi{10.1086/345593}

\bibitem[{{Gunn} \& {Gott}(1972)}]{Gunn1972}
{Gunn}, J.~E., \& {Gott}, J.~Richard, I. 1972, \apj, 176, 1,
  \dodoi{10.1086/151605}

\bibitem[{{Haynes} {et~al.}(1984){Haynes}, {Giovanelli}, \&
  {Chincarini}}]{Haynes1984}
{Haynes}, M.~P., {Giovanelli}, R., \& {Chincarini}, G.~L. 1984, \araa, 22, 445,
  \dodoi{10.1146/annurev.aa.22.090184.002305}

\bibitem[{{Healy} {et~al.}(2020){Healy}, {Blyth}, {Verheijen}, {Hess}, {Serra},
  {van der Hulst}, {Jarrett}, {Yim}, \& {Jozsa}}]{Healy2020}
{Healy}, J., {Blyth}, S.-L., {Verheijen}, M.~A.~W., {et~al.} 2020, arXiv
  e-prints, arXiv:2011.06285.
\newblock \doarXiv{2011.06285}

\bibitem[{{Helfer} {et~al.}(2003){Helfer}, {Thornley}, {Regan}, {Wong},
  {Sheth}, {Vogel}, {Blitz}, \& {Bock}}]{Helfer2003}
{Helfer}, T.~T., {Thornley}, M.~D., {Regan}, M.~W., {et~al.} 2003, \apjs, 145,
  259, \dodoi{10.1086/346076}

\bibitem[{{Herrera} {et~al.}(2020){Herrera}, {Pety}, {Hughes}, {Meidt},
  {Kreckel}, {Querejeta}, {Saito}, {Lang}, {Jim{\'e}nez-Donaire}, {Pessa},
  {Cormier}, {Usero}, {Sliwa}, {Faesi}, {Blanc}, {Bigiel}, {Chevance}, {Dale},
  {Grasha}, {Glover}, {Hygate}, {Kruijssen}, {Leroy}, {Rosolowsky},
  {Schinnerer}, {Schruba}, {Sun}, \& {Utomo}}]{Herrera2020}
{Herrera}, C.~N., {Pety}, J., {Hughes}, A., {et~al.} 2020, \aap, 634, A121,
  \dodoi{10.1051/0004-6361/201936060}

\bibitem[{{Hester}(2006)}]{Hester2006}
{Hester}, J.~A. 2006, \apj, 647, 910, \dodoi{10.1086/505614}

\bibitem[{{Heyer} \& {Dame}(2015)}]{Heyer2015}
{Heyer}, M., \& {Dame}, T.~M. 2015, \araa, 53, 583,
  \dodoi{10.1146/annurev-astro-082214-122324}

\bibitem[{{Hughes} \& {Cortese}(2009)}]{Hughes2009}
{Hughes}, T.~M., \& {Cortese}, L. 2009, \mnras, 396, L41,
  \dodoi{10.1111/j.1745-3933.2009.00658.x}

\bibitem[{Hunter(2007)}]{Hunter2007}
Hunter, J.~D. 2007, Computing in Science \& Engineering, 9, 90,
  \dodoi{10.1109/MCSE.2007.55}

\bibitem[{{Iono} {et~al.}(2005){Iono}, {Yun}, \& {Ho}}]{Iono2005}
{Iono}, D., {Yun}, M.~S., \& {Ho}, P. T.~P. 2005, \apjs, 158, 1,
  \dodoi{10.1086/429093}

\bibitem[{{J{\'a}chym} {et~al.}(2013){J{\'a}chym}, {Kenney},
  {R{\v{z}}ui{\v{c}}ka}, {Sun}, {Combes}, \& {Palou{\v{s}}}}]{Jachym2013}
{J{\'a}chym}, P., {Kenney}, J.~D.~P., {R{\v{z}}ui{\v{c}}ka}, A., {et~al.} 2013,
  \aap, 556, A99, \dodoi{10.1051/0004-6361/201220495}

\bibitem[{{J{\'a}chym} {et~al.}(2019){J{\'a}chym}, {Kenney}, {Sun}, {Combes},
  {Cortese}, {Scott}, {Sivanandam}, {Brinks}, {Roediger}, {Palou{\v{s}}}, \&
  {Fumagalli}}]{Jachym2019}
{J{\'a}chym}, P., {Kenney}, J. D.~P., {Sun}, M., {et~al.} 2019, \apj, 883, 145,
  \dodoi{10.3847/1538-4357/ab3e6c}

\bibitem[{{Jaff{\'e}} {et~al.}(2016){Jaff{\'e}}, {Verheijen}, {Haines}, {Yoon},
  {Cybulski}, {Montero-Casta{\~n}o}, {Smith}, {Chung}, {Deshev},
  {Fern{\'a}ndez}, {van Gorkom}, {Poggianti}, {Yun}, {Finoguenov}, {Smith}, \&
  {Okabe}}]{Jaffe2016}
{Jaff{\'e}}, Y.~L., {Verheijen}, M. A.~W., {Haines}, C.~P., {et~al.} 2016,
  \mnras, 461, 1202, \dodoi{10.1093/mnras/stw984}

\bibitem[{{Kaiser}(1986)}]{Kaiser1986}
{Kaiser}, N. 1986, \mnras, 222, 323, \dodoi{10.1093/mnras/222.2.323}

\bibitem[{{Kaneko} {et~al.}(2018){Kaneko}, {Kuno}, \& {Saitoh}}]{Kaneko2018}
{Kaneko}, H., {Kuno}, N., \& {Saitoh}, T.~R. 2018, \apjl, 860, L14,
  \dodoi{10.3847/2041-8213/aac895}

\bibitem[{{Karachentsev} {et~al.}(2014){Karachentsev}, {Tully}, {Wu}, {Shaya},
  \& {Dolphin}}]{Karachentsev2014}
{Karachentsev}, I.~D., {Tully}, R.~B., {Wu}, P.-F., {Shaya}, E.~J., \&
  {Dolphin}, A.~E. 2014, \apj, 782, 4, \dodoi{10.1088/0004-637X/782/1/4}

\bibitem[{{Kenney} \& {Young}(1986)}]{Kenney1986}
{Kenney}, J.~D., \& {Young}, J.~S. 1986, \apjl, 301, L13,
  \dodoi{10.1086/184614}

\bibitem[{{Kenney} {et~al.}(1992){Kenney}, {Wilson}, {Scoville}, {Devereux}, \&
  {Young}}]{Kenney1992}
{Kenney}, J. D.~P., {Wilson}, C.~D., {Scoville}, N.~Z., {Devereux}, N.~A., \&
  {Young}, J.~S. 1992, \apjl, 395, L79, \dodoi{10.1086/186492}

\bibitem[{{Kenney} \& {Young}(1989)}]{Kenney1989}
{Kenney}, J. D.~P., \& {Young}, J.~S. 1989, \apj, 344, 171,
  \dodoi{10.1086/167787}

\bibitem[{{Kennicutt} \& {Evans}(2012)}]{Kennicutt2012}
{Kennicutt}, R.~C., \& {Evans}, N.~J. 2012, \araa, 50, 531,
  \dodoi{10.1146/annurev-astro-081811-125610}

\bibitem[{{Kim} {et~al.}(2014){Kim}, {Rey}, {Jerjen}, {Lisker}, {Sung}, {Lee},
  {Chung}, {Pak}, {Yi}, \& {Lee}}]{Kim2014}
{Kim}, S., {Rey}, S.-C., {Jerjen}, H., {et~al.} 2014, \apjs, 215, 22,
  \dodoi{10.1088/0067-0049/215/2/22}

\bibitem[{{Koopmann} \& {Kenney}(2004)}]{Koopmann2004}
{Koopmann}, R.~A., \& {Kenney}, J. D.~P. 2004, \apj, 613, 866,
  \dodoi{10.1086/423191}

\bibitem[{{Koopmann} {et~al.}(2001){Koopmann}, {Kenney}, \&
  {Young}}]{Koopmann2001}
{Koopmann}, R.~A., {Kenney}, J. D.~P., \& {Young}, J. 2001, \apjs, 135, 125,
  \dodoi{10.1086/323532}

\bibitem[{{Koyama} {et~al.}(2017){Koyama}, {Koyama}, {Yamashita},
  {Morokuma-Matsui}, {Matsuhara}, {Nakagawa}, {Hayashi}, {Kodama}, {Shimakawa},
  {Suzuki}, {Tadaki}, {Tanaka}, \& {Yamamoto}}]{Koyama2017}
{Koyama}, S., {Koyama}, Y., {Yamashita}, T., {et~al.} 2017, \apj, 847, 137,
  \dodoi{10.3847/1538-4357/aa8a6c}

\bibitem[{{Kroupa}(2001)}]{Kroupa2001}
{Kroupa}, P. 2001, \mnras, 322, 231, \dodoi{10.1046/j.1365-8711.2001.04022.x}

\bibitem[{{Krumholz} {et~al.}(2009){Krumholz}, {McKee}, \&
  {Tumlinson}}]{Krumholz2009}
{Krumholz}, M.~R., {McKee}, C.~F., \& {Tumlinson}, J. 2009, \apj, 693, 216,
  \dodoi{10.1088/0004-637X/693/1/216}

\bibitem[{{Kutner} \& {Ulich}(1981)}]{Kutner1981}
{Kutner}, M.~L., \& {Ulich}, B.~L. 1981, \apj, 250, 341, \dodoi{10.1086/159380}

\bibitem[{{Larson} {et~al.}(1980){Larson}, {Tinsley}, \&
  {Caldwell}}]{Larson1980}
{Larson}, R.~B., {Tinsley}, B.~M., \& {Caldwell}, C.~N. 1980, \apj, 237, 692,
  \dodoi{10.1086/157917}

\bibitem[{{Lee} \& {Chung}(2018)}]{Lee2018}
{Lee}, B., \& {Chung}, A. 2018, \apjl, 866, L10,
  \dodoi{10.3847/2041-8213/aae4d9}

\bibitem[{{Lee} {et~al.}(2017){Lee}, {Chung}, {Tonnesen}, {Kenney}, {Wong},
  {Vollmer}, {Petitpas}, {Crowl}, \& {van Gorkom}}]{Lee2017}
{Lee}, B., {Chung}, A., {Tonnesen}, S., {et~al.} 2017, \mnras, 466, 1382,
  \dodoi{10.1093/mnras/stw3162}

\bibitem[{{Leroy} {et~al.}(2008){Leroy}, {Walter}, {Brinks}, {Bigiel}, {de
  Blok}, {Madore}, \& {Thornley}}]{Leroy2008}
{Leroy}, A.~K., {Walter}, F., {Brinks}, E., {et~al.} 2008, \aj, 136, 2782,
  \dodoi{10.1088/0004-6256/136/6/2782}

\bibitem[{{Leroy} {et~al.}(2009){Leroy}, {Walter}, {Bigiel}, {Usero}, {Weiss},
  {Brinks}, {de Blok}, {Kennicutt}, {Schuster}, {Kramer}, {Wiesemeyer}, \&
  {Roussel}}]{Leroy2009}
{Leroy}, A.~K., {Walter}, F., {Bigiel}, F., {et~al.} 2009, \aj, 137, 4670,
  \dodoi{10.1088/0004-6256/137/6/4670}

\bibitem[{{Leroy} {et~al.}(2011){Leroy}, {Bolatto}, {Gordon}, {Sandstrom},
  {Gratier}, {Rosolowsky}, {Engelbracht}, {Mizuno}, {Corbelli}, {Fukui}, \&
  {Kawamura}}]{leroy2011}
{Leroy}, A.~K., {Bolatto}, A., {Gordon}, K., {et~al.} 2011, \apj, 737, 12,
  \dodoi{10.1088/0004-637X/737/1/12}

\bibitem[{{Leroy} {et~al.}(2013){Leroy}, {Walter}, {Sandstrom}, {Schruba},
  {Munoz-Mateos}, {Bigiel}, {Bolatto}, {Brinks}, {de Blok}, {Meidt}, {Rix},
  {Rosolowsky}, {Schinnerer}, {Schuster}, \& {Usero}}]{Leroy2013}
{Leroy}, A.~K., {Walter}, F., {Sandstrom}, K., {et~al.} 2013, \aj, 146, 19,
  \dodoi{10.1088/0004-6256/146/2/19}

\bibitem[{{Leroy} {et~al.}(2019){Leroy}, {Sandstrom}, {Lang}, {Lewis}, {Salim},
  {Behrens}, {Chastenet}, {Chiang}, {Gallagher}, {Kessler}, \&
  {Utomo}}]{Leroy2019}
{Leroy}, A.~K., {Sandstrom}, K.~M., {Lang}, D., {et~al.} 2019, \apjs, 244, 24,
  \dodoi{10.3847/1538-4365/ab3925}

\bibitem[{{Leroy} {et~al.}(2021{\natexlab{a}}){Leroy}, {Schinnerer}, {Hughes},
  {Rosolowsky}, {Pety}, {Schruba}, {Usero}, {Blanc}, {Chevance}, {Emsellem},
  {Faesi}, {Herrera}, {Liu}, {Meidt}, {Querejeta}, {Saito}, {Sandstrom}, {Sun},
  {Williams}, {Anand}, {Barnes}, {Behrens}, {Belfiore}, {Benincasa},
  {Be{\v{s}}li{\'c}}, {Bigiel}, {Bolatto}, {den Brok}, {Cao}, {Chandar},
  {Chastenet}, {Chiang}, {Congiu}, {Dale}, {Deger}, {Eibensteiner}, {Egorov},
  {Garc{\'\i}a-Rodr{\'\i}guez}, {Glover}, {Grasha}, {Henshaw}, {Ho}, {Kepley},
  {Kim}, {Klessen}, {Kreckel}, {Koch}, {Kruijssen}, {Larson}, {Lee}, {Lopez},
  {Machado}, {Mayker}, {McElroy}, {Murphy}, {Ostriker}, {Pan}, {Pessa},
  {Puschnig}, {Razza}, {S{\'a}nchez-Bl{\'a}zquez}, {Santoro}, {Sardone},
  {Scheuermann}, {Sliwa}, {Sormani}, {Stuber}, {Thilker}, {Turner}, {Utomo},
  {Watkins}, \& {Whitmore}}]{Leroy2021a}
{Leroy}, A.~K., {Schinnerer}, E., {Hughes}, A., {et~al.} 2021{\natexlab{a}},
  arXiv e-prints, arXiv:2104.07739.
\newblock \doarXiv{2104.07739}

\bibitem[{{Leroy} {et~al.}(2021{\natexlab{b}}){Leroy}, {Hughes}, {Liu}, {Pety},
  {Rosolowsky}, {Saito}, {Schinnerer}, {Schruba}, {Usero}, {Faesi}, {Herrera},
  {Chevance}, {Hygate}, {Kepley}, {Koch}, {Querejeta}, {Sliwa}, {Will},
  {Wilson}, {Anand}, {Barnes}, {Belfiore}, {Beslic}, {Bigiel}, {Blanc},
  {Bolatto}, {Boquien}, {Cao}, {Chandar}, {Chastenet}, {Chiang}, {Congiu},
  {Dale}, {Deger}, {den Brok}, {Eibensteiner}, {Emsellem},
  {Garc{\i}a-Rodr{\i}guez}, {Glover}, {Grasha}, {Groves}, {Henshaw}, {Jimenez
  Donaire}, {Kim}, {Klessen}, {Kreckel}, {Kruijssen}, {Larson}, {Lee},
  {Mayker}, {McElroy}, {Meidt}, {Mok}, {Pan}, {Puschnig}, {Razza},
  {Sanchez-Blazquez}, {Sandstrom}, {Santoro}, {Sardone}, {Scheuermann}, {Sun},
  {Thilker}, {Turner}, {Ubeda}, {Utomo}, {Watkins}, \& {Williams}}]{Leroy2021b}
{Leroy}, A.~K., {Hughes}, A., {Liu}, D., {et~al.} 2021{\natexlab{b}}, arXiv
  e-prints, arXiv:2104.07665.
\newblock \doarXiv{2104.07665}

\bibitem[{{Lin} {et~al.}(2017){Lin}, {Li}, {He}, {Xiao}, \& {Wang}}]{Lin2017}
{Lin}, L., {Li}, C., {He}, Y., {Xiao}, T., \& {Wang}, E. 2017, \apj, 838, 105,
  \dodoi{10.3847/1538-4357/aa657a}

\bibitem[{{Lin} {et~al.}(2020){Lin}, {Li}, {Du}, {Wang}, {Xiao}, {Bureau},
  {Fraser-McKelvie}, {Masters}, {Lin}, {Wake}, \& {Hao}}]{Lin2020}
{Lin}, L., {Li}, C., {Du}, C., {et~al.} 2020, \mnras, 499, 1406,
  \dodoi{10.1093/mnras/staa2913}

\bibitem[{{Livio} {et~al.}(1980){Livio}, {Regev}, \& {Shaviv}}]{Livio1980}
{Livio}, M., {Regev}, O., \& {Shaviv}, G. 1980, \apjl, 240, L83,
  \dodoi{10.1086/183328}

\bibitem[{{Liz{\'e}e} {et~al.}(2021){Liz{\'e}e}, {Vollmer}, {Braine}, \&
  {Nehlig}}]{Lizee2021}
{Liz{\'e}e}, T., {Vollmer}, B., {Braine}, J., \& {Nehlig}, F. 2021, \aap, 645,
  A111, \dodoi{10.1051/0004-6361/202038910}

\bibitem[{{Martin} {et~al.}(2005){Martin}, {Fanson}, {Schiminovich},
  {Morrissey}, {Friedman}, {Barlow}, {Conrow}, {Grange}, {Jelinsky},
  {Milliard}, {Siegmund}, {Bianchi}, {Byun}, {Donas}, {Forster}, {Heckman},
  {Lee}, {Madore}, {Malina}, {Neff}, {Rich}, {Small}, {Surber}, {Szalay},
  {Welsh}, \& {Wyder}}]{Martin2005}
{Martin}, D.~C., {Fanson}, J., {Schiminovich}, D., {et~al.} 2005, \apjl, 619,
  L1, \dodoi{10.1086/426387}

\bibitem[{{McLaughlin}(1999)}]{McLaughlin1999}
{McLaughlin}, D.~E. 1999, \apjl, 512, L9, \dodoi{10.1086/311860}

\bibitem[{{McMullin} {et~al.}(2007){McMullin}, {Waters}, {Schiebel}, {Young},
  \& {Golap}}]{CASA2007}
{McMullin}, J.~P., {Waters}, B., {Schiebel}, D., {Young}, W., \& {Golap}, K.
  2007, in Astronomical Society of the Pacific Conference Series, Vol. 376,
  Astronomical Data Analysis Software and Systems XVI, ed. R.~A. {Shaw},
  F.~{Hill}, \& D.~J. {Bell}, 127

\bibitem[{{Mei} {et~al.}(2007){Mei}, {Blakeslee}, {C{\^o}t{\'e}}, {Tonry},
  {West}, {Ferrarese}, {Jord{\'a}n}, {Peng}, {Anthony}, \& {Merritt}}]{Mei2007}
{Mei}, S., {Blakeslee}, J.~P., {C{\^o}t{\'e}}, P., {et~al.} 2007, \apj, 655,
  144, \dodoi{10.1086/509598}

\bibitem[{{Mok} {et~al.}(2017){Mok}, {Wilson}, {Knapen}, {S{\'a}nchez-Gallego},
  {Brinks}, \& {Rosolowsky}}]{Mok2017}
{Mok}, A., {Wilson}, C.~D., {Knapen}, J.~H., {et~al.} 2017, \mnras, 467, 4282,
  \dodoi{10.1093/mnras/stx345}

\bibitem[{{Mok} {et~al.}(2016){Mok}, {Wilson}, {Golding}, {Warren}, {Israel},
  {Serjeant}, {Knapen}, {S{\'a}nchez-Gallego}, {Barmby}, {Bendo}, {Rosolowsky},
  \& {van der Werf}}]{Mok2016}
{Mok}, A., {Wilson}, C.~D., {Golding}, J., {et~al.} 2016, \mnras, 456, 4384,
  \dodoi{10.1093/mnras/stv2958}

\bibitem[{{Moore} {et~al.}(1996){Moore}, {Katz}, {Lake}, {Dressler}, \&
  {Oemler}}]{Moore1996}
{Moore}, B., {Katz}, N., {Lake}, G., {Dressler}, A., \& {Oemler}, A. 1996,
  \nat, 379, 613, \dodoi{10.1038/379613a0}

\bibitem[{{Moore} {et~al.}(1998){Moore}, {Lake}, \& {Katz}}]{Moore1998}
{Moore}, B., {Lake}, G., \& {Katz}, N. 1998, \apj, 495, 139,
  \dodoi{10.1086/305264}

\bibitem[{{Moore} {et~al.}(1999){Moore}, {Lake}, {Quinn}, \&
  {Stadel}}]{Moore1999}
{Moore}, B., {Lake}, G., {Quinn}, T., \& {Stadel}, J. 1999, \mnras, 304, 465,
  \dodoi{10.1046/j.1365-8711.1999.02345.x}

\bibitem[{{Moretti} {et~al.}(2018){Moretti}, {Paladino}, {Poggianti},
  {D'Onofrio}, {Bettoni}, {Gullieuszik}, {Jaff{\'e}}, {Vulcani}, {Fasano},
  {Fritz}, \& {Torstensson}}]{Moretti2018}
{Moretti}, A., {Paladino}, R., {Poggianti}, B.~M., {et~al.} 2018, \mnras, 480,
  2508, \dodoi{10.1093/mnras/sty2021}

\bibitem[{{Moretti} {et~al.}(2020{\natexlab{a}}){Moretti}, {Paladino},
  {Poggianti}, {Serra}, {Roediger}, {Gullieuszik}, {Tomi{\v{c}}i{\'c}},
  {Radovich}, {Vulcani}, {Jaff{\'e}}, {Fritz}, {Bettoni}, {Ramatsoku}, \&
  {Wolter}}]{Moretti2020a}
---. 2020{\natexlab{a}}, \apj, 889, 9, \dodoi{10.3847/1538-4357/ab616a}

\bibitem[{{Moretti} {et~al.}(2020{\natexlab{b}}){Moretti}, {Paladino},
  {Poggianti}, {Serra}, {Ramatsoku}, {Franchetto}, {Deb}, {Gullieuszik},
  {Tomi{\v{c}}i{\'c}}, {Mingozzi}, {Vulcani}, {Radovich}, {Bettoni}, \&
  {Fritz}}]{Moretti2020b}
---. 2020{\natexlab{b}}, \apjl, 897, L30, \dodoi{10.3847/2041-8213/ab9f3b}

\bibitem[{{Morokuma-Matsui} {et~al.}(2021){Morokuma-Matsui}, {Kodama},
  {Morokuma}, {Nakanishi}, {Koyama}, {Yamashita}, {Koyama}, \&
  {Okamoto}}]{Morokuma-Matsui2021}
{Morokuma-Matsui}, K., {Kodama}, T., {Morokuma}, T., {et~al.} 2021, arXiv
  e-prints, arXiv:2103.05867.
\newblock \doarXiv{2103.05867}

\bibitem[{{Mu{\~n}oz-Mateos} {et~al.}(2009){Mu{\~n}oz-Mateos}, {Gil de Paz},
  {Zamorano}, {Boissier}, {Dale}, {P{\'e}rez-Gonz{\'a}lez}, {Gallego},
  {Madore}, {Bendo}, {Boselli}, {Buat}, {Calzetti}, {Moustakas}, \&
  {Kennicutt}}]{MunozMateos2009}
{Mu{\~n}oz-Mateos}, J.~C., {Gil de Paz}, A., {Zamorano}, J., {et~al.} 2009,
  \apj, 703, 1569, \dodoi{10.1088/0004-637X/703/2/1569}

\bibitem[{{Mun} {et~al.}(2021){Mun}, {Hwang}, {Lee}, {Chung}, {Yoon}, \&
  {Lee}}]{Mun2021}
{Mun}, J.~Y., {Hwang}, H.~S., {Lee}, M.~G., {et~al.} 2021, Journal of Korean
  Astronomical Society, 54, 17, \dodoi{10.5303/JKAS.2021.54.1.17}

\bibitem[{{Muraoka} {et~al.}(2016){Muraoka}, {Sorai}, {Kuno}, {Nakai},
  {Nakanishi}, {Takeda}, {Yanagitani}, {Kaneko}, {Miyamoto}, {Kishida},
  {Hatakeyama}, {Umei}, {Tanaka}, {Tomiyasu}, {Saita}, {Ueno}, {Matsumoto},
  {Salak}, \& {Morokuma-Matsui}}]{Muraoka2016}
{Muraoka}, K., {Sorai}, K., {Kuno}, N., {et~al.} 2016, \pasj, 68, 89,
  \dodoi{10.1093/pasj/psw080}

\bibitem[{{Narayanan} {et~al.}(2011){Narayanan}, {Krumholz}, {Ostriker}, \&
  {Hernquist}}]{narayanan11}
{Narayanan}, D., {Krumholz}, M., {Ostriker}, E.~C., \& {Hernquist}, L. 2011,
  \mnras, 418, 664, \dodoi{10.1111/j.1365-2966.2011.19516.x}

\bibitem[{{Narayanan} {et~al.}(2012){Narayanan}, {Krumholz}, {Ostriker}, \&
  {Hernquist}}]{narayanan12}
{Narayanan}, D., {Krumholz}, M.~R., {Ostriker}, E.~C., \& {Hernquist}, L. 2012,
  \mnras, 421, 3127, \dodoi{10.1111/j.1365-2966.2012.20536.x}

\bibitem[{{Nehlig} {et~al.}(2016){Nehlig}, {Vollmer}, \& {Braine}}]{Nehlig2016}
{Nehlig}, F., {Vollmer}, B., \& {Braine}, J. 2016, \aap, 587, A108,
  \dodoi{10.1051/0004-6361/201527021}

\bibitem[{{Nulsen}(1982)}]{Nulsen1982}
{Nulsen}, P.~E.~J. 1982, \mnras, 198, 1007, \dodoi{10.1093/mnras/198.4.1007}

\bibitem[{{Nulsen} \& {Bohringer}(1995)}]{Nulsen1995}
{Nulsen}, P.~E.~J., \& {Bohringer}, H. 1995, \mnras, 274, 1093,
  \dodoi{10.1093/mnras/274.4.1093}

\bibitem[{{Olsen} {et~al.}(2016){Olsen}, {Greve}, {Brinch}, {Sommer-Larsen},
  {Rasmussen}, {Toft}, \& {Zirm}}]{olsen16}
{Olsen}, K.~P., {Greve}, T.~R., {Brinch}, C., {et~al.} 2016, \mnras, 457, 3306,
  \dodoi{10.1093/mnras/stw162}

\bibitem[{{Pappalardo} {et~al.}(2012){Pappalardo}, {Bianchi}, {Corbelli},
  {Giovanardi}, {Hunt}, {Bendo}, {Boselli}, {Cortese}, {Magrini}, {Zibetti},
  {di Serego Alighieri}, {Davies}, {Baes}, {Ciesla}, {Clemens}, {De Looze},
  {Fritz}, {Grossi}, {Pohlen}, {Smith}, {Verstappen}, \&
  {Vlahakis}}]{Pappalardo2012}
{Pappalardo}, C., {Bianchi}, S., {Corbelli}, E., {et~al.} 2012, \aap, 545, A75,
  \dodoi{10.1051/0004-6361/201219689}

\bibitem[{{Poggianti} {et~al.}(2017){Poggianti}, {Jaff{\'e}}, {Moretti},
  {Gullieuszik}, {Radovich}, {Tonnesen}, {Fritz}, {Bettoni}, {Vulcani},
  {Fasano}, {Bellhouse}, {Hau}, \& {Omizzolo}}]{Poggianti2017b}
{Poggianti}, B.~M., {Jaff{\'e}}, Y.~L., {Moretti}, A., {et~al.} 2017, \nat,
  548, 304, \dodoi{10.1038/nature23462}

\bibitem[{{Price-Whelan} {et~al.}(2018){Price-Whelan}, {Sip{\H{o}}cz},
  {G{\"u}nther}, {Lim}, {Crawford}, {Conseil}, {Shupe}, {Craig}, {Dencheva},
  {Ginsburg}, {VanderPlas}, {Bradley}, {P{\'e}rez-Su{\'a}rez}, {de Val-Borro},
  {Paper Contributors}, {Aldcroft}, {Cruz}, {Robitaille}, {Tollerud},
  {Coordination Committee}, {Ardelean}, {Babej}, {Bach}, {Bachetti}, {Bakanov},
  {Bamford}, {Barentsen}, {Barmby}, {Baumbach}, {Berry}, {Biscani}, {Boquien},
  {Bostroem}, {Bouma}, {Brammer}, {Bray}, {Breytenbach}, {Buddelmeijer},
  {Burke}, {Calderone}, {Cano Rodr{\'\i}guez}, {Cara}, {Cardoso}, {Cheedella},
  {Copin}, {Corrales}, {Crichton}, {D{\textquoteright}Avella}, {Deil},
  {Depagne}, {Dietrich}, {Donath}, {Droettboom}, {Earl}, {Erben}, {Fabbro},
  {Ferreira}, {Finethy}, {Fox}, {Garrison}, {Gibbons}, {Goldstein}, {Gommers},
  {Greco}, {Greenfield}, {Groener}, {Grollier}, {Hagen}, {Hirst}, {Homeier},
  {Horton}, {Hosseinzadeh}, {Hu}, {Hunkeler}, {Ivezi{\'c}}, {Jain}, {Jenness},
  {Kanarek}, {Kendrew}, {Kern}, {Kerzendorf}, {Khvalko}, {King}, {Kirkby},
  {Kulkarni}, {Kumar}, {Lee}, {Lenz}, {Littlefair}, {Ma}, {Macleod},
  {Mastropietro}, {McCully}, {Montagnac}, {Morris}, {Mueller}, {Mumford},
  {Muna}, {Murphy}, {Nelson}, {Nguyen}, {Ninan}, {N{\"o}the}, {Ogaz}, {Oh},
  {Parejko}, {Parley}, {Pascual}, {Patil}, {Patil}, {Plunkett}, {Prochaska},
  {Rastogi}, {Reddy Janga}, {Sabater}, {Sakurikar}, {Seifert}, {Sherbert},
  {Sherwood-Taylor}, {Shih}, {Sick}, {Silbiger}, {Singanamalla}, {Singer},
  {Sladen}, {Sooley}, {Sornarajah}, {Streicher}, {Teuben}, {Thomas},
  {Tremblay}, {Turner}, {Terr{\'o}n}, {van Kerkwijk}, {de la Vega}, {Watkins},
  {Weaver}, {Whitmore}, {Woillez}, {Zabalza}, \& {Contributors}}]{astropy:2018}
{Price-Whelan}, A.~M., {Sip{\H{o}}cz}, B.~M., {G{\"u}nther}, H.~M., {et~al.}
  2018, \aj, 156, 123, \dodoi{10.3847/1538-3881/aabc4f}

\bibitem[{{Regan} {et~al.}(2001){Regan}, {Thornley}, {Helfer}, {Sheth}, {Wong},
  {Vogel}, {Blitz}, \& {Bock}}]{Regan2001}
{Regan}, M.~W., {Thornley}, M.~D., {Helfer}, T.~T., {et~al.} 2001, \apj, 561,
  218, \dodoi{10.1086/323221}

\bibitem[{{Saintonge} \& {Spekkens}(2011)}]{Saintonge2011a}
{Saintonge}, A., \& {Spekkens}, K. 2011, \apj, 726, 77,
  \dodoi{10.1088/0004-637X/726/2/77}

\bibitem[{{Saintonge} {et~al.}(2011){Saintonge}, {Kauffmann}, {Kramer},
  {Tacconi}, {Buchbender}, {Catinella}, {Fabello}, {Graci{\'a}-Carpio}, {Wang},
  {Cortese}, {Fu}, {Genzel}, {Giovanelli}, {Guo}, {Haynes}, {Heckman},
  {Krumholz}, {Lemonias}, {Li}, {Moran}, {Rodriguez-Fernandez}, {Schiminovich},
  {Schuster}, \& {Sievers}}]{Saintonge2011b}
{Saintonge}, A., {Kauffmann}, G., {Kramer}, C., {et~al.} 2011, \mnras, 415, 32,
  \dodoi{10.1111/j.1365-2966.2011.18677.x}

\bibitem[{{Saintonge} {et~al.}(2017){Saintonge}, {Catinella}, {Tacconi},
  {Kauffmann}, {Genzel}, {Cortese}, {Dav{\'e}}, {Fletcher},
  {Graci{\'a}-Carpio}, {Kramer}, {Heckman}, {Janowiecki}, {Lutz}, {Rosario},
  {Schiminovich}, {Schuster}, {Wang}, {Wuyts}, {Borthakur}, {Lamperti}, \&
  {Roberts-Borsani}}]{Saintonge2017}
{Saintonge}, A., {Catinella}, B., {Tacconi}, L.~J., {et~al.} 2017, \apjs, 233,
  22, \dodoi{10.3847/1538-4365/aa97e0}

\bibitem[{{Salim} {et~al.}(2007){Salim}, {Rich}, {Charlot}, {Brinchmann},
  {Johnson}, {Schiminovich}, {Seibert}, {Mallery}, {Heckman}, {Forster},
  {Friedman}, {Martin}, {Morrissey}, {Neff}, {Small}, {Wyder}, {Bianchi},
  {Donas}, {Lee}, {Madore}, {Milliard}, {Szalay}, {Welsh}, \& {Yi}}]{Salim2007}
{Salim}, S., {Rich}, R.~M., {Charlot}, S., {et~al.} 2007, \apjs, 173, 267,
  \dodoi{10.1086/519218}

\bibitem[{{Salpeter}(1955)}]{Salpeter1955}
{Salpeter}, E.~E. 1955, \apj, 121, 161, \dodoi{10.1086/145971}

\bibitem[{{S{\'a}nchez Almeida}(2020)}]{SanchezAlmeida2020}
{S{\'a}nchez Almeida}, J. 2020, \mnras, 495, 78, \dodoi{10.1093/mnras/staa1108}

\bibitem[{{Sandstrom} {et~al.}(2013){Sandstrom}, {Leroy}, {Walter}, {Bolatto},
  {Croxall}, {Draine}, {Wilson}, {Wolfire}, {Calzetti}, {Kennicutt}, {Aniano},
  {Donovan Meyer}, {Usero}, {Bigiel}, {Brinks}, {de Blok}, {Crocker}, {Dale},
  {Engelbracht}, {Galametz}, {Groves}, {Hunt}, {Koda}, {Kreckel}, {Linz},
  {Meidt}, {Pellegrini}, {Rix}, {Roussel}, {Schinnerer}, {Schruba}, {Schuster},
  {Skibba}, {van der Laan}, {Appleton}, {Armus}, {Brandl}, {Gordon}, {Hinz},
  {Krause}, {Montiel}, {Sauvage}, {Schmiedeke}, {Smith}, \&
  {Vigroux}}]{sandstrom2013}
{Sandstrom}, K.~M., {Leroy}, A.~K., {Walter}, F., {et~al.} 2013, \apj, 777, 5,
  \dodoi{10.1088/0004-637X/777/1/5}

\bibitem[{{Schindler} {et~al.}(1999){Schindler}, {Binggeli}, \&
  {B{\"o}hringer}}]{Schindler1999}
{Schindler}, S., {Binggeli}, B., \& {B{\"o}hringer}, H. 1999, \aap, 343, 420.
\newblock \doarXiv{astro-ph/9811464}

\bibitem[{{Schruba} {et~al.}(2011){Schruba}, {Leroy}, {Walter}, {Bigiel},
  {Brinks}, {de Blok}, {Dumas}, {Kramer}, {Rosolowsky}, {Sandstrom},
  {Schuster}, {Usero}, {Weiss}, \& {Wiesemeyer}}]{Schruba2011}
{Schruba}, A., {Leroy}, A.~K., {Walter}, F., {et~al.} 2011, \aj, 142, 37,
  \dodoi{10.1088/0004-6256/142/2/37}

\bibitem[{{Sheth} {et~al.}(2005){Sheth}, {Vogel}, {Regan}, {Thornley}, \&
  {Teuben}}]{Sheth2005}
{Sheth}, K., {Vogel}, S.~N., {Regan}, M.~W., {Thornley}, M.~D., \& {Teuben},
  P.~J. 2005, \apj, 632, 217, \dodoi{10.1086/432409}

\bibitem[{{Shetty} {et~al.}(2011){Shetty}, {Glover}, {Dullemond}, {Ostriker},
  {Harris}, \& {Klessen}}]{shetty11}
{Shetty}, R., {Glover}, S.~C., {Dullemond}, C.~P., {et~al.} 2011, \mnras, 415,
  3253, \dodoi{10.1111/j.1365-2966.2011.18937.x}

\bibitem[{{Shibata} {et~al.}(2001){Shibata}, {Matsushita}, {Yamasaki},
  {Ohashi}, {Ishida}, {Kikuchi}, {B{\"o}hringer}, \& {Matsumoto}}]{Shibata2001}
{Shibata}, R., {Matsushita}, K., {Yamasaki}, N.~Y., {et~al.} 2001, \apj, 549,
  228, \dodoi{10.1086/319075}

\bibitem[{{Smith} {et~al.}(2012){Smith}, {Eales}, {Gomez}, {Roman-Duval},
  {Fritz}, {Braun}, {Baes}, {Bendo}, {Blommaert}, {Boquien}, {Boselli},
  {Clements}, {Cooray}, {Cortese}, {De Looze}, {Ford}, {Gear}, {Gentile},
  {Gordon}, {Kirk}, {Lebouteiller}, {Madden}, {Mentuch}, {O'Halloran}, {Page},
  {Schulz}, {Spinoglio}, {Verstappen}, {Wilson}, \& {Thilker}}]{smith12}
{Smith}, M.~W.~L., {Eales}, S.~A., {Gomez}, H.~L., {et~al.} 2012, \apj, 756,
  40, \dodoi{10.1088/0004-637X/756/1/40}

\bibitem[{{Solomon} \& {Vanden Bout}(2005)}]{Solomon2005}
{Solomon}, P.~M., \& {Vanden Bout}, P.~A. 2005, \araa, 43, 677,
  \dodoi{10.1146/annurev.astro.43.051804.102221}

\bibitem[{{Sorce} {et~al.}(2016){Sorce}, {Gottl{\"o}ber}, {Hoffman}, \&
  {Yepes}}]{Sorce2016}
{Sorce}, J.~G., {Gottl{\"o}ber}, S., {Hoffman}, Y., \& {Yepes}, G. 2016,
  \mnras, 460, 2015, \dodoi{10.1093/mnras/stw1085}

\bibitem[{{Speagle} {et~al.}(2014){Speagle}, {Steinhardt}, {Capak}, \&
  {Silverman}}]{Speagle2014}
{Speagle}, J.~S., {Steinhardt}, C.~L., {Capak}, P.~L., \& {Silverman}, J.~D.
  2014, \apjs, 214, 15, \dodoi{10.1088/0067-0049/214/2/15}

\bibitem[{{Stark} {et~al.}(1986){Stark}, {Knapp}, {Bally}, {Wilson}, {Penzias},
  \& {Rowe}}]{Stark1986}
{Stark}, A.~A., {Knapp}, G.~R., {Bally}, J., {et~al.} 1986, \apj, 310, 660,
  \dodoi{10.1086/164717}

\bibitem[{{Stevens} {et~al.}(2019){Stevens}, {Diemer}, {Lagos}, {Nelson},
  {Obreschkow}, {Wang}, \& {Marinacci}}]{Stevens2019}
{Stevens}, A. R.~H., {Diemer}, B., {Lagos}, C. d.~P., {et~al.} 2019, \mnras,
  490, 96, \dodoi{10.1093/mnras/stz2513}

\bibitem[{{Stevens} {et~al.}(2021){Stevens}, {Lagos}, {Cortese}, {Catinella},
  {Diemer}, {Nelson}, {Pillepich}, {Hernquist}, {Marinacci}, \&
  {Vogelsberger}}]{Stevens2021}
{Stevens}, A. R.~H., {Lagos}, C. d.~P., {Cortese}, L., {et~al.} 2021, \mnras,
  502, 3158, \dodoi{10.1093/mnras/staa3662}

\bibitem[{{Sun} {et~al.}(2018){Sun}, {Leroy}, {Schruba}, {Rosolowsky},
  {Hughes}, {Kruijssen}, {Meidt}, {Schinnerer}, {Blanc}, {Bigiel}, {Bolatto},
  {Chevance}, {Groves}, {Herrera}, {Hygate}, {Pety}, {Querejeta}, {Usero}, \&
  {Utomo}}]{Sun2018}
{Sun}, J., {Leroy}, A.~K., {Schruba}, A., {et~al.} 2018, \apj, 860, 172,
  \dodoi{10.3847/1538-4357/aac326}

\bibitem[{{Trujillo} {et~al.}(2020){Trujillo}, {Chamba}, \&
  {Knapen}}]{Trujillo2020}
{Trujillo}, I., {Chamba}, N., \& {Knapen}, J.~H. 2020, \mnras, 493, 87,
  \dodoi{10.1093/mnras/staa236}

\bibitem[{{Tully} \& {Shaya}(1984)}]{Tully1984}
{Tully}, R.~B., \& {Shaya}, E.~J. 1984, \apj, 281, 31, \dodoi{10.1086/162073}

\bibitem[{{Urban} {et~al.}(2011){Urban}, {Werner}, {Simionescu}, {Allen}, \&
  {B{\"o}hringer}}]{Urban2011}
{Urban}, O., {Werner}, N., {Simionescu}, A., {Allen}, S.~W., \&
  {B{\"o}hringer}, H. 2011, \mnras, 414, 2101,
  \dodoi{10.1111/j.1365-2966.2011.18526.x}

\bibitem[{Virtanen {et~al.}(2020)Virtanen, Gommers, Oliphant, Haberland, Reddy,
  Cournapeau, Burovski, Peterson, Weckesser, Bright, {van der Walt}, Brett,
  Wilson, Millman, Mayorov, Nelson, Jones, Kern, Larson, Carey, Polat, Feng,
  Moore, {VanderPlas}, Laxalde, Perktold, Cimrman, Henriksen, Quintero, Harris,
  Archibald, Ribeiro, Pedregosa, {van Mulbregt}, \& {SciPy 1.0
  Contributors}}]{2020SciPy-NMeth}
Virtanen, P., Gommers, R., Oliphant, T.~E., {et~al.} 2020, Nature Methods, 17,
  261, \dodoi{10.1038/s41592-019-0686-2}

\bibitem[{{Vulcani} {et~al.}(2018){Vulcani}, {Poggianti}, {Gullieuszik},
  {Moretti}, {Tonnesen}, {Jaff{\'e}}, {Fritz}, {Fasano}, \&
  {Bettoni}}]{Vulcani2018}
{Vulcani}, B., {Poggianti}, B.~M., {Gullieuszik}, M., {et~al.} 2018, \apjl,
  866, L25, \dodoi{10.3847/2041-8213/aae68b}

\bibitem[{{Wang} {et~al.}(2016){Wang}, {Koribalski}, {Serra}, {van der Hulst},
  {Roychowdhury}, {Kamphuis}, \& {Chengalur}}]{Wang2016}
{Wang}, J., {Koribalski}, B.~S., {Serra}, P., {et~al.} 2016, \mnras, 460, 2143,
  \dodoi{10.1093/mnras/stw1099}

\bibitem[{{Wang} {et~al.}(2014){Wang}, {Fu}, {Aumer}, {Kauffmann}, {J{\'o}zsa},
  {Serra}, {Huang}, {Brinchmann}, {van der Hulst}, \& {Bigiel}}]{Wang2014}
{Wang}, J., {Fu}, J., {Aumer}, M., {et~al.} 2014, \mnras, 441, 2159,
  \dodoi{10.1093/mnras/stu649}

\bibitem[{{W}es {M}c{K}inney(2010)}]{mckinney-proc-scipy-2010}
{W}es {M}c{K}inney. 2010, in {P}roceedings of the 9th {P}ython in {S}cience
  {C}onference, ed. {S}t\'efan van~der {W}alt \& {J}arrod {M}illman, 56 -- 61,
  \dodoi{10.25080/Majora-92bf1922-00a}

\bibitem[{{White} {et~al.}(1997){White}, {Jones}, \& {Forman}}]{White1997}
{White}, D.~A., {Jones}, C., \& {Forman}, W. 1997, \mnras, 292, 419,
  \dodoi{10.1093/mnras/292.2.419}

\bibitem[{{Wilson} {et~al.}(2009){Wilson}, {Warren}, {Israel}, {Serjeant},
  {Bendo}, {Brinks}, {Clements}, {Courteau}, {Irwin}, {Knapen}, {Leech},
  {Matthews}, {M{\"u}hle}, {Mortier}, {Petitpas}, {Sinukoff}, {Spekkens},
  {Tan}, {Tilanus}, {Usero}, {van der Werf}, {Wiegert}, \& {Zhu}}]{Wilson2009}
{Wilson}, C.~D., {Warren}, B.~E., {Israel}, F.~P., {et~al.} 2009, \apj, 693,
  1736, \dodoi{10.1088/0004-637X/693/2/1736}

\bibitem[{{Wolfire} {et~al.}(2010){Wolfire}, {Hollenbach}, \&
  {McKee}}]{Wolfire2010}
{Wolfire}, M.~G., {Hollenbach}, D., \& {McKee}, C.~F. 2010, \apj, 716, 1191,
  \dodoi{10.1088/0004-637X/716/2/1191}

\bibitem[{{Wright} {et~al.}(2010){Wright}, {Eisenhardt}, {Mainzer}, {Ressler},
  {Cutri}, {Jarrett}, {Kirkpatrick}, {Padgett}, {McMillan}, {Skrutskie},
  {Stanford}, {Cohen}, {Walker}, {Mather}, {Leisawitz}, {Gautier}, {McLean},
  {Benford}, {Lonsdale}, {Blain}, {Mendez}, {Irace}, {Duval}, {Liu}, {Royer},
  {Heinrichsen}, {Howard}, {Shannon}, {Kendall}, {Walsh}, {Larsen}, {Cardon},
  {Schick}, {Schwalm}, {Abid}, {Fabinsky}, {Naes}, \& {Tsai}}]{Wright2010}
{Wright}, E.~L., {Eisenhardt}, P. R.~M., {Mainzer}, A.~K., {et~al.} 2010, \aj,
  140, 1868, \dodoi{10.1088/0004-6256/140/6/1868}

\bibitem[{{Yajima} {et~al.}(2021){Yajima}, {Sorai}, {Miyamoto}, {Muraoka},
  {Kuno}, {Kaneko}, {Takeuchi}, {Yasuda}, {Tanaka}, {Morokuma-Matsui}, \&
  {Kobayashi}}]{Yajima2021}
{Yajima}, Y., {Sorai}, K., {Miyamoto}, Y., {et~al.} 2021, \pasj, 73, 257,
  \dodoi{10.1093/pasj/psaa119}

\bibitem[{{Yoon} {et~al.}(2017){Yoon}, {Chung}, {Smith}, \&
  {Jaff{\'e}}}]{Yoon2017}
{Yoon}, H., {Chung}, A., {Smith}, R., \& {Jaff{\'e}}, Y.~L. 2017, \apj, 838,
  81, \dodoi{10.3847/1538-4357/aa6579}

\bibitem[{{York} {et~al.}(2000){York}, {Adelman}, {Anderson}, {Anderson},
  {Annis}, {Bahcall}, {Bakken}, {Barkhouser}, {Bastian}, {Berman}, {Boroski},
  {Bracker}, {Briegel}, {Briggs}, {Brinkmann}, {Brunner}, {Burles}, {Carey},
  {Carr}, {Castander}, {Chen}, {Colestock}, {Connolly}, {Crocker}, {Csabai},
  {Czarapata}, {Davis}, {Doi}, {Dombeck}, {Eisenstein}, {Ellman}, {Elms},
  {Evans}, {Fan}, {Federwitz}, {Fiscelli}, {Friedman}, {Frieman}, {Fukugita},
  {Gillespie}, {Gunn}, {Gurbani}, {de Haas}, {Haldeman}, {Harris}, {Hayes},
  {Heckman}, {Hennessy}, {Hindsley}, {Holm}, {Holmgren}, {Huang}, {Hull},
  {Husby}, {Ichikawa}, {Ichikawa}, {Ivezi{\'c}}, {Kent}, {Kim}, {Kinney},
  {Klaene}, {Kleinman}, {Kleinman}, {Knapp}, {Korienek}, {Kron}, {Kunszt},
  {Lamb}, {Lee}, {Leger}, {Limmongkol}, {Lindenmeyer}, {Long}, {Loomis},
  {Loveday}, {Lucinio}, {Lupton}, {MacKinnon}, {Mannery}, {Mantsch}, {Margon},
  {McGehee}, {McKay}, {Meiksin}, {Merelli}, {Monet}, {Munn}, {Narayanan},
  {Nash}, {Neilsen}, {Neswold}, {Newberg}, {Nichol}, {Nicinski}, {Nonino},
  {Okada}, {Okamura}, {Ostriker}, {Owen}, {Pauls}, {Peoples}, {Peterson},
  {Petravick}, {Pier}, {Pope}, {Pordes}, {Prosapio}, {Rechenmacher}, {Quinn},
  {Richards}, {Richmond}, {Rivetta}, {Rockosi}, {Ruthmansdorfer}, {Sandford},
  {Schlegel}, {Schneider}, {Sekiguchi}, {Sergey}, {Shimasaku}, {Siegmund},
  {Smee}, {Smith}, {Snedden}, {Stone}, {Stoughton}, {Strauss}, {Stubbs},
  {SubbaRao}, {Szalay}, {Szapudi}, {Szokoly}, {Thakar}, {Tremonti}, {Tucker},
  {Uomoto}, {Vanden Berk}, {Vogeley}, {Waddell}, {Wang}, {Watanabe},
  {Weinberg}, {Yanny}, {Yasuda}, \& {SDSS Collaboration}}]{York2000}
{York}, D.~G., {Adelman}, J., {Anderson}, John~E., J., {et~al.} 2000, \aj, 120,
  1579, \dodoi{10.1086/301513}

\bibitem[{{Zabel} {et~al.}(2019){Zabel}, {Davis}, {Smith}, {Maddox}, {Bendo},
  {Peletier}, {Iodice}, {Venhola}, {Baes}, {Davies}, {de Looze}, {Gomez},
  {Grossi}, {Kenney}, {Serra}, {van de Voort}, {Vlahakis}, \&
  {Young}}]{Zabel2019}
{Zabel}, N., {Davis}, T.~A., {Smith}, M. W.~L., {et~al.} 2019, \mnras, 483,
  2251, \dodoi{10.1093/mnras/sty3234}

\bibitem[{{Zabel} {et~al.}(2020){Zabel}, {Davis}, {Sarzi}, {Nedelchev},
  {Chevance}, {Kruijssen}, {Iodice}, {Baes}, {Bendo}, {Corsini}, {De Looze},
  {de Zeeuw}, {Gadotti}, {Grossi}, {Peletier}, {Pinna}, {Serra}, {van de
  Voort}, {Venhola}, {Viaene}, \& {Vlahakis}}]{Zabel2020}
{Zabel}, N., {Davis}, T.~A., {Sarzi}, M., {et~al.} 2020, \mnras, 496, 2155,
  \dodoi{10.1093/mnras/staa1513}

\bibitem[{{Zahid} {et~al.}(2012){Zahid}, {Dima}, {Kewley}, {Erb}, \&
  {Dav{\'e}}}]{Zahid2012}
{Zahid}, H.~J., {Dima}, G.~I., {Kewley}, L.~J., {Erb}, D.~K., \& {Dav{\'e}}, R.
  2012, \apj, 757, 54, \dodoi{10.1088/0004-637X/757/1/54}

\end{thebibliography}
\bibliographystyle{aasjournal}


\end{document}